%% file: Thesis.tex
\documentclass[oneside]{ecsthesis}     
\usepackage{color}
\usepackage{calligra}
\usepackage{pdfpages}
\graphicspath{{../Figures/}}   
\usepackage[mathscr]{euscript}
\hypersetup{colorlinks=false}   
\begin{document}
\frontmatter
\title      {Experimental studies of superconducting gap structure and quantum fluctuations in novel superconductors and heavy fermion compounds}
\authors    {\texorpdfstring
            {\href{mailto:kartik.panda@gm.rkmvu.ac.in}{Kartik Panda}}
             {Kartik Panda}
            }
\addresses  {\Department of Physics\\\Ramakrishna Mission Vivekananda Educational and Research Institute}
\date       {\month\today}
\subject    {}
\maketitle

\include{Declaration}

\include{Dedication}
\addcontentsline{toc}{chapter}{\bf Acknowledgments}
\include{Acknowledgments}

\addcontentsline{toc}{chapter}{\bf Publication}
\include{publication}
\tableofcontents
\thispagestyle{empty}
\listoffigures
\listoftables
\doublespacing

\mainmatter
\doublespacing
\include{chapter1}

\include{chapter2}

\include{chapter3}
\include{chapter4}
\include{chapter5}

\include{chapter6}

\include{chapter7}
\include{chapter8}

\include{biblio}

\end{document}

%% file: Declaration.tex
{\bf {\LARGE {Declaration}}}\\~\\~\\~\\
\begin{Large}
This thesis is a presentation of my original research work carried out under the guidance of Dr. Amitava Bhattacharyya at Ramakrishna Mission Vivekananda Educational and Research Institute, Howrah, India. This work has not been submitted for any other degree to this or any other University or body. Wherever contributions of others are involved, every effort has been made to indicate this clearly, with due reference to the literature.\\~\\~\\
\begin{flushright}
.................................................\\
Kartik Panda
\end{flushright}

In my capacity as the supervisor of the candidate’s thesis, I certify that the above statements are true to the best of my knowledge.\\~\\

...............................................................\\
Dr. Amitava Bhattacharyya
\end{Large}

%% file: Dedication.tex
\vspace*{250px}
\begin{Large}
{\bf{If we knew what we were doing it wouldn't be called research.}}\\
\hspace*{8cm} \Large{................... Albert Einstein}
\end{Large}

\newpage
\vspace*{250px}
\begin{center}
\begin{Huge}
\calligra{This thesis is affectionately dedicated to my \newline Maa \& Baba}
\end{Huge}
\end{center}

%% file: Acknowledgments.tex
{\bf {\LARGE {Acknowledgements}}}\\

This PhD has been a truly life-changing experience for me, and I could not have done it without the help and support of many persons. I have no valuable words to express my thanks, but my heart is still full of the favours received from every person.

First and foremost, I want to express my gratitude to my supervisor, {\bf Dr. Amitava Bhattacharyya}, for all of his support, inspiration, and motivation over the past few years. I am thankful to him for his valuable lesson in research. He instilled in me the value of hard work and dedication and always pushed me to do better with my full effort.

{\bf Prof. D. T. Adroja}, for a fascinating and fruitful experimental collaboration, I am tremendously grateful. {\bf Dr. Pabitra Biswas, Prof. A. D. Hillier, Dr. J. Lord, and Dr. K. Yokoyama} from the ISIS Neutron and Muon Source, RAL, UK, deserve special thanks for their invaluable knowledge and unwavering support during my experiments.

{\bf Prof. Tanmoy Das} and his team have been a continual source of theoretical assistance and useful discussion for me. {\bf P. P. Ferreira, Prof. T. T. Dorini, and Prof. L. T. F. Eleno} are also to be thanked for their theoretical assistance.

{\bf Prof. N. Kase, Prof. A. J.S. Machado, Prof. Y. J. Sato, Prof. Y. Shi, and Prof. D. Aoki} are also to be thanked for giving me some high-quality samples for the experiment. {\bf Prof. A. M. Strydom, Prof. M. R. Lees, and Prof. Z. Fisk} for their unique contributions to the publications.

I would like to thank {\bf Prof. Yoshikazu Mizuguchi} from Tokyo Metropolitan University, Japan and {\bf Prof. V. P. S. Awana} from CSIR-National Physical Laboratory, New Delhi, India for kindly refereeing this thesis and enriching it with their valuable suggestions.

I owe a debt of gratitude to {\bf Swami Atmapriyanandaji Maharaj, Aditya Maharaj, and Abhijit Bandyopadhyay, Bobby Ezhuthachan, Sanjoy Biswas,} and other faculty members for their continuous support and encouragement.

I am extremely thankful to all of my teachers specially {\bf Prabir Jana, Soumen Parua, Manidatta Majhi and others} for forming me into the person I am today. Having a very poor economic background, it was not possible without their help at that moment. I owe a debt of gratitude to the monks of the Ramakrishna Mission Calcutta Students Home, Belgharia, for their constant support and guidance throughout my academic and personal lives.

To my labmate-{\bf Arnab, Koustav} we had great times. To the old gang - {\bf Abhik, Pradeepta, and Samit} - you're the best. 

I want to acknowledge DST INSPIRE for providing my research fellowship and India UK Newton-Bhabha Fund for funding support through to perform the experiments at ISIS neutron and muon source, RAL, UK.

Finally, I  must emphasize my eternal gratitude to {\bf my parents}, to whom this thesis is dedicated, for their unfailing love, support and encouragement. I'm thankful to my {\bf Didi, Jamaibabu, Dola} and others for their constant support. {\bf Atri}, I thank you for your patience and unselfish love - especially in times of distress.

%% file: publication.tex
\chapter*{List of Publications}
\begin{enumerate}
\item  A. Bhattacharyya, D. T. Adroja, {\bf K. Panda}, Surabhi Saha, Tanmoy Das, A. J. S. Machado, T. W. Grant, Z. Fisk, A. D. Hillier, and P. Manfrinetti, \textit{`Evidence of Nodal Line in the Superconducting Gap Symmetry of Noncentrosymmetric ThCoC$_{2}$'}, \href{https://journals.aps.org/prl/abstract/10.1103/PhysRevLett.122.147001} {\textcolor{blue}{Physical Review Letters {\bf 122}, 147001 (2019)}}.

\item  {\bf K. Panda}, A. Bhattacharyya, D. T. Adroja, N. Kase, P. K. Biswas, Surabhi Saha, Tanmoy Das, M. R. Lees, and A. D. Hillier, \textit{`Probing the superconducting ground state of ZrIrSi: A muon spin rotation and relaxation study'}, \href{https://journals.aps.org/prb/abstract/10.1103/PhysRevB.99.174513} {\textcolor{blue}{Physical Review B {\bf 99}, 174513 (2019)}}.

\item  A. Bhattacharyya, {\bf K. Panda}, D. T. Adroja, N. Kase, P. K. Biswas, Surabhi Saha, Tanmoy Das, M. R. Lees, and A. D. Hillier, \textit{`Investigation of Superconducting Gap Structure in HfIrSi using muon spin relaxation/rotation'}, \href{https://iopscience.iop.org/article/10.1088/1361-648X/ab549e/meta} {\textcolor{blue}{Journal of Physics: Condensed Matter {\bf 32}, 085601 (2020)}}.

\item  A. Bhattacharyya, D. T. Adroja, J. S. Lord, L. Wang, Y. Shi, {\bf K. Panda}, H. Luo, and A. M. Strydom, \textit{`Quantum fluctuations in the quasi-one-dimensional non-Fermi liquid system CeCo$_{2}$Ga$_{8}$ investigated using $\mu$SR'}, \href{https://journals.aps.org/prb/abstract/10.1103/PhysRevB.101.214437} {\textcolor{blue}{Physical Review B {\bf 101}, 214437 (2020)}}.

\item  D. T. Adroja, A. Bhattacharyya, Y. J. Sato, M. R. Lees, P. K. Biswas, {\bf K. Panda}, V. K. Anand, Gavin B. G. Stenning, A. D. Hillier, and D. Aoki, \textit{`Pairing symmetry of an intermediate valence superconductor CeIr$_{3}$ investigated using $\mu$SR measurements'}, \href{https://journals.aps.org/prb/abstract/10.1103/PhysRevB.103.104514} {\textcolor{blue}{Physical Review B {\bf 103}, 104514 (2021)}}.

\item  A. Bhattacharyya, P. P. Ferreira, {\bf K. Panda}, F. B. Santos, D. T. Adroja, K. Yokoyama, T. T. Dorini, L. T. F. Eleno, A. J. S. Machado, \textit{`Electron-phonon superconductivity in C-doped topological nodal-line semimetal Zr$_{5}$Pt$_{3}$: A muon spin rotation and relaxation ($\mu$SR) study'}, \href{https://iopscience.iop.org/article/10.1088/1361-648X/ac2bc7/meta} {\textcolor{blue}{Journal of Physics: Condensed Matter {\bf 34}, 035602 (2020)}}.

\item A. Bhattacharyya, M. R. Lees, K. Panda, V. K. Anand, J. Sannigrahi, P. K. Biswas, R. Tripathi, and D. T. Adroja, \textit{`Investigation of superconducting ground state of Sc$_{5}$Co$_{4}$Si$_{10}$: Muon spin rotation and relaxation study'}\href{https://journals.aps.org/prmaterials/abstract/10.1103/PhysRevMaterials.6.064802}{\textcolor{blue} {Physical Review Materials {\bf 6}, 064802 (2022).}}
\newpage

{\bf{\underline{Conference Proceedings}}}
\begin{itemize}
\item [3.] \begin{large}
\it {Ground state magnetic properties of Ho$_{15}$Si$_{9}$C compound}\\
\end{large}
\textbf {K. Panda} and A. Bhattacharyya\\ 
\begin{large}
Proceedings of NCFMP-2020\\
\end{large}\
\item [2.] \begin{large}
\it {Inelastic Neutron Scattering Study of Newly Discovered Ising type ferromagnet CeRu$_{2}$Al$_{2}$B}\\
\end{large}
\textbf {Kartik Panda}, A. Bhattacharyya, and D. T. Adroja\\ 
\begin{large}
Proceedings of AMDP-2017
\end{large}\\
\item [1.] \begin{large}
\it{Structural Changes in Ferromagnetic Shape Memory Alloys Due to The Effect of Heat Treatments}\\
\end{large}
\textbf {Kartik Panda}, Md. Sarowar Hossain, and P K Mukhopadhyay\\ 
\begin{large}
Proceedings of ICMAGMA-2017
\end{large}
\end{itemize}

\end{enumerate}

%% file: chapter1.tex
\chapter{Introduction} \label{chapter:1}
The nature of the matter in our everyday world, i.e. solid, liquid, and gaseous phases, is defined by the interaction of the fundamental particles electrons and nuclei. Not only do electrons from the ``quantum glue" that holds the nuclei together in solid, liquid, and molecular phases do this, but excited electrons also determine the numerous properties of the materials. The properties of matter are naturally divided into two categories: (i) those determined by electrons in their ground state, such as cohesive energy, equilibrium crystal structure, phase transitions between structures, elastic constants, charge density, magnetic order, static dielectric and magnetic susceptibilities; and (ii) those determined by electrons in their excited states, such as low-energy excitation in metals involved in specific heat, Pauli spin susceptibility, transport, and so on. Many properties of materials may now be determined directly from the electron's fundamental equation, bringing new insights into fundamental physics, chemistry, and material science challenges. The electronic structure of materials lies at the heart of the many-body problem of interacting electrons, which is one of the great challenges in theoretical physics. It is obvious from the hamiltonian for the system of electrons and nuclei that electron-electron interactions are not negligible. The strong electron-phonon interaction in the materials causes superconductivity, one of the most fascinating properties of material.

Superconductivity is a phenomenon in which the electrical resistance of the material completely disappears and the magnetic field is expelled from the interior below a certain temperature ($T_\mathrm{C}$), is one of the significant scientific discoveries of the last century. Till now not only the room temperature superconductors yet to be discovered but the theory of high temperature superconductivity is not clear. In April of 1911, Dutch physicist Heike Kamerlingh Onnes of Leiden university first discovered the superconductivity in pure mercury (Hg)~\cite{onnes1911}. He and his co-workers were cooling the mercury in liquid helium and observed that the resistance of mercury abruptly vanishes at 4.2 K, which was more precisely lower than the measurement uncertainty. Later, in 1913, he was awarded the Nobel Prize in Physics for this significant discovery in low-temperature physics. Since then the quest for new superconducting materials and the mechanism behind the superconductivity has attracted considerable attention of theoretical and experimental communities in condensed-matter physics. After two decades, German physicists Meissner and Ochsenfeld observed another hallmark property of superconductors~\cite{meissner1933}. Below $T_{C}$, the magnetic field is expelled from the interior of superconductors, exhibits a perfect diamagnetic state, which has come to know as Meissner Effect. After the next few decades, several attempts had been made to develop a microscopic theory of superconductivity. With the introduction of the London theory in 1935~\cite{london1935} and the Ginzberg-Landau theory in 1950~\cite{ginzburg1950}, major progress was made toward such a theory for superconductivity. However, American scientists John Bardeen, Leon Cooper, and John Schrieffer reported the first true microscopic theory of superconductivity in 1957, winning them a Nobel Award in 1972~\cite{BCS}. When an electron passes through a crystal lattice, the positive crystal lattice is moved toward it, which leads to the creation of phonons, which can be thought of as a series of quantum oscillators. This charge might be enough to attract an electron and create a ``Cooper pair''. As electrons were fermions (particles with non-integer spins), it obeys Pauli's exclusion principle. In a Cooper pair, however, electrons with opposite spins add up to an integer spin of 0. The electron pair is then promoted to the Boson state. This theory successfully explains the zero resistivity, Meissner effect, isotope effect and the specific heat anomaly related to the energy gap of the superconductor. However, this theory is inadequate to explain the origin of superconductivity in some novel superconductors. 

Superconductivity had only been reported in metals and metal alloys before the mid-1980s, with the maximum recorded transition temperature being 23 K in Nb$_{3}$Ge~\cite{matthias1965}. In the realm of superconductivity, the 1980s were a decade of unmatched breakthroughs. In 1979, German physicist F. Steglich discovered superconductivity in heavy fermion (HF) compounds CeCu$_{2}$Si$_{2}$~\cite{steglich1979}, one of the most awe-inspiring and elusive properties of HF materials. The discovery established a new reality of electronically mediated superconductivity. Interestingly, in this sample magnetism is lurking in the background-introduced doubt that this conforms to the simple BCS scenario. In 1980, the first organic superconductor(carbon based) (TMTSF)$_{2}$PF$_{6}$ with $T_{C}$ = 0.9 K, at an external pressure of 11 kbar,  was successfully grown by K. Bechgaard and his teams ~\cite{bechgaard1980}. The possibility of organic superconductors is first suggested by Bill Little in 1964~\cite{Little}.

\begin{figure}
\centering
 \includegraphics[height=0.9\linewidth]{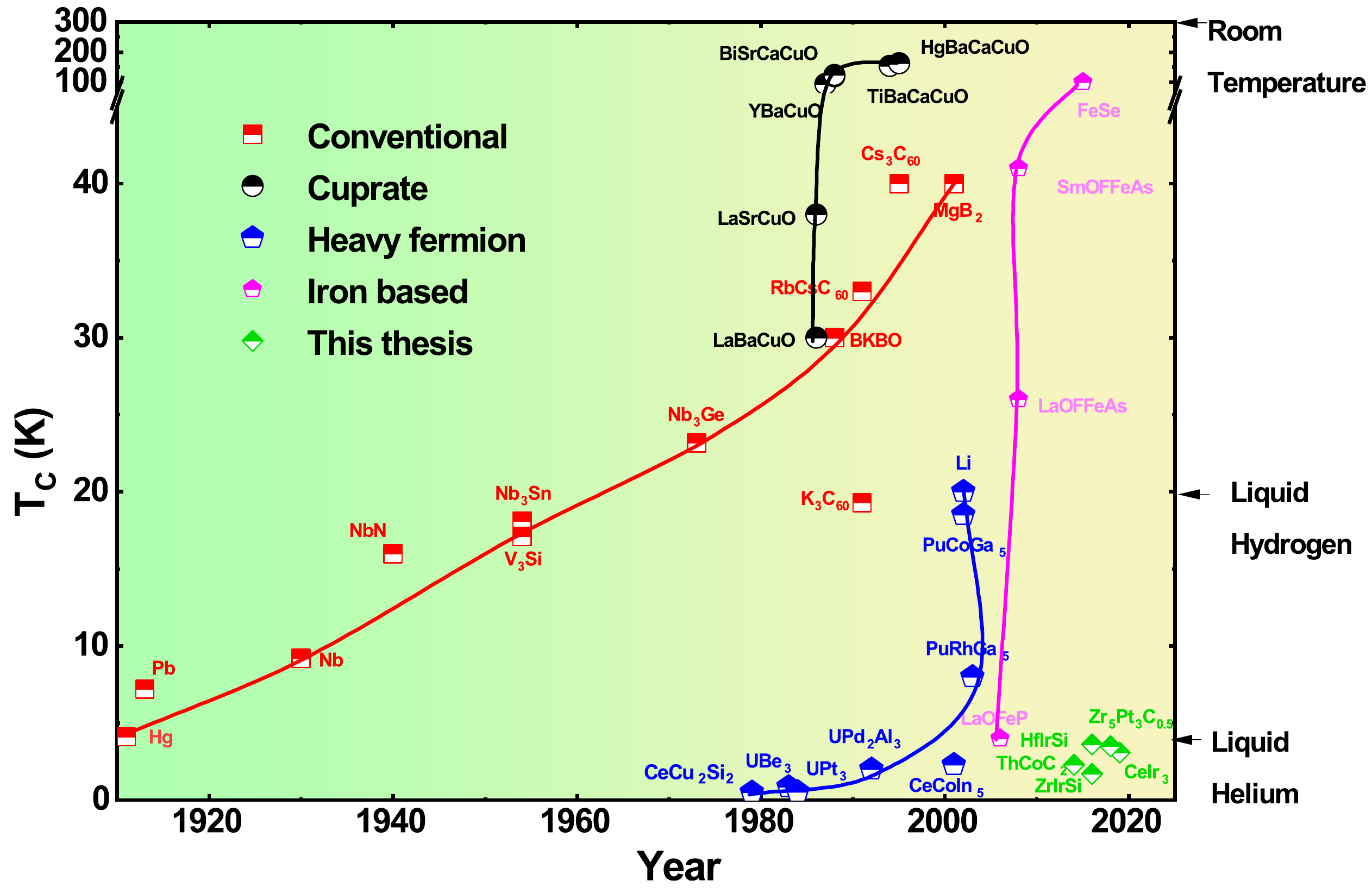}
\caption {The history of some of the discovered superconducting compounds. In this figure the BCS superconductors (red squares), the cuprates (black circle),  heavy Fermion superconductors (blue pentagon), iron based superconductor (pink pentagon), and superconductors of this thesis (green diamond) are shown.}
\label{musr}
\end{figure} 

A true milestone is touched in superconducting research in 1986 when German physicist Georg Bednorz and Alex M\"{u}ller reported that the lanthanum copper oxide (La$_{2}$CuO$_{4}$), when doped with Barium (LBCO) shows superconducting transition at 36 K~\cite{bednorz1986}. They shared the Nobel prize in physics in 1987 for their discovery. After the discovery of superconductivity in LBCO other cuprate materials $T_{C}$ more than the boiling point of liquid nitrogen (77 K) was reached. By 1993 the $T_{C}$ was pushed to 135 K in HgBa$_{2}$Ca$_{2}$Cu$_{3}$O$_{8+x}$ (Hg-1223)~\cite{Schilling}. The undoped parent compounds of the cuprates are insulators, the novel superconductors are doped materials. Doping leads to conductivity and, for larger concentrations, to superconductivity. Superconductivity in the cuprates is considered unconventional and is not explained by BCS theory. Several experimental techniques such as angle-resolved photoemission spectroscopy (ARPES)~\cite{Damascelli}, Josephson tunneling~\cite{Sun}, and nuclear magnetic resonance (NMR)~\cite{Jha} confirm that the cuprates have $d-$ wave pairing symmetry. The pairing in cuprates is highly anisotropic with a line of nodes in the superconducting gap and also the electron-electron interactions are more important than electron-phonon interactions. In contrast to conventional superconductors, which are well characterized by BCS theory, most cuprates show weak isotope effects. 

Akimitsu et al. revealed in 2001 that MgB$_{2}$, a simple binary compound known since the 1950s, is a superconductor with a transition temperature of 39 K~\cite{prassides2001}. It is, however, more commonly known as the conventional superconductor with the highest $T_{C}$, the higher $T_{C}$ coming from the presence of two distinct bands in the Fermi surface. Noncentrosymmetric superconductors have been known for quite some time. However, after the discovery of superconductivity in the heavy fermion compound CePt$_{3}$Si [$T_{C}$ = 0.75 K] in 2004, it attracted a lot of attention, due to the presence of various unusual properties in the superconducting state such as the appearance of spontaneous magnetism in the superconducting ground state~\cite{smidman2017,bauer2012}. Superconductivity was observed in the iron pnictide La(O$_{1-x}$F$_{x}$)FeAs at 26 K by Hosono et al. in 2006~\cite{kamihara2008}. In the similar iron pnictide Sm[O$_{1-x}$F$_{x}$]FeAs~\cite{Ren}, the critical temperature could be rapidly increased to as high as 55 K, which are considered the second family of high temperature superconductors. The discovery of superconductivity in Fe-based materials came as a big surprise as there is a historically antagonistic relationship between magnetism and superconductivity. The findings of iron-based superconductors suggest that there may be other families of high temperature superconductors besides cuprates and iron pnictides. The Fermi surface, the boundary between occupied and non-occupied electron states in the zero temperature limit which ultimately determines many physical properties, was soon the target of much experimental and theoretical effort. Quantum oscillations in high magnetic fields~\cite{Carrington} quickly provided a general confirmation of the density-functional theory prediction of quasi-2D hole and electron pockets around the Brillouin zone center and corners respectively (with some modification); it was also soon realised that this band structure favoured an unusual ``$s_\pm$'' superconducting gap function in a spin-fluctuation theory~\cite{Shun}, while the magnetic state in the parent compounds could, from a purely itinerant point of view, be understood in terms of a spin-density wave transition driven by nesting between the hole and electron pockets.\\
The technique of muon spin rotation and relaxation ($\mu$SR) is at the heart of the measurement methods presented in this thesis. $\mu$SR is a remarkably effective magnetic probe that may be used in a wide range of fields such as condensed matter physics, material science, and chemistry. The fundamental idea behind this technique is to implant 100\% spin polarized muons into the sample and then analyze the asymmetry spectra of the observed positrons evolving over time. These muons are ideal for studying magnetic fields at the atomic scale within matter, such as those created by different types of magnetism and/or superconductivity. The temperature dependence of the superfluid density determined from TF-$\mu$SR measurement directly indicates the symmetry of the superconducting gap structure. As the muon has an intrinsic magnetic moment of 3.18 $\mu_\mathrm{p}$, $\mu$SR measurement (zero field) are very sensitive to the weak internal field in a sample, down to the order of 0.1 G. Search for time reversal symmetry breaking in superconductors started with $\mu$SR measurements. The chiral $p$-wave superconductivity is reported in Sr$_{2}$RuO$_{4}$, which also exhibit TRS breaking, detected using $\mu$SR measurement ~\cite{ray2014}. 

\section{Thesis overview}
This thesis aims to understand the superconducting gap structure of novel superconductors and quantum fluctuations in heavy fermion compounds through sample preparation, X-ray, resistivity, magnetization, heat capacity and muon spin spectroscopy measurements. We have prepared high quality polycrystalline sample of HfIrSi, ZrIrSi, ThCoC$_{2}$, and CeIr$_{3}$ using arc melting method and single crystal of CeCo$_{2}$Ga$_{8}$ using flux grown technique. Various characterization techniques have been employed to understand the physical and microscopic properties of these systems. The thesis is structured as follows:

\begin{itemize}
\item{\textbf{ Chapter \ref{chapter:2} Introduction:}}\\
Electrons are fermions. But when is a fermion not a fermion? When there are two of them coupled in a Cooper pair. Chapter \ref{chapter:2}  introduces the fundamental theory of superconductors.
 
\item{\textbf{ Chapter\ \ref{chapter:3} Experimental details:}}\\
Here we introduce all the experimental techniques used throughout this thesis. In this chapter, we have discussed sample preparation, X-ray diffraction measurement. The resistivity, dc-magnetization and heat capacity measurement methods are discussed here. A detailed muon spin relaxation and rotation measurement techniques are also presented in this chapter.

\item{\textbf{Chapter\ \ref{chapter:4} Studies of ternary equiatomic superconductors:}} \\
In this chapter, we present the superconducting properties of the ternary equiatomic superconductors HfIrSi and ZrIrSi. Good quality polycrystalline samples of HfIrSi and ZrIrSi have been prepared successfully using the arc melting method. These compounds have been characterized through X-ray diffraction, transport, magnetization and specific heat measurements. We have discussed a transverse-field muon spin rotation (TF-$\mu$SR) study on these compounds. The temperature dependence of the magnetic penetration depth is compatible with an isotropic $s$-wave model. We have also performed ZF-$\mu$SR studies on these systems found that time reversal symmetry is preserved.

\item{\textbf {Chapter \ \ref{chapter:5} Nodal line in the superconducting gap symmetry of noncentrosymmetric superconductor:}}\\
The superconducting properties of noncentrosymmetric superconductor ThCoC$_{2}$, which exhibits numerous types of unconventional behaviour in the field-dependent heat capacity data, have been investigated using low-temperature $\mu$SR and magnetization measurements. The presence of a nodal line $d$ wave gap has been suggested from the TF-$\mu$SR measurement. This finding is in agreement with the field dependent heat capacity, $\gamma(H) \sim \sqrt{H}$. 

\item{\textbf {Chapter \ \ref{chapter:6} Studies of intermediate valence superconductor:}}\\ 
The role of strong spin orbit coupling in superconductivity is discussed in this chapter. Here we have thoroughly presented a systematic resistivity, magnetization, heat capacity and $\mu$SR measurement on CeIr$_{3}$ with $T_{C}$ = 3.4 K, which is the second-highest $T_{C}$ among the Ce-based intermetallic compounds. It has shown that the density of states (DOS) at the Fermi surface principally arises from the $5d$ states of the Ir atoms, suggesting that CeIr$_{3}$ is indeed an Ir $5d$-band superconductor and that the $5d$ electrons play a crucial role in the superconductivity. The ZF-$\mu$SR data show no strong evidence for spontaneous internal fields developing at or below $T_{C}$, and weak temperature dependence in the ZF-$\mu$SR relaxation rate below 3 K is taken as evidence for the presence of spin fluctuations rather than the breaking of time-reversal symmetry.

\item {\textbf {Chapter \ \ref{chapter:7} Quantum fluctuations in the non-fermi liquid system:}}\\
Reduced dimensionality offers a piece of crucial information in deciding the type of the quantum ground state in heavy fermion materials. To accommodate the deeper understandings of the nature of quantum critical points, we have discussed the non Fermi liquid properties of CeCo$_{2}$Ga$_{8}$ evidence from transport, magnetic and muon spin relaxation measurements. All the experimental data confirms that the stoichiometric CeCo$_{2}$Ga$_{8}$ compound exhibits a non Fermi liquid ground state without any doping. 

\item{\textbf{Chapter \ \ref{chapter:8} Summary and Conclusions:}} \\
The important findings and its importance for further research are summarized.
\end{itemize}


%% file: chapter2.tex
\chapter{Fundamental properties of superconductivity} \label{chapter:2}
\section{Introduction}
The theory of superconductivity has been the subject of numerous attempts during the twentieth century. There has been some progress, but the ultimate theory has yet to be discovered. C. Gorter and H. Casimir made the first attempt to explain the origin of superconductivity in 1934~\cite{gorter}. Later London brothers i.e. H. London and F. London presented the phenomenological theory in 1935~\cite{london1935}, followed by Ginzburg, Landau's theory in 1950~\cite{ginzburg1950}. All of these efforts aided in gaining a better understanding of superconductivity. Several empirical connections have been postulated, indicating that crystal lattice structure and superconductivity have a close relationship. Although Ginzburg-Landau theory does not include the microscopic details contained within BCS theory, it describes many properties of superconductors, and in particular, it can describe properties of unconventional superconductors where BCS theory fails to do so. It does, however, have limitations in that it cannot explain the origins of superconductivity at the microscopic level. In 1957, Bardeen, Cooper, and Schrieffer (BCS) presented the first microscopic theory of superconductivity~\cite{BCS}. This theory gives a satisfactory explanation of the key properties of elemental superconductors. However, this theory fails to explain the origin of superconductivity in heavy fermion superconductors, cuprate superconductors and some other novel superconductors. This chapter gives an introduction to superconductivity. The BCS and Ginzburg-Landau theories, and their key results, are presented, with a focus on the parameters that can be measured using $\mu$SR experiments. 

\begin{figure*}
\centering
 \includegraphics[width=0.45\linewidth]{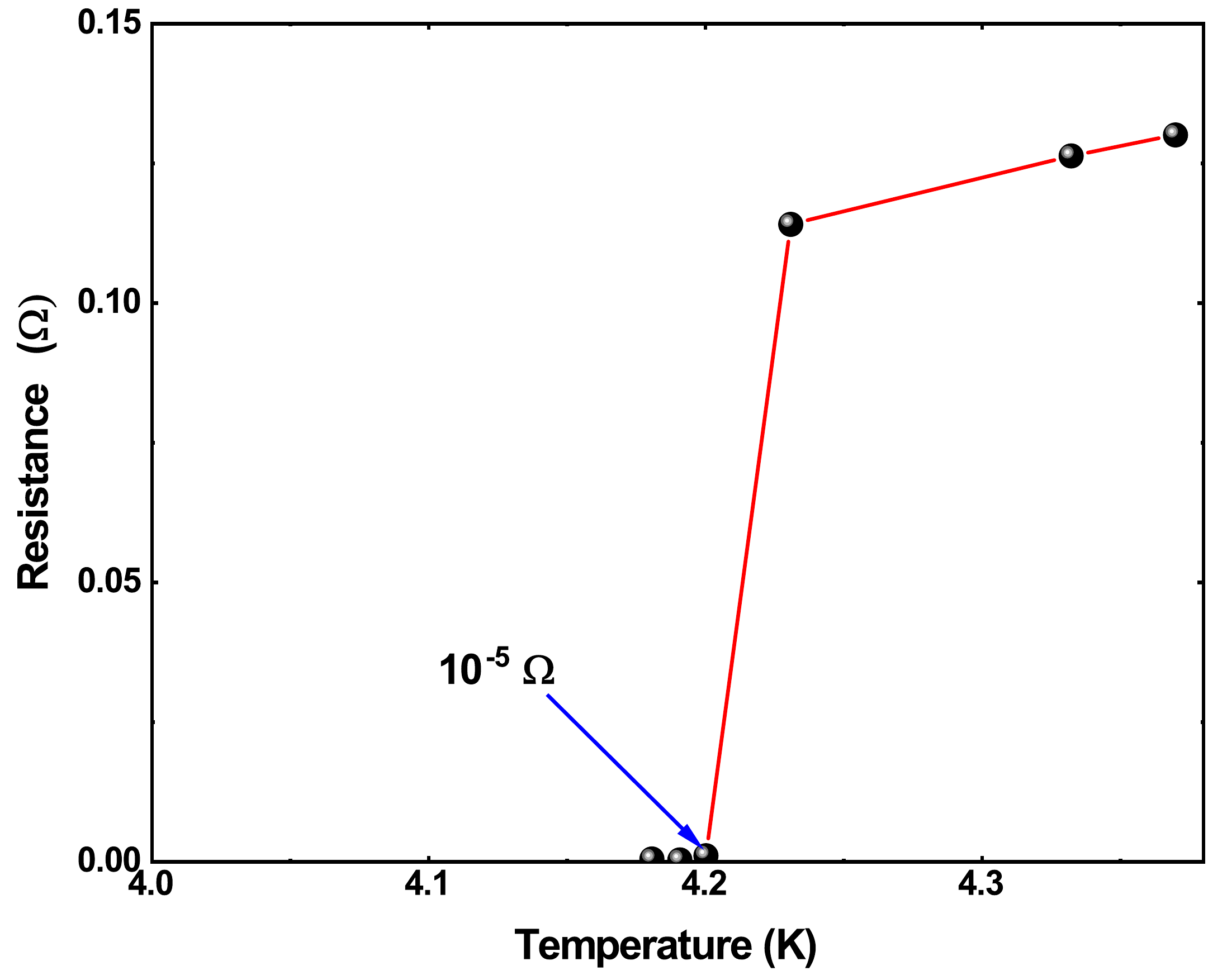}
 \includegraphics[height = 0.4\linewidth, width=0.45\linewidth]{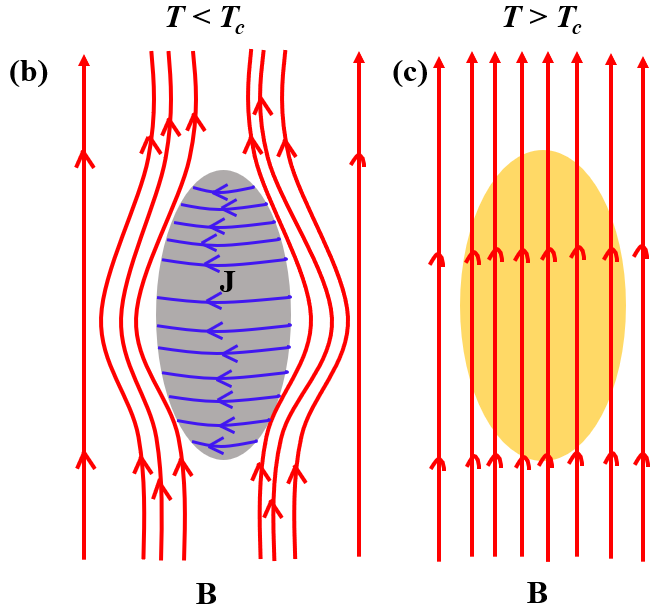}
\caption {(a) Kamerlingh Onnes’s measurements of resistance versus temperature in mercury from 1911, showing the superconductivity below 4.2 K. Adapted from Onnes~\cite{onnes1911}. Illustration of magnetic field lines for a superconductor in the (b) Meissner state, where the shielding currents (shown in blue) make it so the resulting magnetic field inside the sample is zero and, (c) normal state, where the flux lines go through the sample.}
\label{Onnes}
\end{figure*}

\section{Hallmark properties of superconductor}
\subsection*{Zero Resistance}
If a material exhibit both perfect conductivity and perfect diamagnetism below a certain temperature, known as superconducting transition temperature ($T_{C}$), is called a superconductor. For superconductivity, below $T<T_{C}$ the electrical resistivity ($\rho$) goes to zero and the conductivity becomes infinite below $T_{C}$. So from Maxwell's quation, we can write
\begin{eqnarray}
\bf{\nabla . E} &=& 4\pi \rho \\
\implies  \bf{E} &=& 0
\end{eqnarray}
The electric field is zero inside the superconductor, which leads to the finite current density ($\bf{J_{s}}$), which implies current flow in the superconductor without an electric field. Persistent currents in superconducting rings provide the most persuasive evidence for this. A supercurrent is induced in the ring is cooled down to $T_{C}$ in the presence of a magnetic field. As $\bf{E} = 0$ everywhere along the ring, so
\begin{equation}
\frac{d\phi}{dt} = 0
\end{equation}  
As a result, the magnetic flux across the ring remains constant over time, is equal to  $\phi_{0} = \frac{h}{2e}$ {$\approx~$2.068$\times$10$^{-15}$ Wb }. When the external magnetic field is removed, then also the supercurrent flows continuously because the resistance in the superconducting state is zero. As a result, the overall flux passing through the ring does not change. Depending on the geometry of the wire and temperature, persistent electricity can theoretically last longer than the life of the universe. However, it has been demonstrated experimentally that the supercurrent can sustain for 100,000 years, confirming the superconductor's zero resistance.

\subsection*{Meissner Effect}
Only zero resistivity, $\rho$ = 0, is no longer considered to be the true signature of superconductivity. The Meissner-Ochsenfeld effect~\cite{meissner1933} is the actual hallmark of superconductivity. When diamagnetic materials are placed in an external magnetic field it partly opposes the magnetic field from their interior. Superconductors are the perfect example of diamagnetic material, which completely expels the magnetic field everywhere. The creation of whirling supercurrents on the surface of the superconductor causes magnetization to happen. The expulsion of flux from the interior of the superconductor is known as the Meissner effect, the actual hallmark of superconductivity. 

The three vectors {\bf M, H, B} are related by the following equation:
\begin{equation}
{\bf B} = \mu_{0}{(\bf{H+M})}
\end{equation} 

In the Meissner state ${\bf B} = 0$, which leads to {\bf{ M = -H}}, the susceptibility of the superconductor is equal to
\begin{equation}
\chi=\frac{M}{H}=-1
\end{equation}

\subsection{Critical field}
Although superconductors exhibit perfect diamagnetism in a weak magnetic field, superconductivity can be destroyed in a strong magnetic field. The Meissner state in most elemental superconductors is abruptly disrupted at a magnetic field known as the critical field, $H_{C}$, and above $H_{C}$ these materials behave like a normal metal. When an external magnetic field is applied, the free energy of the superconductor increases. The Helmholtz free energy difference between the superconducting state and the normal state is given by 
\begin{equation}
\epsilon_{cond} = F_{sc}(T)-F_{n}(T) = -\frac{H_{C}^2}{8\pi}
\end{equation}
$\epsilon_{cond}$ is known as condensation energy. 

\begin{figure*}
\centering
 \includegraphics[width=0.9\linewidth]{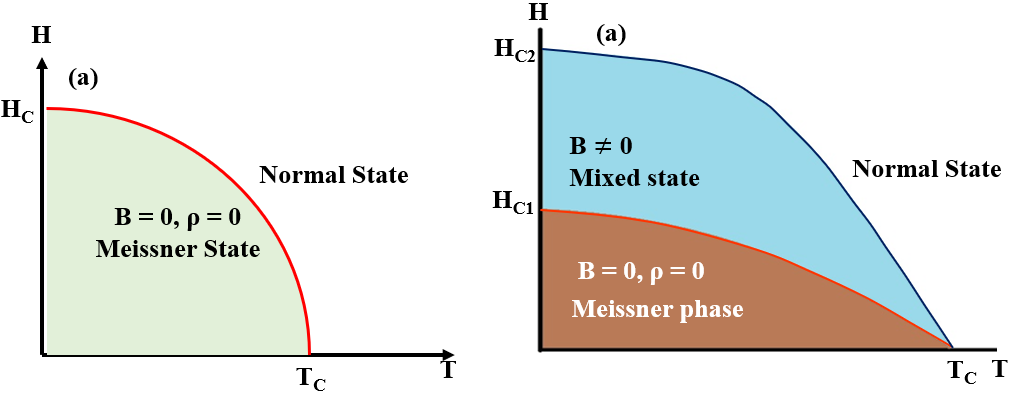}
\caption {(a) Schematic magnetic phase diagram presents Meissner state for a type-I superconductor. (b) Magnetic phase diagram for a type-II superconductor. The magnetic field penetrates the superconductor above the lower critical field $H_{C1}$, generating the mixed state until superconductivity is eliminated at the upper critical field $H_{C2}$.}
\label{fig22}
\end{figure*}

\subsection{Two kinds of superconductivity}
\begin{figure*}
\centering
 \includegraphics[width=0.9\linewidth]{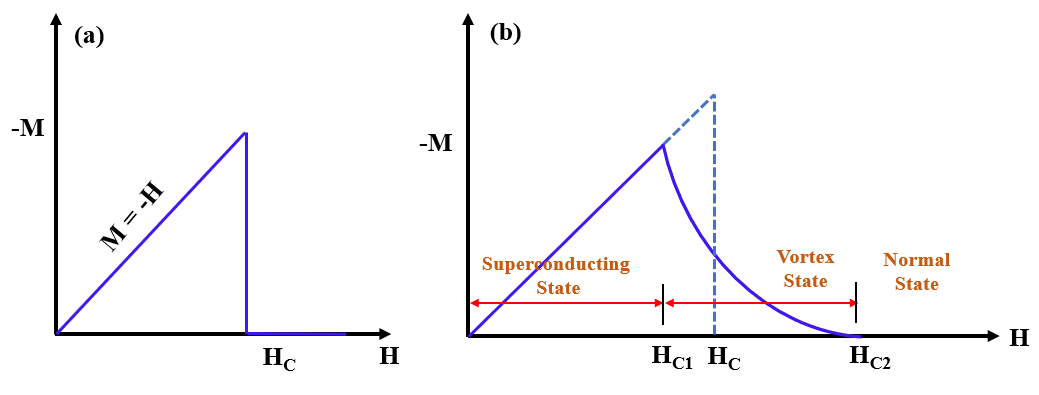}
\caption {(a) The field dependence of the magnetization in the case of a type-I superconductor. (b) The field dependence of the magnetization in the case of a type-II superconductor.}
\label{fig23}
\end{figure*}

Based on their behaviour in an external magnetic field, the superconductors are divided into two categories: type-I and type-II superconductors. Below $T_{C}$ type I superconductors fully expel the magnetic field, up to some critical applied field ($H_{C}$) [known as Meissner effect]. As the field is totally expelled, the magnetisation of the sample is proportional to the applied field; this relationship has been plotted in Fig. \ref{fig23}(a).  Type II superconductors also exhibit the Meissner state up to some critical applied field, known as the lower critical field ($H_{C1}$) then above $H_{C1}$ (but below a second critical field ($H_{C2}$) quantized flux lines from the applied field are trapped in the sample, forming the flux line lattice (FLL) [see Fig.\ref{fig22} (b]. These flux lines, also known as Abrikosov vortices, are small pockets of the normal state surrounded by the superconductor; this state is therefore known as the mixed state. As the applied field increases, the density of vortices increase, until a critical flux density is reached at $H_{C2}$, and superconductivity is destroyed. A plot of the field dependence of the magnetization of a type-II superconductor is presented in Fig. \ref{fig23} (b).

\subsection*{The London equation}
Inspired by the two fluid model of superfluid $^{4}$He of Gorter and Casimir in 1934~\cite{gorter}, London brother assumed that the superconductor is a two fluid system (i) a fluid made up of ordinary electrons ($n_{n}$) (ii) another fluid made up of superconducting electrons ($n_{s}$).
\begin{equation}
n = n_{n} + n_{s} 
\end{equation}  

To describe the observed zero resistance property and Meissner effect two brothers Fritz and Heinz London in 1935 ~\cite{london1935} proposed a pair of equations, 
\begin{equation}
\frac{\partial {\bf J_{s}}}{\partial t} = \frac{n_{s}e^2}{m_{e}}\bf{E}
\label{London1}
\end{equation}
\begin{equation}
 {\bf \nabla} \times  {\bf J_{s}}= \frac{n_{s}e^2}{m_{e}}\bf{B}
 \label{London2}
\end{equation}

\begin{figure*}
\centering
 \includegraphics[width=0.9\linewidth]{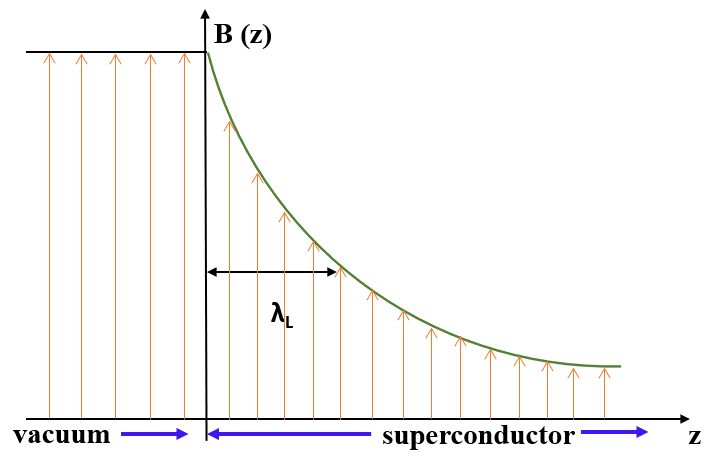}
\caption {The Meissner effect is depicted schematically in this figure based on London's equation. Inside the superconductor, the magnetic field decays exponentially. }
\label{fig21}
\end{figure*}

Here {\bf B} and {\bf E} are the microscopic magnetic field and electric field within the superconductor, $J_{s}$ is the local supercurrent density, $n_{s}$ is the superfluid density related to the number density of superconducting electrons, $m_{e}$ is the mass of the electron, respectively. Eq. \ref{London1} and \ref{London2} relates the variation of current to the electric field and magnetic field with time. Eq. \ref{London1} represents the phenomena of zero resistivity of the superconducting state. In presence of an external electric field, the superconducting electrons are accelerated contrary to Ohm's law i.e. constant drift velocity inside the conductor.

The second Eq. \ref{London2} can be written by combining with Maxwell equation, $\bf{\nabla \times B} = \mu_{0}\bf{J}$, we get

\begin{eqnarray}
\bf{\nabla \times (\nabla \times B)} & = & -\mu_{0}\frac{n_{s}e^2}{m_{e}} {\bf B}\\
\bf{\nabla \times (\nabla \times B)} & = & -\frac{1}{\lambda^{2}} {\bf B}
\end{eqnarray} 
where the dimensions of $\lambda$ is similar to the length, and is known as London penetration depth, 
\begin{equation}
\lambda_{L}= \sqrt{\frac{m_{e}c^2}{4\pi n_{s}e^2}}
\end{equation}
It is the distance inside the surface over which an external magnetic field is screened out to zero, given that $B = 0$ in the bulk.

The Meissner effect can be understood easily assuming a superconductor is placed in an external field $B_{a} = B_{0}\hat{z}$ parallel to the surface of the superconductor. By putting this value of {\bf B} and solving it, it can be easily shown that any external magnetic field is screened out inside the superconductor [see Fig.~\ref{fig21}], as
\begin{equation}
B(z) = B_{0}~\exp(-z/\lambda_{L})
\label{London6}
\end{equation}
Therefore the current flow at the surface of the superconductor exponentially decays inside the superconductor. Although the London equations are not derived from fundamental properties, they successfully explain the zero resistance property and Meissner effect in a superconductor.

The superfluid density $n_{s}$ is a measure of the number of Cooper pairs in the system, and so as $T$ is increased, $n_{s}$ is expected to drop and therefore $\lambda_{L}$ increases as the external field can penetrate further into the sample. This theory, however, is unable to account for the experimentally observed change in penetration depth as a function of temperature. The London penetration depth can be calculated using Gorter-Casimir's formula of the temperature dependence of superconducting carrier density~\cite{gorter}.
\begin{equation}
\frac{\lambda(T)}{\lambda(0)} = \left[1-t^4\right]^{-1/2}
\label{London7}
\end{equation}
where $t=T/T_{C}$. It can be shown from Eq. \ref{London7} that when $T\rightarrow T_{C}$, the London penetration depth tends to infinity. Using BCS microscopic calculation, $\lambda(T) = \lambda(0)/\sqrt{1-t^p}$ with $p$ = 2 for s-wave and $p$ = 4/3 for d-wave superconductors~\cite{Prozorov2006}.

\subsection*{Isotope effect}
In 1950 it is reported that the $T_{C}$ of Hg is related to its isotopic mass~\cite{Tinkham} and follows the relationship,
\begin{equation}
T_{C} \propto M^{-\alpha}
\end{equation}
where $\alpha$ = 1/2 for Hg. Most conventional superconductors agree very well with this value. The isotope effect indicates the role played by electron-phonon interaction, in the occurrence of superconductivity in conventional superconductors.

\section{BCS Model of superconductivity}
In 1957, Bardeen, Cooper, and Schrieffer (BCS) published their most famous theory in condensed matter physics to describe the unique phase of materials known as superconductivity~\cite{BCS}. As per their hypothesis, even a modest gross attractive attraction between two electrons induced by second order interaction mediated by phonon results in the formation of a Cooper pair, which is a bound pair. As a result, at the Fermi energy, electrons form bound Cooper pairs, which then condense into a phase coherent macroscopic quantum state, resulting in superconductivity. In this section, I will give a brief overview of this elegant theory and its novelty to describe the different properties of superconductors.

\subsection{Formation of Cooper pair}
In 1950, despite popular belief, Frohlich ~\cite{Frohlich} demonstrated that there might be a resulting attractive attraction between two electrons mediated by phonon. The following way illustrates how the attractive interaction works physically: when an electron passes through the lattice, it causes a positive charge imbalance by pulling positive ions towards it, and this surplus positive charge draws another electron. Electrons will scatter in a particular way and the electron-electron interaction in presence of phonon is given by~\cite{Tinkham},
\begin{equation}
V_{s}(q,\omega) = \frac{4\pi e^2}{q^2+k_{s}^2}+\frac{4\pi e^2}{q^2+k_{s}^2}\frac{\omega_{q}^2}{\omega^2-\omega_{q}^2}
\end{equation}
where the first term is the screened Coulomb potential which is always positive, repulsive in nature. The second term is caused by electron phonon interaction, which is negative when $\omega < \omega_{q}$. When the attractive force exceeds the Coulomb repulsive force, the result is a net attractive interaction. This is possible for the two electrons if they can construct the wave function which selectively chooses the attractive frequency range.  Cooper~\cite{cooper1956} in 1956  demonstrated that two electrons with opposite momenta and spin might form a bound pair known as a Cooper pair in the presence of net attractive interactions, no matter how tiny they are. It can be shown employing the uncertainty principle that the size of a Cooper pair is of the order of BCS coherence length, $\xi_{BCS} = \hbar v_{F}/\pi \Delta$ which is much larger than the inter particles distance. As a result, pairs are extensively overlapping and form a collective state.

\subsection{BCS wavefunction}
It was shown by L. N. Cooper in the same work that the Fermi sea is unstable against the formation of bound pairs induced by net attractive interaction~\cite{cooper1956}. In this situation, the Cooper pairs should condense until they reach a point of equilibrium when the binding energy for the additional pair goes to zero and gives rise to a macroscopic quantum state.

If the two electrons have opposite momenta ($k$,$-k$), their binding energy will be maximum, and their spin should be opposite to each other to minimize the exchange correlation energy. So, the angular  momentum of the Cooper pair is zero ($L$ = 0 ) and spin ($S$ = 0), is known as spin singlet state with $k\uparrow$ and $k\downarrow$ electrons. To describe the state BCS proposed a ground state wave function of the form~\cite{Tinkham}

\begin{eqnarray}
|{\Psi_{BCS}}>=\prod_{k} \hspace{1mm}  (|u_k| + |v_k| c^\dagger_{k\uparrow} c^\dagger_{-k\downarrow}) |{0}>
\end{eqnarray}

where, $|u_{k}|^2 + |v_{k}|^2  = 1$. Here $c^{\dagger}_{k\uparrow}$( $c^{\dagger}_{k\downarrow}$) is creation operator which creates one electron (hole) of momentum {\bf $k$}({\bf{$- k$}}) and  spin up (down). $|0>$ is the vacuum state.  The probability of the pair ($k\uparrow$, $k\downarrow$) being occupied is $v_{k}^2$, and the probability that it is not occupied is $u_{k}^2 = 1 - v_{k}^2 $.

The coefficients $u_{k}$ and $v_{k}$ can be determined by minimizing the ground state energy $E = <\psi_{BCS}|H|\psi_{BCS}>$ where $H$ is the BCS pairing Hamiltonian or reduced Hamiltonian given by,

\begin{equation}
H = \sum_{{\bf k},\sigma} n_{\sigma,{\bf k}}\epsilon_{\bf{k}}+\sum_{\bf{k,l}}V_{\bf{k,l}}c^{*}_{{\bf k},\uparrow}c^{*}_{{\bf -k},\downarrow}c_{{\bf l},\uparrow}c_{{\bf -l},\downarrow}
\end{equation}

The first term corresponds to the kinetic energy of noninteracting electron gas and second term is the pairing interaction via second order phonon scattering. Here,$V_{\bf{k,l}}$ is the electron-phonon scattering matrix element which is approximately constant and can be replaced by –V where V is
a positive quantity. Now applying variational method~\cite{Tinkham}, the ground state energy can be
determined as follows, 
\begin{equation}
-\epsilon_{cond} = F_{sc}(0) - F_{n}(0) = -\frac{1}{2}N(0)\Delta^{2}(0)
\end{equation}
This is the condensation energy at T = 0, which is equal to $\frac{H_{C}^2}{8\pi}$, $H_{C}$ is the thermodynamic critical field.F is the free energy per unit volume , N(0) is the normal state electronic DOS at the Fermi level and $\Delta$(0) is the SC energy gap equivalent to the pairing energy given by,
\begin{equation}
\Delta_{0} = 2\hbar \omega_{D}\exp(-1/N(0)V)
\label{Eq20}
\end{equation}
$\omega_{D}$ is the Debye frequency.

\subsection{Excitation energies and the energy gap}
The elementary excitation in superconductors is electron and hole like quasi particles originating from the breaking of the Cooper pairs. Using self-consistent approach~\cite{Tinkham} energy of these quasi particles can be determined as,

\begin{figure*}
\centering
 \includegraphics[height= 0.5\linewidth,width=0.6\linewidth]{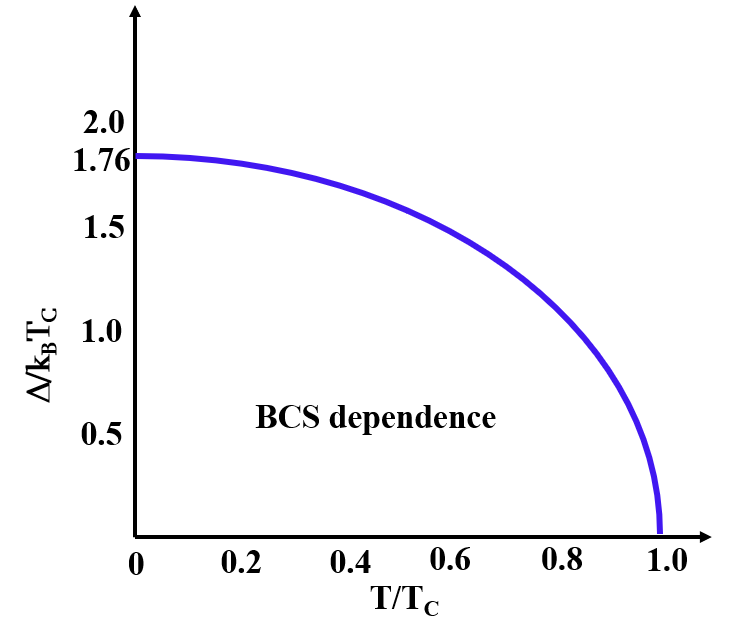}
\caption { The normalized temperature dependence of BCS energy gap.}
\label{fig25}
\end{figure*}

\begin{eqnarray}
E_k=\sqrt{\xi_{k}^2+\Delta^2},
\end{eqnarray}

In finite temperature limit: 
\begin{equation}
k_{B}T_{C} = 1.14\hbar \omega_{D}\exp(-1/N(0)V)
\label{Eq22}
\end{equation}

Dividing \ref{Eq20} by \ref{Eq22}, we get
\begin{equation}
\frac{\Delta_{0}}{k_{B}T_{C}}= \frac{2}{1.14} = 1.76
\end{equation}
which leads to $\frac{2\Delta_{0}}{k_{B}T_{C}}$ = 3.52, gap to $T_{C}$ ratio for BCS weak coupling superconductor. 

The temperature dependence of the energy gap can be computed from:
\begin{eqnarray}
\int_{0}^{\infty} \mathrm{d}E \left[\frac{\tanh \left(\frac{1}{2T}\sqrt{E^2+\Delta^2}\right)}{\sqrt{E^2+\Delta^2}}-\frac{1}{E}\tanh \left(\frac{E}{2T_{C}}\right)\right]=0,
\end{eqnarray}

Near $T_{C}$, $\Delta(T)$ drops to zero, approximated as
\begin{equation}
\frac{\Delta(T)}{\Delta(0)} \approx 1.74\left(1-\frac{T}{T_{C}}\right)^{1/2}
\end{equation}

The temperature variation of $\Delta$ is shown in Fig. \ref{fig25}. 

\section{Ginzburg-Landau model}
The microscopic BCS theory provides excellent explanations of various properties of superconductors such as prediction of $T_{C}$, isotope effect, energy gap, elementary excitation etc where energy gap $\Delta$ is constant over space. However, in inhomogeneous superconductors where $\Delta$ changes spatially and fluctuations are involved, BCS theory becomes very complicated. In such a situation, another exciting theoretical description proposed by Ginzburg and Landau (GL) in 1950~\cite{ginzburg1950} much before the development of BCS theory, provides an elegant description of superconductivity close to $T_{C}$. GL theory concentrates entirely on superelectrons rather than excitation and is generalized to deal with spatially varying and time dependent order parameters.
\subsection{Ginzburg Landau theory of the bulk phase transition}
Based on Landau's theory of second order phase transition in 1930, Ginzburg and Landau assumed that there exists an order parameter $\psi$ (quantum mechanical wave function), which characterize the superconducting state~\cite{Tinkham}. The order parameter is assumed to have some physical value below $T_{C}$ which characterizes the state, and above $T_{C}$ in the normal state it is zero i.e.
\[\psi=\begin{cases}
\text{0}, & \text{if $T>T_{C}$}\\
\text{$\psi(T) \neq 0$}, & \text{if $T<T_{C}$}
\end{cases}
\]
Considering it's a macroscopic wave function, they assumed that the wave function should be a complex number. They have introduced the wave function as,
\begin{equation}
\psi(r) = |\psi(r)|~e^{i\varphi(r)}
\end{equation}
The complex scalar is the Ginzburg-Landau order parameter. (i) $|\psi(r)|^{2}$ represent the amplitude of the wave function, which is related to local superconductor electron density. (ii) The phase factor $\varphi(r)$ is related to the supercurrent that flows through the material below $T_{C}$. 

\begin{figure*}
\centering
 \includegraphics[width=0.9\linewidth]{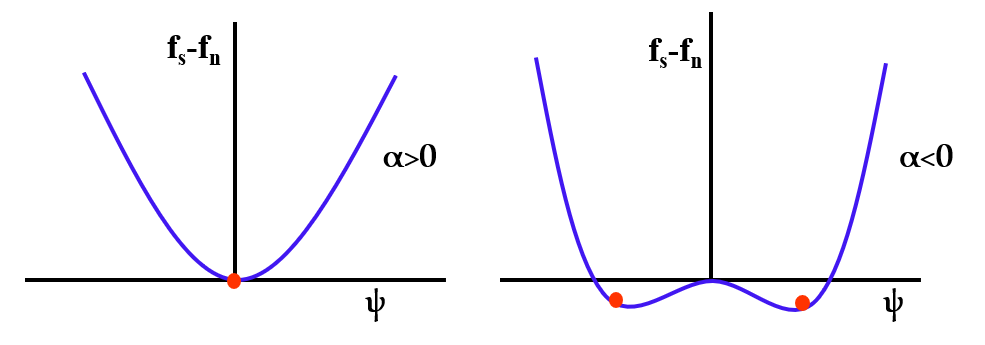}
\caption {Ginzburg Landau free energy functions for $T>T_{C}~ (\alpha > 0)$ and for $T<T_{C} ~(\alpha < 0)$. Red dots indicates equilibrium position. For simplicity, $\psi$ has taken to be real. }
\label{fig26}
\end{figure*}

The free energy of a superconductor in absence of magnetic field and spatial variation of $n_{s}$ can be written as:
\begin{equation}
F_{sc} = F_{n}+\alpha |\psi|^2+\frac{\beta}{2}|\psi|^{4}
\end{equation} 

The parameters $\alpha$ and $\beta$ are temperature dependent expansion coefficients. $\beta$ is positive throughout the transition but $\alpha$ changes sign to keep the free energy of the system minimum. The temperature dependence of $\alpha$ can be approximated by:
\begin{equation}
\alpha = \alpha_{0}(T/T_{C} -1)........\rm{\alpha>0}
\end{equation}
Assuming $n_{s} \propto |\psi|^{2}$ the equilibrium value obtained from 
\begin{equation}
\frac{\partial(F_{sc}-F_{n})}{\partial n_{s}} = 0 = \alpha + \beta |\psi|^2
\end{equation}

It is found that:
(i) for $\alpha>0$ minimum must be when $|\psi|^2$ = 0 \\
(ii) for $\alpha<0$ minimum is when $|\psi|^2 = -\frac{\alpha}{\beta} = |\psi_{\infty}|^2$

So in the superconducting state at equilibrium we have
\begin{equation}
F_{sc} - F_{n} = -\frac{\alpha^2}{2\beta}
\end{equation} 
We already know from condensation energy 
\begin{equation}
F_{sc} - F_{n} = - \frac{1}{2}\mu_{0}H_{C}^2
\end{equation}
so $\mu_{0}H_{C}^2 = \frac{\alpha^2}{\beta}$

\subsection{The full G-L free energy}

In presence of the external magnetic field spatial gradients the free energy superconducting ground state can be written as~\cite{Tinkham}
\begin{equation}
F_{sc} = F_{n}+\alpha |\psi|^2+\frac{\beta}{2}|\psi|^{4} + \frac{1}{2m}|\left(\frac{\hbar}{i}{\bf \Delta}-\frac{e}{c}{\bf A} \psi\right)|^{2}+\frac{1}{2}\mu_{0}H{(r)}^2
\label{GL2}
\end{equation}

Here $\bf{A}$ is the magnetic vector potential which is related with $\bf{H}$ ($\mu_{0}\bf{H} = {\bf \nabla} \times {\bf A}$). The third term denotes kinetic energy associated with the fact that $\psi$ is not uniform in space, but has a gradient. The last term is the magnetic energy associated with the magnetization in a local field $H(r)$. e, and m are the charge and mass of the superelectron.  

The minimization of free energy, $F$ with respect to $\psi$, leads to the well known GL equation
\begin{equation}
\alpha|\psi|+\beta|\psi|^2\psi+\frac{1}{2m}\left(\frac{\hbar}{i}{\bf \Delta} - \frac{e}{c} \bf{A}\right)^2\psi =0
\end{equation}

In a superconductor, the Ginzburg-Landau equations predicted two additional characteristic lengths. (i) The coherence length, $\xi$, the distance between the two electrons in the Cooper pair and (ii) penetration depth, $\lambda_{GL}$, previously introduced by the London brothers in their theory. When $T \rightarrow$ $T_{C}$ both the quantities goes to infinity, however their ratio 
\begin{equation}
\kappa = \frac{\lambda_{GL}(T)}{\xi(T)}
\end{equation}
From critical field measurements, both $\lambda_{GL}$ and $\xi_{GL}$ can be estimated experimentally. The upper critical field is directly related with $\xi$ by the following equations:
\begin{equation}
\mu_{0}H_{C2} = \frac{\phi_{0}}{2\pi\xi^2(T)}
\label{GL12}
\end{equation}
$H_{C2}$ can be determined in the lab using magnetization, resistivity, or heat capacity measurements. The following equation relates the effective penetration depth to the lower critical field:
\begin{equation}
\mu_{0}H_{C1} = \frac{\phi_{0}}{4\pi\lambda^2(T)} \ln~\kappa
\end{equation}

$H_{C1}$ can be measured using isothermal field dependence of magnetization. However, it is difficult to measure the point at which the magnetic line of force begins to penetrate it. So it is convenient to measure the thermodynamic critical field, from the London penetration depth, which can be easily measurable using different experimental techniques. The GL theory describes the relationship between the two characteristic length scales and the thermodynamic critical field as follows:
\begin{equation}
\mu_{0}H_{{C}} = \frac{\phi_{0}}{2\sqrt{2}\pi\lambda\xi}
\label{GL14}
\end{equation}
From Eqs. \ref{GL12}  and Eq. \ref{GL14}, the $H_{C2}$ is related to $H_{{C}}$ by 
\begin{equation}
H_{C2} = \kappa \sqrt{2}H_{{C}}
\end{equation}

\begin{itemize}
\item {If $\kappa < 1/\sqrt{2}$, implies $H_{C2} < H_{C}$. The superconducting state arises at and below $H_{{C}}$ with total expulsion of the magnetic flux, indicating that the superconductor is of type I.}
\item {If $\kappa>1/\sqrt{2}$, implies $H_{{C2}} > H_{{C}}>H_{{C1}}$. The superconducting state appears below $H_{{C2}}$. As some magnetic flux penetrate in this state, the superconductor is known as type-II superconductor.}
\end{itemize}

Through solving the GL equations in a magnetic field, Abrikosov discovered that the order parameter’s spatial distribution meant that, in a type II superconductor in the mixed state, forms into an ordered periodic structure of vortices.

\subsection{Superconducting gap structure}
To understand the pairing mechanism of the superconductor it is is crucial to investigate the temperature dependence of the superconducting gap, related to the superfluid density. The temperature evolution of the normalized superfluid density can be described as:
\begin{equation}
\tilde{n_{s}}=\frac{n_{s}(T)}{n_{s}(0)}=\frac{\lambda^{-2}(T)}{\lambda^{-2}(0)} = 1 + \frac{1}{\pi} \int_{0}^{2\pi}\int_{\Delta(\phi,T)}^{\infty}\frac{\partial f}{\partial E} \frac{E dE d\phi}{\sqrt{E^2-\Delta^2(\phi,T)}}
\end{equation}
here f is the Fermi function 
\begin{equation}
f = \frac{1}{1+\exp(E/k_{B}T)}
\end{equation}
The temperature dependence of the gap can be described using the approximate form
\begin{equation}
\Delta(\phi,T) = \Delta (T) \tanh(1.82[1.018\lbrace{T_{c}/T-1}\rbrace]^{0.51}) g(\phi)
\end{equation}

\begin{figure*}
\centering
 \includegraphics[width=0.9\linewidth]{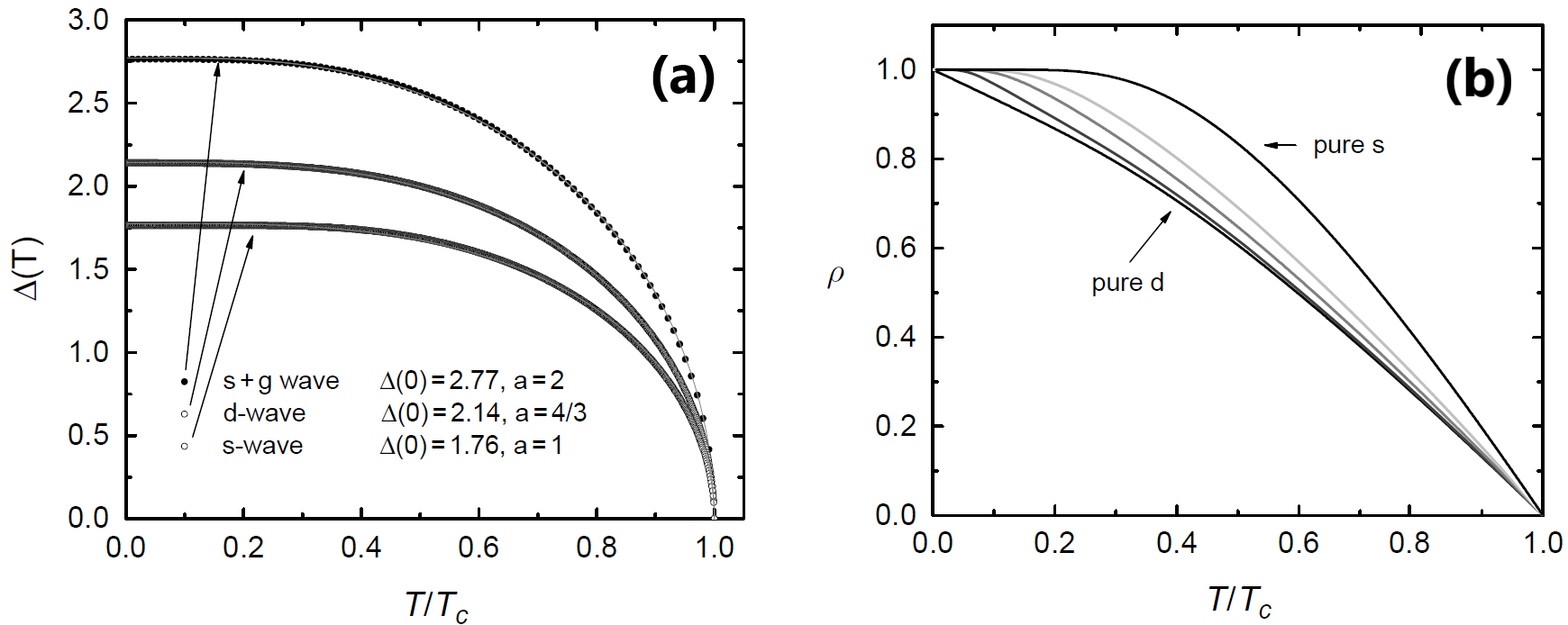}
\caption {(a) The temperature dependence of superconducting gap in case of $s$, $d$ and $s+g$ symmetries (b) Superfluid density for a mixed gap, $d_{x^{2}-y^{2}} +is$ for a component content of 0, 10, 20, 30, 40 and 100\% (from bottom up). Similar curves are obtained for $d_{x^2-y^2} +id_{xy}$ type mixed gap. Adapted from Ref.~\cite{Poole}}
\label{fig27}
\end{figure*}

here $\Delta (T)$ is the temperature dependence of the superconducting gap, is given by the following equations:
\begin{equation}
\Delta_{0}(T) = \Delta_{0}(0)\tanh\left(\frac{\pi T_{C}}{\Delta_{0}(0)}\sqrt{a\left(\frac{T_{C}}{T}-1\right)}\right)
\end{equation}
here $\Delta (0)$ is the gap magnitude at zero temperature, is a constant, $a$ is a parameter dependent upon the particular pairing state. $g({\phi}$) represents the angular dependence of the superconducting gap. Table~\ref{tb21} summarizes $g({\phi}$) for several different gap functions~\cite{Poole}.

\begin{table}[h]
 \caption{Some typical pairing state gap functions}
 \centering
  \begin{tabular}{lc}
\hline \hline
    Notation & $g(\phi$) \\  \hline
    isotropic $s$ wave & 1 \\~\\
    $\frac{\epsilon}{1+\epsilon-\cos(4 \varphi)}$ & Abrikosov's anisotropic $s$-wave \\~\\
    $\frac{1}{\sqrt{1-\epsilon\cos^{2}(\theta)}}$ & Spheroidal anisotropic $s$-wave\\~\\
    $\frac{1+\epsilon|\cos(6\varphi)|}{1-\epsilon}$  & Anisotropic 6-fold $s$-wave\\~\\
    $\frac{1-\sin^{4}(\theta)\cos(4\varphi)}{2}$ & $s+g$ pairing \\~\\
    $\cos(2\varphi)$ & $d_{x^2-y^2}$ \\~\\
    $\sin(2\varphi)$ & $d_{xy}$ \\~\\
    $\frac{cos(2\varphi)}{\sqrt{1-\epsilon\cos^2(\theta)}}$ & Anisotropic $d$-wave \\~\\
    $\sin \theta$ & $p$ wave (point) \\~\\
    $\sin \theta \cos \phi$ & $p$ wave (line) \\~\\
    \hline \hline
\end{tabular}
\label{tb21}
\end{table}

\begin{figure*}
\centering
 \includegraphics[width=0.9\linewidth]{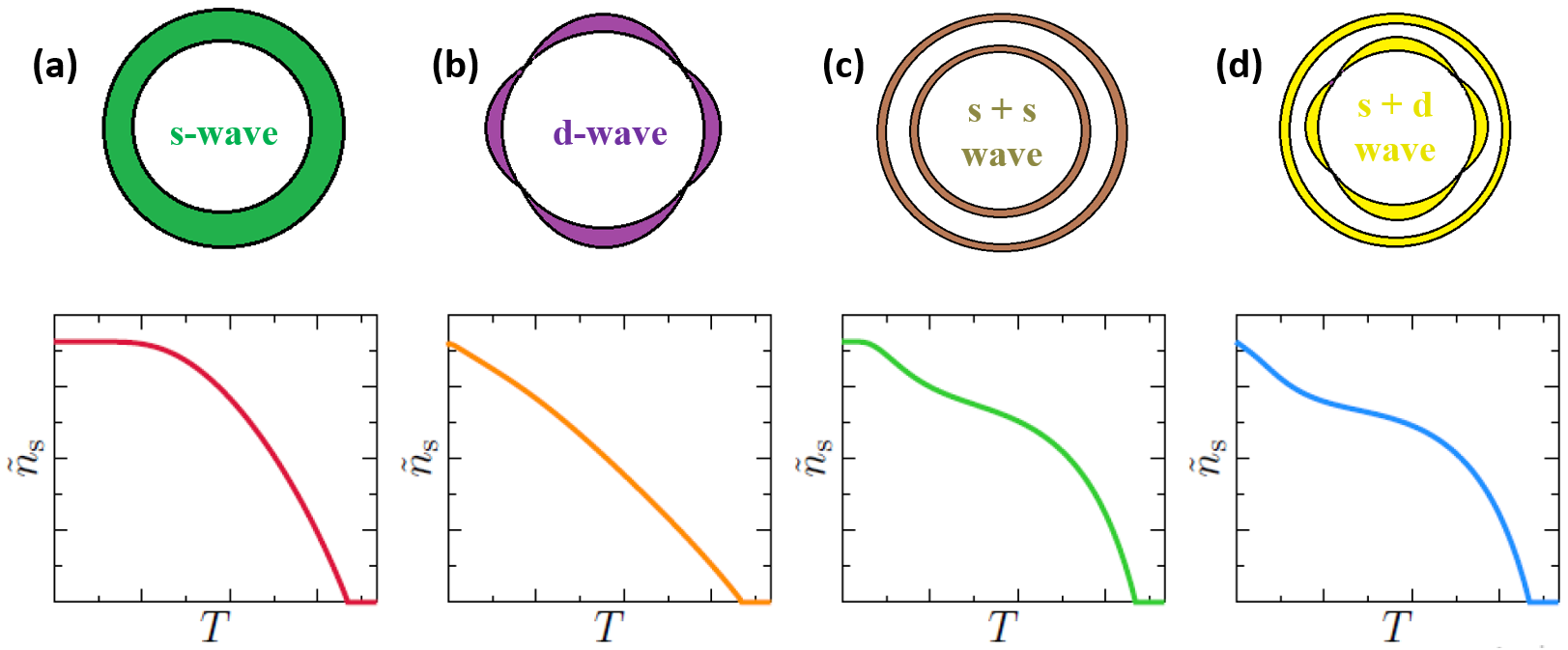}
\caption {Top: A schematic of the superconducting gap symmetry and bottom: the T dependence of the superfluid density of (a) $s$-wave, (b) $d$-wave, (c) $s+s$-wave, and (d) $s + d$-wave superconductors.}
\label{fig28}
\end{figure*}

The temperature evolution of $\sim^n_{s}(t)$ for several gap topologies is given in Fig. ~\ref{fig28}. By examining the low temperature behaviour of $\sim^n_{s}(t)$, it becomes relatively straightforward to tell whether a gap is nodal or not. For a fully gapped superconductor, the low temperature region has a constant value of superfluid density. This is because Cooper pairs need to reach a certain temperature $T \sim \Delta/k_{B}$ in order to have sufficient thermal energy to cross the gap and break up. Nodal superconductors, on the other hand, exhibit a linear dependence of $\sim^n_{s}(t)$ at low temperatures. The absence of a gap in nodal systems means low-energy excitations are possible as there is always sufficient energy for some Cooper pairs to break up.

It is worth noting that $\mu$SR experiments are not sensitive to the sign of the gap function. The sign of a gap can be deduced by examining the evolution of $\Delta(\phi)$ as one moves across the Fermi surface. A $d$-wave gap simply refers to a gap with line nodes. BCS theory holds well for sufficiently weak electron interactions. By inspecting the ratio $2\Delta(0)/k_{B}T_{C}$ (which = 3.53 for weakly-coupled BCS superconductors), it is possible to determine whether the superconductor is a strong or weakly coupled superconductor. For strongly-coupled superconductors, $2\Delta(0)/k_{B}T_{C} > 3.53$~\cite{Tinkham}.

\section{Novel superconductors}
The classical superconductors consist of elements, intermetallic compounds, alloys, and ionic compounds. They are all $s$-waves with features easily explained by typical isotropic BCS theory. But the properties of newly discovered cuprate and other high temperature superconductors can not be explained by conventional BCS theory, often referred to as ``Unconventional superconductor". Deviation from conventional BCS theory first reported in the heavy fermion compound CeCu$_{2}$Si$_{2}$ by German physicist F. Steglich et al~\cite{steglich1979}. In this compound, magnetism is lurking in the superconducting state. In the following section, we will discuss the heavy fermion superconductor and noncentrosymmetric superconductor. 

\subsection{Heavy fermion superconductors}
Heavy-fermion systems (HFS) are perhaps the most remarkable manifestation of strongly correlated electron systems. These are the intermetallic compounds in which the effective mass of the electron is high ($m^{*} \sim 100 m_{e}$) and specific heat $\gamma >$ 400 mJ/mol K$^2$~\cite{Stewart}. Superconductivity in all samples could be understood within the BCS paradigm for approximately 60 years, from Onnes' original discovery in 1911 until the late 1970s. The discovery of superconductivity in heavy Fermion compound CeCu$_{2}$Si$_{2}$ in 1979 by Steglich et al.~\cite{steglich1979} was a huge surprise at this time. Superconductivity and magnetism were thought to be mutually exclusive at the time. A little amount of magnetic impurities is enough to suppress superconductivity in any superconductor known before 1979. Surprisingly magnetism is lurking in the superconducting state of this material. In Ce based compounds there is an interplay between two competing mechanisms. Due to strong RKKY interaction in CeCu$_{6}$ and CeRu$_{2}$Si$_{2}$, it shows a collapse of long range magnetism. When pressure is applied to CeIn$_{3}$, CeCu$_{2}$Ge$_{2}$, CeRh$_{2}$Si$_{2}$, and CePd$_{2}$Si$_{2}$, the Neel temperature $T_{N}$ decreases, and a quantum critical point is reached which makes the way for heavy fermion superconductivity to emerge. At the quantum critical point, a small change in pressure destroys the magnetic ordering of atoms.

In UPt$_{3}$, UBe$_{13}$, URu$_{2}$Si$_{2}$, UPd$_{2}$Al$_{3}$, UNi$_{2}$Al$_{3}$, and U$_{2}$PtC$_{2}$, a transition from to an antiferromagnetic state to a superconducting state takes place. The number of known heavy fermion superconductors has grown from five in 1984 to eleven in 1998 and around forty today. For these materials, a variety of unconventional pairing mechanisms have been proposed, including antiferromagnetic and ferromagnetic spin fluctuations, ferromagnetic magnons, and even electric quadrupole fluctuations. The angular momentum of the Cooper pair is one i.e. $L$ = 1 or $p$-wave state.
 
\begin{figure*}
\centering
 \includegraphics[width=0.9\linewidth]{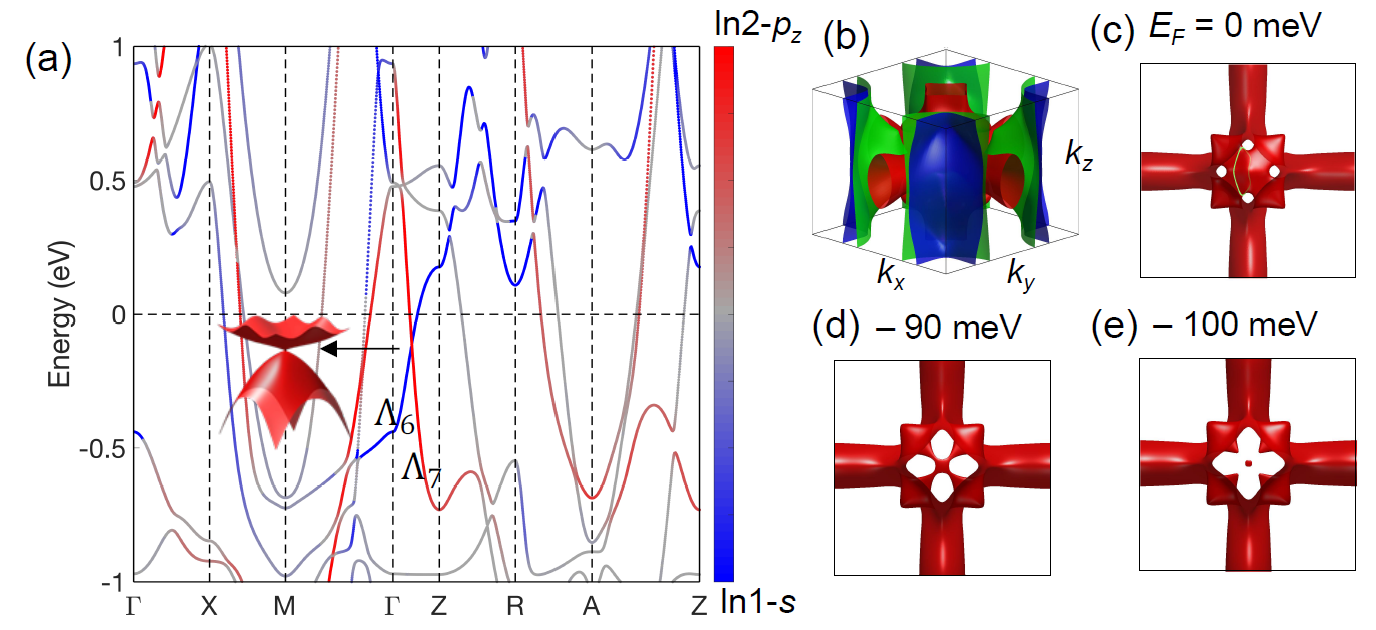}
\caption {(a)Electronic structure of CeCoIn$_{5}$. (a) The band structure. Red and blue colors represent the contribution of In2-p$_{z}$ and In1-s orbitals, respectively.  (b) The whole Fermi surface in the Brillouin zone. Red, blue and green indicate different bands. (c)-(e) Fermi surfaces at different energies viewed from the top of the ($k_{x}$; $k_{y}$)-plane. For clarity, only the red part of the Fermi surface is shown. The Dirac pocket is located at the center of the zone. The green circle in (c) represents a quantum orbit caused by the Dirac pocket. Adapted from Ref.~\cite{Shirer}}
\label{fig211}
\end{figure*}

Several features of superconducting materials are influenced by their large effective mass. At the Fermi level, the electron density of states can be estimated using
\begin{equation}
D(E_{F}) = \frac{1}{2\pi^2}(\frac{2m^{*}}{\hbar^{2}})^{3/2}E^{1/2}_{F}
\label{Eq28}
\end{equation}

As indicated by the data in table \ref{tb22}, the density of states in heavy-electron compounds relate to  $m^{*} \approx 200 m_{e}$ values. Experimental observations in the normal condition provide a lot of evidence for the large effective mass. The contribution of conduction electrons to specific heat, for example, is

\begin{equation}
\gamma = \frac{1}{3}\pi^{2}D(E_{F})k_{B}^{2}
\end{equation}
is proportional to the density of states in Eq. \ref{Eq28}. As a result, for heavy-fermion compounds, $\gamma$ is unusually large. Table \ref{tb22} shows that heavy electron superconductors have electronic specific heat coefficients $\gamma$ that are on average 10 times higher than typical superconducting materials. As can be seen from the same table, the specific heat capacity jump at the transition temperature is considerably high for heavy-electron compounds, therefore the ratio ($C_s$-$\gamma T_C$)/$\gamma T_C$ is near to the normal BCS value of 1.43.

\begin{table}[h]
 \caption{Properties of several heavy fermion superconductors}
 \centering
  \begin{tabular}{lcccc}
\hline \hline

      Compounds & $T_{C}$ & $\gamma_{n}$ (mJ/mole K$^{2}$) & ($C_{s}$-$C_{n}$)/$\gamma T_{C}$ & $m^{*}$/m \\ \hline
      CeCu$_{2}$Si$_{2}$ & $\approx$ 0.8 &340 & 3.5 &220 \\~\\
      UBe$_{13}$ & 0.85 &1100 & & 260  \\~\\
      UPt$_{3}$ & 0.43 & 460 & $\approx$ 0.9 & 187 \\~\\
      UNi$_{2}$Al$_{3}$ & 1.0 &120 & &48  \\~\\
      UPd$_{2}$Al$_{3}$ & $\approx$2.0 &145-210 & & 666\\~\\
      URu$_{2}$Si$_{2}$ & 1.3 & 65 & 0.42 &140 \\~\\
      \hline \hline

\end{tabular}
\label{tb22}
\end{table}

Anisotropic features of heavy-electron superconductors can be seen in measurements including critical fields, ultrasonic attenuation, electrical resistivity, NMR relaxation, and thermal conductivity. At 0 K, the lower critical fields $\mu_{0}H_{C1}$ are a few mT, but the upper critical fields $\mu_{0}H_{C2}$ approach 1 or 2 Tesla. The critical field derivatives $dH_{C2}/dT$ are extremely large in absolute terms, with values as low as -10 T/K for CeCu$_{2}$Si$_{2}$ and -44 T/K for UBe$_{13}$~\cite{Stewart}.

A large value (several thousand angstroms) of London penetration depth ($\lambda_L$) is compatible with the huge effective mass $m^{*}$, which enters the classical expression as a square root factor.
\begin{equation}
\lambda_{L} = \left(\frac{m^{*}}{\mu_{0}n_{s}e^2}\right)^{1/2}
\end{equation}

\subsection{Noncentrosymmetric superconductors}
\begin{figure*}
\centering
 \includegraphics[width=0.9\linewidth]{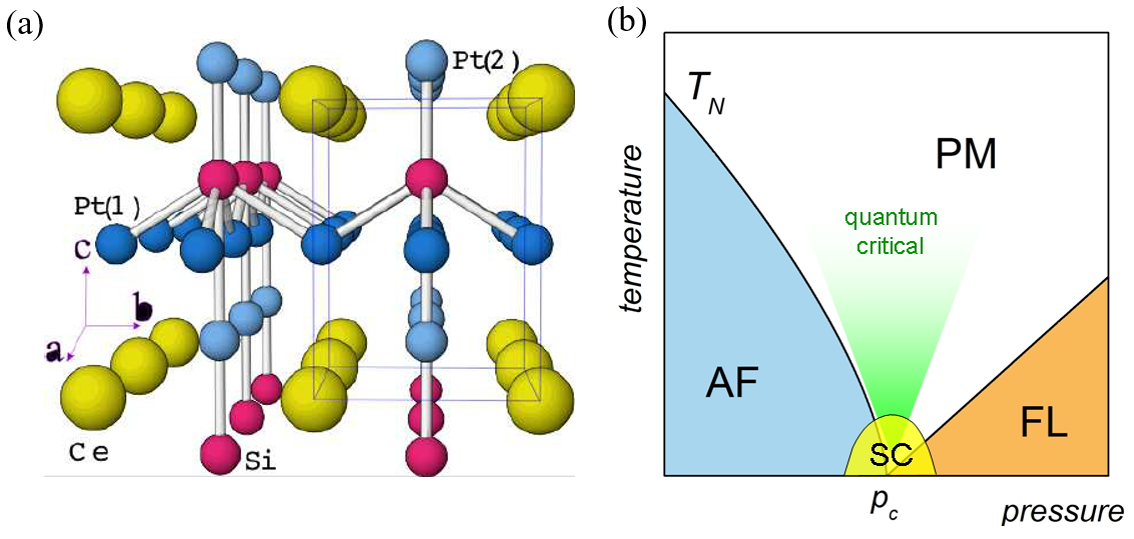}
\caption {(a) Crystal Structure of CePt$_{3}$Si. Adapted from Ref.~\cite{BauerCePt3Si}(b) Temperature-pressure phase diagram of heavy fermion compounds.}
\label{fig212}
\end{figure*}

Since the discovery of superconductivity in the exotic heavy-fermion compound CePt$_3$Si [$T_{C}$ = 0.75 K], noncentrosymmetric compounds which lacks a centre of inversion have been a hot topic of research [see Fig. \ref{fig212}(a)]~\cite{BauerCePt3Si,Metoki}. This is the only known heavy fermion non centrosymmetric compound that exhibits superconductivity at ambient pressure, as opposed to CeIrSi$_{3}$, CeRhSi$_{3}$, CeCoGe$_{3}$, CeIrGe$_{3}$ and UIr$_{3}$, which all display superconductivity at applied pressure. Large Sommerfeld coefficient $\gamma \approx$ 0.39 J/K$^{2}$ mol of this compound confirms heavy fermion character. Different experimental techniques such as zero field muon-spin relaxation and neutron scattering confirm the coexistence of antiferromagnetism and superconductivity. The relaxation rate $1/T_{1}$ of nuclear magnetic resonance (NMR), low temperature thermal conductivity, and previous penetration depth studies all show power law behaviour, indicating nodes in the superconducting gap. Large upper critical field and no change in the Knight shift has been seen over the $T_C$ transition temperature. A triplet superconducting order parameter is thought to be responsible for these properties. However, towards $T_{C}$, a peak in the NMR spectra was noticed, which was interpreted as $s$-wave singlet-like characteristics. Because of these findings, it has been suggested that the order parameter is a combination of singlet and triplet pairing. It's also been proposed that the nodal structures in the gap are the consequence of the superposition of singlet and triplet order parameters~\cite{smidman2017}. At a Neel temperature ($T_{N}$) of 2.25 K, which is higher than the superconducting transition temperature, CePt$_3$Si also orders antiferromagnetically, however, the impact of this ordering on superconducting characteristics is unclear. Antiferromagnetic fluctuations, which favour sign changes in order parameters in momentum space, might potentially be the cause of nodal behaviour~\cite{smidman2017}.

Another example of noncentrosymmetric superconductor is Li$_{2}$Pt$_{3}$B which shows superconductivity at $T_{C}$ = 2.7 K~\cite{Takeya}. NMR measurements and heat capacity measurement suggest the presence of a line node in the superconducting gap structure. When Pt is replaced with Pd, the transition temperature rises steadily with increasing Pd content, eventually reaching 7 K for Li$_{2}$Pd$_{3}$B, which is reported to be a fully gapped $s-$ wave superconductor. These findings suggest the mixture of singlet and triplet superconducting order parameter Li$_{2}$Pt$_{1-x}$Pd$_{x}$B for any value of $x$, but with a pure s-wave singlet in the $x$ = 1 limit.   

CeTX$_3$ [T = Transition element, X = group IV element], a novel class of noncentrosymmetric superconductor, is also a heavy fermion compound~\cite{Muro}. One striking feature of these compounds is their large value of the upper critical field, which is also typically concave upward as a function of temperature.  The antiferromagnetic ordering temperature of CeIrSi$_3$ $T_{N}$ is 5 K, and greater pressure is required for superconductivity. $T_N$ is suppressed when pressure is raised, with superconductivity forming at 18 kbar and reaching its maximal transition temperature near 25 kbar. Heat capacity measurement under applied pressure shows a sharp superconducting jump at $T_{C}$ with $\Delta C/\gamma T_{C}$ = 5.7, much higher than the BCS weak coupling value of 1.43, the highest reported heat capacity jump at $T_{C}$, indicates strong coupling superconductivity. CeCoGe$_3$, while structurally similar to CeIrSi$_3$ and CeRhSi$_3$, exhibits a distinct set of characteristics. The antiferromagnetic ordering temperature $T_{N1}$ = 21 K of CeCoGe$_{3}$ is greater than that of the preceding compounds. It possesses two more antiferromagnetic transitions than the other compounds, according to reports. The antiferromagnetic phase is suppressed with applied pressure before superconductivity is observed between 54 and 75 kbar, with a maximum $T_{C}$ = 0.69 K. Due to a modest increase in the Sommerfeld coefficient $\gamma_{n}$ = 49 mJ/mol $K^2$, the Kondo interaction is also shown to be decreased.     

In these noncentrosymmetric compounds, the lack of inversion symmetry causes an asymmetric electrical field gradient in the crystal lattice, resulting in a Rashba-type spin orbit coupling (RSOC, energy scale $\sim$ 10-100 meV). The spin of the Cooper pair is not a good quantum number in the presence of RSOC; except for time-reversal invariant momenta, the conduction band's spin degeneracy is lifted, allowing for the admixture of spin-singlet and spin-triplet states to create Cooper pairs. The superconducting gap displays line or point nodes in the order parameter if the triplet part is strong. Below $T_{C}$ = 2.7 K, the noncentrosymmetric compound LaNiC$_{2}$ exhibits USC with broken time-reversal symmetry (TRS) as pointed by $\mu$SR measurements, a nodal energy gap as observed by very low temperature magnetic penetration depth measurements, and probable multigap superconductivity due to the tiny value of the RSOC.

\subsection{Electronic band structure of cuprate and iron based superconductor}

\begin{figure*}
\centering
 \includegraphics[ width=0.9\linewidth]{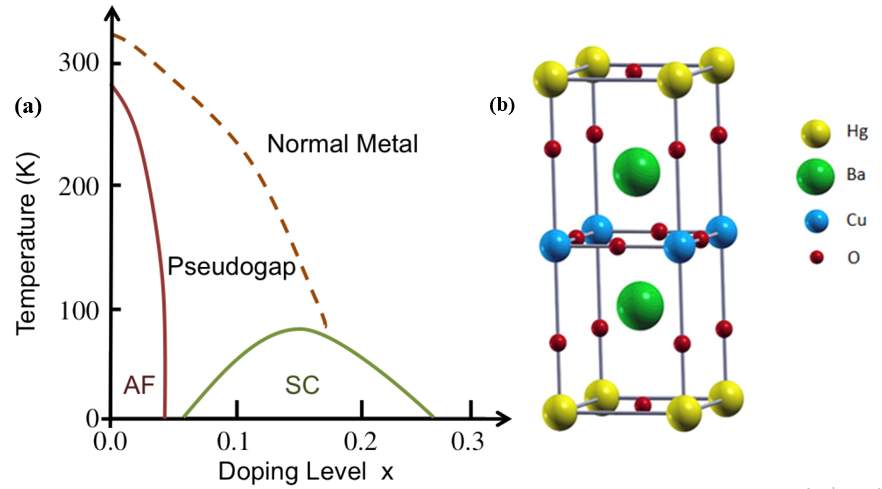}
\caption {(a) Phase diagram of cuprate superconductors. (b)Unit cell of Hg-Ba-Cu-O superconductor. Adapated from~\cite{Yu} }
\label{fig210}
\end{figure*}

The parent stoichiometric compounds of the cuprate superconductors are antiferromagnetic Mott insulators, which then become superconducting upon doping. Fig.~\ref{fig210}(a) shows a schematic phase diagram for a typical cuprate superconductor. One or more CuO$_{2}$ layers forms the crystal structure [see Fig.~\ref{fig210}(b)]. The tetragonal cuprate superconductors show a small difference in lattice parameters $a$ and $b$ close to 0.38 nm because of common structural features of the CuO$_{2}$ planes. Many theories were initially proposed to explain how high temperature superconductivity arises in cuprates but to date, none of them provides a complete generalised microscopic description that can explain all experimental evidence and that most scientists agree on. Two main theories however have the most support. The first is the resonating valence bond theory, proposed by P. Anderson~\cite{anderson1959} and attributes superconductivity in cuprates to their atomic structure. It argues that electrons from neighbouring copper atoms can form pairs via valence bonds which then become mobile and behave as cooper pairs as the material is doped. The other theory known as spin fluctuation instead holds antiferromagnetic correlations as responsible for electron pairing. Unlike traditional BCS-type superconductors which exhibit an isotropic superconducting gap in momentum space (called $s$-wave superconductors), electron pairing in these novel superconductors display $d$-wave symmetry ~\cite{tsuei2000}.

\begin{figure*}
\centering
 \includegraphics[ width=\linewidth]{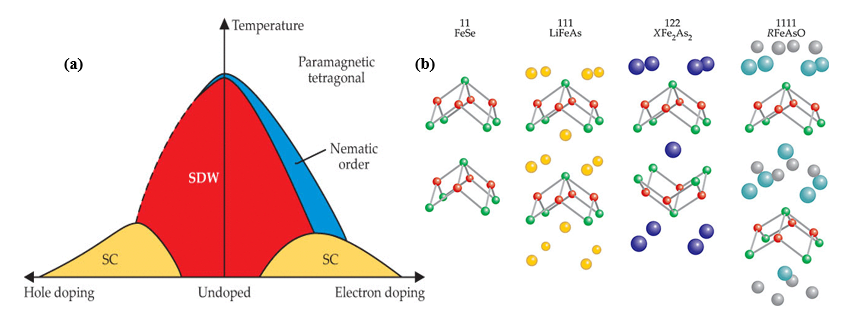}
\caption {(a) The phase diagram of iron based superconductor. (b) The five tetragonal structure of Fe-based suprconductors. Adapted from~\cite{Paglione} }
\label{fig29}
\end{figure*}

The discovery of iron based superconductors opened a new avenue of research in high temperature superconductivity. These compounds shared a common layered structure based on a planar layer of iron atoms joined by tetrahedrally coordinated pnictogen (P, As) or chalcogen (S, Se, Te) anions arranged in a stacked sequence separated by alkali, alkaline-earth or rare-earth and oxygen/fluorine ‘blocking layers'. Here the parent compounds are in spin density wave (SDW) state, upon doping superconductivity arises in these compounds, like cuprates. However there is some characteristics difference between iron based superconductors and cuprate superconductors are found (i) in the arrangement of pnictogen/chalcogen anions above and below the planar iron layer (see Fig.\ref{fig29}(e)) as opposed to the planar copper–oxygen structure of the cuprates; (2) in the ability to substitute or dope directly into the active pairing layer; and (3) in the metallic (rather than insulating) multiband nature of the parent compounds. It is these traits, together with the similar interplay of magnetism and superconductivity, that mark the iron pnictides and cuprates as distinct, but closely related, superconducting families. Considering first the electronic structure of the high-symmetry paramagnetic phase, the typical picture arising both from Density Functional Theory calculations and experiment consists of up to three quasi-2D hole pockets around the centre of the Brillouin zone with two electron-like pockets centred at a corner of the Brillouin zone, as obseved in BaFe$_{2}$As$_{2}$. The band structure reflects the layered crystal structure having poor inter-layer charge transport giving rise to the quasi-2D Fermi surfaces and means that the gross features of the electronic structure are determined by the FeAs plane and do not depend strongly on what is in between the FeAs (or FeSe etc) layers. This general picture was experimentally verified for LaFePO soon after the discovery of Fe-based superconductors and forms the basis for further discussions. One immediately apparent property of this unusual quasi-2D Fermi surface is that there is good ``nesting" of the hole and electron Fermi surfaces that is, that a wavevector of {\bf $Q$} = ($\pi$/a, $\pi$/a) (2-Fe cell) would map hole-like sections of the Fermi surface onto the electron sheets. More formally, the Lindhard dielectric response function $\chi$(q) will be strongly peaked around $\bf {q = Q}$ and this will give rise to spin-density waves with ordering vector $Q$.

\subsection{Superhydride superconductors}
Searching for superconductivity with $T_{c}$ near room temperature is of great interest both for fundamental science \& many potential applications. Superhydrides have received growing attentions recently because of high temperature superconductivity. In 1968, Ashcroft predicted that the metallic hydrogen could be come a high-temperature superconductor. Up till now, the closest realization of Ashcroft’s prediction is the hydrogen-rich compounds. In 2014, a H$_{3}$S compound was predicted to be a conventional superconductor at 200 GPa with $T_{c} \approx$ 200 K ~\cite{39}. In 2015, the H$_{3}$S compound was observed in a high-pressure experiment~\cite{39}. It was confirmed that it is a conventional superconductor with $T_{C}$ = 203 K at 155 GPa~\cite{39}. These researches have opened a new direction of searching for new hydrogen-rich compounds which exhibit a high value of $T_{c}$ under high pressure.

A large number of metal superhydrides share commonhost-guest clathrate structures. In these structures, hydrogen atoms form a hydrogen cage which hosts a metal atom inside. The first example is ThH$_{10}$ with an Fm-3m structure. It was predicted that $T_{c}$ = 176-241 K at 100 GPa and $T_{c}$ = 166-228 K at 200 GPa. The Fm-3m ThH$_{10}$ was confirmed by an experimental observation, and it was reported that $T_{c}$ = 159-161 K at 170-175 GPa ~\cite{45} in excellent agreement with the previous prediction. The most famous examples are the prediction of superconducting LaH$_{10}$ and YH$_{10}$, as the predicted $T_{c}$ is approaching the room temperature. In LaH$_{10}$, it was predicted that $T_{c}$ = 274-286 K at 210 GPa and it was reported by experiments that LaH$_{10}$ is indeed a conventional superconductor with $T_{C}$ of 260K at 180-200 GPa, and the latest report has shown the evidence that its T$_{c}$ could be as high as 280 K. In YH$_{10}$, it was predicted that $T_{C}$ could be as high as 305-326 K at 250GPa. This theoretical prediction has been very fascinating as, at around 300K, it is the so -called room temperature. The YH$_{x}$ compounds have been synthesized under high pressure. The existence of superconducting YH$_{6}$ and YH$_{9}$ compounds have been confirmed with $T_{c}$=227 K at 237 GPa~\cite{51} and $T_{c}$ = 243 K at 201 GPa~\cite{51}, respectively. However, YH$_{10}$ and its superconductivity are yet to be confirmed. A recent article reported the discovery of superconductivity with maximum critical temperature($T_{c}$) above 210 K in calcium superhydrides, the new alkali earth hydrides experimentally showing superconductivity above 200 K in addition to sulfur hydride and rare-earth hydride system~\cite{Ca}. I conclude this section with a Table \ref{s} which summarizes $T_{c}$ of several superhydride superconductors.

\begin{table}
\caption{$T_{c}$ of some selected superhydride superconductors}
\label{s}
\centering
\begin{tabular}{lcccc}
\hline\hline
Superconductors & P (GPa) & $T_{C}$ & Ref.\\ \hline
ThH$_{10}$   & 175 & 161 & ~\cite{45} \\~\\
H$_{3}$S   & 155 & 203 &~\cite{39} \\~\\
YH$_{6}$   & 166 & 224 &~\cite{50}  \\~\\
YH$_{6}$   & 237 & 227 & ~\cite{51} \\~\\
YH$_{9}$   & 201 & 243 & ~\cite{51}\\~\\
LaH$_{10}$   & 180-200 & 260,280 & ~\cite{48}\\~\\
CaH$_{6}$   & 160-190 & 200 & ~\cite{Ca}\\~\\
\hline\hline
\end{tabular}
\end{table}

\section{Experimental measurable quantities of superconductor}
\subsection{Resistivity}
From Drude -Lorent's theory we can say that resistivity is proportional to the scattering rate ($\tau^{-1}$). The electrical conductivity depends on temperature mainly via the different scattering processes which enter into the mean lifetime $\tau$. In a typical metal there will be three main scattering processes, scattering by impurities ($\tau_{imp}$), by electron-electron interactions ($\tau_{el-el}$) and by electron-phonon ($\tau_{el-ph}$) collisions.
\begin{equation}
    \frac{1}{\tau} =  \frac{1}{\tau_{imp}} + \frac{1}{\tau_{el-el}} + \frac{1}{\tau_{el-ph}}
\end{equation}
where $\frac{1}{\tau_{imp}}$ is independent of temperature, $\frac{1}{\tau_{el-el}}$ is propotional to $T^{2}$, and $\frac{1}{\tau_{el-ph}}$ is propotional to $T^{5}$. The resistivity of the metal can be written in the form:
\begin{equation}
    \rho = \rho_{0} + aT^{2}+bT^{5}+..............
\end{equation}

According to Matthiessen's rule, in room temperature region there is two contribution in the total resistivity ($\rho_{T}$) of the metal (i) temperaturature independent part ($\rho_{0})$ and (ii) temperature dependent part ($\rho_{iT}$). So the total resistivity can be written as~\cite{Terence} 
\begin{equation}
    \rho_{T} = \rho_{0}+\rho_{iT}
\end{equation}
In low temperature the resistivity of a typical metal is linearly proportional to  $T^{5}$ and become linear at high temperate. Considering band structures, Fermi surfaces, and phonon
spectra of the metals Bloch-Gr{\"u}neisen suggested the following relation~\cite{Terence}: 
\begin{equation}
\rho_{iT} = \left(\frac{C}{M\theta_{D}}\right) \left(\frac{T}{\theta_{D}}\right)^{5}\int_{0}^{\Theta_{D}/T} \frac{z^5}{(e^{z}-1)(1-e^{-z})}dz
\end{equation}
where M is the atomic weight and C is a constant. With the given limits, the integral itself reduces to $\frac{1}{4}(\theta_{D}/T)^4$ at high temperature and to a constant at low temperature.

\begin{figure*}
\centering
 \includegraphics[width=0.9\linewidth]{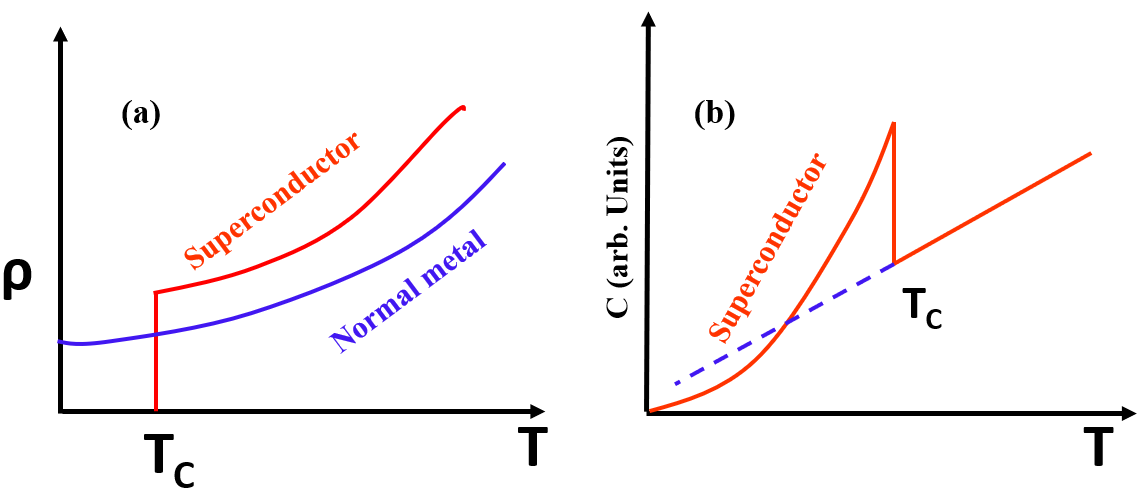}
\caption {(a) Resistivity as a function of temperature for normal metal and superconductor. (b) Temperature dependence of heat capacity of superconductor and a normal metal.}
\label{fig213}
\end{figure*}

\subsection{Heat Capacity}
Specific heat of superconductors is one of the most researched characteristics. Because the whole sample reacts, it offers a 'bulk' measurement that considers the entire sample.

Above the critical transition temperature $T_{C}$ the specific heat of high-temperature superconductors tends to follow the Debye theory. We know that specific heat ($C_{n}$) of a normal metal is the sum of the linear term $C_{e} = \gamma T$ emerging due to the conduction electrons, a lattice vibrations or phonon terms $C_{ph} = \beta T^{3}$
\begin{equation}
C_{n}= \gamma T + \beta T^{3}
\end{equation}

As per the free electron theory, the electronic contribution to the specific heat per mole of conduction electrons are as follows:
\begin{equation}
C_{e} = \gamma T = \frac{1}{2}\pi^{2}R\left(\frac{T}{T_{F}}\right)
\end{equation}
In the Debye theory, phonon contribution to the specific heat per gram atom is;
\begin{equation}
C_{ph} = \beta T^3 = \left(\frac{12\pi^4}{5}\right)R\left(\frac{T}{\Theta_{D}}\right)^{3}
\end{equation}
Using this equation the Debye temperature can be determined. 

In the absence of an applied magnetic field, the transition from the normal to the superconducting state is referred to as a second-order phase transition. The BCS theory predicts that at $T_{C}$, electronic specific heat leaps from the normal-state value $\gamma T_{C}$ to superconducting-state value $C_{s}$ in a ratio of
\begin{equation}
\frac{C_{s}-\gamma T_{c}}{\gamma T_{c}} = 1.43
\end{equation}

The electronic contributions to the specific heat $C_{e}$ will depend exponentially with temperature, according to the BCS theory for $T<$T$_{C}$

\begin{equation}
C_{e} \approx a~e^{ -\frac{\Delta}{k_{B}T}}
\end{equation}

Here 2$\Delta$ is the energy gap in the superconducting state. For BCS superconductor the gap to $T_{C}$ ratio ($2\Delta/k_{B}T_{C}$) is  3.53~\cite{Tinkham}.
\begin{figure*}
\centering
 \includegraphics[width=0.9\linewidth]{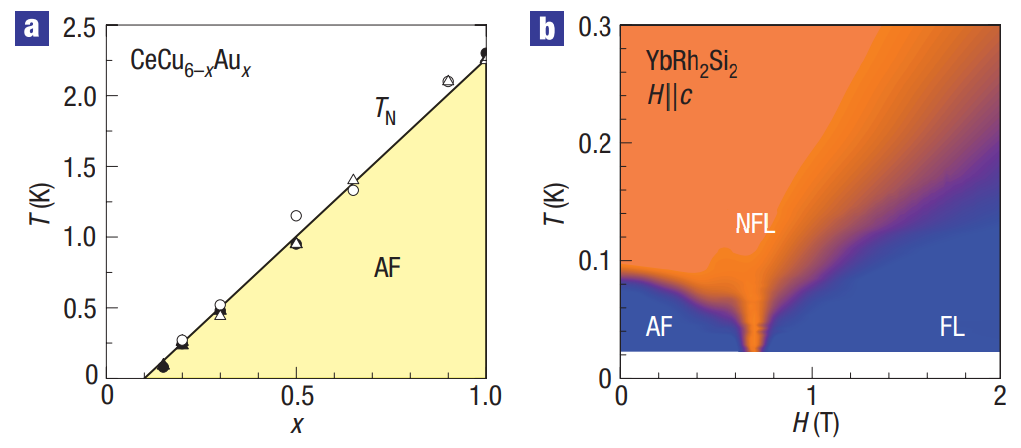}
\caption {(a) QCP in heavy fermion materials. Adapted from  \cite{GEGENWART}. }
\label{fig214}
\end{figure*}

\section{Quantum fluctuation}
One of the most cherished goals in physics is to achieve room-temperature superconductivity. Several attempts have been to achieve it. The first success comes with the discovery of a cuprate superconductor by German physicist Bednorz and M\"{u}ller~\cite{bednorz1986}. It is reported that the atomic quantum fluctuations are the pairing glue in these superconductors rather than conventional electron-phonon interactions.

In the classical universe, any oscillation freezes at absolute zero temperature. To decrease the potential energy of the system by lowering the temperature, all macroscopic system forms an ordered pattern at the microscopic level. However, quantum mechanics allows fluctuation even at absolute zero temperature, as predicted by Heisenberg's uncertainty principle. Such fluctuations are known as quantum fluctuation, which can be tuned through the variation of a non-thermal control parameter, such as applied pressure. When the strength of the quantum fluctuation is sufficiently strong the system experiences a quantum phase transition to a new ground state. However, attaining quantum transition in a real system is not as straightforward as it appears. At zero temperature, a continuous phase transition occurs around the quantum critical point (QCP). Recently, QCP is experimentally observed in heavy fermion compounds and rare earth based intermetallic compounds~\cite{StewartNFL,Lohneysen}. These systems are potential candidates to study the QCP as their ground state can easily be tuned through the application of pressure, chemical doping or magnetic field.

Two prominent examples of heavy fermion compounds to study the QCP are CeCu$_{6-x}$Au$_{x}$ and YbRh$_{2}$Si$_{2}$. In CeCu$_{6-x}$Au$_{x}$~\cite{LohneysenPRL,Schroder}, the paramagnetic state of CeCu$_{6}$ tuned into antiferromagnetic state. As the doping concentration increases, $T_{N}$ also increases. A Similar thing is also observed in low temperature staggered moment-spatially modulated magnetization in antiferromagnetic vector, confirming the presence of antiferromagnetic QCP. In YbRh$_{2}$Si$_{2}$ the magnetic transition occurs at 70 mK~\cite{Trovarelli}, which is continuously suppressed by the application of a small magnetic field [see Fig.~\ref{fig214}]. 

The explicit identification of QCP is crucial to understand several properties that are important to understand the physics of a strongly correlated electron system. At QCP, Fermi liquid breaks may down, leading to non Fermi liquid (NFL) state. There exist only a few undoped or stoichiometric materials which exhibit NFL states at ambient pressure, such as UBe$_{13}$, CeNi$_{2}$Ge$_{2}$, CeCu$_{2}$Si$_{2}$, and CeRhBi. Moreover, quantum critically may lead to novel quantum phases such as unconventional superconductivity. 

%% file: chapter3.tex
\chapter{Experimental techniques} \label{chapter:3}
This chapter introduces the experimental techniques that will be employed throughout the rest of the thesis. This is accomplished by explaining the ideas that underpin each sort of experiment, a description of the experimental apparatus and methods. The techniques used here are divided into the microscopic technique of muon spin rotation and relaxation ($\mu$SR), which investigates the environment local to an implanted muon, and the bulk techniques of resistivity, magnetic susceptibility and heat capacity measurements, which determine the signal by the response of the entire sample to an external stimulus. First I've presented the resistivity, magnetic susceptibility, and heat capacity measurement techniques. The $\mu$SR technique is described after that.

\section{Sample growth techniques}
\subsection{Polycrystalline sample}
It is best, to begin with, a polycrystalline sample when developing a novel material. We can take some bulk measurements with this sample as well. We have synthesized the high-quality polycrystalline sample using two methods:
\begin{itemize}
\item{\bf{Arc Melting Method:}}\\
Using a single arc furnace, we have prepared a high-quality polycrystalline sample of HfIrsi, ZrIrSi, ThCoC$_2$, and CeIr$_3$. A high DC voltage is applied to the electrode, and the current was stabilized by contacting the tip of the electrodes to the base of the copper hearth. The current arc is kept in a stable state by drawing the electrodes away from the sample and placing them above the sample. The high purity elements are placed in a stoichiometric ratio on a water cooled copper hearth before being sealed and evacuated using a rotary pump. Before the electrodes were powered on, the chamber was flushed with Argon gas and re-evacuated at least three times to decrease the influence of oxygen contamination. The melting takes place in an Ar-rich environment. To increase the sample homogeneity, the sample is flipped and remelted several times. We then weigh the prepared sample to check if any weight has been lost during the melting process. This initial melt was usually followed by a lengthier period of high-temperature annealing. The samples were sealed in quartz tubes that had been evacuated or had been exposed to a partial atmosphere of Argon before being transferred to a box furnace. 
\begin{figure*}
\centering
 \includegraphics[width=0.6\linewidth]{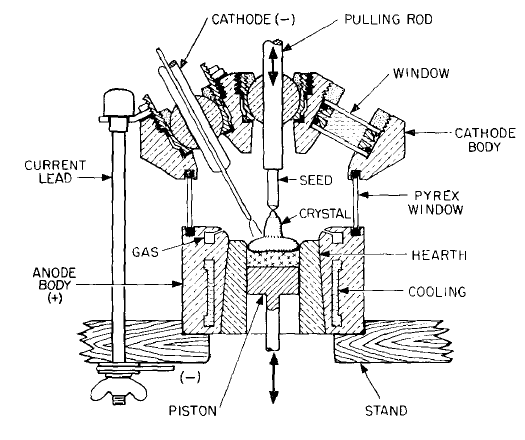}
\caption {Schematic diagram of single arc furnace.}
\label{fig31}
\end{figure*}

\item{\bf{Solid State Reaction Method:}}\\
In the solid state synthesis process, a mortar pestle is used to mix all of the powders in the stoichiometric ratio. The powders are thoroughly grounded. To improve the contact surface area of the reactants, the homogeneous mixture is palletized using a hydraulic pellet press. The pellets are placed on an alumina boat or crucible and transferred into a box furnace at a controlled temperature. To reduce any reactions with air, especially oxygen, the pellet was sometimes encapsulated in a quartz tube under a high vacuum. To ensure that the sample is uniform, the grinding and heating operation is usually performed numerous times.
\end{itemize}
\begin{figure*}
\centering
 \includegraphics[width=0.9\linewidth]{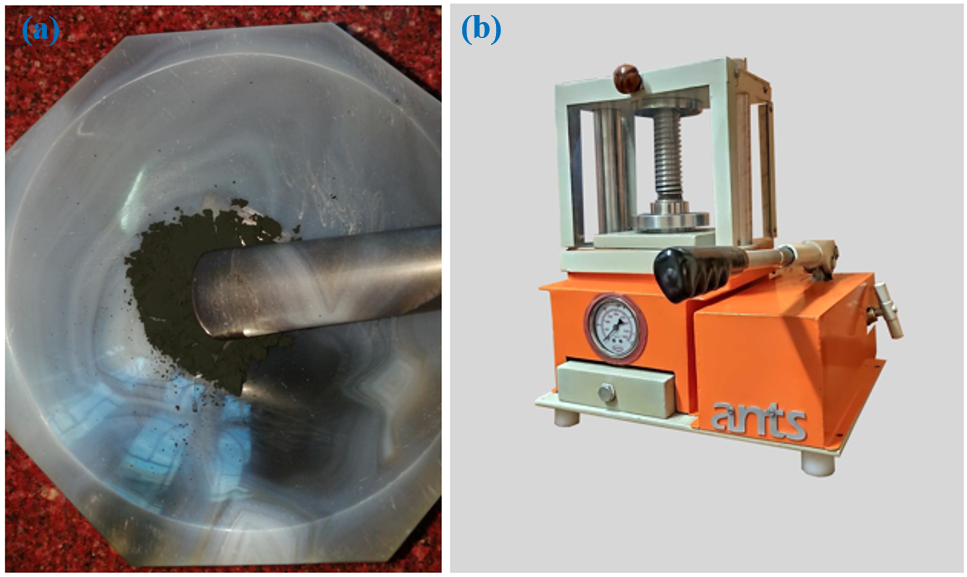}
\caption {The in house solid state synthesis method at RKMVERI. (a) Powders are grounded thoroughly in Agate mortal pestle. (b) Pallet press die to prepare the pallet of the sample. (c) Hydraulic press in our lab.}
\label{fig30}
\end{figure*}

\subsection{Single crystal }
Because of their continuous, homogeneous, and highly-ordered structure, single crystals are one of the most important categories of materials, allowing them to have unique features. Single crystal materials are preferable to polycrystalline materials in many ways, and many features found in single crystals are not reproduced in polycrystals. Even with current scientific advancements in sophisticated polycrystalline materials developed for specific uses, electrical, optical, thermal, mechanical, and other characteristics of single crystal are still remain outstanding. For these reasons, it's no surprise that single crystals and the methods for preparing them are a hot issue among scientists. To grow single crystals, we have the used flux grown method:

\begin{itemize}
\item{\bf Flux Growth Method:}
In general, the flux growth approach is quite straightforward. The material is melted after being combined with a suitable solvent in a polycrystalline form. The material crystallises with spontaneous nucleation without any favored nucleation sites when the solution cools slowly. Yet, there are some disadvantages of producing a single crystal using this method. The crystals formed by this process are often quite tiny, and separating the crystals from the solvent is generally challenging. Contamination from the crucible material and the solvent must also be avoided. Crystals can be grown even without any external flux, where part of the materials works as a self-flux.
 
The single crystal of CeCo$_{2}$Ga$_{8}$ described in this thesis was grown in an alumina crucible utilising the Ga-self flux technique. The crucible was enclosed within a quartz tube that was completely evacuated. The crucible was heated to 1100$^{0}$C for 10 hours, then gradually cooled to 630$^{0}$C, after which the Ga flux was spun out in a centrifuge and quenched in cold water~\cite{wang2017heavy}.

\end{itemize}

\section{Characterization techniques}

\subsection{Powder X-ray diffraction}
The wavelength of X-ray is of the same order as the interatomic spacing in crystals. Hence, an X-ray can be efficiently diffracted by the crystal. W.L. Bragg in 1913, employ X-rays to study the internal structure of the crystal. Every lattice plane in a crystal behaves like a diffraction grating, on the exposure of X-rays. For details please see Ref. ~\cite{jenkis1996introduction}. 
Suppose a monochromatic beam of X-ray incidents at a glancing angle $\theta$ which gets scattered by the atoms such as A and B, in the arbitrary direction. Constructive interference patterns take place only among these scattered waves which are reflected spectacularly having $n\lambda$ path difference. Here n is an integer, $\lambda$ is the wavelength of the X-ray. The path difference for the wave reflected from the adjacent plane is AB+BC = 2$d\sin \theta$.
For constructive interference ,
\begin{equation}
2d\sin\theta = n\lambda~~~    \rm{(n= 1,2,3....)}
\end{equation} 
The intensity ($I_{hkl}$) of scattered waves as a function of scattering angle (2$\theta$) is used to generate a diffraction pattern. The arrangement of atoms in the diffracting planes determines the amplitude of light dispersed by a crystal.

\begin{equation}
I_{hkl}\propto \vert F_{hkl} \vert^{2}
\end{equation}

\begin{equation}
F_{hkl}=\sum_{j=1}^{n}N_{j} f_{j} \exp [2\pi i (hx_{j}+ky_{j}+lz_{j})]
\end{equation}

Where $F_{hkl}$ denotes structure factor that measures the amplitude of the light scattered by a crystal, $x_{j}$, $y_{j}$, $z_{j}$ are fractional coordinates, a fraction of each corresponding position occupied by $j^{th}$ atom is denoted by $N_{j}$, $f_{j}$ is the scattering factor while $hkl$ being the miller indices. The scattering factor, f, is comprised of several parameters that describe how an X-ray interacts with electrons in the vicinity of an atom;

\begin{equation}
\vert f \vert^{2} = \left[ f_{0} \exp \left(\frac{-B \sin^{2} \theta}{\lambda^{2}}\right) +  \Delta f^{\prime}   \right]^{2}+(\Delta f^{\prime\prime})^{2}
\end{equation}

here $\lambda$ is the X-ray wavelength, $f_{0}$ at 0$^{0}$ is equivalent to the number of electrons around the atom, B (=~8$\pi^{2}U^{2}$) is the Debye-Waller temperature factor where $U$ is the mean-square amplitude of thermal vibration.

\begin{figure*}
\centering
 \includegraphics[width=0.9\linewidth]{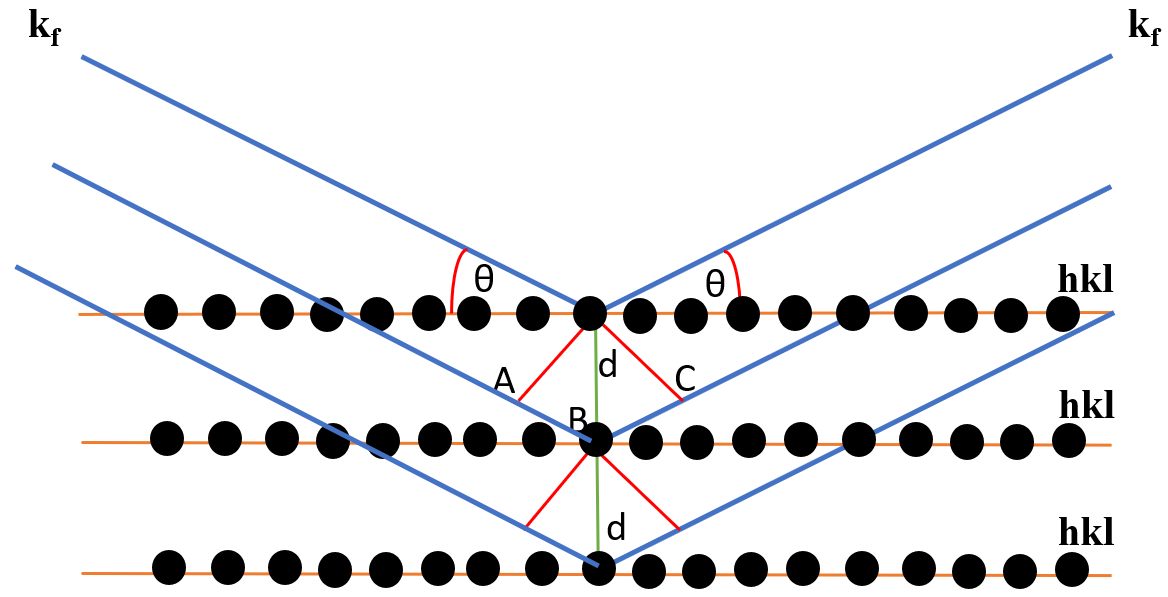}
\caption {Schematic diagram of the diffraction process as described by Bragg's scattering model where planes of atoms are uniformly spaced.}
\label{fig32}
\end{figure*}

From the X-ray diffraction pattern using Rietveld refinement, we can determine the lattice parameters, phase quantities, space group, bond length, bond angle, microstrain in the crystals and the position of the atoms etc. In Rietveld refinement, we minimize the difference between calculated and observed patterns by the least square method. At the starting of the refinement, we check if the peaks are at the right positions which dictate that phase was identified correctly. But in most cases difference exists as one can see observing peak width, slightly tilted peal positions, intensities etc. Traditionally Pseudo Voigt (V) curves are used in X-ray Rietveld analysis, which is an admixture of Lorentzian (L) and Gaussian (G) curves;

\begin{equation}
V(x)=m~ L(x)+(1-m)~ G(x)
\end{equation}

where m is an integer changing whose value refinement is done. Sometimes instead of pseudo voigt (PV) function Pearson VII, Thompson-Cox-Hastings PV, Split PV, Finger-Cox-Jephcoat PV etc are also used. The parameters will be tweaked during the refinement process until the computed and observed patterns are as close as possible to each other. The system will avoid diverging or achieving a false minimum if the initial model is closer to the correct one. The residual values, which are presented below, can be used to assess the goodness of fit ($\chi^{2}$) of the least-square refinement or if a real minimum has been obtained;

\begin{equation}
R_{P}=\dfrac{\sum \left(Y_{i,Obs}-Y_{i,Calc}\right)^{2}}{\sum Y_{i,Obs}}, \,\, \rm{R_{P} \equiv Pattern}
\end{equation}

\begin{equation}
R_{F}=\dfrac{\sum \left|\sqrt{I_{n,Obs}}-\sqrt{I_{n,Calc}}\right|}{\sum I_{n,Obs}}, \,\, \rm{R_{F} \equiv Structure factor}
\end{equation}

\begin{equation}
R_{WP}=\left| \dfrac{\sum w_{i} \left(Y_{i,Obs}-Y_{i,Calc}\right)^{2}}{\sum w_{i} (Y_{i,Obs})^{2}}\right|, \,\, \rm{R_{WP} \\equiv Weighted \,\, Pattern}
\end{equation}

Where, at the end of the refining cycle, $I_{n, Obs}$ and $I_{n, Calc}$ are the observed and simulated intensities for the $n^{th}$ Bragg reflection. Among different software, Fullprof is used for structural refinement of the samples.

\subsection{Energy dispersive X-ray spectroscopy}
Energy dispersive X-ray spectroscopy (EDX), often known as EDS or EDAX, is an X-ray method that is used in combination with scanning electron microscopy (SEM) to determine the elemental composition of the sample. When an electron beam collides with a sample, it may interact and excite an electron from a lower orbit to a higher orbit, causing a hole in the lower orbit to form. Then an electron from a higher orbit falls into a lower orbit, releasing energy as X-rays, which is a property of the energy difference among two energy levels and hence related to the electronic structure of the element. Energy spectra with peaks corresponding to the components that make up the real composition of the material under investigation are created by EDX analysis. See Ref.~\cite{bell2003energy} for further information. This technique was utilised to examine the composition of the various samples in this thesis.

\section{Magnetic and physical properties measurements}

\subsection{Magnetization}
In this thesis, a superconducting quantum interference device (SQUID) magnetometer is used to perform bulk magnetization measurement over the temperature range 1.5 K – 300 K in an external applied magnetic field. A SQUID magnetometer is a highly sensitive magnetic probe, with a sensitivity of up to 5$\times$10$^{-8}$ emu. This instrument can operate both in both a dc mode (for probing a static magnetization) and an ac mode (for examining spin dynamics). However, in this thesis, we have presented only dc measurements.  

\begin{figure*}
\centering
 \includegraphics[width=0.9\linewidth]{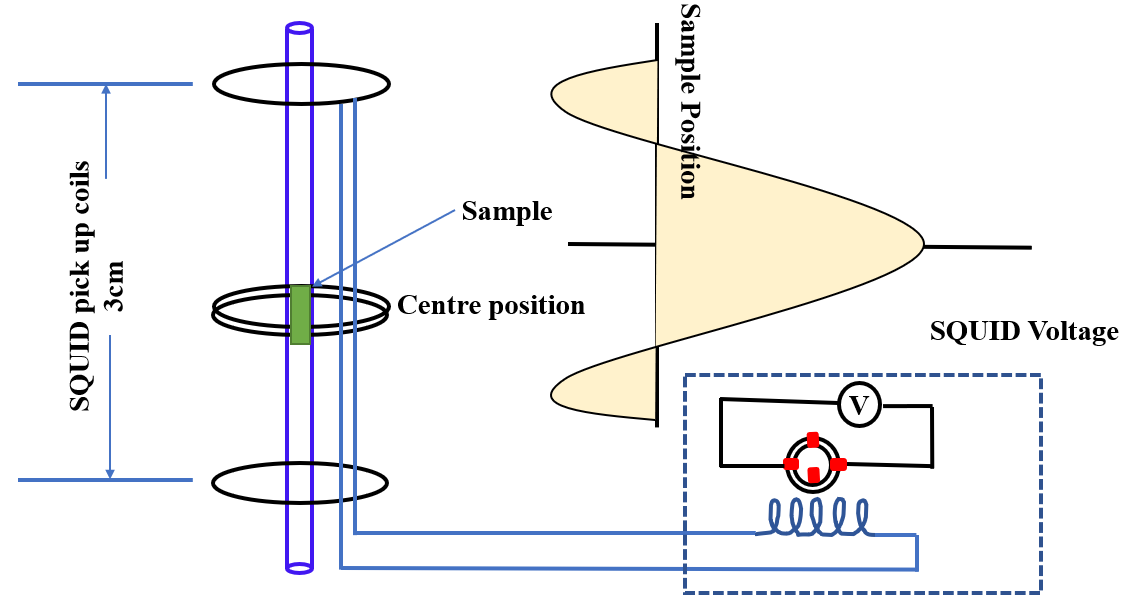}
\caption {schematic of a SQUID magnetometer. The sample is moved through the gradiometer coils, which induces a current. The current flows to the SQUID, which in turn produces a voltage response. The magnetization is then found by fitting the voltage response.}
\label{fig33}
\end{figure*}

A SQUID magnetometer consists of a second-order gradiometer (a set of conducting pick-up coils) connected to a superconducting loop consisting of two parallel Josephson junctions. The sample is placed in a non-magnetic sample holder, and the sample holder is inserted inside the pick-up coils. As the sample moves in the gradiometer, Faraday’s law dictates that a current must be induced in the coils. The SQUID then converts the change in magnetic flux measured by the coils to a dipole response voltage. The SQUID voltage is measured for several sample positions along the length of the gradiometer and is then fitted with the expected signal from a point source magnetic dipole. Thus, the magnetic moment of the sample can be extracted. A schematic of a SQUID magnetometer is shown in Fig. \ref{fig33}.

There are two main dc configurations in a SQUID: field-cooled (FC) and zero field-cooled (ZFC). ZFC measurements involve cooling the sample down, then applying an external magnetic field $H$. FC measurements apply the field before cooling occurs. These two configurations allow for a range of magnetic phenomena to be observed. Both the magnetization ($M$) of the sample and the magnetic susceptibility can be measured using a SQUID. DC susceptibility can be measured using the following equation:
\begin{equation}
\chi =\lim_{H \to 0} \frac{M}{H}
\end{equation}

It is particularly useful for identifying different magnetic regimes and the locations of the phase transitions between them. Often a SQUID is used to characterise the magnetic phases of a sample in preparation for a more detailed $\mu$SR experiment. What must be kept in mind, however, is that the SQUID is a bulk probe, and therefore local phenomena that can be detected using $\mu$SR (for example time reversal symmetry breaking, short-range ordered magnetism) may not be detectable using a SQUID~\cite{nakatsuji2005spin,yaouanc2008short}.

\subsection{Resistivity}
Resistivity measurement is one of the most sensitive measurements of the electrical transport properties of the material. A Physical Properties Measurement System is also known as PPMS consisting of alternating-current transport (ACT) mode is used to measure the resistivity of the sample. A sample probe is positioned inside a wide helium bath, which is enclosed by a nitrogen jacket. It was possible to achieve a temperature range of 1.8 K to 400K, as well as a magnetic field up to 9 T can be applied. 
 
A more accurate method, especially for materials having low resistance, is provided by the "four point arrangement" method. Here the resistance now depends on the length ($L$) between the voltage probe and the area of cross section ($A$) of the sample. Vernier callipers is used to measure these values. The following equation is then used to compute the resistivity, $\rho$.
\begin{equation}
\rho = \frac{VA}{IL}
\end{equation}

Two outside wires are used to deliver current to the sample, and two different inner wires are utilised to assess the potential difference (across the sample), as shown in Fig. \ref{fig31}. The current flowing through the high impedance voltmeter is extremely low, which reduces the impact of cable and contact resistance. 

\begin{figure*}
\centering
 \includegraphics[ width=0.8\linewidth]{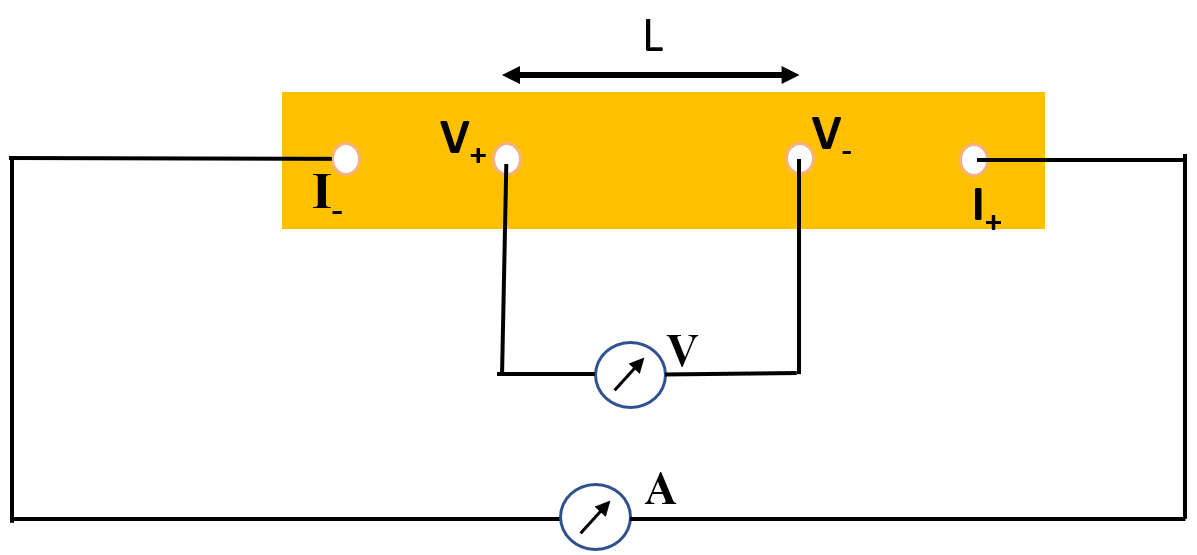}
\caption {Schematic diagram of a sample configured for a four-probe resistivity measurement, where L is the distance between the two voltage wires.}
\label{fig34}
\end{figure*}
Samples were usually sliced into bar forms with a homogeneous cross-sectional area and four silver wires with diameters of 0.05 mm were joined with silver paste. Because the current leads create an electric field throughout the length of the bar shaped sample, the voltage wires must be connected to the current wires in order to appropriately detect the related potential difference. It's also crucial that all the voltage and current wires have unique contact points; or else, contact resistance will no longer be nullified, and the voltage measurement would be influenced. The wired sample was next soldered to the proper connections and glued to a sample puck using GE varnish. A loading rod was then used to place the puck in the sample chamber.

\subsection{Heat capacity}
The Physical Property Measurement System (PPMS) is an automated instrument similar to the MPMS. A multifunctional probe with a series of different sample platforms makes it very versatile. It can perform a variety of material property measurements as a function of field (up to 14 T) and temperature (0.4 K to 400 K). These include heat capacity, thermal transport, DC and AC magnetometry, resistivity and AC transport~\cite{design2000physical}. For the experiments presented here the heat capacity option was used.

Heat capacity measurements give information on the entropy change of a sample as a function of temperature and thus make it possible to extract information on the lattice, magnetic and electronic properties. The results obtained from heat capacity allow a direct comparison between experiment and theory, making them a valuable tool in condensed matter physics. The heat capacity is measured as a function of temperature ($T$) at constant pressure:
\begin{equation}
C_{p} = \left(\frac{\partial Q}{\partial T}\right)_{p}
\end{equation}

\begin{figure*}
\centering
 \includegraphics[ width=0.8\linewidth]{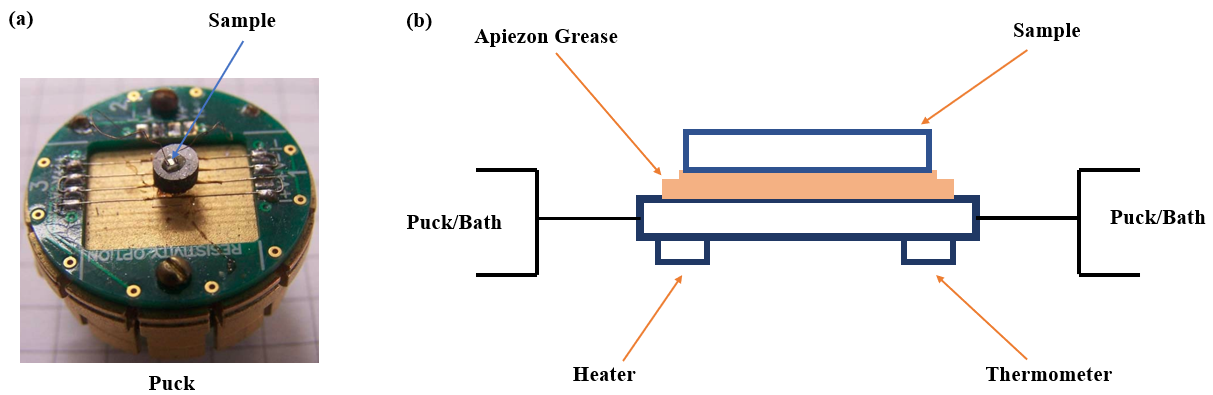}
\caption {Schematic of sample platform with a mounted sample.}
\label{fig314}
\end{figure*}
where $dQ$ is the change in the heat energy in the system. To realize the experiment the sample goes through the heating and cooling periods of fixed time duration. The temperature profile of the sample is monitored to extract the thermal response and thus the heat capacity. The sample is mounted on a quartz platform with a thin layer of apiezon grease. A thermometer and heater are attached to the platform and two small wires provide electrical contact as well as a well-defined thermal link. Samples between 1 mg to 200 mg are measured with this system. The sample has to be relatively at to allow good thermal contact and ensure that the time to reach thermal equilibrium is relatively small.

The 'addenda' is produced for calibration, by measuring the response of the puck and grease without a sample. After each measurement, the provided software fits the whole temperature range to model the relaxation of the sample, sample platform and bath:
\begin{equation}
C_{\rm {total}}\frac{dT}{dt} = -K_{W}(T-T_{B}) + P(t)
\end{equation}

$C_{total}$ in this expression describes the heat capacity of sample and platform, $K_{W}$ represents the thermal conduction of the wires, $T_{B}$ is the temperature of the bath and $P(t)$ is the heater power.

\section{Muon spin spectroscopy}
The heat capacity, resistivity and magnetization measurements are the conventional techniques that measure the bulk properties of the system. Muon spin relaxation/rotation/resonance ($\mu$SR) technique is particularly sensitive for observing the magnetic field distribution inside a material at a microscopic level. This technique is especially helpful for studying the pairing mechanism of novel superconductors. All of the $\mu$SR results presented in this thesis were collected at the ISIS Rutherford Appleton Laboratory in the United Kingdom.
\subsection{Fundamental concepts}
The implantation of 100\% spin polarized muon inside the sample is the main theme of $\mu$SR technique. This technique is extremely sensitive to weak fields in the atomic level, makes it interesting for the investigation of structure and dynamics on the atomic scale. Muons are a local probe of internal fields, and may be used as sensitive magnetometers~\cite{blundell1999muon}. The principle of this technique is that the implanted muons in the system precess in the low magnetic field in the material. In contrast to other techniques such as neutrons and X-rays, $\mu$SR does not involve any scattering. The advantage of this technique over Electron Spin Resonance (ESR) and Nuclear Magnetic Resonance (NMR), both of which are also local probe techniques, is that a muon may be implanted into the material with 100\% spin polarization. The variation in the spin polarization of muons can be used to study the spatial and temporal properties of internal fields, allowing researchers to look into difficulties in solid state physics that would otherwise be difficult to analyze .  \\
\begin{table}[h]
 \caption{Characteristics properties of electron ($e$), muon ($\mu^{+}$) and proton ($p$)}
 \centering
  \begin{tabular}{lccc}
\hline
\hline
    & $e$ &  $\mu^{+}$ & $p$ \\  \hline 
    Charge  & $-e$  & $+e$ & $+e$ \\~\\
    Spin  & 1/2 & 1/2 & 1/2\\~\\
  Mass &  $m_{e}$ & 207 $m_{e}$ & 1836 $m_{e}$\\~\\
$\gamma/2\pi$(MHzT$^{-1}$) &28025 & 135.5 & 42.58  \\~\\
Magnetic moment & 657$\mu_{p}$ & 3.18 $\mu_{p}$ & $\mu_{p}$  \\~\\
Lifetime & $>$4.6$\times$10$^{26}$ yr &  2.197 $\mu$s & $>$2.1 $\times$ 10$^{29}$ yr \\~\\
\hline \hline
\end{tabular}
  \label{T1}
\end{table}

\begin{figure*}
\centering
 \includegraphics[height=0.5\linewidth, width=\linewidth]{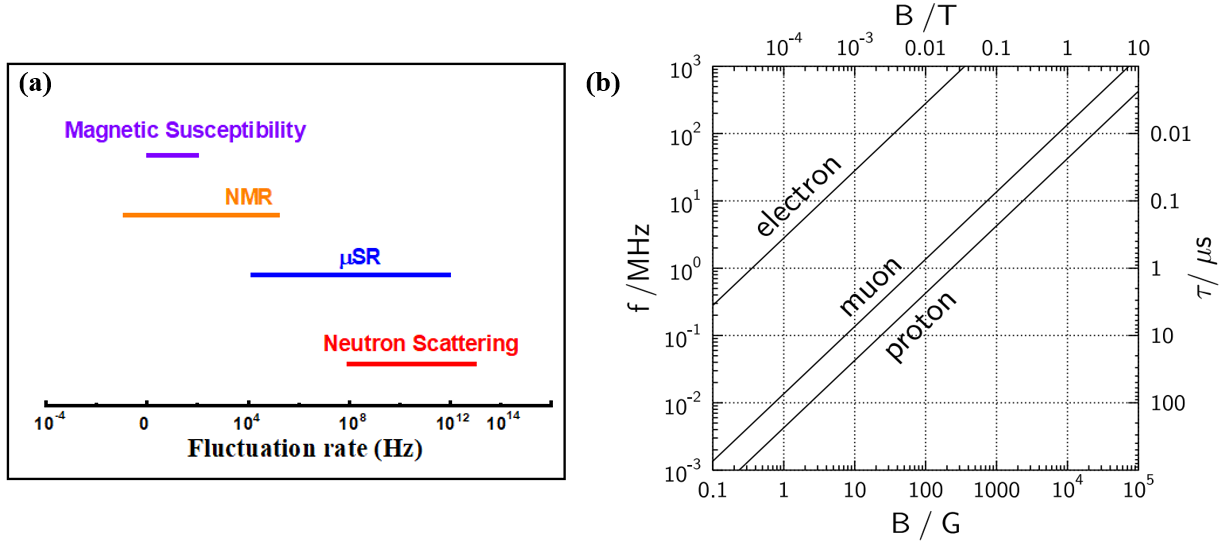}
\caption {(a) Range of fluctuation rates that can be measured with different techniques. (b) Larmor precession frequencies for electron, muon and proton. Adapted from Ref.~\cite{blundell1999spin}}
\label{fig35}
\end{figure*}

In Table \ref{T1}, the characteristics of the muon are listed alongside those of an electron and a proton. The uniqueness of these characteristics is what makes $\mu$SR such a valuable tool in condensed matter physics. Muons are the elementary particle with a mass of 207 times the mass of an electron, in between the mass of electron and proton. As muon is a spin 1/2 particle so no quadrupolar interaction does not occur, in contrast to NMR of S $>$ 1/2 where this can lead to broadening and weakening of the observed resonances. A muon has a lifetime of  2.19 $\mu$s, which is comparatively long for a condensed matter process. This enables the measurement of fluctuation rates in the range of 10$^{4}$ to 10$^{12}$ Hz. This time window connects the gap between NMR and neutron scattering as illustrated in Fig. \ref{fig35}.\\

Although muons can be negatively or positively charged, only the positive muon is of interest in condensed matter physics. This is because this technique is based on the implantation of the muon at a specific site in the studied material. A negatively charged muon is attracted to the atomic nuclei and thus is not very sensitive to the magnetic properties of the system. However, the positive muons are attracted to negative charge density, such as large areas of electron density. So they are implanted at interstitial sites in inorganic materials or bond to organic molecules~\cite{blundell1999spin}. Most importantly the large gyromagnetic ratio of the muon, which is proportional to its magnetic moment, results in a technique that is extremely sensitive and can detect fields down to 10$^{-5}$ T.

\subsection{Muon production and implantation}
The $\mu$SR experiment is only possible in cyclotron or synchrotron . They provide high-energy protons (about 800 MeV), which are required for muon synthesis. These protons collide with a target (usually graphite) that contains both protons and neutrons, resulting in pions: 
\begin{eqnarray}
p+p &\rightarrow& p+n+\pi^{+}\\
p+n &\rightarrow& n+n+\pi^{+}\\
p+n &\rightarrow& p+\pi^{-}
\end{eqnarray}

where $p$, $n$ and $\pi^{\pm}$ are a proton, a neutron and a positive or negative pion, respectively. Although there exist different techniques for muon production, for the one most commonly used the pions decay into muons $\mu^{\pm}$ and neutrinos, $\nu_{\mu}$ and $\bar{\nu_{\mu}}$, while at rest in the surface layer of the target via the reactions:
\begin{eqnarray}
\pi^{+} &\rightarrow& \mu^{+}+\nu_{\mu}\\
\pi^{-} &\rightarrow& \mu^{-}+\bar{\nu_{\mu}}
\end{eqnarray}
Conservation of angular and linear momentum and parity violation ensures that the muons are spin polarized as follows: consider a pion with spin 0 decaying from rest in the target; the decay is a two-body process thus the neutrino and muon must have equal and opposite momentum/spin and, as the neutrino has negative helicity, it follows that the muon spin must also be antiparallel to its momentum. The muons described here are termed surface muons, as they are created very close to the surface of the target and typically have energies around 4 MeV~\cite{blundell2001magnetism}. 

\begin{figure*}
\centering
 \includegraphics[height=0.5\linewidth, width=0.8\linewidth]{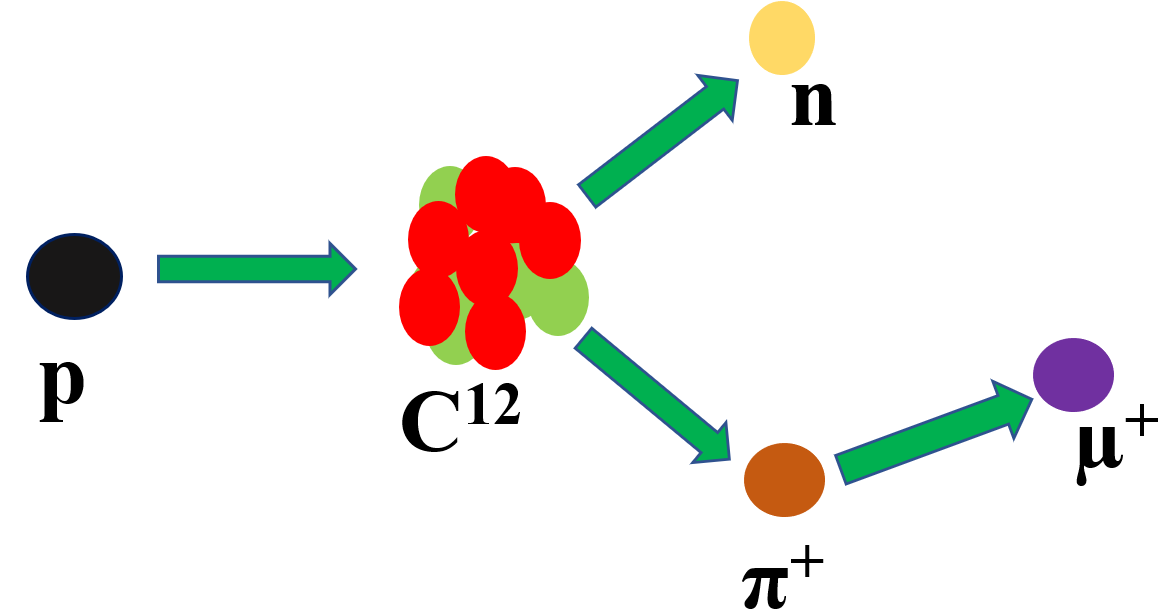}
\caption {Production of muon from high energy proton beam with collision with graphaite target }
\label{fig36}
\end{figure*}

This muon beam produced is then guided to the sample through a combination of steering and focussing quadrupole magnets. Contaminating uncharged and negatively charged particles in the beam are separated off by a system of opposing electric and magnetic fields. Following implantation in the sample, the muons rapidly lose energy down to a few hundred eV, due to stopping processes such as ionization of atoms, scattering with electrons and series of electron capture and loss, whilst still retaining their polarization.

\subsection{Interactions and decay of muons}
The $\mu^{+}$ implanted in the sample decays into an electron-neutrino ($\nu_{e}$), positron ($e^+$), and a muon anti-neutrino ($\bar{\nu_{\mu}}$) with a lifetime of 2.19 $\mu$s:
\begin{equation}
\mu^{+} \rightarrow e^{+} + \nu_{e} + \bar{\nu_{\mu}}
\end{equation}

The decay takes place via weak interaction and violates parity, which means that the positrons created in this reaction are emitted in the muon spin direction.  The angular distribution of emitted positrons with respect to the muon spin at the time of decay is given by;
\begin{equation}
N(\theta)\rightarrow N_{0}(1+\beta \cos\theta)
\end{equation}
Here $\theta$ is the angle formed by the muon spin and the positron's emission direction. The asymmetry factor $\beta$ reflects the positron energy, where $\beta$ = 1 for the maximum possible positron energy (52.8 MeV) and $\beta = \frac{1}{3}$ when averaged over all possible energies~\cite{garwin1957observations}. Thus, by observing a substantially large number of decay events, the time-dependence of the spin-polarization of an ensemble of muons may be deduced.

\begin{figure*}
\centering
 \includegraphics[height=0.5\linewidth, width=0.6\linewidth]{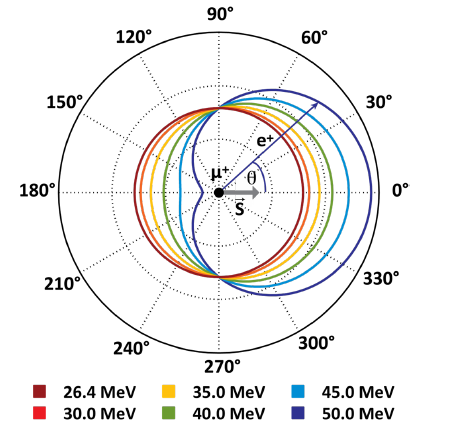}
\caption {Angular distribution of positrons emitted when the muon decays. The emission distribution depends on the positron energy. The black horizontal arrow denotes the direction of the muon spin. Larger energies result in a more anisotropic distribution. }
\label{fig37}
\end{figure*}

\subsection{Muon precession and relaxation}
In a $\mu$SR experiment, the quantity of interest is the time-dependent muon polarization. Consider a muon initially polarized along the z-axis in a magnetic field $B$. Its spin will precess around $B$ with Larmor frequency $\omega=\gamma_{\mu}B$, having $\gamma_{\mu} = 2\pi \times 135.5$ MHzT$^{-1}$ is the muon gyromagnetic ratio. Note how $\omega$ depends on the magnitude of $B$ and not its direction with respect to the muon polarization which only affects the amplitude of the frequencies. This makes $\mu$SR a powerful technique for both single and polycrystalline samples. The time evolution of its $z$-component of polarization $P_{z}$(t) is then given by
\begin{equation}
P_{z}(t) = \cos^2 \theta +~ \sin^2 \theta \cos \omega t
\end{equation}
where $\theta$ is the angle between $B$ and the z-axis (see Fig.2.2). Taking the powder average yields

\begin{equation}
P_{z}(t) = \frac{1}{3} + \frac{2}{3}\cos(\gamma_{\mu}Bt)
\end{equation}
where the 1/3 denotes the fraction of spins that are parallel to the initial muon polarization and therefore do not precess (this is the origin of the `1/3-tail' often observed in ordered polycrystalline samples).
\begin{figure*}
\centering
 \includegraphics[height=0.5\linewidth, width=0.6\linewidth]{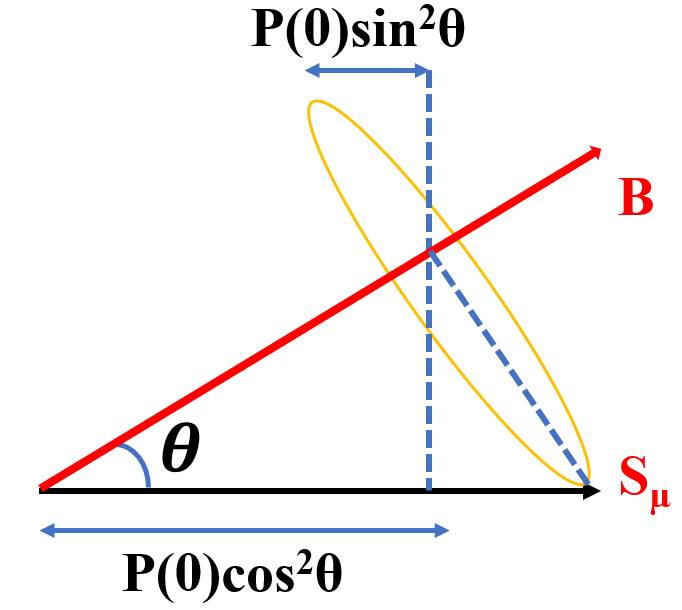}
\caption {Larmor precession of muon polarization around a local field $B$. }
\label{fig39}
\end{figure*}

An arrangement of scintillation detectors is placed around the sample space to track the direction of emitted positrons. There are two possible geometrical configurations for these detectors; longitudinal and transverse, as shown in Fig.~\ref{fig315}. The former is used for transverse-field (TF, or muon spin rotation) whilst the latter is used longitudinal-field (LF) and zero-field (ZF) measurements (muon spin relaxation). A more detailed discussion of these types of measurements and their applications will be addressed in the following sections, although for the discussion one needs only to consider the LF arrangement. This involves placing one bank of detectors behind the sample (the forward detector) and the other to intersect the muon beam in between the sample and the entry window (the backward detector). The polarization is measured directly by observing the relative counts of the backward and forward detectors via the time-dependent asymmetry function

\begin{equation}
A(t) = \frac{N_{F}(t)-\alpha N_{B}(t)}{N_{F}(t)+\alpha N_{B}(t)}
\end{equation}

where $N_{B}$ and $N_{F}$ denote the number of counts in the forward and backward detectors, respectively. The constant $\alpha$ is experimentally determined to account for any difference in detector efficiency. Note how without the normalising denominator, $A(t)$ would reflect the radioactive nature of the muons as an exponential decay curve. It is the asymmetry spectrum that carries information on the internal field distribution. Oscillations in that spectrum will dephase with any spatial or time-dependent variations in the local field, which is observed as a relaxation of the ensemble spin polarization. In a $\mu$SR experiment, one interprets the data by fitting the asymmetry to a proposed model for the internal field distribution of the sample using the least-squares fitting routine. In this thesis, we fitted all the $\mu$SR data using the WIMDA (Windows Muon Data Analysis) package~\cite{pratt2000wimda}. Many different functions that are used to describe a wide range of physical scenarios may be studied with $\mu$SR, each with its approximations. The following sections will discuss some of the cases relevant to the research presented in this thesis.
\begin{figure*}
\centering
 \includegraphics[height=0.5\linewidth, width=0.8\linewidth]{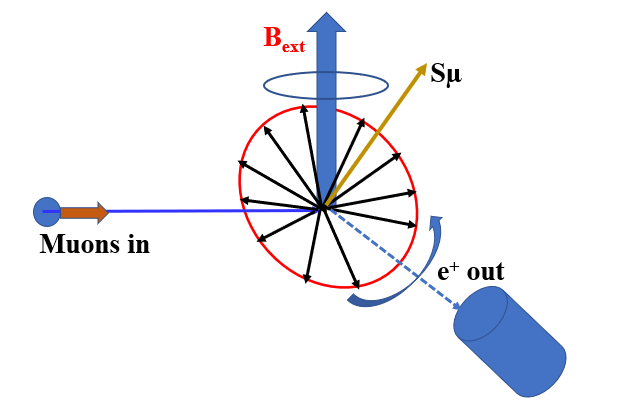}
\caption {Schematic picture of spin precession in a magnetic field. }
\label{fig310}
\end{figure*}

\subsection{Zero-field and longitudinal field muon spin relaxation}
Weak internal magnetism may be studied using ZF $\mu$SR. This is the simplest setup in which the local $B$ originates from intrinsic magnetic (or nuclear) moments near the muon site. If long-range magnetic order is present, a significant number of muons will stop in sites of equivalent local fields and this results in a coherent precession of muon spins which manifests as a cosinusoidal asymmetry function. Typically, the maximum value of $A(t)$ is $\sim$ 25\% which results from the intrinsic asymmetry of the muon decay and the finite solid angle subtended by the detectors.

If the local field has an inhomogeneous distribution and thus varies from site to site, the signal is damped (relaxed) as muons oscillating at different frequencies dephase over time. In the ordered case, there may be a slight variation in the field at each muon site which will result in weakly damped oscillations. Another common example arises from nuclear moments, in which the variation between muon sites is so large that no oscillations are observed. Instead, the asymmetry may be described employing a Gaussian Kubo-Toyabe relaxation function $G_{KB}$($\Delta$, t)~\cite{kubo1967stochastic,hayano1979zero}.

\begin{equation}
G_{KB}(\Delta,T) = \frac{1}{3}+\frac{2}{3} \exp(-\Delta^2t^2/2)(1-\Delta^{2}t^{2})
\label{Eq323}
\end{equation}

This is observed as an initially Gaussian depolarization which recovers to a constant long-time `1/3 tail' (note that is not 1/3 of the total asymmetry, as mentioned earlier, but 1/3 of its normalised value). This model is based on the central limit theorem; that is, in a statistically disordered system the magnetic field sensed by the muon at any given point would take a random value and this may be approximated by a normal distribution of width $\Delta/\gamma_{\mu}$  about the mean (in this case zero). Note that this distribution is appropriate for concentrated moments. Dilute distributions are better modelled by a Lorentzian function, giving rise to the Lorentzian Kubo-Toyabe function, as first predicted by Walstedt and Walker~\cite{walstedt1974nuclear}.

\begin{figure*}
\centering
 \includegraphics[width=\linewidth]{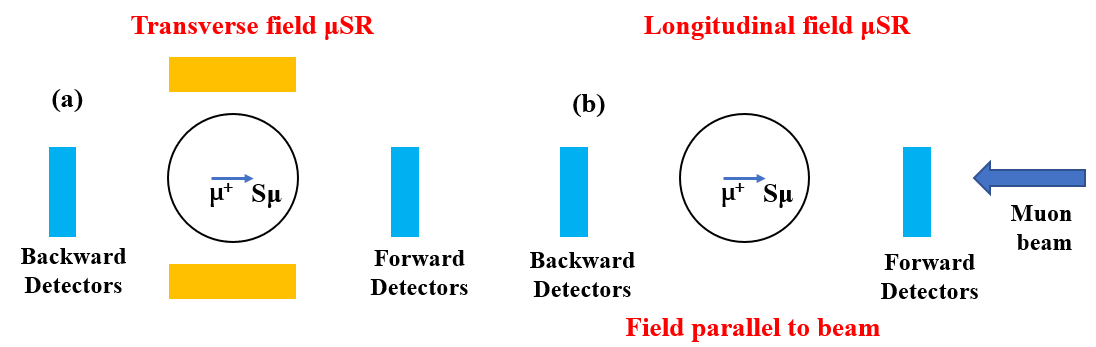}
\caption {Schematic diagram of (a) transverse field muon spin rotation, where the magnetic field is applied perpendicular to the perpendicular to the muon spin direction (b) longitudinal field muon spin relaxation, where the magnetic field is parallel to the muon spin direction. }
\label{fig315}
\end{figure*}
So far, all of these examples have involved static moments. Another cause of damping is from a time-dependent field distribution that arises from either fluctuating moments or muon diffusion (muon hopping) and this manifests as an exponential time relaxation~\ref{Eq323}. The effects of muon hopping on the measured relaxation are shown in Fig 2.4(b) for a range of fluctuation (or hopping) rates $\nu$. In the case of $\nu$ = 0, the relaxation takes the form of the zero-field Kubo-Toyabe curve described by Eq. 2.8. If the dynamics are comparatively slow ($\nu < \Delta$), then this is observed as a relaxation in the 1/3 tail at late times, and the form of the tail may be approximated by
\begin{equation}
P_{z}(\nu,t) = \frac{1}{3}\exp(-\frac{2}{3}\nu t)
\end{equation}
If the dynamics are fast ($\nu >> \Delta$), the asymmetry is dominated by the exponential relaxation and, rather counter-intuitively, the relaxation rate decreases as the fluctuation or hopping rate increases. In this case, the relaxation is well represented by a simple exponential
\begin{equation}
P_{z}(t)= e^{-\lambda t}
\end{equation}
in which the relaxation rate is given by $\lambda = 2 \Delta^2/\nu$.

Of course, these are simplified examples and often the spectra display a range of both static and dynamic phenomena. One of the ways of distinguishing between these is to use LF $\mu$SR, in which a magnetic field $B_{L}$ is applied in the initial direction of muon polarization~\cite{blundell1999spin}. Initially, the muons precess in both applied and internal fields, but with increasing $B_{L}$ the vector sum of the local and applied field becomes increasingly aligned until the muons are completely decoupled from the intrinsic moments (the applied field pins the muon spin along its initial direction). The polarization is restored, which in the case of a Kubo-Toyabe relaxation, for example, means a recovery of the 1/3 tail to its maximum value. Since static moments decouple in much lower fields, a combination of ZF and LF $\mu$SR can be used to infer the origin of the relaxation.

\begin{figure*}
\centering
 \includegraphics[width=0.6\linewidth]{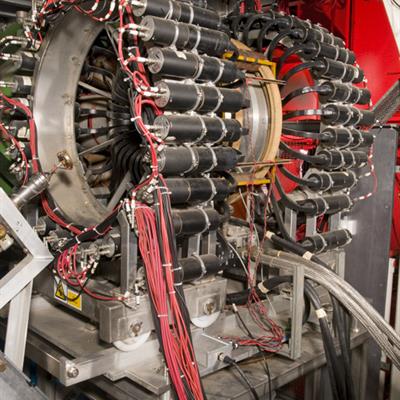}
\caption {The MUSR spectrometer at ISIS ISIS Neutron and Muon Source, Rutherford Appleton Laboratory. Adapted from ISIS neutron and muon source. }
\label{fig311}
\end{figure*}
\subsection{Transverse-field muon spin rotation}
In TF $\mu$SR (also known as transverse-field muon spin rotation), the detectors are placed around the sample and the field is applied perpendicular to the initial muon spin. In this case, the ensemble spin rotates around the applied field $B_{TF}$ and on implantation, any variation in the local field distribution is observed as an envelope in $A(t)$. A typical relaxation function has the form; 
\begin{equation}
A(t) = A_{0}\cos(\omega t)f(t)
\end{equation}

where initial asymmetry is $A_{0}$, $\omega = \gamma_{\mu}B_{\rm{loc}}$ is the angular frequency of the oscillations which are dependent on the local transverse field $B_{loc}$, and $f(t)$ is a function describing the relaxing envelope.
One of the principal applications of TF-$\mu$SR is for studying type-II superconductors. These systems form vortex lattices when cooled to their Shubnikov state in an applied field, and the magnetic field distribution inside the vortex lattice can be mapped out using $\mu$SR. Depending on the superconducting penetration depth, the local field measured by each muon can range from zero in the superconducting regions to $B_{TF}$ in the normal vortex cores. Muons will settle in any set of crystallographically equivalent sites but the vortex lattice is generally out of sync with the crystal lattice and of a significantly larger scale, the full range of field distribution is sampled. It is slightly broadened by nuclear spin contributions, as well as any disorder in the vortex lattice, for example, due to pinning. A typical field distribution is shown schematically in Fig. 2.5. In these instances, TF-$\mu$SR can be used to deduce the penetration depth and its temperature dependence.

\begin{figure*}
\centering
 \includegraphics[width=0.8\linewidth]{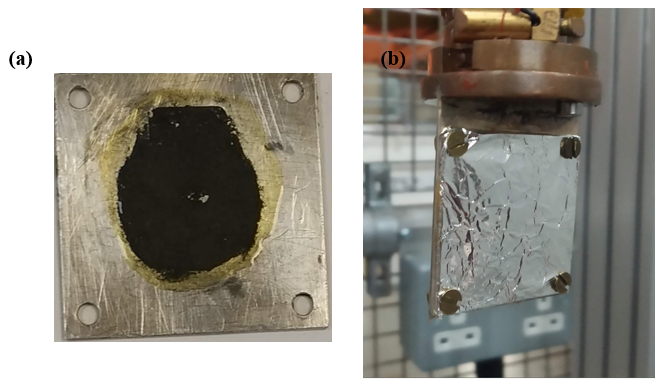}
\caption {(a)Powder sample was placed inside a silver (99.995\%) sample holder using GE varnish  diluted with ethanol. (b) The sample holder wrapped with silver foil placed in a cryostat inside the MuSR spectrometer. Taken in ISIS Neutron and muon source, UK during experiment.}
\label{fig312}
\end{figure*}

\subsection{MUSR spectrometer}
All the $\mu$SR experiments presented in this thesis have been performed in MuSR spectrometer at ISIS facility, Rutherford Appleton Laboratory, UK. The MuSR spectrometer consists of 64 scintillator detectors that are usually grouped into forward and backward directions for longitudinal configuration, or 8 individual groups for TF mode. The latter is accessed by rotating the instrument through 90$^{0}$ with respect to the beamline. Available sample environments include TF fields of up to 60 mT and longitudinal field 0.3 T and variable-temperature cryostats which allow temperatures down to 40 mK. 

Generally, the powder sample mounted on the silver(Ag) or haematite (Fe$_{2}$O$_{3}$) sample holder using GE varnish diluted with ethanol. Silver has a negligible relaxation rate since it is weakly diamagnetic, does not superconduct under normal circumstances, and has very small nuclear moments ($\sim 0.1 \mu_{N}$). As a result, silver does not affect muon spin ensemble precession or depolarization. Haematite is occasionally utilized for the exact opposite reason: enormous, random internal magnetic fields depolarize any muons that land in it almost instantaneously. The silver sample holder is wrapped with silver (Ag) foil placed in the cryostat. Choosing a suitable cryostat for the experiment is a crucial part. There are different cryostats with different temperature ranges such as (i) Dilution fridge for 40 mK to 4 K, (ii) Sorb for 300 mK to 50 K, (iii) Variox for 1.5 K to 300 K (iv) CCR for 4 K (10 K) to 700 K and (v) Furnace for 300 K to 1500 K.  \\

In summary, it can be seen that $\mu$SR has a wide range of applications in condensed matter physics. In this thesis, we have used TF-$\mu$SR and ZF/LF-$\mu$SR measurements. From TF-$\mu$SR we can estimate the temperature dependence of the London penetration depth, related with superfluid density. Using different gap models we can estimate the superconducting gap structure. Moreover, we can estimate the different superconducting parameters such as London penetration depth ($\lambda_{L}$), the effective mass of the quasiparticle and superconducting carrier density etc. Though, the ZF-$\mu$SR technique is extremely sensitive to the magnetic field at the atomic level, helps us to detect the appearance of the magnetic field in the superconducting state, crucial to detecting the broken time reversal symmetry. ZF-$\mu$SR is also helpful to understand the nature of ordering i.e. long-range or short-range ordering in magnetic materials. 


%% file: chapter4.tex
\chapter{Studies of ternary equiatomic superconductors ZrIrSi and HfIrSi} 
\label{chapter:4}
\section{Introduction}
Recently, a number of studies have been carried out in ternary metal phosphide, silicide and arsenide with general formula  $TrT'X$ ($Tr$ and $T'$ are either 4$d$ or 3$d$ transition  elements and whereas $X$ is either a group IV or member)~\cite{barz1980, muller1983, seo1997, ching2003, keiber1984, shirotani2001, shirotani1998,shirotani2000,shirotani1999,shirotani1995}. These systems have attracted considerable attention due to their relatively high superconducting transition temperature ($T_{C}$), for example 15.5 K for hexagonal $h-$MoNiP~\cite{shirotani2000}, 13 K for $h-$ZrRuP~\cite{shirotani1993} and 12 K for $h-$ZrRuAs~\cite{meisner1983}. These ternary equiatomic systems have provided a playground to investigate the role of spin orbit (SO) coupling in superconductivity, which were not so well studied in these systems. Compounds with Ir are often characterized with a strong SO coupling effect, due to presence of the Ir $5d$-orbitals. Superconductivity is observed in a number of Ir-based compounds such as Li$_{2}$IrSi$_{3}$ ($T_{C}$ = 4.2 K) ~\cite{pyon2014, lu2015}, IrGe ($T_{C}$ = 4.7 K) ~\cite{matthias1963}, RIr$_{3}$ [$T_{C}$ = 3.1 K (La), $T_{C}$ = 3.3 K (Ce)] ~\cite{haldolaarachchige2017, sato2018,bhattacharyya2019},  CaIr$_{2}$($T_{C}$ = 5.8 K) ~\cite{haldolaarachchige2015}, HfIrSi ($T_{C}$ = 3.1 K) ~\cite{kase2016} and ScIrP ($T_{C}$ = 3.4 K) ~\cite{pfannenschmidt2011}.  Cuamba {\it et.al.}~\cite{pyon2014,lu2015,matthias1963,cuamba2016} suggests that the presence strong spin orbit (SO) coupling and a significant contribution to the total density of states (DOS) comes from the Ir-atom in most of the Ir-based compounds. Recently, we have reported time reversal symmetry (TRS) breaking superconductivity on the transition metal based caged type R$_{5}$Rh$_{6}$Sn$_{18}$ (R = Lu, Sc, and Y) ~\cite{bhattacharyya2015broken,bhattacharyya2015unconventional,bhattacharyya2015unconventional} compounds due to  strong spin orbit coupling.

\begin{figure*}
\centering
 \includegraphics[width=0.6\linewidth]{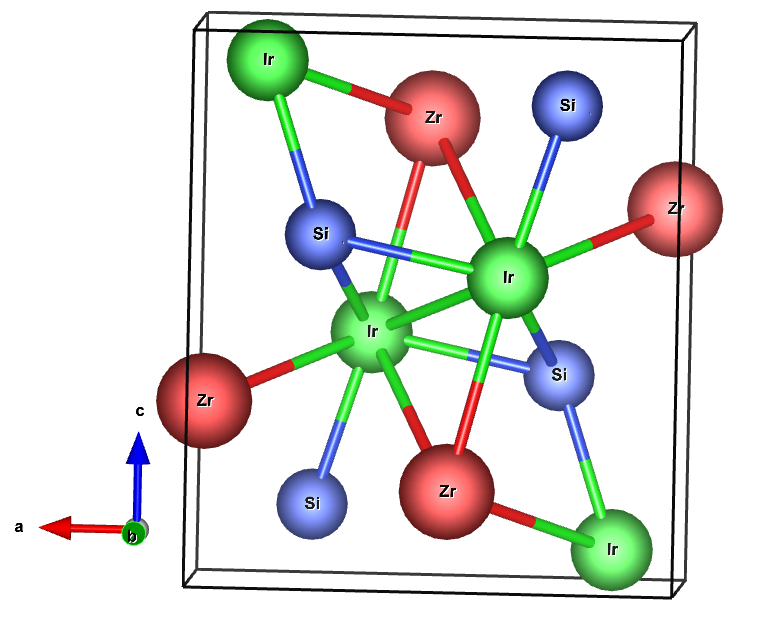}
\caption {The crystal structure of ZrIrSi which crystallizes in the orthorhombic structure with the space group Pnma, where the spheres represent the Zr (red), Ir (green), and Si (blue) atoms.}
\label{fig41}
\end{figure*}

In ternary equiatomic compounds, superconductivity has only been found in two types of crystal structures: the first one is the hexagonal Fe$_{2}$P-type (space group $P\bar{{6}}m$2),~\cite{ching2003,shirotani2000, shirotani2001} and the second one is the orthorhombic Co$_{2}$Si-type (space group $Pnma$)~\cite{shirotani1998,shirotani2000, ching2003, muller1983}. It is interesting to note that in these systems $T_{C}$ is strongly associated with crystal structure. Furthermore, $h-$Fe$_{2}$P-type structure exhibit higher $T_{C}$ than $o-$Co$_{2}$Si-type structure, for example: $h-$ZrRuP  shows $T_{C}$ at 13.0 K whereas $o-$ZrRuP exhibit $T_{C}$ at 3.5 K. In case of $h-$Fe$_{2}$P-type structure each layer is fiiled up with either $Tr$ and $T'$ or $X$ and $X$ elements. In case of $o-$ZrRuP,  Shirotani  {\it et al.}~\cite{shirotani1999} reported the formation of two dimensional triangular  Ru$_{3}$ clusters and in the basal plane they are connected through Ru-P ionic bonds. It also connected through Zr-Ru bonds where Zr atoms occupy the $ z=1/2$ plane. If phosphorus is replaced by the more electronegative silicon then nearest neighbour Ru-Ru bond length enlarged to 2.87 \AA. Surprisingly $o-$MoRuP shows superconductivity at 15.5 K which is as high as isoelectronic $h-$ZrRuP ($T_{C}$ = 13 K)  and $h-$MoNiP ($T_{C}$ = 13 K).  Ching {\it  et. al.}~\cite{ching2003} shown that in $o-$MoRuP and $o-$ZrRuP higher value of DOS at the FL is directly related to higher $T_{C}$, as suggested for BCS superconductor. In these systems, the density of states are governed by Mo-$4d$ orbitals. Ching {\it  et. al.}~\cite{ching2003} have calculated the values of density of states which are 0.46 states per eV atom and 0.33 states per eV atom for $o-$MoRuP and $o-$ZrRuP, respectively.

Therefore, to investigate the superconducting pairing mechanism in ZrIrSi and HfIrSi, in this chapter we have presented a systematic $\mu$SR study. Zero field (ZF)-$\mu$SR is a powerful technique to know the time reversal symmetry (TRS) breaking in the superconducting state~\cite{sonier2000musr}. ZF-$\mu$SR data indicates the absence of any spontaneous magnetic fields below $T_{C}$, thus implying that TRS is not broken in the superconducting state. The superfluid density as a function of temperature is determined from the depolarization rate of the transverse field (TF)-$\mu$SR which is well described by an isotropic $s-$wave model. These results are further supported by {\it ab-initio} electronic structure calculation.
\begin{figure*}
\centering
 \includegraphics[width=0.6\linewidth]{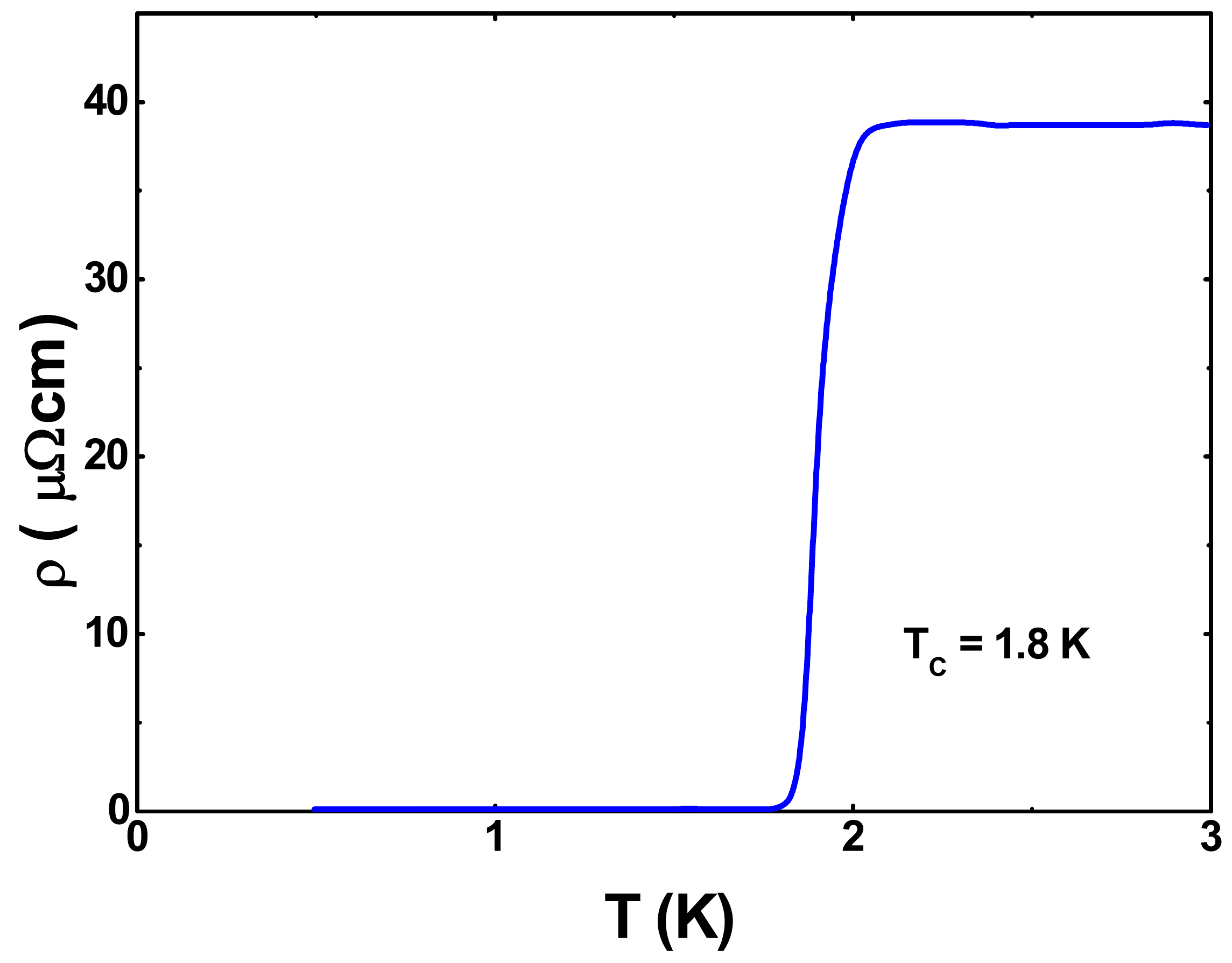}
\caption {Temperature variation of resistivity in zero field.}
\label{fig42}
\end{figure*}

\section{Studies of ZrIrSi}
\subsection{Sample preparation}
For this study, a polycrystalline sample of ZrIrSi was synthesized using a typical arc melting process on a water-cooled copper hearth using Zr (99.99\%), Ir (99.99\%), and Si (99.999\%) in a stoichiometric ratio. The arc melted ingot was remelted several times to confirm the phase homogeneity. After that, the sample was annealed at 1000$^{0}$C for a week in a sealed vacuum quartz tube. X-ray diffraction was carried out using Cu-$K_{{\alpha}}$ radiation. X-ray powder diffraction data revealed that ZrIrSi crystallizes in the orthorhombic structure (space group $Pnma$) as displayed in Fig.~\ref{fig41}. The calculated lattice parameters are $a$ = 6.557~\AA, ~ $b$ = 3.942~\AA ~ and $c$ = 7.413~\AA, which are in agreement with previous report~\cite{kase2016}.

\subsection{Resistivity measurements}
Electrical resistivity measurement was done in a standard dc-four probe technique down to 0.5 K. The temperature ($T$) variation of the electrical resistivity $\rho (T)$ in zero applied magnetic field is presented in Fig. \ref{fig42}. The electrical resistivity data reveals superconductivity at $T_{{C}}$ = 1.7 K. Kase {\it et. al.}~\cite{kase2016} have estimated the  Ginzburg Landau coherence length ($\zeta$) = 23.1 nm. It is interesting to note that $T-$ dependence of upper critical field shows a convex curvature~\cite{kase2016}, which might suggest the presence of SO coupling. Similar curvature is also found in R$_{3}$T$_{4}$Sn$_{13}$ (R = La, Sr; T = Rh, Ir) which is a SO coupled superconductor~\cite{kase2011superconducting}.

\begin{figure*}
\centering
 \includegraphics[width=0.9\linewidth]{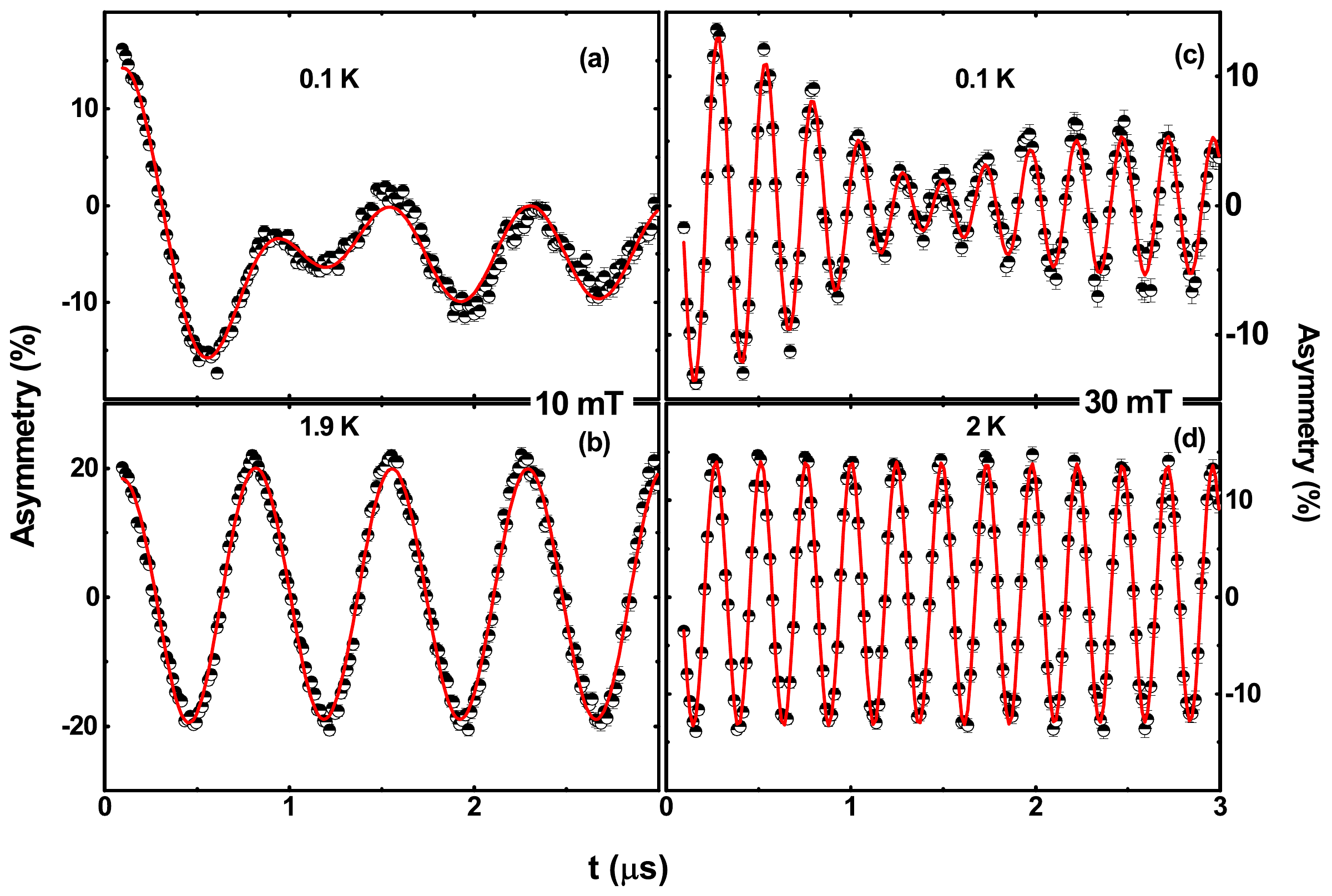}
\caption {Time dependence TF-$\mu$SR asymmetry spectra for ZrIrSi recorded at (c) T = 0.1 K and (d) T = 1.9 K in the presence of the applied field $\mu_{0}H$ = 10 mT and at (e) T = 0.1 K and (f) T = 2.0 K in the applied field $\mu_{0}H$ = 30 mT. The red solid line shows the fittings to the data using Eq. \ref{Eq41} described in the text.}
\label{fig43}
\end{figure*}
 
\subsection{TF-$\mu$SR measurements}
\begin{figure*}
\centering
 \includegraphics[width=0.6\linewidth]{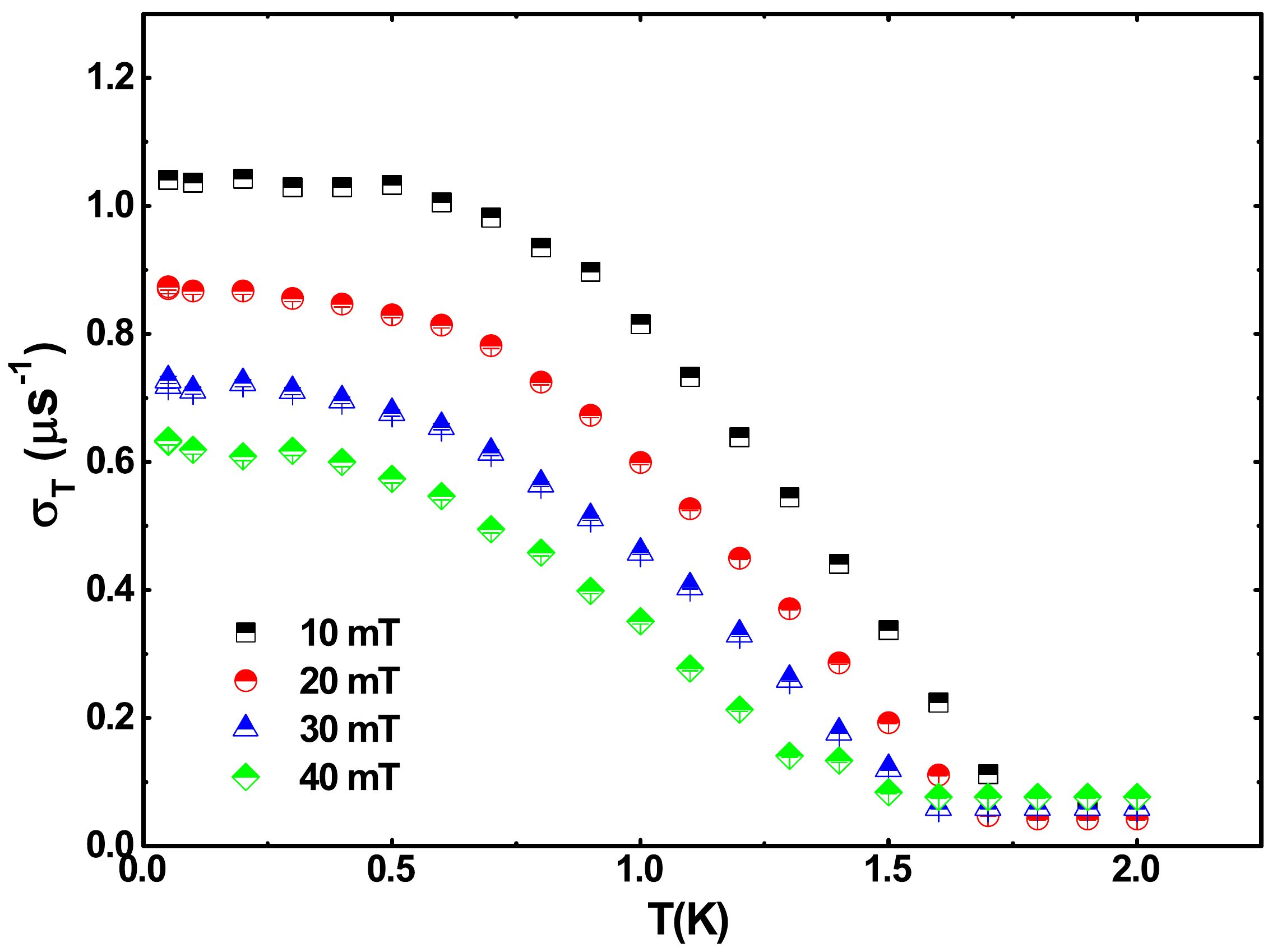}
\caption {The superconducting depolarization rate $\sigma_{T}$ as a function of temperature in the presence of an applied field of 10 mT, 20 mT, 30 mT, and 40 mT.}
\label{fig44}
\end{figure*}

\begin{figure*}
\centering
 \includegraphics[width=0.6\linewidth]{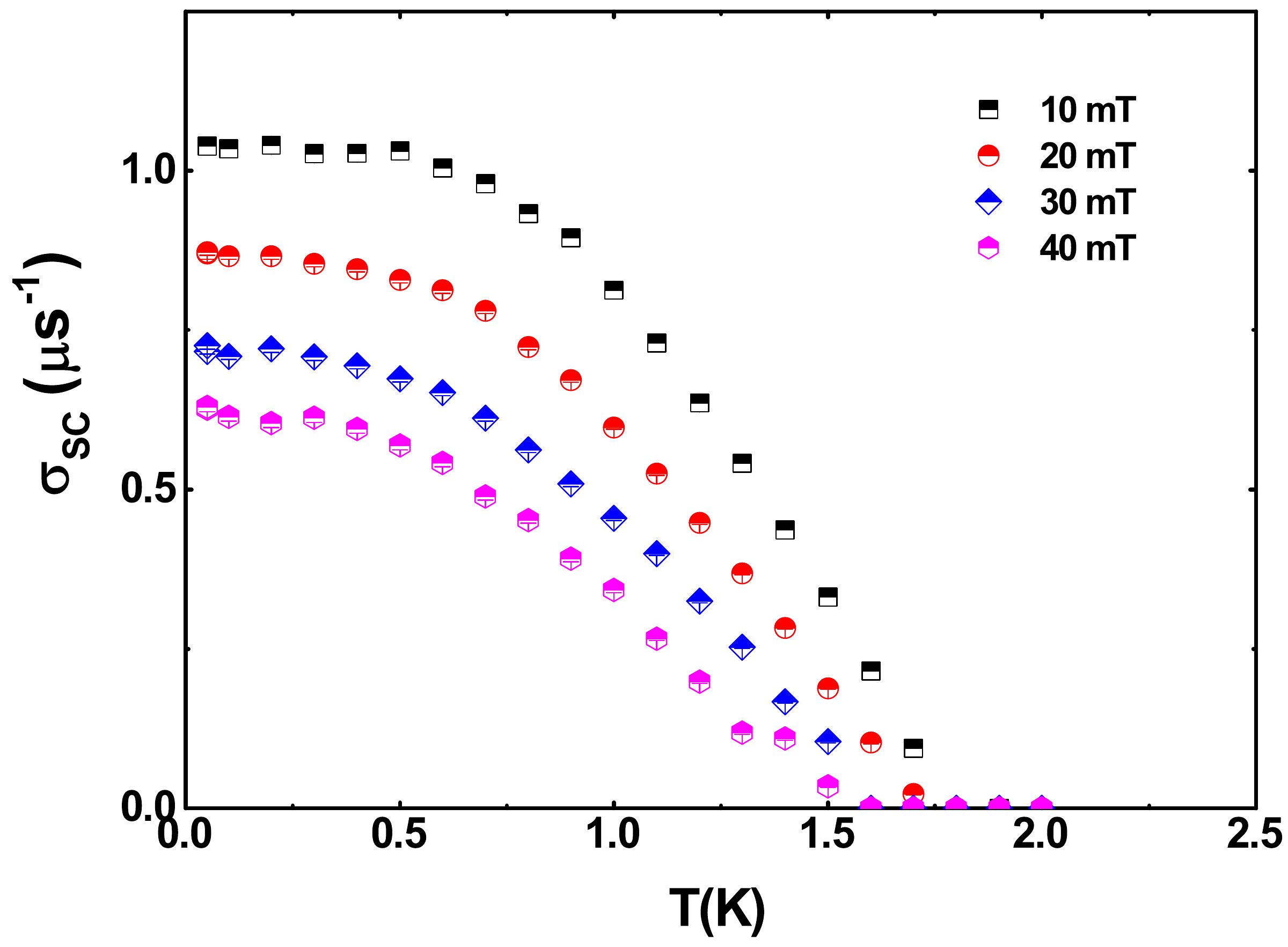}
\caption {The superconducting depolarization rate $\sigma_{sc}$ as a function of temperature in the presence of an applied field of 10 mT, 20 mT, 30 mT, and 40 mT.}
\label{fig45}
\end{figure*}

\begin{figure*}
\centering
 \includegraphics[width=0.6\linewidth]{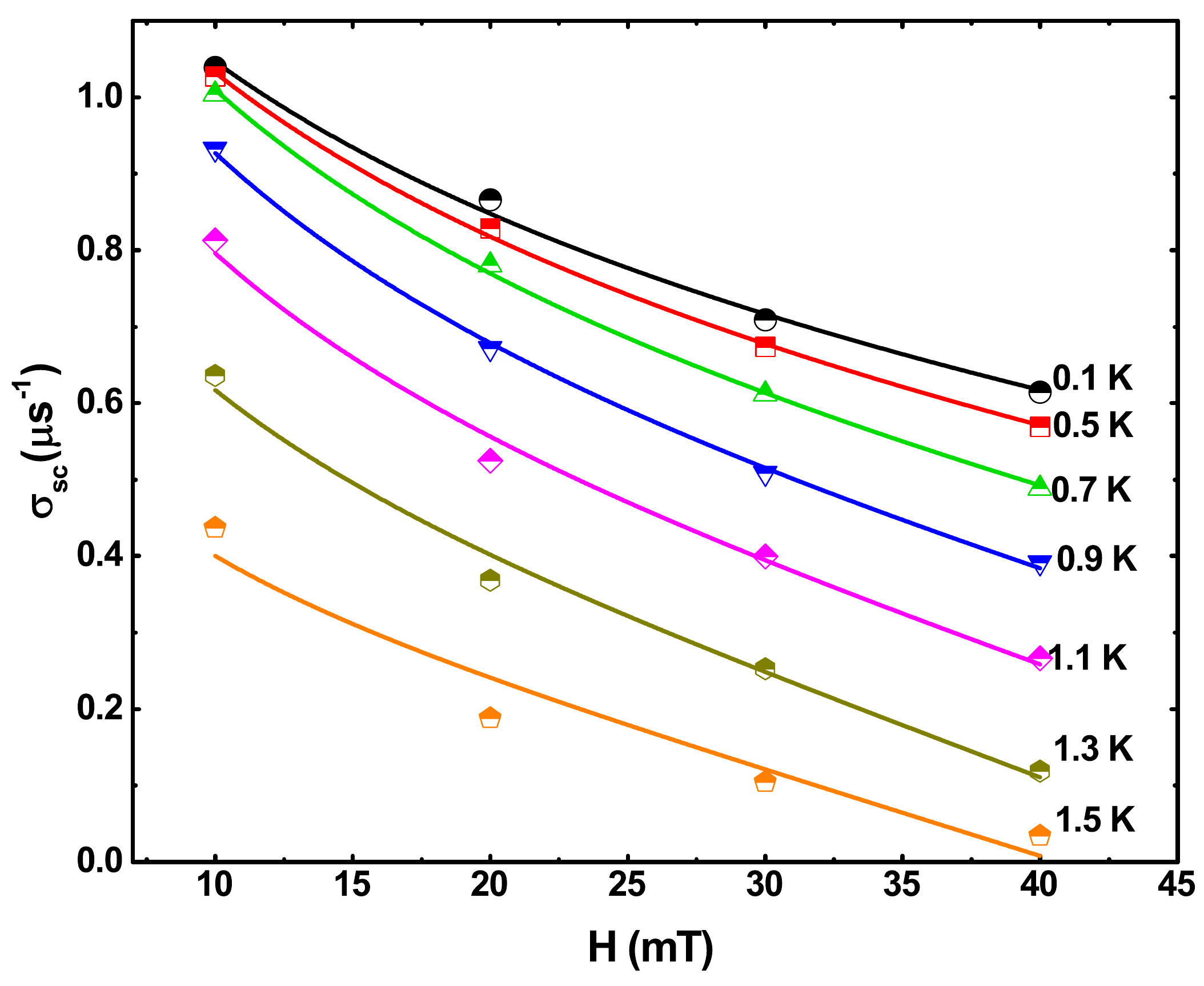}
\caption {The magnetic field dependence of the muon spin depolarization rate is shown for a range of different temperatures. The solid lines are the results of fitting the data using Brandt’s equation as discussed in Eq. \ref{eqn2}.}
\label{fig46}
\end{figure*}

To explore the pairing mechanism and gap structure of the superconducting state of ZrIrSi, TF-$\mu$SR measurements were performed down to 0.05 K. Fig.~\ref{fig43} (a)-(d) represent the TF-$\mu$SR asymmetry time spectra in presence of 10 mT and 30 mT applied magnetic field at above and below $T_{C}$.  Below $T_{C}$ the spectra depolarize strongly because of the inhomogeneous field distribution in the vortex state. TF-$\mu$SR data fitted using two Gaussian oscillatory, functions~\cite{bhattacharyya2018brief, adroja2017multigap, bhattacharyya2019evidence}:

\begin{equation}
G_{TF}(t) = \sum_{i=1}^{2} A_{i}\cos(\omega_{i}t+\phi)\exp(-\frac{\sigma_{i}^{2}t^{2}}{2})
\label{Eq41}
\end{equation}

where $A_{i}$, $\sigma_{i}$, $\omega_{i}$, $\phi$ is the initial asymmetry, Gaussian relaxation rate, muon spin precession frequency and the initial phase of the offset, respectively. In this fit $\sigma_{i}$  for the $2^{nd}$ part is equal to zero, which corresponds to the background term. This term comes from those muons which missed the sample and directly hit the silver sample holder and therefore the depolarization of this oscillating term is zero, i.e. $\sigma_{2}$ = 0 as silver has a minimal nuclear moment. $\sigma_{1}$ can be expressed as $\sigma_{1} = \sqrt{\sigma_{sc}^{2}+\sigma_{n}^2}$, where $\sigma_{sc}$ comes from superconducting part and $\sigma_{n}$ comes from nuclear magnetic dipolar moment which is fixed in the entire temperature range, supported by the ZF-$\mu$SR later. The total depolarization rate $\sigma_{T}$ as a function of temperature in different applied fields is shown in Fig. ~\ref{fig44}.\\
The temperature variation of $\sigma_{sc}$ is depicted in Fig.~\ref{fig45}. As $H_{C2}$ value is low in this sample, $\sigma_{{sc}}$ depends on the applied field as displayed in Fig.~\ref{fig46}. Brandt~\cite{brandt1988magnetic, brandt2003properties} has reported that for a superconductor with $H_{ext}/H_{C2}$ $\leq$0.25, $\sigma_{sc}$ is associated with the London penetration depth [$\lambda(T)$] by the following equation:

\begin{equation}
\sigma_{sc}[\mu s^{-1}] = 4.83 \times 10^{4}(1-H_{ext}/H_{C2}) \\ 
\times [1+1.21(1-\sqrt{{(H_{ext}/H_{C2}})} )^{3}]\lambda^{-2}[nm]
\label{eqn2}
\end{equation}

This equation is a good approximation for $\kappa \geq 5$, which is valid for our case as $\kappa = 40.5$ for ZrIrSi~\cite{kase2016}. From this relation we have determined the temperature dependence of $\lambda(T)$ and $\mu_{0}H_{C2}(T)$. Isothermal cuts perpendicular to the temperature axis of $\sigma_{{sc}}$ data sets were used to determine the $H$-dependence of the depolarization rate $\sigma_{{sc}}(H)$ as displayed in Fig.~\ref{fig46}. We have estimated the London penetration depths $\lambda$ = 254.4(3) nm, using $s$-wave model.

We have plotted the temperature variation of normalized $\lambda^{-2}(T)/\lambda^{-2}(0)$, which is directly proportional to the superfluid density. $\lambda^{-2}(T)/\lambda^{-2}(0)$ data were fitted using the following equation~\cite{prozorov2006magnetic, adroja2015superconducting, adroja2017nodal, das2018multigap, bhattacharyya2017superconducting}:

\begin{eqnarray}
\frac{\sigma_{sc}(T)}{\sigma_{sc}(0)} &=& \frac{\lambda^{-2}(T)}{\lambda^{-2}(0)}\\
 &=& 1 + \frac{1}{\pi}\int_{0}^{2\pi}\int_{\Delta(T)}^{\infty}(\frac{\delta f}{\delta E}) \times \frac{EdEd\phi}{\sqrt{E^{2}-\Delta(T,\phi})^2} \nonumber
\end{eqnarray}

here $f$ is the Fermi function which can be expressed as $f= [1+\exp(E/k_{B}T)]^{-1}$. $\Delta(T,\phi) = \Delta(0)\delta(T/T_{C})g(\phi)$ whereas $g(\phi)$ is the angular dependence of the gap function, $\phi$ is the azimuthal angel in the direction of FS. The temperature variation of the superconducting gap is approximated by the relation $\delta(T/T_{C}) = \tanh \{{1.82[1.018 (T_{C}/T -1)]^{0.51}} \}$. The spatial dependence $g(\phi$) is substituted by (a) 1 for $s-$wave gap, (b) $\vert\cos(2\phi)\vert$ for $d-$wave gap with line nodes.

\begin{figure*}
\centering
 \includegraphics[width=0.6\linewidth]{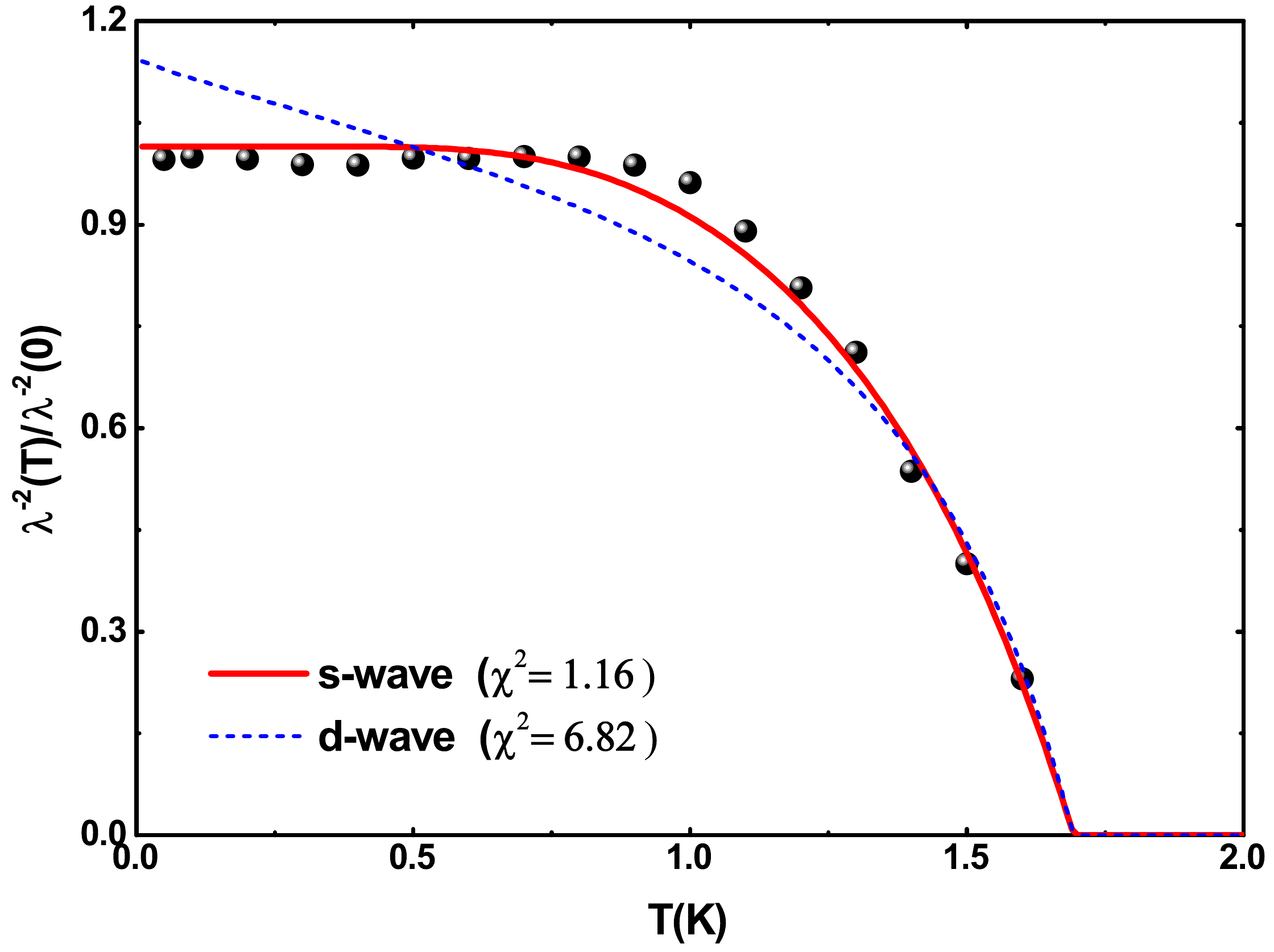}
\caption {The inverse magnetic penetration depth squared as a function of temperature is shown here. The lines show the fits using $s$-wave (solid) and $d$-wave (dashed) gap functions.}
\label{fig47}
\end{figure*}

Fig.~\ref{fig47} represents the fits to the $\lambda^{-2}(T)/\lambda^{-2}(0)$ data of ZrIrSi  using a single gap $s-$wave and nodal $d-$wave models. It is clear that the data can be well described by the isotropic $s-$wave model with a gap value 0.374 meV. This model gives a gap to $T_{C}$ ratio, $2\Delta(0)/k_{B}T_{C}$ = 5.10. The higher value of gap compare to BCS gap (3.53) suggest the presence of strong SO coupling. Similar high gap value was obtained for Ir-based superconductors, for example: IrGe [$2\Delta(0)/k_{B}T_{C}$ = 5.14]~\cite{cuamba2016, matthias1963}, CaIrSi$_{3}$  [$2\Delta(0)/k_{B}T_{C}$ = 5.4]~\cite{singh2014probing}. On the other hand $d-$wave model is clearly not suitable for this system as the $\chi^{2}$ value increased significantly for this fit ($\chi^{2}$ = 6.82). As ZrIrSi is a type II superconductor, supposing that approximately all the normal states carriers ($n_{e}$) contribute to the superconductivity ($n_{s} \approx$ n$_{e}$), superconducting carrier density $n_{s}$, the effective-mass enhancement $m^{*}$ have been estimated to be $n_{s}$ = 6.9(1) $\times$ 10$^{26}$ carriers $m^{-3}$, and $m^{*}$ = 1.474(3) $m_{e}$ respectively for ZrIrSi. Detail calculations can be found in Ref.~\cite{hillier1997classification,adroja2005probing, anand2014physical}.

\subsection{ZF-$\mu$SR measurements}
\begin{figure*}
\centering
 \includegraphics[width=0.6\linewidth]{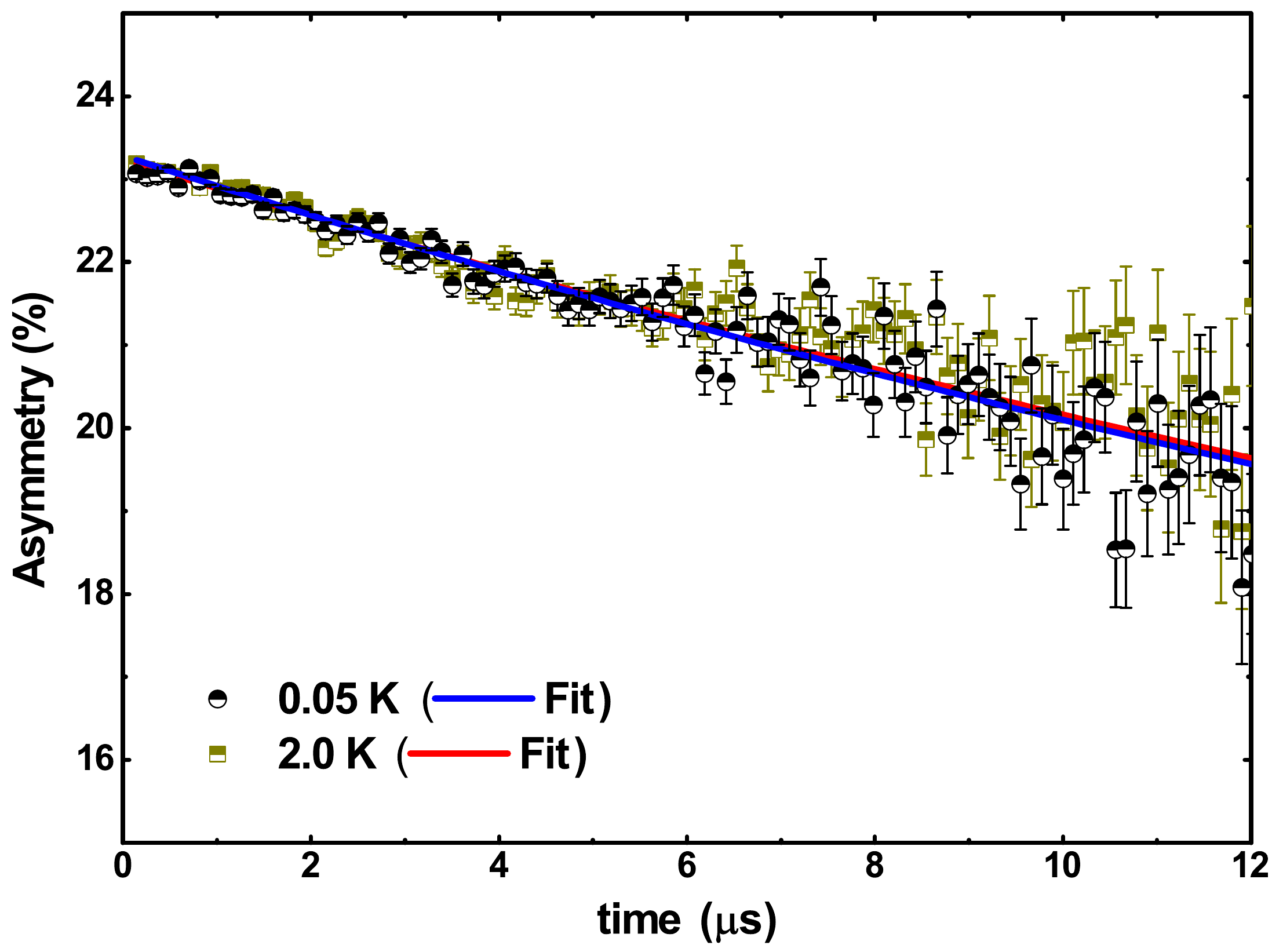}
\caption {ZF-$\mu$SR asymmetry time spectra for ZrIrSi at 0.05 K (black circles) and 2 K (dark yellow squares) are shown together. The lines are the least-squares fits to the data using Eq.~\ref{Eq4}. }
\label{fig48}
\end{figure*}

In order to investigate the pairing mechanism in the superconducting ground state, we used the ZF-$\mu$SR study. The time evolution of asymmetry spectra is shown in Fig.~\ref{fig48} for $T$ = 0.05 K $< T_{C}$ and $T$ = 2 K $> T_{C}$. The spectra below and above $T_{C}$ are found identical, ruling out the presence of any magnetic field. This reveals that the TRS is preserved in the superconducting state of ZrIrSi. This ZF data were fitted by a Lorentzian function: 

\begin{equation}
G_{ZF}(t) = A_{0}(t)\exp{(-\lambda t)}+A_{bg}
\label{Eq4}
\end{equation} 

$A_{0}$ and $A_{bg}$ are the sample and background asymmetry respectively, which are nearly temperature independent. $\lambda$ is the relaxation rate that comes from nuclear moments. The red and blue lines in Fig.~\ref{fig48} indicate the fits to the ZF-$\mu$SR data. The fitting parameters of the ZF-$\mu$SR asymmetry data are as follows: $\lambda$ = 0.030(9) $\mu {s}^{-1}$ at  0.05 K  and $\lambda$ = 0.026(3) $\mu {s}^{-1}$ at 2 K.  The change in the relaxation rate is within the error bar, indicating no clear evidence of TRS breaking in ZrIrSi.

\subsection{Summary and Theoretical Calculations}
In conclusion, we have performed ZF and TF-$\mu$SR measurements in the mixed state of ZrIrSi. Using Brandt's equation we have determined the temperature dependence of the magnetic penetration depth. The superfluid density $n_{s} \propto 1/\lambda^{2}$ well described by an isotropic $s-$wave model. The obtained gap value is $2\Delta(0)/k_{B}T_{C}$ = 5.1, which suggest ZrIrSi to be a strongly coupled BCS superconductor. {\it Ab-initio} electronic structure calculation indicates BCS superconductivity, which supports our experimental results. The low-energy bands are dominated by the $4d-$orbitals of the transition metal Zr, with a substantially lesser weight from the $5d-$orbitals of the Ir-atoms. ZF-$\mu$SR reveals, there is no spontaneous magnetic field below $T_{C}$, which suggest the absence of TRS breaking. The present results pave the way to develop a realistic theoretical model to interpret the origin of superconductivity in ternary systems.Ab initio electronic structure calculation indicates BCS superconductivity, which supports our experimental results. The low-energy bands are dominated by the 4d orbitals of the transition metal Zr, with a substantially less weight from the 5d orbitals of the Ir atoms.\footnote{Theoretical calculations have been performed by Prof. Tanmoy Das and Surabhi Saha from Department of Physics, Indian Institute of Science, Bangalore 560012, India}

ZrIrSi unit-cell has a $\it{mmm}$ point group symmetry and it belongs to the $\it{Pnma(62)}$ space group (orthorhombic crystal structure). We have used the Vienna Ab-initio Simulation Package(VASP) for ab-initio electronic structure calculation. The projector augmented wave (PAW) pseudo-potentials are used to describe the core electrons and for the exchange-correlation functional Perdew-Burke-Ernzerhof (PBE) form is used. We have used a local density application (LDA) functional with a cut-off energy for the plane wave basis set of 500 eV.  The Monkhorst-Pack $k$-mesh is set to $14\times14\times14$ in the Brillouin zone for the self-consistent calculation. The optimized lattice parameter by energy minimization were found to be as follows $a$ = 3.9643~$\AA$, $b$ = 6.5893~$\AA$, and $c$ = 7.4070~$\AA$~and $\alpha=\beta=\gamma=90^{0}$. To deal with the strong correlation effect of the $d$-electrons of the Ir atoms, we employed the LDA+U method with $U$ = 2.8 eV. For the Fermi surface calculation, we have used a larger $k$-mesh of $31\times31\times31$.

\begin{figure*}
\centering
 \includegraphics[width=0.6\linewidth]{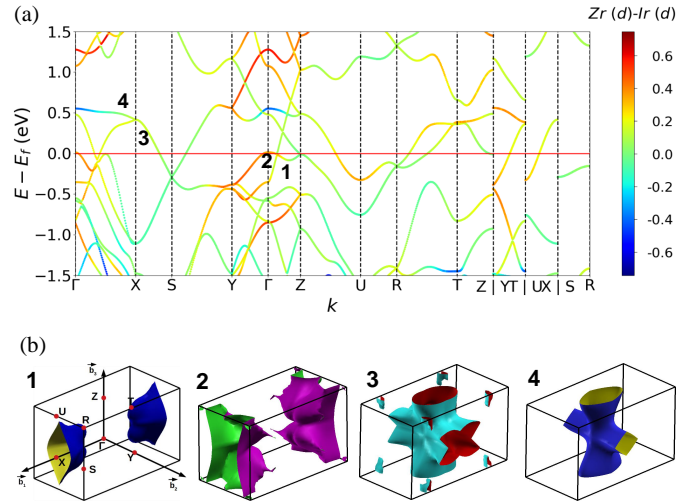}
 \caption{(a) Computed DFT band structure is plotted along with the standard high-symmetric directions for the orthorhombic crystal structure of ZrIrSi. The bands are colored with the difference in the $d$-orbital weight of the Zr and Ir-atoms, where red color gives stronger Zr-$d$ orbital weight while blue color dictates Ir-$d$ orbital weight. Four bands that pass through the Fermi level are indicated by 1,2,3,4 numbers. (b) Corresponding Fermi surfaces are plotted in the full three-dimensional Brillouin zone. We find that Fermi surface 1 and 2 form pockets around the X-point, while Fermi surface 3 and 4 constitute pockets centering the $\Gamma$-point. The colors on the Fermi surface only aid visualization.}
\label{DFT_Zr}
\end{figure*}

In Fig. \ref{DFT_Zr}, we show the band structure and Fermi surface plots.  We find that there are four bands passing through the Fermi level, forming four Fermi surface pockets. Two Fermi pockets are centered around the $\Gamma$-point, while the other two pockets are centered around the X-point. Unlike the multi-gap superconductivity in MgB$_2$ \cite{liu2001beyond} and Mo$_8$Ga$_{41}$\cite{sirohi2019multiband} which are driven by the presence of multiple Fermi surfaces, we do not find any evidence of multigap superconductivity in ZrIrSi, which is in agreement with the TF-$\mu$SR results. This is presumably because of the absence of $E_{2g}$-phonon mode which could enable inter-band scattering, while in the present case phonon modes cause intra-band electron-phonon coupling. Moreover, we observe substantial three-dimensionality in all four Fermi surfaces. This substantially weakens Fermi surface nesting strength. Thus the possibility of an inter-band nesting driven unconventional $s^{\pm}$-pairing symmetry is suppressed as compared to a two-dimensional iron-pnictide family with similar Fermi surface topology (see, e.g., Ref.~\cite{das2012pairing}).

Finally, we study the orbitals' contributions to the low-energy electronic states. We find that the low-energy bands are dominated by the $4d$-orbitals of transition metal Zr, with substantially lesser weight from the $5d$-orbitals of the Ir-atoms. Due to the presence of $d$-orbitals in the low-energy states, it is natural to anticipate the involvement of strong correlation in these systems and hence strong coupling superconductivity which changes from typical electron-phonon to quasiparticle-phonon mechanism within the Eliashberg theory\cite{das2015superconducting}. However, to our surprise, we find a substantially lower effective mass $1.5 m_e$ (where $m_e$ is the bare electron's mass) which is in remarkable agreement with the results from the TF-$\mu$SR measurements. This is well captured within our LDA+U calculation without essentially including dynamical correlations. We repeated the calculations for isostructural compounds TiIrSi and HfIrSi and find that the essential Fermi surface topology and three-dimensionality remain the same in all three materials (not shown). Therefore, we conclude that the superconductivity in ZrIrSi and its isostructural materials (such as TiIrSi and HfIrSi) can be well understood within the conventional BCS theory. Although our estimates of the BCS ratio of 5.1 is slightly higher than the BCS estimate of 3.5, however, we believe this slight increment is caused by the spin-orbit coupling of the Ir-atoms, and the Fermi surface anisotropy.

The present results pave the way to develop a realistic theoretical model to interpret the origin of superconductivity in ternary systems.

\newpage

\section{Studies of HfIrSi}

\subsection{Sample preparation}
 A polycrystalline sample of orthorhombic HfIrSi was prepared by melting stoichiometric quantities of high purity Hf, Ir, and Si on a water-cooled copper hearth, under an argon atmosphere in an arc furnace. The as-cast ingot was turned and remelted several times to improve the phase homogeneity. The sample was then annealed in an evacuated quartz tube for 168~hrs at 1273~K. Powder X-ray diffraction data were collected using a RAD-2X Rigaku X-ray diffractometer.

\subsection{Magnetization measurements}
\begin{figure*}
\centering
 \includegraphics[height = 0.5\linewidth, width=1.0\linewidth]{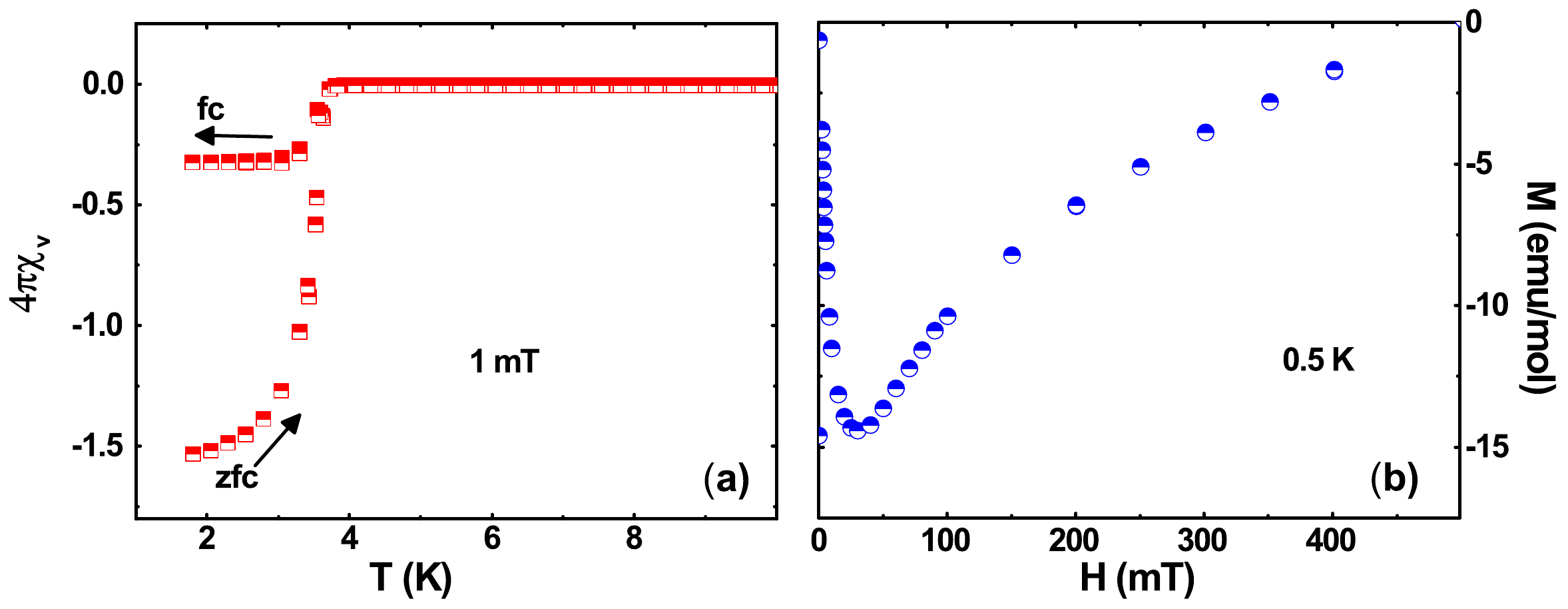}
\caption {(a) Magnetic susceptibility as a function of temperature, $\chi$(T), collected in zero-field-cooled (ZFC) and field-cooled (FC) modes in an applied field of 1 mT in SI units. (b) Isothermal field dependence of the magnetization of HfIrSi at 0.5 K.}
\label{fig49}
\end{figure*}

The superconducting properties of the sample were characterized via dc magnetic susceptibility measurements made using a Quantum Design, Magnetic Property Measurement System (MPMS), over the temperature range 1.5 to 10~K in an applied magnetic field 1 mT. The temperature ($T$) dependence of the magnetic susceptibility, $\chi(T)$, of HfIrSi in an applied magnetic field of 1 mT is shown in Fig.~\ref{fig49}(a). $\chi(T)$ reveals a clear signature of superconductivity below a superconducting transition temperature, $T_{C} = 3.6(1)$~K.  The magnetization, $M$, versus field, $H$, curve shown in Fig.~\ref{fig49}(b) at 0.5~K is typical of type II superconductivity. The lower critical field was estimated from the $M(H)$ curve at 0.5~K to be $\approx 1 $~mT. From the field dependence of the resistivity data~\cite{kase2016} the upper critical field $H_{{C2}}(0)$ was estimated to be 22.3(1)~kOe, while the Pauli paramagnetic limit 18.4$T_{C} = 66(2)$~kOe.

\subsection{Heat capacity measurement}
\begin{figure*}
\centering
 \includegraphics[width=0.6\linewidth]{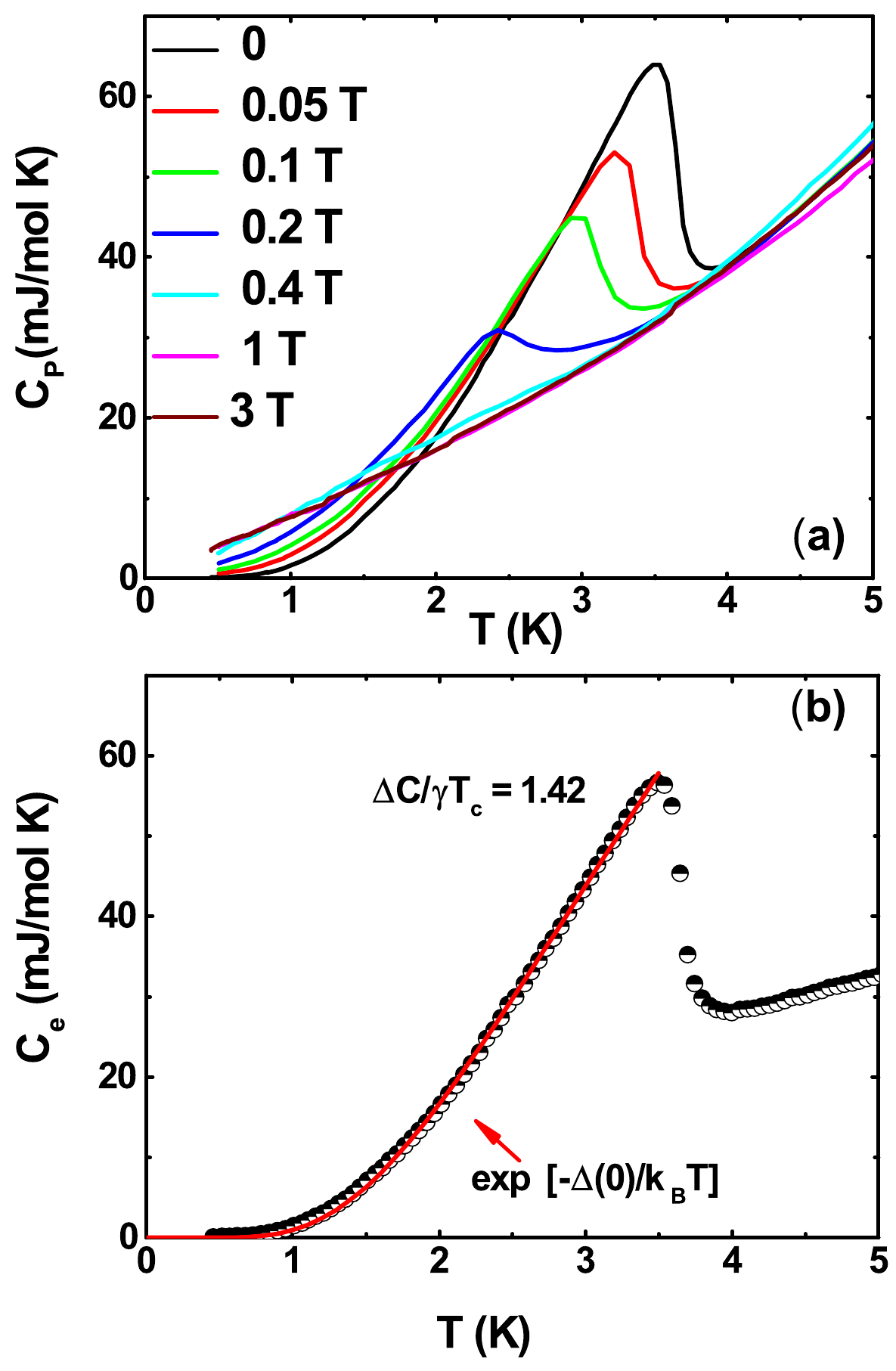}
\caption {(a) Temperature dependence of heat capacity, $C_{P}$(T), for 0.45 $\leq$ T$\leq$ 5 K measured in different applied magnetic fields. (b) Electronic contribution to the zero-field heat capacity, Ce, as a function of temperature. The solid line indicates a fit to $C_{e}$(T) made with an isotropic BCS expression.}
\label{fig410}
\end{figure*}

Heat capacity measurements were performed down to 0.3~K using a Quantum Design Physical Property Measurement System (PPMS) with a $^3$He insert. Heat capacity, $C_{{P}}$, as a function of temperature for $0.45 \leq T \leq 5$~K is shown in Fig.~\ref{fig410}(a) in different applied magnetic fields. In the normal state $C_{ P}\left(T\right)$ was found to be independent of the external magnetic field. Above $T_{{C}}$ in the normal state, the $C_{P}\left(T\right)$ data can be described using $C_{P}(T) = \gamma T + \beta T^{3}$, where $\gamma $ is the electronic heat capacity coefficient, and $\beta T^{3}$ is the lattice (phonon) contribution to the specific heat. Fitting gives $\gamma = 5.56(1)$~mJ mol$^{-1}$K$^{-2}$ and $\beta = 0.17(2)$~mJ mol$^{-1}$K$^{-4}$. Using the Debye model, the Debye temperature is given by $\Theta_{{D}} = (\frac{12\pi^{4}}{5\beta}nR)^{1/3}$, where $R = 8.314$~J mol$^{-1}$K$^{-1}$ is the gas constant and $n = 3$ is the number of atoms per formula unit in HfIrSi. Using this relationship, $\Theta_{{D}}$ is estimated to be 325(12)~K. The jump in the heat capacity $\Delta C_{P}(T_{{C}}) = 28.5(1)$~mJ mol$^{-1}$K$^{-1}$ and $T_{C} = 3.6(1)$~K, yields $\Delta C/\gamma T_{C} = 1.42(5)$~\cite{kase2016}. This value is close to 1.43 expected for weak-coupling BCS superconductors~\cite{Tinkham}.

Fig.~\ref{fig410}(b) shows the temperature dependence of the electronic specific heat, $C_{e}(T)$ , obtained by subtracting the phonon contribution from $C_{P}(T)$. $C_{e}(T)$  can be used to investigate the superconducting gap symmetry. From the fit to the exponential temperature dependence of $C_{e}(T)$  shown in fig. ~\ref{fig410}(b), we find $\Delta (0) = 0.50(2)$  meV which is close to 0.51(1) meV obtained from the analysis of the TF-$\mu$ SR data presented below. $\Delta (0) = 0.50(2)$  meV gives 2$\Delta(0)/k_{B}T_{ C} = 3.2(2)$ , which is close to the value of 3.53 expected for weak-coupling BCS superconductors~\cite{BCS}.

\subsection{TF-$\mu$SR measurements}
\begin{figure*}
\centering
 \includegraphics[width=0.5\linewidth]{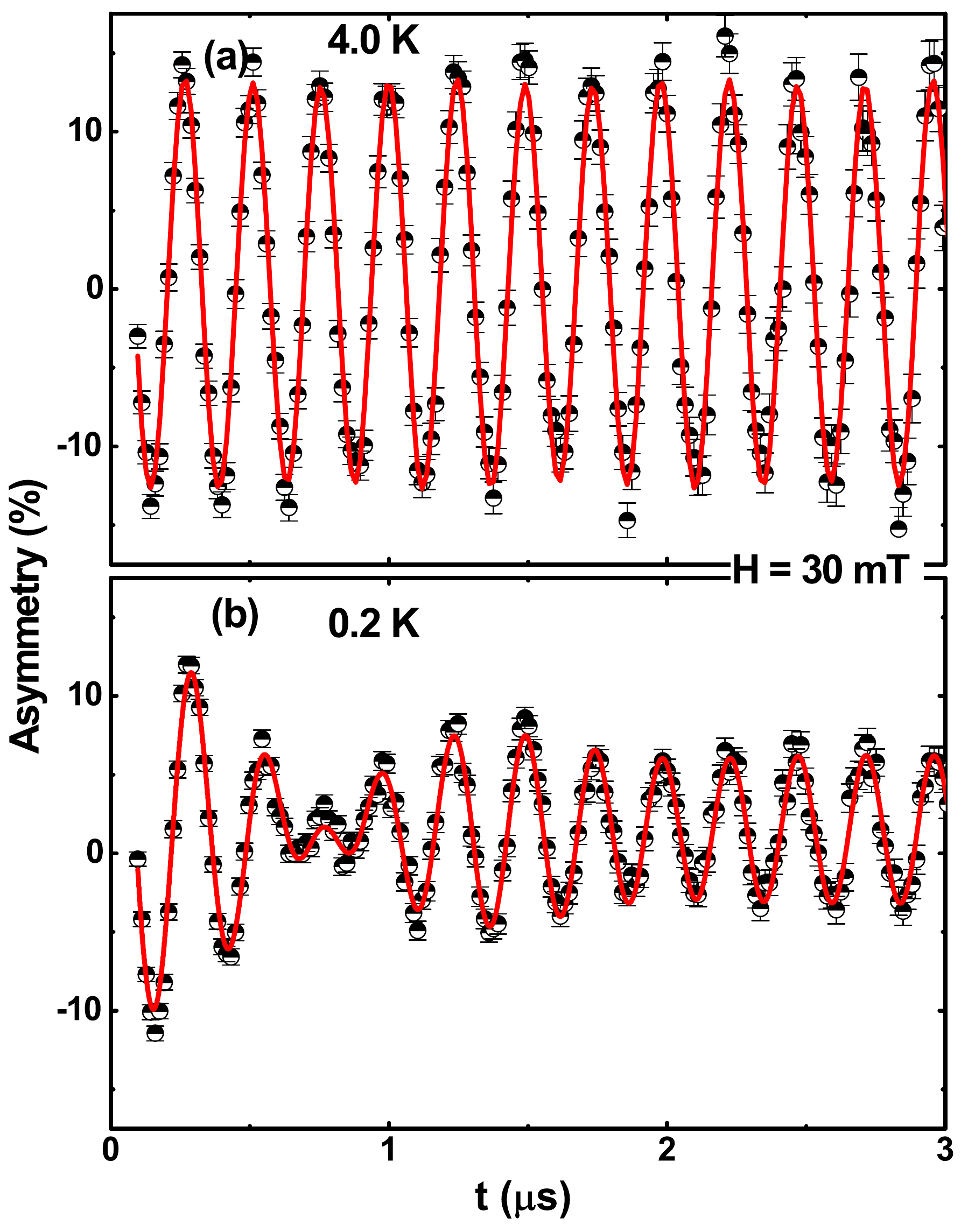}
\caption {Time evolution of TF-$\mu$SR asymmetry spectra for HfIrSi recorded at (a) T = 4.0 K and (b) T = 0.2 K in a transverse magnetic field H = 30 mT. The solid red line is a fit to the data using equation \ref{GaussRel} as described in the text.}
\label{fig411}
\end{figure*}

To investigate the superconducting gap structure in $o$-HfIrSi, we have performed TF-$\mu$SR measurements. Figs.~\ref{fig411}(a) and~\ref{fig411}(ab display the TF-$\mu$SR asymmetry spectra taken at temperatures above and below $T_{{C}}$ in an applied magnetic field of 300~Oe. The presence of a flux-line lattice (FLL) in the superconducting state results in an inhomogeneous field distribution within the sample, which in turn induces a faster decay in the asymmetry spectra below $T_{C}$ (Figs.~\ref{fig411}(b)). The time evolution of the TF-$\mu$SR data at all temperatures above and below $T_{{C}}$ is best described by a sinusoidal oscillatory function damped with a Gaussian relaxation and an oscillatory background term~\cite{bhattacharyya2018brief, adroja2017multigap, bhattacharyya2019evidence, anand2014physical}:

\begin{equation}
G_{z1}(t) = A_{1}\cos(\omega_{1}t+\varphi)\exp(-\frac{\sigma^{2}t^{2}}{2})+A_{2}\cos(\omega_{2}t+\varphi).
\label{GaussRel}
\end{equation}

\begin{figure*}
\centering
 \includegraphics[width=0.6\linewidth]{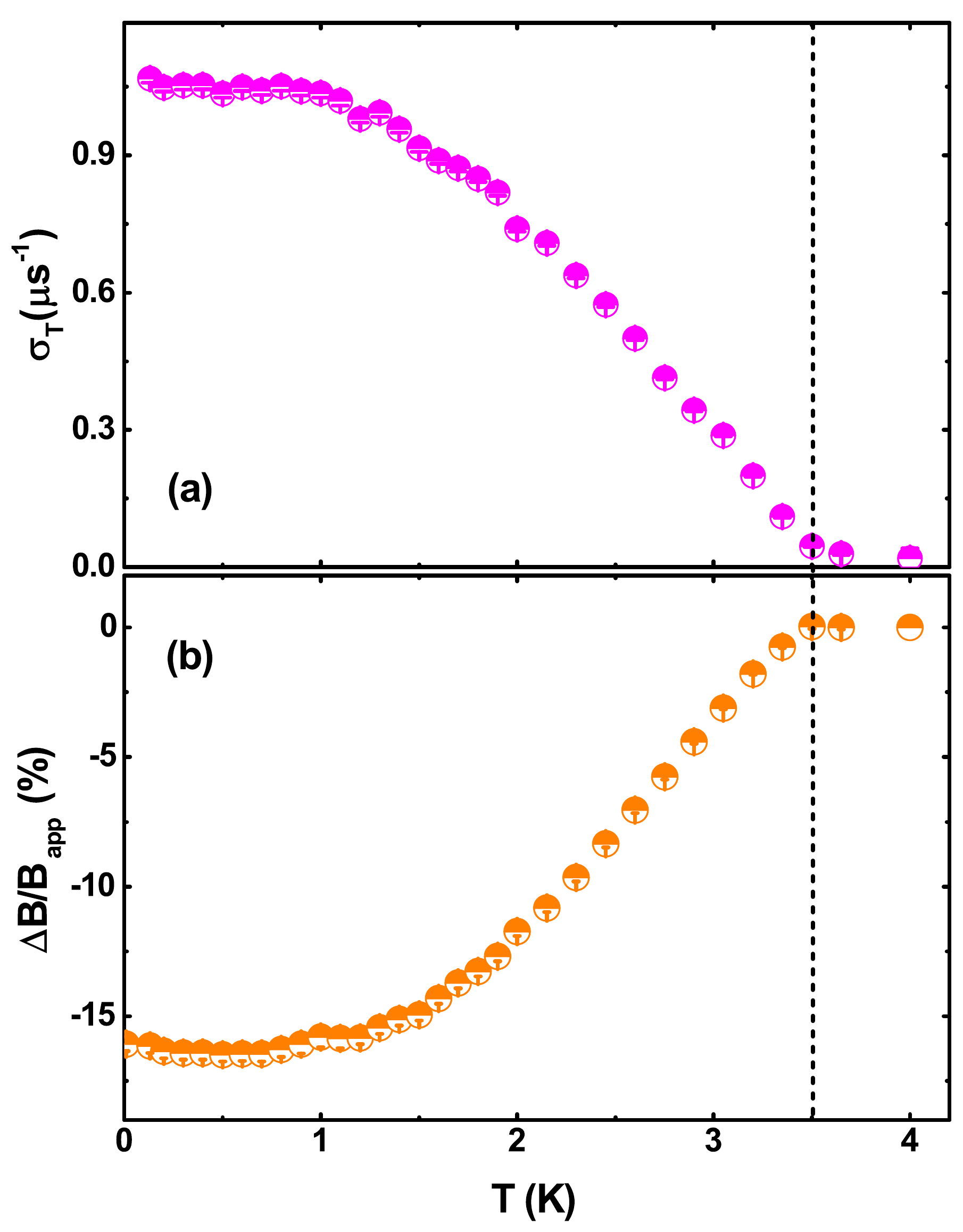}
\caption {(a) Temperature dependence of the total depolarization rate $\sigma_{T}(T)$ in the presence of an applied magnetic field of 30 mT. (b)  The relative change of the internal field normalized to external applied field as a function of temperature, where $\Delta B = B_{SC} - B_{app}$.}
\label{fig412}
\end{figure*}

Here $A_{{1}}$= 64.20\% and are the initial asymmetry of the sample while $A_{{2}}$ = 35.80\% is a background that arises from the muons implanted directly into the silver sample holder that do not depolarize, $\omega_{{1}}$ and $\omega_{{2}}$ are the muon precession frequencies within the sample and the sample holder, respectively, $\varphi$ is an initial phase of the offset, and $\sigma$ is the total muon spin relaxation rate. The temperature dependence of total depolarization rate ($\sigma_{T}$) is shown in fig. \ref{fig412} (a). The field shift is $\Delta B = B_{SC} - B_{app}$, where $B_{SC}$ indicates the superconducting field induced by the vortex lattice and $B_{app}$ is the applied field, as shown in fig.~\ref{fig412}(b). As the sample goes through the transition into the superconducting state, there is a strong negative shift in the peak field, which is a unique characteristic of the vortex lattice. $\sigma$ consists of two contributions: one is due to the inhomogeneous field variation across the superconducting vortex lattice, $\sigma_{sc}$, and the other is a normal state contribution, $\sigma_{{n}}$ = 0.029 $\mu s^{-1}$, which is taken to be temperature independent over the entire temperature range studied and was obtained from spectra measured above $T_{C}$. Using $\sigma^{2}$ = $\sigma_{sc}^{2}$ + $\sigma_{n}^2$ we obtain the superconducting contribution $\sigma_{sc}$. The temperature variation of the penetration depth/superfluid density was modeled using~\cite{prozorov2006magnetic, adroja2015superconducting, adroja2017nodal, das2018multigap, bhattacharyya2017superconducting}

\begin{eqnarray}
\frac{\sigma_{sc}(T)}{\sigma_{sc}(0)} &=& \frac{\lambda^{-2}(T,\Delta_{0,i})}{\lambda^{-2}(0,\Delta_{0,i})},\\ \nonumber
 &=& 1 + \frac{1}{\pi}\int_{0}^{2\pi}\int_{\Delta(T)}^{\infty}(\frac{\delta f}{\delta E}) \times \frac{EdEd\phi}{\sqrt{E^{2}-\Delta(T,\phi})^2}, 
\end{eqnarray}

where $f= [1+\exp(E/k_{B}T)]^{-1}$ is the Fermi function and $\Delta(T,\phi) = \Delta(0)\delta(T/T_{C}){g}(\phi)$. $\Delta(0)$ is the value of superconducting gap. The temperature dependence of the superconducting gap is approximated by the relation $\delta(T/T_{C}) = \tanh[1.82[1.018(T_{C}/T-1)]^{0.51}]$ where ${g}(\phi$) refers to the angular dependence of the superconducting gap function. ${g}(\phi$) is replaced by (a) 1 for an $s$-wave gap, and (b) $\vert\cos(2\phi)\vert$ for a $d$-wave gap with line nodes \cite{pang2015evidence, annett1990symmetry}. The data is best modeled using a single isotropic $s$-wave gap of 0.51(1)~meV, which yields a gap to $T_{C}$ ratio, 2$\Delta/k_{B}T_{C} = 3.38(2)$ that is very close to the 3.3(2) obtained from the heat capacity data presented earlier, and indicates weak-coupling superconductivity in HfIrSi. The muon spin depolarization rate attributable to the superconducting state ($\sigma_{sc}$) is related with penetration depth via $\sigma_{sc}(T) = 0.0431\times\frac{\gamma_{\mu}\phi_{0}}{\lambda^{2}(T)}$, where $\phi_0 = 2.609 \times 10^{-15}$ Wb is the magnetic flux quantum. This gives $\lambda_{L}(0)$ = 259(4)~nm for the $s$-wave fit. The London model provides a direct relation between $\lambda(T)$ and ($m^{*}/n_{s}$)($\lambda_{L}^2=m^{*}c^{2}/4\pi n_{s} e^2)$ where m$^{*} = (1+\lambda_{{e-ph}})m_{{e}}$ is the effective mass in units of the electron rest mass $m_{{e}}$, and $n_{{s}}$ the carrier density. $\lambda_{{e-ph}}$ is calculated from $\Theta_{D}$ and $T_{C}$ using McMillan equation $\lambda_{e-ph} = \frac{1.04+\mu^{*}\ln(\Theta_{D}/1.45T_{C})}{(1-0.62\mu^{*})\ln(\Theta_{D}/1.45T_{C})-1.04}$. The superconducting carrier density is then estimated to be $n_{s} = 6.6(1) \times 10^{26}$ carriers m$^{-3}$ and the effective-mass enhancement $m^{*} = 1.57(3)m_{e}$, for HfIrSi. Details of similar calculations can be found in Refs.~\cite{brandt2003properties, chia2004probing, amato1997heavy, mcmillan1968transition, bhattacharyya2019, das2018multigap} 
\begin{figure*}
\centering
 \includegraphics[width=0.6\linewidth]{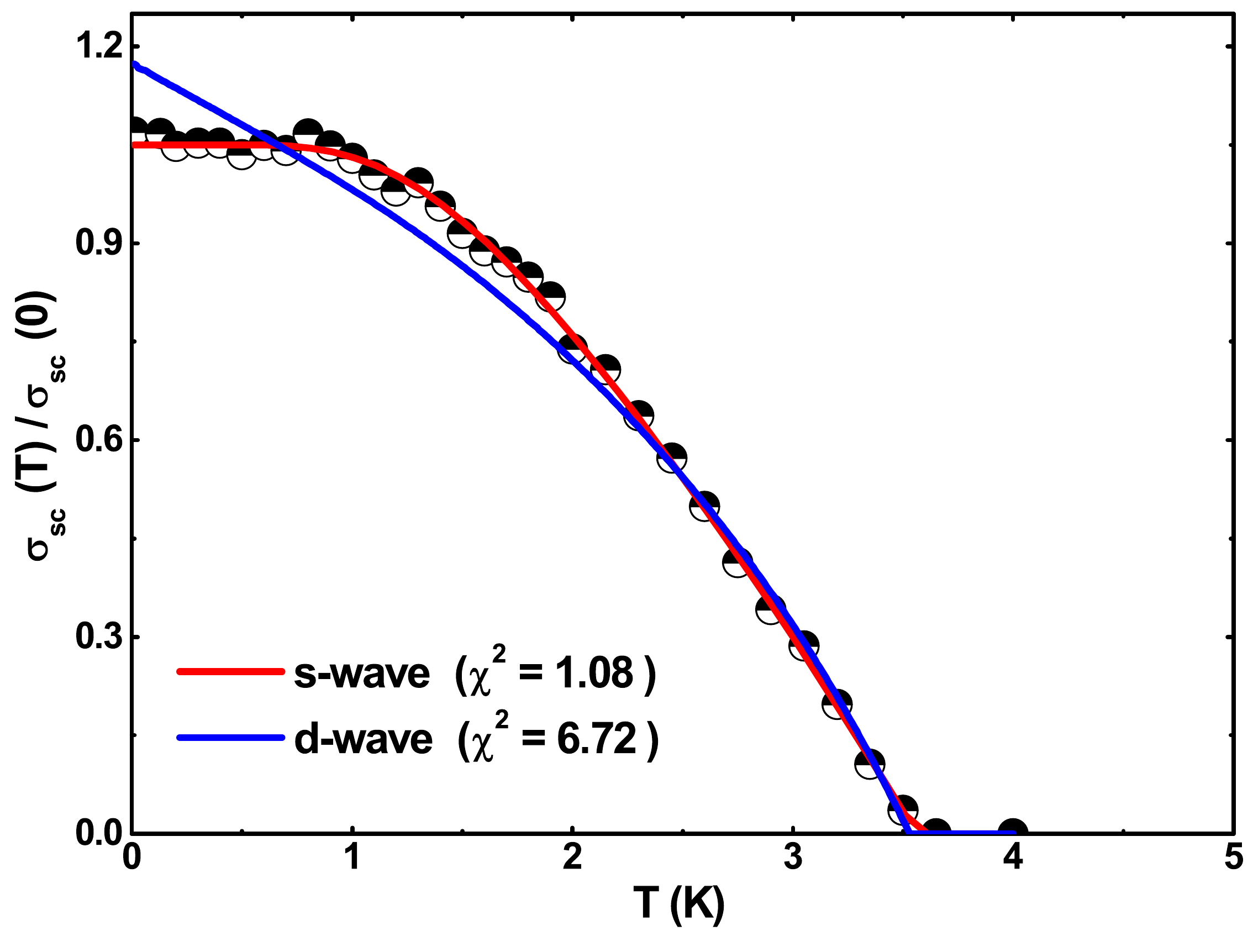}
\caption {Temperature dependence of the normalized superconducting depolarization rate $\sigma_{sc}(T)/\sigma_{sc}(0)$ in the presence of an applied magnetic field of 30 mT.}
\label{fig413}
\end{figure*}

\subsection{ZF-$\mu$SR measurements}
\begin{figure*}
\centering
 \includegraphics[width=0.6\linewidth]{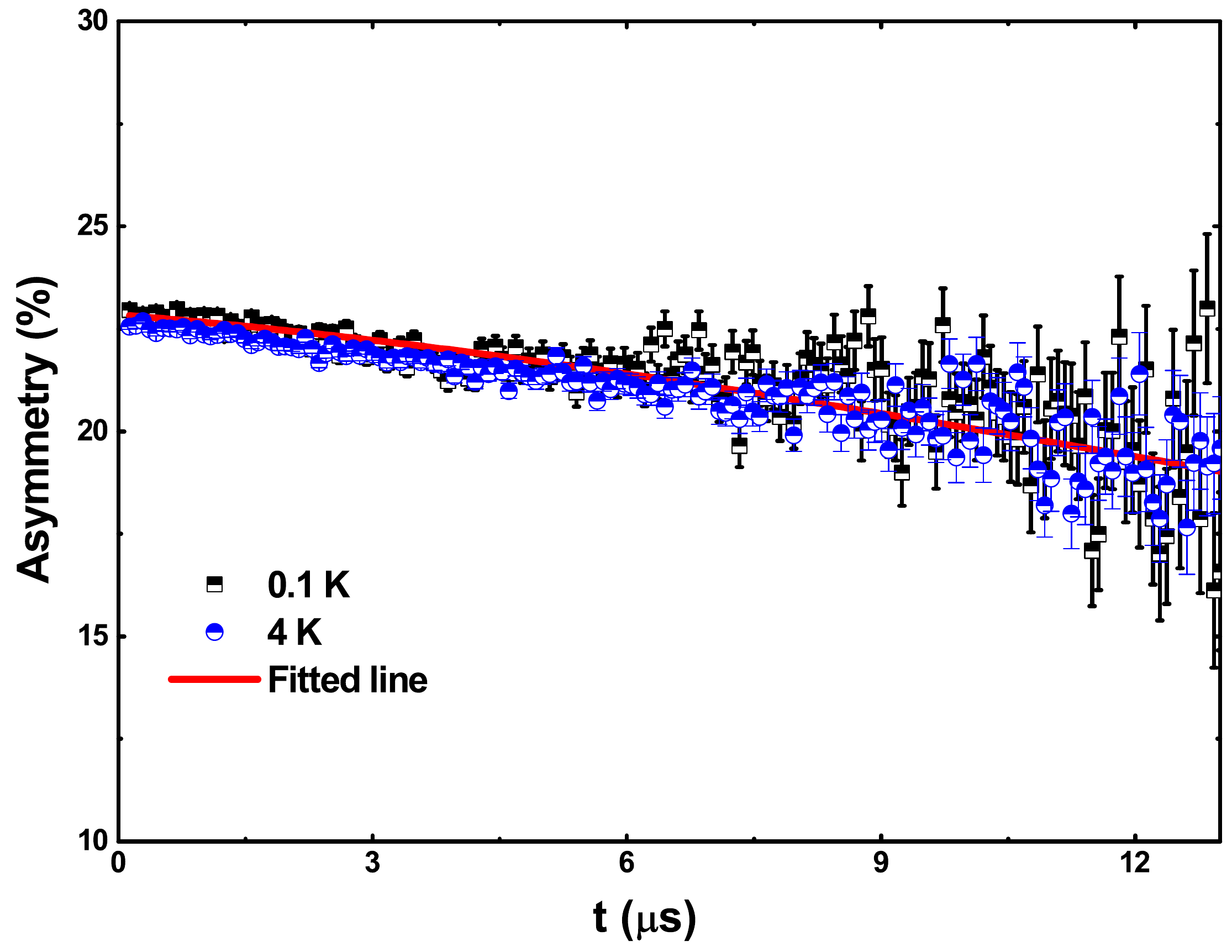}
\caption {Zero-field $\mu$SR time spectra for HfIrSi collected at 0.1 K (green squares) and 4 K (blue circles) are shown together with a line that is a least squares fit to the data collected at 4 K using equation \ref{GKTfunction}(please see text for details).}
\label{fig414}
\end{figure*}

ZF-$\mu$SR measurements were performed to check for the appearance of spontaneous magnetic fields in the superconducting state of HfIrSi. The time evolution of the zero-field asymmetry spectra above (4~K) and below (0.1~K) $T_{C}$ are shown in Fig.~\ref{fig414}. The ZF-$\mu$SR data are well described using a damped Gaussian Kubo-Toyabe (KT) function, and a background term arising from muons that miss the sample and stop in the silver sample plate, 

\begin{equation}
G_{z2}(t) = A_{3}G_{KT}(t)e^{-\lambda_{\mu}t}+A_{bg}.
\label{GKTfunction}
\end{equation}

Here $G_{KT}(t)$ is the Gaussian Kubo-Toyabe function given by $G_{KT}(t) = [\frac{1}{3}+\frac{2}{3}(1-\sigma_{KT}^{2}t^{2})\exp({-\frac{\sigma_{KT}^2t^2}{2}})]$. $A_{3}$ and $A_{bg}$ are the asymmetries arising from the sample and background, respectively. $\sigma_{KT}$ and $\lambda_{\mu}$ are the muon spin relaxation rates due to randomly oriented nuclear moments. The fitting parameters $A_{3}$, $A_{bg}$ are independent of temperature. Fits to the ZF-$\mu$SR asymmetry data using Eq.~\ref{GKTfunction} are shown by the solid lines in Fig.~\ref{fig414} and give $\sigma_{KT} = 0.068(1) ~\mu {s}^{-1}$ and $\lambda_{\mu} = 0.0046(2) ~\mu {s}^{-1}$ at 0.1~K and $\sigma_{KT} = 0.064(1) ~\mu {s}^{-1}$ and $\lambda_{\mu} = 0.0046(1) ~\mu {s}^{-1}$ at 4~K. The  values of $\sigma_{KT} $ and  $\lambda_{\mu}$ above and below $T_{C}$ agree to within error, indicating that TRS is preserved in the superconducting state of HfIrSi.

\subsection{Uemura Classification}
The Uemura classification scheme~\cite{uemura1989universal, hillier1997classification} correlates the $T_{C}$ and the effective Fermi temperature, $T_{F}$ ( $=\frac{\hbar^{2}(3\pi^{2})^{2/3}n_{s}^{2/3}}{2k_{B}m^{*}}$) of a superconductor. The values of $n_{s}$ and $m^{*}$ have been estimated from the TF-$\mu$SR data. In this classification unconventional superconductors lie between $1/10\geq (T_{C} /T_{F}) \ge 1/100$, while for conventional BCS superconductors $T_{C} / T_{F} \leq 1/1000$.  The position of HfIrSi and ZrIrSi indicates it is a conventional superconductor with its $T_{C}/T_{F}$ value of 0.00155 for HfIrSi and 0.000758 for ZrIrSi.

\begin{table}
 \caption{Comparison of superconducting parameters of ZrIrSi and HfIrSi}
 \centering
  \begin{tabular}{lcc}
\hline \hline
    Parameters & ZrIrSi &  HfIrSi \\  \hline
    Crystal group & Pnma (no. 62) & Pnma (no.62)\\~\\
    Lattice constants (\AA) & a = 6.557 & a= 6.523 \\~\\
     Lattice constants (\AA) & b = 3.942 & b = 3.912 \\~\\
      Lattice constants (\AA) &  c = 7.413 &  c = 7.353 \\~\\
    Cell volume (\AA$^{3}$) & 191.0 & 187.3 \\~\\
    T$_{C}$ (K)     & 1.7 & 3.6 \\~\\
    H$_{C}$ (mT) &  10.8 & 21.8 \\~\\
     H$_{C1}$(mT)  & 0.7 &  1 \\~\\
  H$_{C2}$(T) &  0.6 &  2.23\\~\\
$\gamma$(0) (mJ/mol K$^{2}$) &  7.63 & 5.56  \\~\\
$\Theta_{D} (K) $ & 368 & 325  \\~\\
$\Delta C/\gamma T_{C}$ & 1.4 &  1.42 \\~\\ 
2$\Delta$/$k_{B} T_{C}$ & 5.10 &  3.2 \\~\\
$\lambda$(nm) & 254.4 & 259 \\~\\
$\lambda_{e-ph}$ & 0.464  & 0.545 \\~\\
$n_{s}$(carriers/m$^{3})$ & 6.9 $\times$ 10$^{26}$ & 6.6 $\times$ 10$^{26}$ \\~\\
\hline \hline

  \end{tabular}

  \label{tab:41}
\end{table}
\subsection{Summary and Theoretical Calculations}
We have performed magnetization, heat capacity, and ZF and TF-$\mu$SR measurements at temperatures above and below $T_{C}$ on the orthorhombic 111 superconductor HfIrSi. The heat-capacity and magnetization data confirm bulk superconductivity in this material with a $T_{{C}} = 3.6(1)$~K. The temperature dependence of muon depolarization rate, $\sigma_{{sc}}(T)$, determined from the TF-$\mu$SR data collected in the field-cooled mode, is the best fit by an isotropic $s$-wave model. The value of $2\Delta(0)/k_{B}T_{{C}} = 3.38(2)$ obtained from the $s$-wave gap fit, suggests weak-coupling BCS-type superconductivity in HfIrSi. The ZF-$\mu$SR measurements reveal no sign of any spontaneous field appearing below $T_{C}$ which suggests that TRS is preserved in HfIrSi. Electronic structure calculations suggest that the Hf and Ir atoms in HfIrSi hybridize strongly along the c-axis, and that this hybridization is responsible for the strong three dimensionalities of the system which screens the Coulomb interaction. As a result, despite the presence of d-electrons in this system, the correlation effects are weakened, meaning that the electron-phonon coupling gains in importance \footnote{Theoretical calculations have been performed by Prof. Tanmoy Das and Surabhi Saha from the Department of Physics, Indian Institute of Science, Bangalore 560012, India}. 
\begin{figure*}
\centering
 \includegraphics[width=0.6\linewidth]{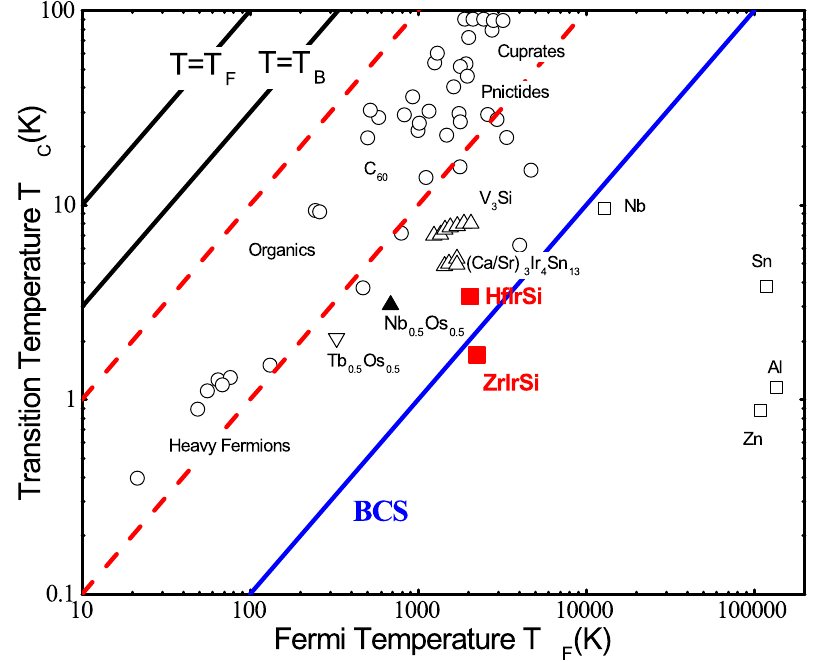}
\caption {The superconducting transition temperature, $T_{C}$, versus the effective Fermi temperature, $T_{F}$, where the position of HfIrSi and ZrIrSi is shown by the solid red square. The unconventional superconductors fall within a band indicated by the red dashed lines for which 1/10 $\geq$ $(T_{C}/T_{F})$ $\geq$ 1/100.}
\label{fig415}
\end{figure*}

The Vienna Ab-initio Simulation Package (VASP)~\cite{kresse1993ab} was used to perform density-functional theory (DFT) electronic structure calculations on HfIrSi. The projected augmented wave (PAW) pseudopotentials were used to describe the core electrons, with the Perdew-Burke-Ernzerhof (PBE) functional~\cite{perdew1996generalized} used for the exchange-correlation potential. The cut-off energy for the plane-wave basis set was fixed at 500~eV. The Monkhorst-Pack $k$-mesh was set to $14\times14\times14$ in the Brillouin zone for the self-consistent calculations.\\

\begin{figure*}
\centering
 \includegraphics[width=0.6\linewidth]{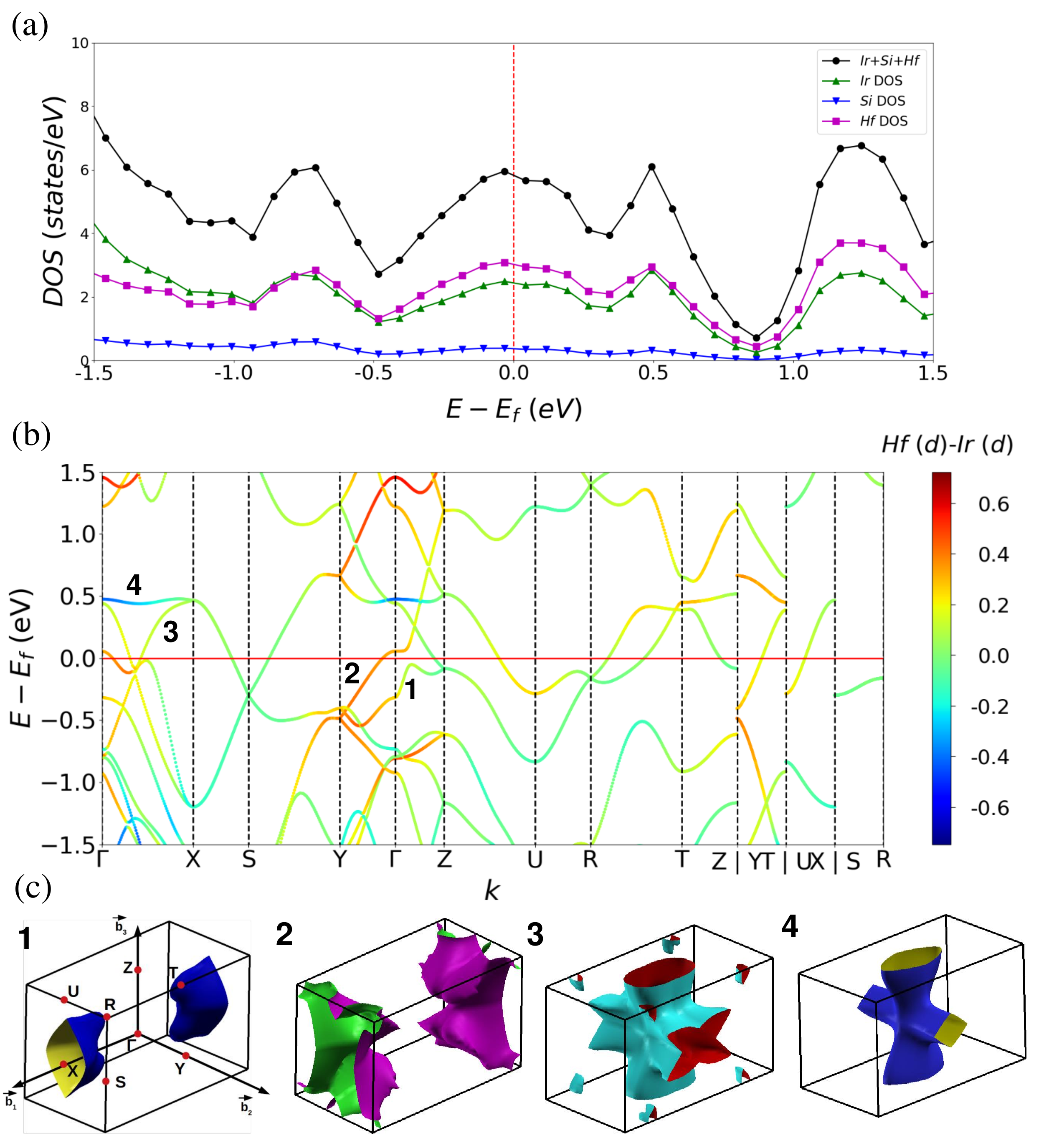}
 \caption{(a) Computed partial DOS for Hf, Ir, and Si atoms, along with the total DOS. We clearly observe that the Hf and Ir atoms contribute dominantly to the low-energy spectrum. For both these atoms, outermost d-orbitals contribute mostly to this energy scale. (b) DFT calculation of the band structure along with the standard high-symmetry directions for the orthorhombic crystal structure of HfIrSi. The band structure is coloured with a gradient colour map where blue indicates Ir-d orbitals and red shows the Hf-d orbitals. (c) Corresponding four Fermi surfaces in the 3D Brillouin zone. The colours on the Fermi surfaces have no significance here. Four Fermi surfaces are denoted by ‘1’, ‘2’, ‘3’, and ‘4’. The result indicates that the bands near the `Z'-point are dominated by Hf-d states while those near the ‘Z’ point gain more weight from the Ir-d states.}
\label{DFT}
\end{figure*}

HfIrSi is orthorhombic ($Pnma$($62$) space group) with $mmm$ point group symmetry. The relaxed lattice parameters obtained are $a=3.944$~\AA, $b=6.499$~\AA, $c=7.376$~\AA~ and $\alpha=\beta=\gamma=90^{\circ}$. Typically, it is observed that band structure calculated with relaxed coordinates give better agreement with experiments than experimental one. It is also true that the difference between the relaxed coordinates and experimental ones is very small and the corresponding band structures are not significantly different. To deal with the strong correlation effects of the $d$-electrons of the Ir atoms, the LDA+U method~\cite{liechtenstein1995density, dudarev1998electron} was employed with $U=2.8$~eV. For the Fermi surface and density of states calculations, a larger $k$-mesh of $31\times31\times31$ was used. The calculations were repeated with SOC but this produced no significant change in the low-energy spectrum.\\

Due to the involvement of the strongly correlated transition metals as well as possible SOC, one may anticipate that the superconductivity in HfIrSi may be exotic. However, the observation of conventional, time-reversal invariant superconductivity leads to a question: how does a phonon mediated attractive potential win out over the strong Coulomb interaction to give conventional superconductivity? Our investigation favors a conventional pairing symmetry based on the observations of fully gapped superconductivity and weak correlation strength in this material.\\

To address this point, we investigate the DFT band structure as shown in Fig.~\ref{DFT}. From the partial-DOS shown in Fig.~\ref{DFT}(a), we find that the transition metals Ir and Hf have nearly equal weight in the low-energy spectrum, and contribute most to the total DOS. Additionally, both transition metals have the corresponding $4d$ and $5d$ orbitals, respectively providing the dominant contribution to the Fermi surfaces. The result indicates that these two atoms both undergo strong hybridization in this system. This is confirmed by the visualization of the orbital weight distributions of the electronic structure as shown in Fig.~\ref{DFT}(b). We plot here the difference in the orbital weights between the Ir-$d$ and Hf-$d$ orbitals. We indeed find that the bands near the $\Gamma$-point are dominated by the Hf-$d$ orbitals while the Ir-$d$  atoms contribute strongly to those near Z (on the $k_z=\pm \pi$-plane). This result indicates that the Hf and Ir atoms hybridize rather strongly along the $c$-axis of the lattice. This hybridization is responsible for the strong three-dimensionality of this system which screens the Coulomb interaction. As a result, despite the presence of  $d$ electrons in this systems, the correlation effect is weakened, leading to the electron-phonon coupling increasing in importance~\cite{chen2021emergence}.\\

The Fermi surface topologies are shown in Fig.~\ref{DFT}(c) confirm the strong three-dimensionality in all four bands.  We consistently find two Fermi pockets near the $\Gamma$-points and two Fermi pockets centering on the X-points. The large Fermi surface volume is consistent with the high carrier density of this system as measured in the $\mu$SR experiments, and the higher value of the Fermi temperature presented in Fig.~3. These results are also consistent with the weak correlation strength in this system~\cite{das2014intermediate}. Strong three-dimensionality can also reduce relativistic effects, weakening the SOC.

To date, a large number of equiatomic $TrT'X$ compounds have been discovered with high superconducting transition temperatures and high critical magnetic fields, but $\mu$SR investigations have been carried out on just a few of these 111 compounds. The present study provides a valuable comparison for future $\mu$SR investigations on this family of compounds. The present results will also help in the development of realistic theoretical models, including the role of strong spin-orbit coupling, that explain the origin of superconductivity in HfIrSi, and also may help us arrive at empirical criteria for the occurrence of superconductivity with strong SOC, high $T_c$, and $H_{{C2}}$ in other $TrT'X$ equiatomic ternary systems.


%% file: chapter5.tex
\chapter{Nodal line in the superconducting gap symmetry of noncentrosymmetric ThCoC$_{2}$} 
\label{chapter:5}
\section{Introduction}
In recent years, noncentrosymmetric (NCS) superconductors have attracted considerable attention in condensed matter physics, both theoretically and experimentally ~\cite {smidman2017, bauer2012}. These NCS compounds lack inversion symmetry, which generates an asymmetric electrical field gradient in the crystal lattice and, thereby, produces a Rashba-type antisymmetric spin-orbit coupling (ASOC, energy scale $\sim$ 10-100 meV)~\cite{rashba1960properties,frigeri2004superconductivity,samokhin2004magnetic,fujimoto2007electron}.  In presence of  ASOC, the spin of a Cooper pair is not a good quantum number, the spin degeneracy of the conduction band is lifted and hence allows for the admixture of spin-singlet and spin-triplet states in Cooper pairs formation~\cite{frigeri2004superconductivity, samokhin2004magnetic}. If the triplet part is large~\cite{frigeri2005superconductors}, the superconducting gap reveals line or point nodes in the order parameter. The discovery of unconventional superconductivity (USc) in CePt$_3$Si with $T_C$ = 0.75 K, has spurred theoretical and experimental works to find out the effect of the lack of inversion symmetry on NSC~\cite{bauer2004heavy,yogi2004evidence,bonalde2005evidence}. These compounds orders antiferromagnetically below 2.25 K, several unconventional properties such as large anomalous width of the superconducting transition. The upper critical field for this compound is also more than the Pauli limit which is 1.4 T with 5 T, which is an indicative signature for spin triplet superconductivity. 

After this discovery, unconventional superconductivity under applied pressure is also reported in other NSC heavy fermion (HF)  compounds~\cite{kimura2005pressure,kimura2007normal,okuda2007condensed}, such as CeTSi$_3$(T = Rh, Ir)~\cite{sugitani2006pressure,kawai2008magnetic,wang2018superconductivity} and  CeTGe$_3$ (T = Co, Rh and Ir)~\cite{honda2010pressure}. These compounds crystallize in the BaNiSi$_{3}$-type tetragonal structure with noncentrosymmetric space group I4mm. At ambient pressure, CeRhSi$_{3}$ orders antiferromagnetically with $T_{N}$ = 1.6 K. The Sommerfeld coefficient is estimated $\gamma_{n}$ = 110 mJ/ mol K$^{2}$, indicates the heavy fermionic nature of this compound. For this compound also the upper critical value exceeds (16 T at p = 26 kbar) the Pauli limit, evidence of spin triplet superconductivity. Heat capacity measurement on CeIrSi$_{3}$ under pressure shows a sharp superconducting jump around $T_{C}$ with a heat capacity jump 5.7, much larger than the BCSS weak coupling value 1.43, which indicates the strong coupling superconductivity. This is one of the possible explanations of the large upper critical field. So, these compounds are crucial to investigate the effects of strong correlations in superconductivity. 

\par

Recently, NCS compounds without $f$ or $d$ electrons have attracted considerable attention due to the presence of several unconventional properties. To date, USc properties, e.g., a nodal gap structure and/or a large upper critical field, have been presented in a few weakly correlated NCS superconductors, including Li$_2$(Pd$_{1-x}$Pt$_x$)$_3$B~\cite{yuan2006s,yuan2008penetration}, Y$_2$C$_3$~\cite{chen2011evidence}, Mo$_3$Al$_2$C~\cite{bauer2010unconventional},  Mg$_{10}$Ir$_{19}$B$_{16}$~\cite{bonalde2009possible}  and with special attention for, LaNiC$_2$~\cite{hillier2009evidence}. Li$_{2}$(Pd${1-x}$Pt$_x$)$_{3}$B is a promising candidate to study the spin-orbit coupling in NCS compounds. The ASOC is raised by doping from Pd to Pt because 4d electrons are replaced with 5d electrons. Experimental measurements such as NMR, thermodynamic measurements suggest that the Li$_{2}$Pd$_{3}$B is a fully gapped s-wave superconductor. On the other hand, nodal superconducting gap structure is reported in Li$_{2}$Pt$_{3}$B. The deviation from the conventional exponential behavior in the heat capacity data of Mo$_{3}$Al$_{2}$C [$T_{C}$ = 9.05 K] suggest the presence of a line nodal gap in the superconducting gap structure. Whereas, Two fully gapped superconductivity is reported in Y$_{2}$C$_{3}$ and La$_{2}$C$_{3}$. The presence of a magnetic field in the superconducting state, which leads to time reversal symmetry breaking in the superconducting ground state is also reported in these NCS compounds. First TRS breaking is observed in LaNiC$_{2}$~\cite{hillier2009evidence}.

\begin{figure*}
\centering
 \includegraphics[width=0.6\linewidth]{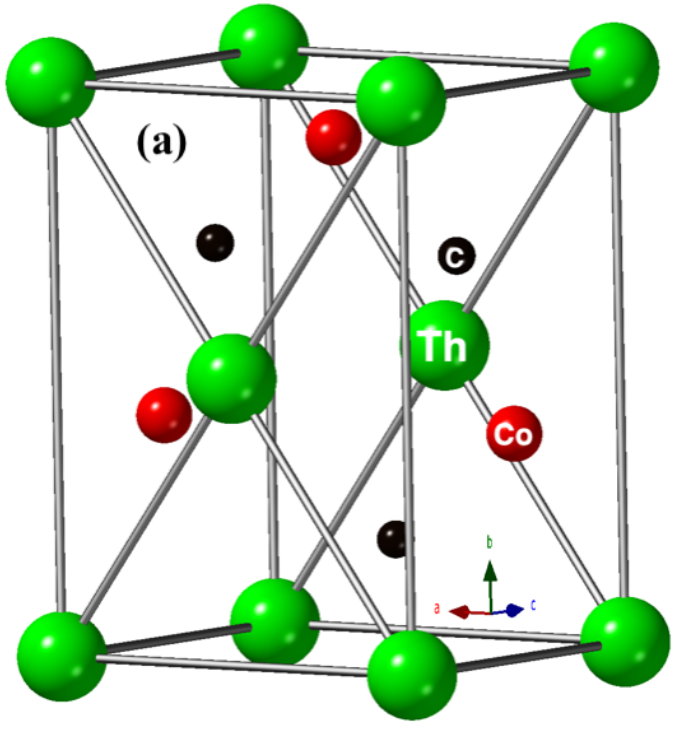}
\caption {The orthorhombic crystal structure of ThCoC$_{2}$, where green, red, and black symbols dictate Th (big in size), Co (medium), and C atoms (small), respectively.}
\label{fig51}
\end{figure*}

ThCoC$_2$ with $T_C$ = 2.3 K, crystallizes in the NCS CeNiC$_{2}-$type orthorhombic structure with space group  $Amm2$ (No. 38), in which a mirror plane is missing along the $c-$axis~\cite{grant2014superconductivity,grant2017tuning}, as displayed in Fig.~\ref{fig51}. The upper critical field ($H_{C2}$) vs temperature phase diagram confirms a positive curvature~\cite{grant2014superconductivity}, and the electronic specific heat coefficient at low temperature exhibits a square root field dependence $\gamma(H)\propto \sqrt{H}$, which indicates presence of nodes in the superconducting order parameter~\cite{chen2013evidence,wright1999low,yang2001low,van1990specific} and a potential USc state in ThCoC$_2$. $\gamma(H)\propto \sqrt{H}$ property, was admitted a general sign on $d-$band  and HF superconductors~\cite{wright1999low, yang2001low,van1990specific}. Moreover, examples of positive curvature in $H_{C2}(T)$ phase diagram includes  LnNi$_2$B$_2$C (Ln=Lu and Y)~\cite{takagi1994superconducting,shulga1998upper}, MgB$_2$~\cite{takano2001superconducting}, the NCS Li$_2$(Pd,Pt)$_3$B~\cite{peets2011magnetic}, and the NCS HF CeRhSi$_3$~\cite{kimura2007extremely}. The isostructural compound  LaNiC$_2$ displays USc below $T_C$ = 2.7 K with broken time-reversal symmetry (TRS) in $\mu$SR measurements, a nodal energy gap from very-low-temperature magnetic penetration depth measurements, and a suggested multigap superconductivity due to the modest value of the ASOC in LaNiC$_2$~\cite{hillier2009evidence,quintanilla2010relativistic}. Motivated by these results, in this chapter, we have investigated the compound ThCoC$_2$ using transverse (TF) and zero field (ZF) $-\mu$SR measurements and probed its superconducting state, which is found to be an unconventional in origin. The presence of nodal line $d-$wave superconductivity is manifested from the TF$-\mu$SR and further supported through our theoretical calculation in ThCoC$_{2}$.

\begin{figure*}
\centering
 \includegraphics[width=0.6\linewidth]{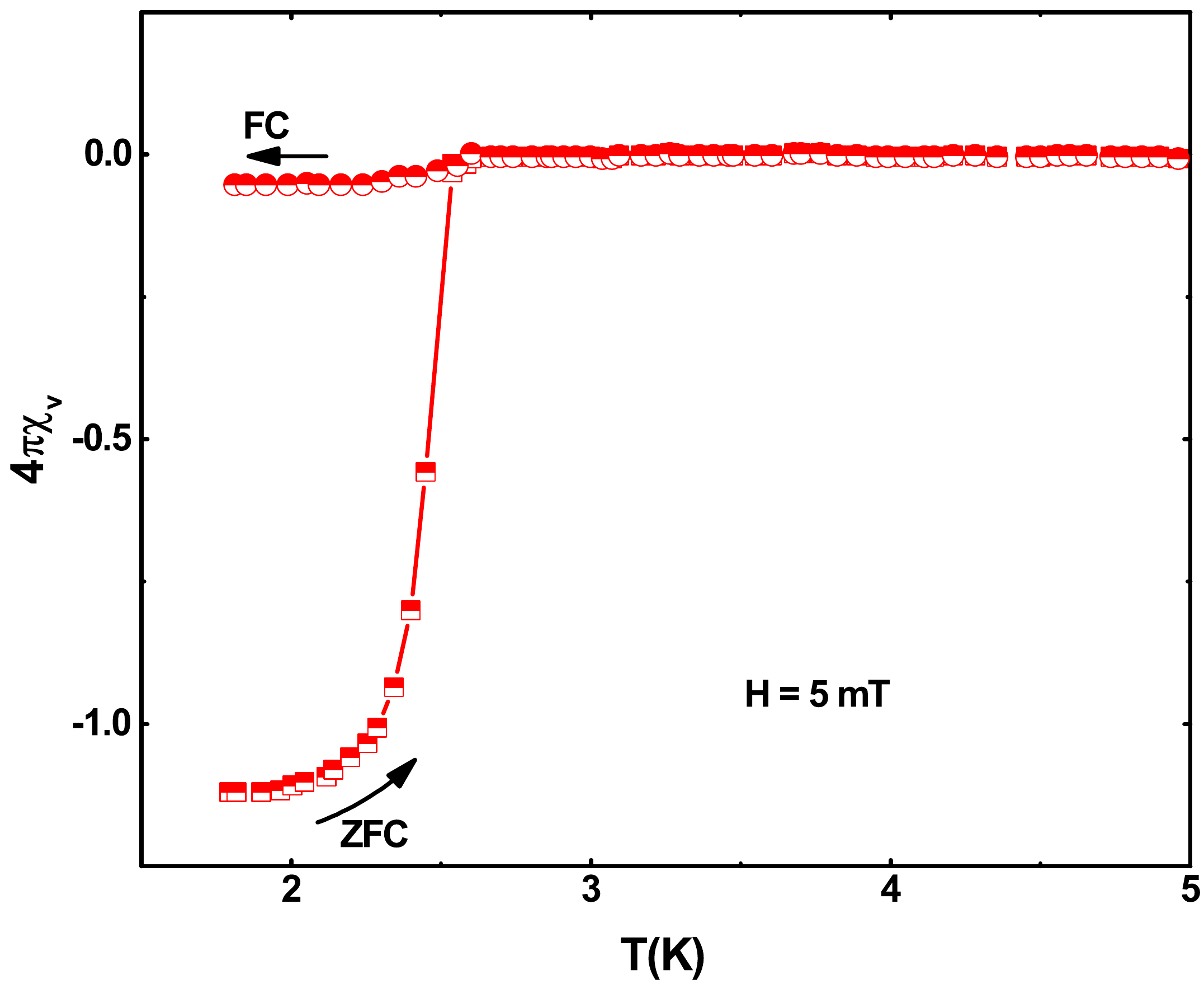}
\caption {Measured temperature dependence data of magnetic susceptibility $\chi(T)$ are shown in ZFC and FC modes.}
\label{fig52}
\end{figure*}

\section{Sample preparation}
A polycrystalline sample of ThCoC$_{2}$ was prepared by arc melting of the stoichiometric amount of Th (3N), Co (4N), and  C (5N) on a water-cooled Cu hearth in a high-purity Ar atmosphere~\cite{grant2014superconductivity}. The ingots were sealed in an evacuated quartz tube and annealed at 1100 $^\circ$C for 336 hr to get better sample quality.

\section{Type-II superconductivity}
 The magnetization data were obtained using a Quantum Design Superconducting Quantum Interference Device. The superconducting transition temperature $T_C$ = 2.3 K was confirmed by the magnetic susceptibility $\chi(T)$, as shown in Fig.~\ref{fig52}. The observed superconducting volume was 100\% of perfect diamagnetism~\cite{sonier2000musr}.
 
 \begin{figure*}
\centering
 \includegraphics[width=0.6\linewidth]{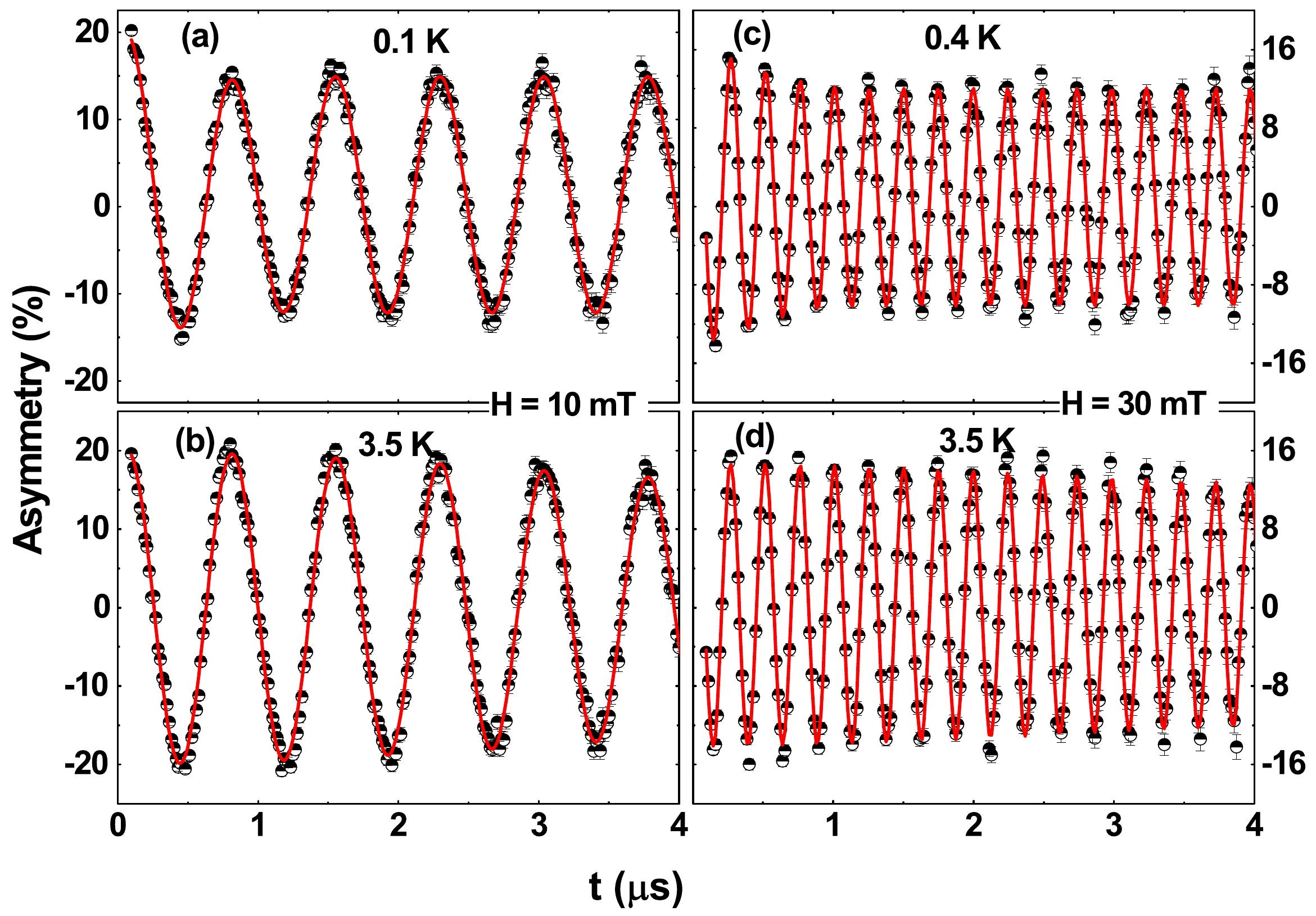}
\caption {Transverse field $\mu$SR asymmetry spectra are shown at (a) T = 0.1 K, (b) T = 3.5 K for H = 10 mT, and (c) T = 0.4 K (d) T = 3.5 K for H = 30 mT. The solid lines are fit using an oscillatory function with a Gaussian relaxation, plus a nondecaying oscillation that originates from muons stopping in the silver sample holder. The effect of the flux line lattice can be seen in the top panel as the strong Gaussian decay envelope of the oscillatory function. Above $T_{C}$, the depolarization is reduced, which is due to the randomly oriented array of nuclear magnetic moments.}
\label{fig53}
\end{figure*}
 
\section{Evidence of nodal line}
\begin{figure*}
\centering
 \includegraphics[width=0.6\linewidth]{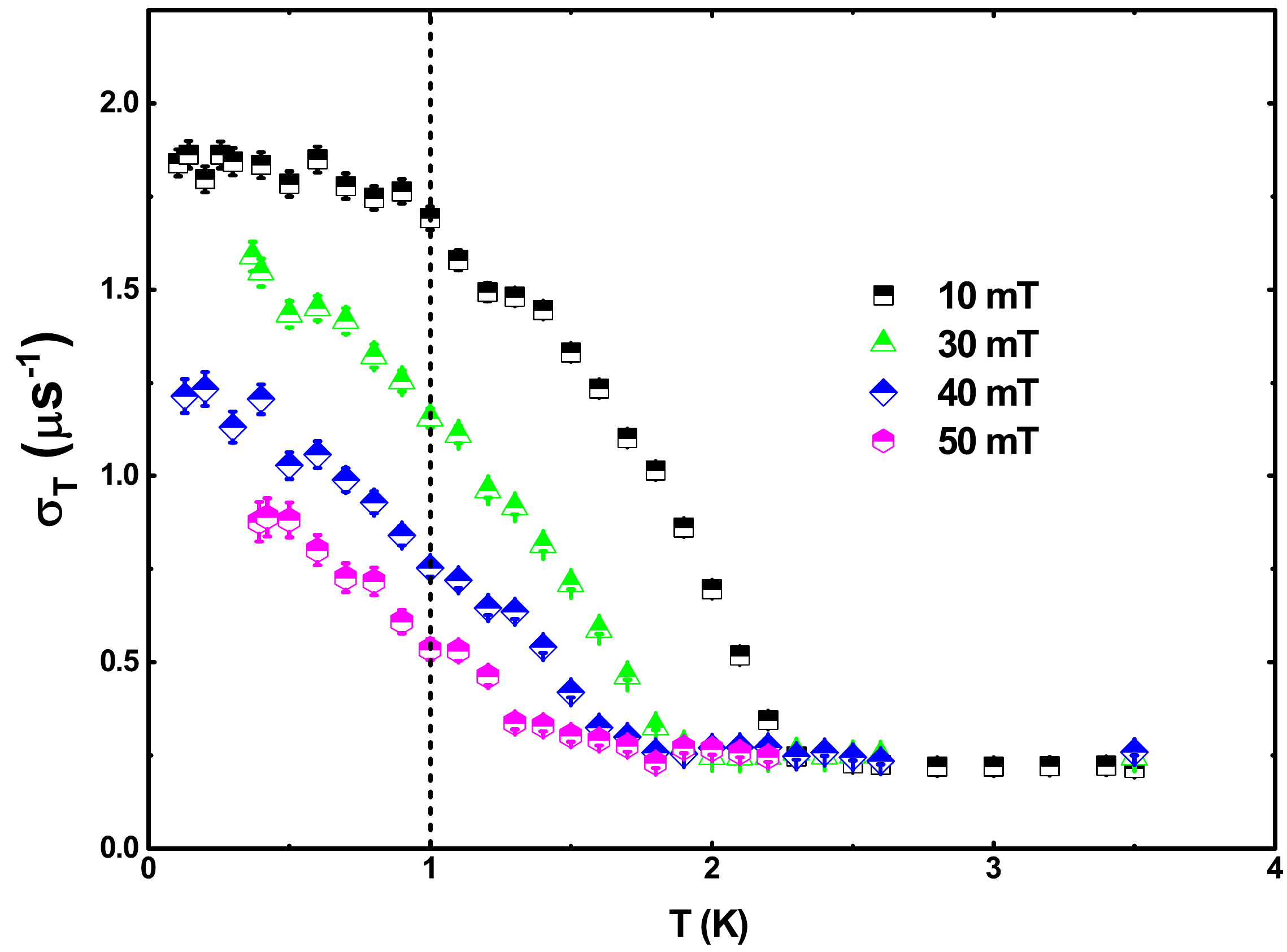}
\caption {Temperature dependence of the TF-$\mu$SR spin-depolarization rate collected in a range of fields 10 mT $\leq$ H $\leq$ 50 mT. }
\label{fig54}
\end{figure*}

\begin{figure*}
\centering
 \includegraphics[width=0.6\linewidth]{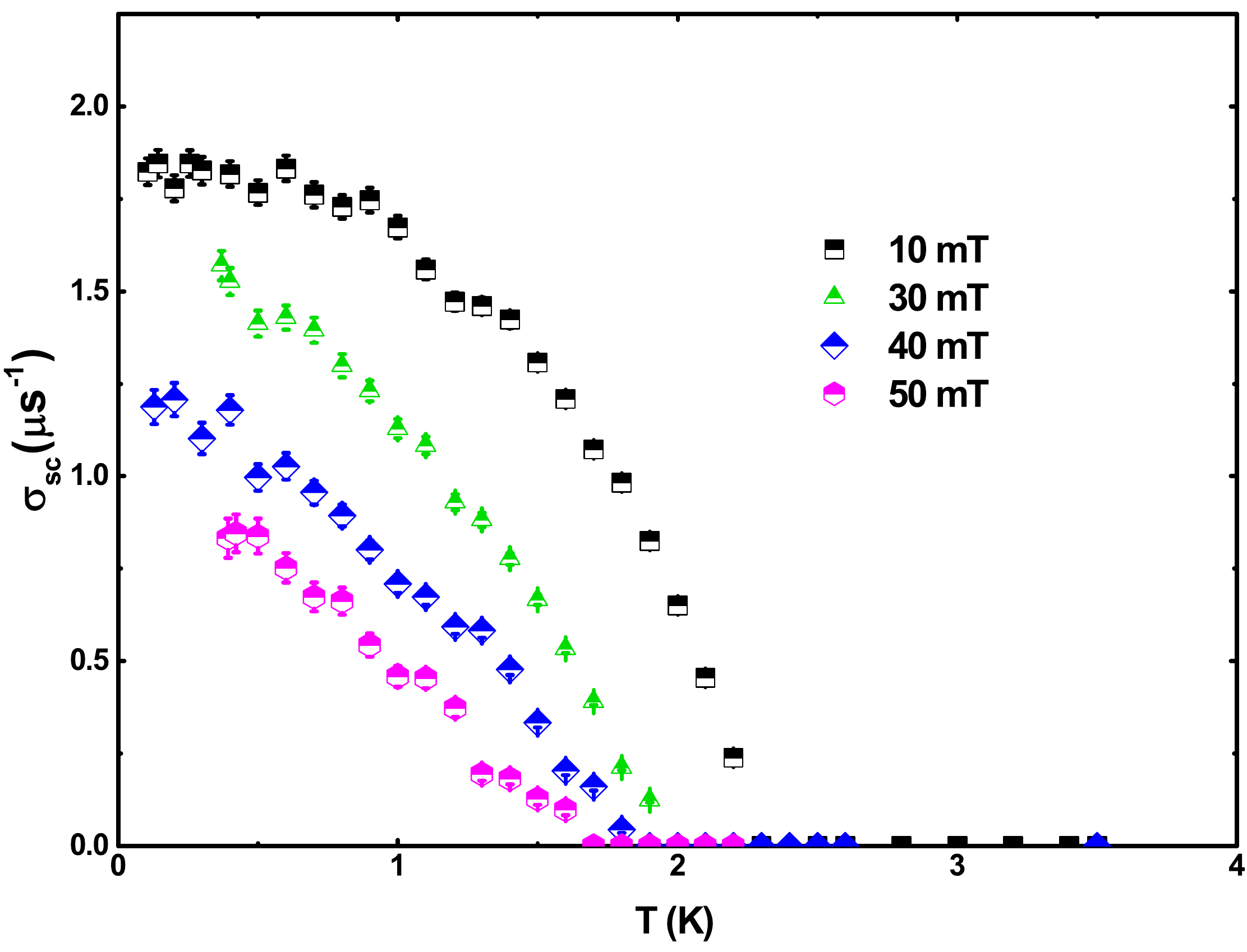}
\caption {Temperature dependence of the TF-$\mu$SR spin-depolarization rate collected in a range of fields 10 mT $\leq$ H $\leq$ 50 mT. The dashed line shows an example of the isothermal cuts used to find the field dependence of $\sigma_{sc}$ at a particular temperature.}
\label{fig55}
\end{figure*}

The flux line lattice (FLL) in the vortex state of a type-II superconductor leads to a distinctive field distribution in the sample, which can be detected in the $\mu$SR relaxation rate.  Fig.~\ref{fig53}(a-d) shows the TF-$\mu$SR asymmetry spectra above and below $T_{C}$ in an applied field of 10 mT and 30 mT. Below T$_{C}$, the TF-$\mu$SR precession signal decays with time due to the distribution of fields associated with the FLL.  The analysis of TF-$\mu$SR asymmetry spectra was carried out in the time domain using a sinusoidal function damped with Gaussian relaxation plus a nondecaying oscillation that originates from muons stopping in the silver sample holder, $G_{z}(t) = A_{1} \cos (2\pi\nu_{1}t + \phi)\exp (-\frac{\sigma^{2}t^{2}}{2}) + A_{2} \cos(2\pi\nu_{2} t+\phi)$, where $A_1$ and $A_2$ are the sample and background asymmetries, $\nu_{1}$ and $\nu_{2}$ are the frequencies of the muon precession signal from the sample and background signal, respectively with phase angle $\phi$ and $\gamma_\mu/2\pi$ = 135.5 MHz T$^{-1}$ is the muon gyromagnetic ratio. The total depolarization rate $\sigma_{T}$ as a function of temperature in different applied fields is shown in ~\ref{fig54}.   $\sigma$ is the overall muon depolarization rate; there are contributions from both the vortex lattice ($\sigma_{sc}$) and nuclear dipole moments ($\sigma_{nm}$, which is assumed to be temperature independent) [where $\sigma = \sqrt{(\sigma_{sc}^2+\sigma_{nm}^2}$]. $\sigma_{sc}$ was determined by quadratically deducting the background nuclear dipolar depolarization rate obtained from spectra measured above T$_{C}$, which is shown in Fig.~\ref{fig55}.

\begin{figure*}
\centering
 \includegraphics[width=0.6\linewidth]{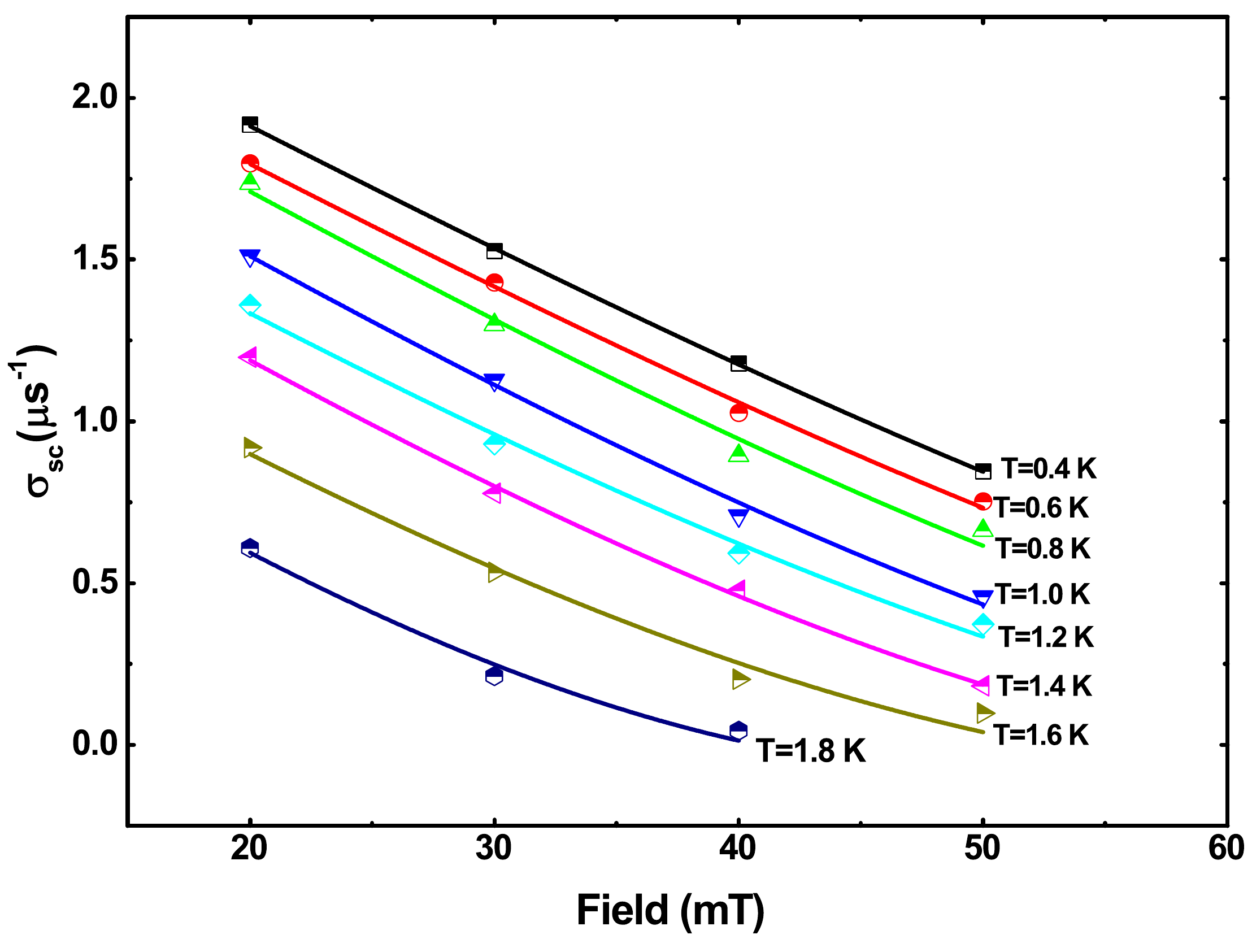}
\caption {Field dependence of the muon spin-depolarization rate is shown for a range of different temperatures. The solid lines are the results of fitting the data using the Brandt equation as discussed in the text.}
\label{fig56}
\end{figure*}

\begin{figure*}
\centering
 \includegraphics[width=0.6\linewidth]{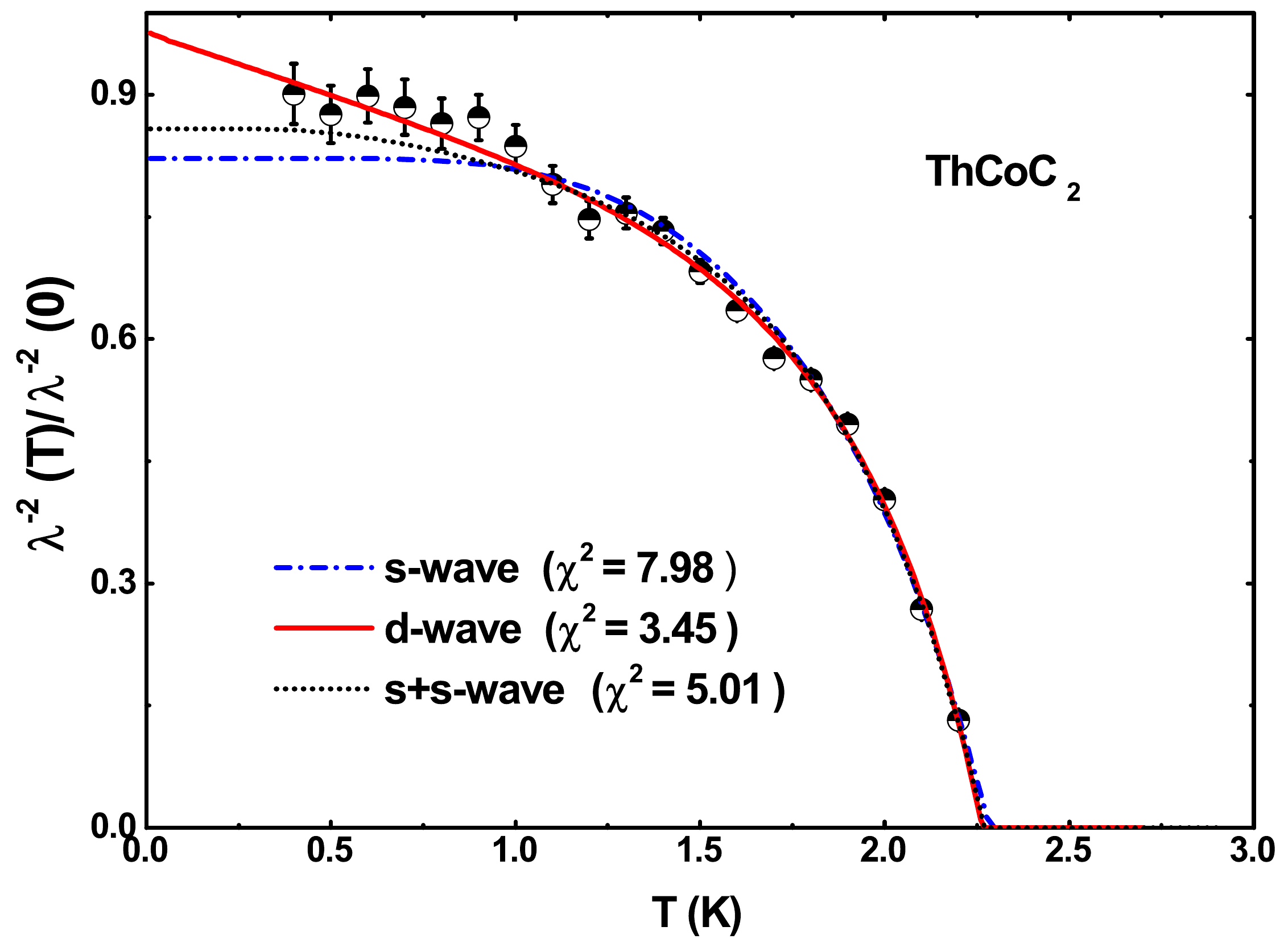}
\caption {Temperature dependence of the inverse magnetic penetration
depth squared is shown here. The lines show the fits using $s$ wave (dashed), $s + s$ wave (dotted), and $d$ wave (solid) gap functions.}
\label{fig57}
\end{figure*}

As can be seen in Fig.~\ref{fig55} that $\sigma_{sc}$ depends on the applied field. Isothermal cuts perpendicular to the temperature axis of the $\sigma_{sc}$ data sets were used to determine the $H$ dependence of the depolarization rate $\sigma_{sc}(H)$ displayed in Fig.~\ref{fig55}. Brandt~\cite{brandt2003properties} suggested a useful relation that describe the field dependence for a hexagonal lattice, which is valid over the field range examined in this experiment, $\sigma_{s}[\mu s^{-1}] = 4.83 \times 10^{4}(1-H_{ext}/H_{C2})\times [1+1.2(1-\sqrt{H_{ext}/H_{C2})^3]\lambda^{-2}[nm]}$, to estimate the temperature dependence of the inverse-squared penetration depth, $\lambda^{-2}$ and $\mu_{0}H_{C2}$. The resulting fits to the $\sigma_{sc}(H)$ data are displayed as solid lines in Fig.~\ref{fig56}. The temperature dependence of $\lambda^{-2}$ is presented in Fig.~\ref{fig57}. We can model the temperature dependence of the $\lambda^{-2}$ by the following equation~\cite{bhattacharyya2017superconducting,adroja2017multigap},
\begin{eqnarray}
\frac{\lambda^{-2}(T,\Delta_{0,i})}{\lambda^{-2}(0,\Delta_{0,i})}= 1 + \frac{1}{\pi}\int_{0}^{2\pi}\int_{\Delta(T)}^{\infty}(\frac{\delta f}{\delta E}) \times \frac{EdEd\phi}{\sqrt{E^{2}-\Delta(T,\phi})^2} 
\end{eqnarray}

, where $f= [1+exp(E/k_{B}T)]^{-1}$ is the Fermi function. The gap is given by $\Delta_{i}(T,\phi)=\Delta_{0,i}\delta(T/T_{C})g(\phi)$, whereas g($\phi$) refer to the angular dependence of the superconducting gap function and $\phi$ is the azimuthal angle along the Fermi surface. The $T$ dependence of the superconducting gap is approximated by the relation $\delta(T/T_{C}) = \tanh\{1.82(T_{C}/T-1)]^{0.51}\}$. g($\phi$) is substituted by (a) 1 for an $s-$ or $s+s-$wave gap and (b) $\vert\cos(2\phi)\vert$ for a $d-$wave gap with line nodes.

We have analyzed the temperature dependence of $\lambda^{-2}$ based on three different models, a single-gap isotropic $s-$wave, a line node $d-$wave and two gap $(s + s)-$wave models. The result of the fits using various gap models are shown by lines (dashed, dotted, and solid) in Fig.~\ref{fig57}. It is clear from Fig.~\ref{fig57}that the $s-$ or $s+s-$ wave models do not fit the data and gave a larger value of $\chi^2$ = 7.98(3) for $s-$wave and $\chi^2$ = 5.01(1) for $s+s-$wave.  As square root field dependence of the electronic heat capacity suggests a presence of nodes in the superconducting gap symmetry, we therefore fitted our data using a $d-$wave model with line nodes in the gap structure. The $d-$wave model gives the best description of the temperature dependence of $\lambda^{-2}$ of ThCoC$_2$ as the $\chi^2$ value reduced significantly ($\chi^2$ = 3.45(1) for $d$-wave). The $d-$wave fit gives $\Delta(0)$ = 0.77(1) meV and $T_{C}$ = 2.3(1) K. This gives a gap to $T_{C}$ ratio 2$\Delta(0)$/$k_BT_{C}$ = 7.8, which is significantly larger than the weak coupling BCS value of 3.53, indicating strong coupling superconductivity in ThCoC$_2$.  Enhanced value of the ratio is also observed in other NCS superconductors such as Re$_6$Zr~\cite{singh2014detection} with 2$\Delta(0)$/$k_B$$T_{C}$ = 4.2, BiPd~\cite{sun2015dirac} with 2$\Delta(0)$/$k_B$$T_{C}$ = 3.8,  La$_7$Ir$_3$~\cite{barker2015unconventional} with 2$\Delta(0)$/$k_B$$T_{C}$ = 3.81, K$_2$Cr$_3$As$_3$~\cite{adroja2015superconducting} with 2$\Delta(0)$/$k_B$$T_{C}$ = 6.4 and  Cs$_2$Cr$_3$As$_3$~\cite{adroja2017nodal} with 2$\Delta(0)$/$k_B T_{C}$ = 6.0. 

\section{Preserved time reversal symmetry}
\begin{figure*}
\centering
 \includegraphics[width=0.6\linewidth]{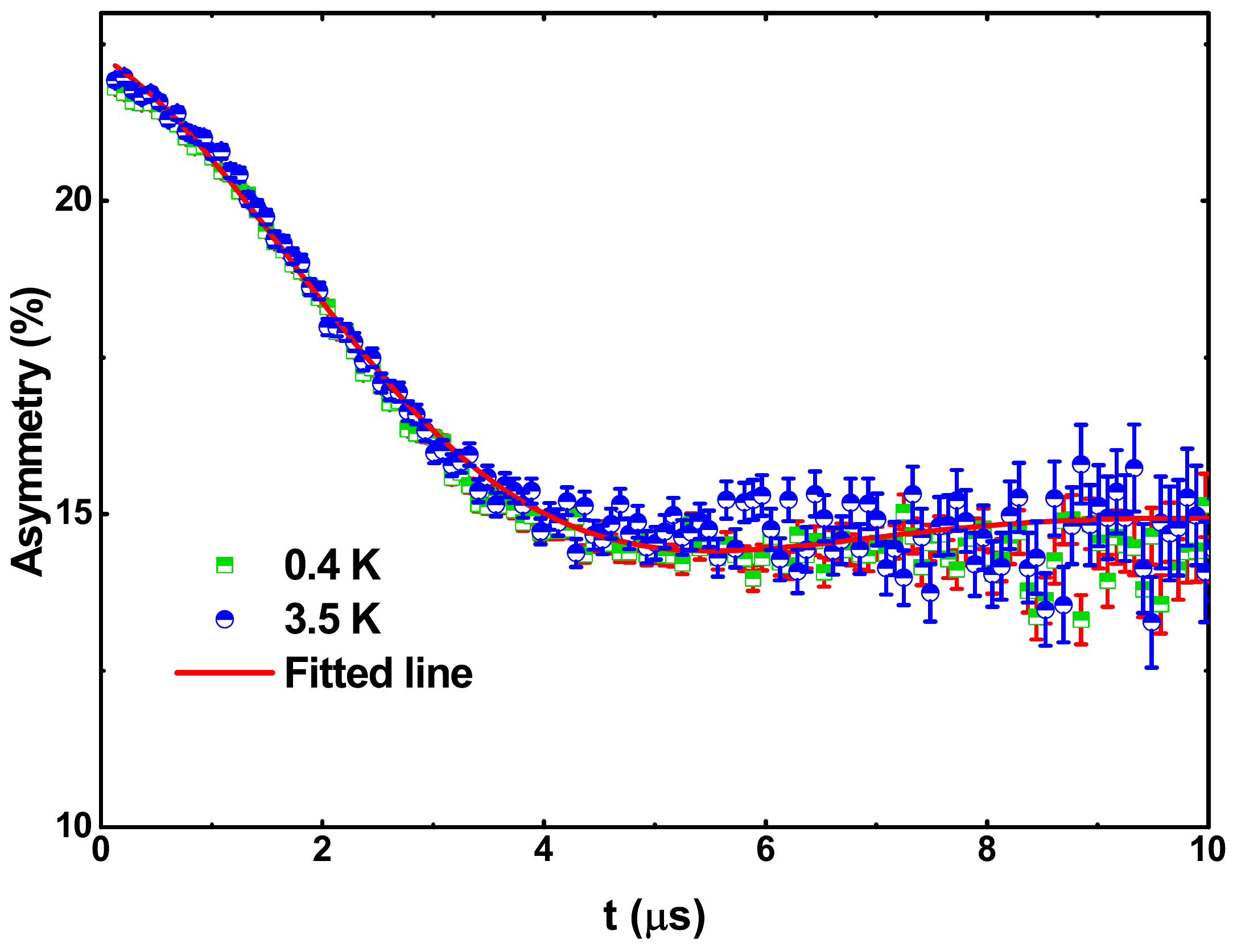}
\caption {(a) Time dependence asymmetry spectra at 0.4 K and 3.5 K measured in zero magnetic field of ThCoC$_{2}$ with $T_{C}$ = 2.3 K. The lines are the results of the fits. The absence of extra relaxation below $T_{C}$ indicates no internal magnetic fields and, consequently, suggests that the
superconducting state preserves time reversal symmetry.}
\label{fig58}
\end{figure*}

The time evolution of the ZF$-\mu$SR of ThCoC$_2$ is shown in Fig.~\ref{fig58} for $T$ = 0.4 K and 3.5 K. In these relaxation experiments, any muons stopped on the silver sample holder gave a time independent background. No signature of precession is visible (either at 0.4 K or 3.5 K), ruling out the presence of a sufficiently large internal magnetic field as seen in magnetically ordered compounds. The only possibility is that the muon$-$spin relaxation is due to static, randomly oriented local fields associated with the electronic and nuclear moments at the muon site. The ZF$-\mu$SR data are well described in terms of the damped Gaussian Kubo-Toyabe (KT) function, $G_{z2}(t) =A_1 G_{KT}(t)e^{-\lambda t}+A_{bg}$ where $G_{KT}(t) = \left[\frac{1}{3}+\frac{2}{3}(1-\sigma_{KT}^2t^2)e^{{\frac{-\sigma_{KT}^2t^2}{2}}}\right]$ is Gaussian Kubo-Toyabe function, $\lambda$ is the electronic relaxation rate, $A_1$ is the initial asymmetry, $A_{bg}$ is the background.  The parameters $\sigma_{KT}$, $A_1$, and $A_{bg}$ are found to be temperature independent.  It is evident from the ZF$-\mu$SR spectra that there is no noticeable change in the relaxation rates at 3.5 K ($\ge T_{\bf c}$) and 0.4 K ($\le T_{C}$). This indicates that the time-reversal symmetry is preserved upon entering the superconducting state. $\sigma_{KT}$ accounts for the Gaussian distribution of static fields from nuclear moments (the local field distribution width $H_\mu$ = $\sigma/\gamma_\mu$, with muon gyromagnetic ratio $\gamma_\mu$ = 135.53 MHz/T). The fits of $\mu$SR spectra in Fig.S1 by the decay function gave $\sigma_{KT}$ = 0.31(7) $\mu s^{-1}$ and $\lambda$ = 0.12(6) $\mu s^{-1}$ at 3.5 K and $\sigma_{KT}$ = 0.30(7) $\mu s^{-1}$ and $\lambda$= 0.11(9) $\mu s^{-1}$ at 0.4 K. The fits are shown by the solid lines in Fig.S1. Since within the error bars both $\sigma_{KT}$ and $\lambda$ at $T < T_{C}$ and $T > T_{C}$ are similar, there is no evidence of time-reversal symmetry breaking in ThCoC$_2$. This result also confirms the absence of any kind of magnetic impurity in our sample.

\section{Summary and Theoretical Calculations}
In summary, we have investigated the gap symmetry of ThCoC$_{2}$ using transverse field $\mu$SR. Our $\mu$SR analysis confirmed that a nodal line $d-$wave model fits better than a single or two gaps isotropic $s-$wave model to the observed temperature dependence of the penetration depth. This finding is in agreement with the field dependent heat capacity, $\gamma(H) \sim \sqrt{H}$. The finding of the nodal line gap is further supported through our theoretical calculations\footnote{Theoretical calculations have been performed by Prof. Tanmoy Das and Surabhi Saha from the Department of Physics, Indian Institute of Science, Bangalore 560012, India }.

\subsection*{DFT calculation}
For the first-principles electronic structure calculation, we used the Vienna Ab-initio Simulation Package (VASP)~\cite{kresse1996efficient} and use the Perdew-Burke-Ernzerhof (PBE) form for the exchange-correlation functional~\cite{perdew1996generalized}. The projector augmented wave (PAW) pseudo-potentials are used to describe the core electrons. Electronic wave-functions are expanded using plane waves up to a cut-off energy of 500 eV. The Monkhorst-Pack k-mesh is set to $14\times14\times14$ in the Brillouin zone for the self-consistent calculation. All atoms are relaxed in each optimization cycle until atomic forces on each atom are smaller than 0.01 $eV \AA$. The lattice constants are obtained by relaxing the structure and with total energy minimization. Our obtained relaxed lattice parameters are $a$ = 3.8493${\AA}$, $b$ = 3.8493$\AA$, $c$ = 3.9462$\AA$, which are close to the experimental values. To deal with the strong correlation effect of the $d$-electrons of the Th and Co atoms, we employed LDA+$U$ method with $U = 5 eV$ on both atoms. We have recalculated the band structure with spin-orbit coupling, and no considerable change is obtained in the low-energy bands of present interests.

\subsection*{Calculation details}

\begin{figure*}
\centering
 \includegraphics[width=0.9\linewidth]{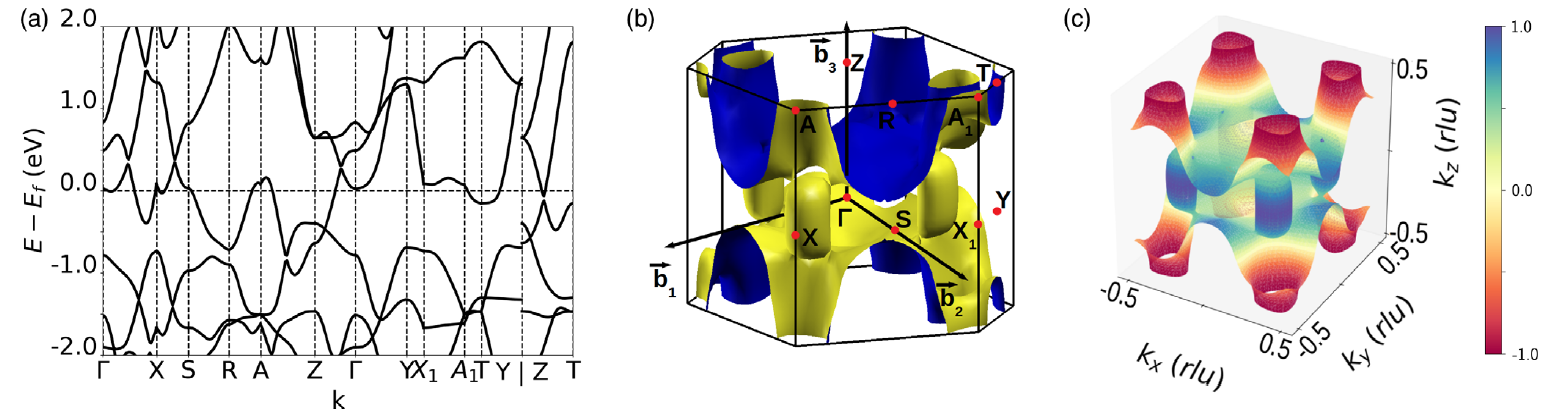}
\caption {Computed result of DFT band structure is plotted along with the standard high-symmetric directions for the orthorhombic crystal structure of ThCoC$_{2}$. The low energy part of the band structure is dominated by Co-d orbitals. (b) Corresponding Fermi surface topology is plotted in the first Brillouin zone of the lattice. The color in this plot does not have any specific meaning. The red dots show the high-symmetry points on the Brillouin zone and b1, b2, and b3 are the reciprocal lattice vectors. (c) Computed pairing eigenfunction plotted in a color map of red (negative) to yellow (nodes) to blue (positive) color. The result is overlaid on the corresponding FS. We immediately observe the presence of nodal lines on the $k_{z} = \pm \pi/2$ planes (visualized by yellow contour).}
\label{fig59}
\end{figure*}

Now we discuss our theoretical calculations. For the first-principles electronic structure calculation, we used the Vienna Ab-initio Simulation Package (VASP)~\cite{kresse1996efficient} and use the Perdew-Burke-Ernzerhof (PBE) form for the exchange-correlation functional~\cite{perdew1996generalized}. To deal with the strong correlation effect of the $d$-electrons of the Th and Co atoms, we employed LDA+$U$ method with $U = 5 eV$ on both atoms. We have recalculated the band structure with spin-orbit coupling, and no considerable change is obtained in the low-energy bands of present interests. The computed band structure is shown in Fig.~\ref{fig59}(a). We notice that only a single band osculates around the Fermi level. This band is dominated by Co-$d$-orbital with hybridization with the C-$p$ orbitals (with about 10\% contributions from the C-$p$ states to the low-energy band structure). The corresponding Fermi surface (FS) is shown in Fig.~\ref{fig59}(b). The three-dimensional FS topology is quite interesting in this system, exhibiting a hole-like pocket around the zone corner (`A' point), but an electron-like topology around the `T'-point. In addition, as we scan across the $k_z$-direction, we notice that the Fermi pocket near the $k_z=\pm\pi$-plane first closes at discrete Fermi-points, and then reopens into another pocket around some non-high-symmetric point. Such an interesting FS topological change in the 3D Brillouin zone (BZ) within a single band model implies the presence of saddle-points in the electronic structure around the Fermi level. Hence one obtains a large density of states near the Fermi level, which is desirable for superconductivity. In addition, due to multiply-pockets features in the FS, one expects large FS nestings, and hence unconventional superconductivity.

\subsection*{Details of pairing eigenvalue calculations}

Our calculation of pairing interaction and pairing eigenvalue originating from spin-fluctuation is done by directly including the DFT band structure in a three-dimensional Brillouin zone (BZ). We start with a single-band Hubbard model, as dictated by the DFT calculation, which is given by
\begin{equation}
H=\sum_{\mathbf{k},\sigma=\uparrow,\downarrow}\xi_{\mathbf{k}}c^{\dagger}_{\mathbf{k}\sigma}c_{\mathbf{k}\sigma}+U\sum_{\mathbf{k},\mathbf{k}^{'},\mathbf{q}}c^{\dagger}_{\mathbf{k}\uparrow}c_{\mathbf{k}+\mathbf{q}\uparrow}c^{\dagger}_{\mathbf{k}^{'}\downarrow}c_{\mathbf{k}^{'}-\mathbf{q}\downarrow},\nonumber\\
\end{equation}
where $c^{\dagger}_{\mathbf{k}\sigma} (c_{\mathbf{k}\sigma})$ is the creation (annihilation) operator for non-interacting electron with momentum ${\bf k}$, and spin $\sigma=\uparrow/\downarrow$, and $\xi_{\bf k}$ is the corresponding band structure, taken directly from the DFT calculation. The second term is the Hubbard interaction, written in the band basis with the onsite potential $U$. Perturbative expansion of the Hubbard term in the spin-singlet and spin-triplet pairing channels yields,

\begin{eqnarray}
H&=&\sum_{\mathbf{k},\sigma=\uparrow,\downarrow}\xi_{\mathbf{k}}c^{\dagger}_{\mathbf{k}\sigma}c_{\mathbf{k}\sigma}\nonumber\\
&&+\sum_{\mathbf{k}\mathbf{k}^{'},\sigma\sigma'}V_{\sigma\sigma'}(\mathbf{k}-\mathbf{k^{'}})c^{\dagger}_{\mathbf{k}\sigma}c^{\dagger}_{-\mathbf{k}\sigma^{'}}c_{-\mathbf{k^{'}}\sigma^{'}}c_{\mathbf{k}'\sigma}.
\label{pairH}
\end{eqnarray}

Here $V_{\sigma\sigma'}$ refers to the spin singlet and triplet case when $\sigma=\mp\sigma$, respectively. The corresponding (static) pairing potentials are given in Eqs.(2-3) in the main text. The bare susceptibility is 
\begin{equation}
\chi^0(\mathbf{q},\omega)=-\sum_{\mathbf{k}}\frac{f(\xi_{\mathbf{k}})-f(\xi_{\mathbf{k}+\mathbf{q}})}{\omega-\xi_{\mathbf{k}}+\xi_{\mathbf{k}+\mathbf{q}}+\iota \delta},
\end{equation}
where $f$ is the corresponding Fermi function, and $\delta$ is an infinitesimal number. Since we are interested in the static limit of the pairing potential, the imaginary part of the above particle-hole correlator does not contribute and we obtain a real pairing potential $V$. In the RPA channel, the spin and charge channels are decoupled, and due to different denominators $1\mp U\chi^0$, the spin channel is strongly enhanced while simultaneously the charge channel is suppressed. The method is a weak to intermediate coupling approach and thus the value of $U$ is restricted by the non-interacting bandwidth. We use $U$ = 400 meV in our numerical calculation and the result is reproduced with different values of $U$. We note that this value of $U$ is defined in the band basis, while the LDA $U$ is for the orbitals. The $k$-dependent of the pairing eigenfunction is dictated by the anisotropy in $V({\bf q})$ which is directly related to the bare FS nesting character, this result remains unaffected by a particular choice of $U$ in the weak to intermediate coupling range. With increasing $U$, the overall strength of $V$ mainly increases and thus the pairing eigenvalue $\lambda$ also increases. Since in this work, we are primarily interested in uncovering the pairing eigenfunction, the value of $U$ remains irrelevant in this coupling strength. 
\begin{figure*}
\centering
 \includegraphics[width=\linewidth]{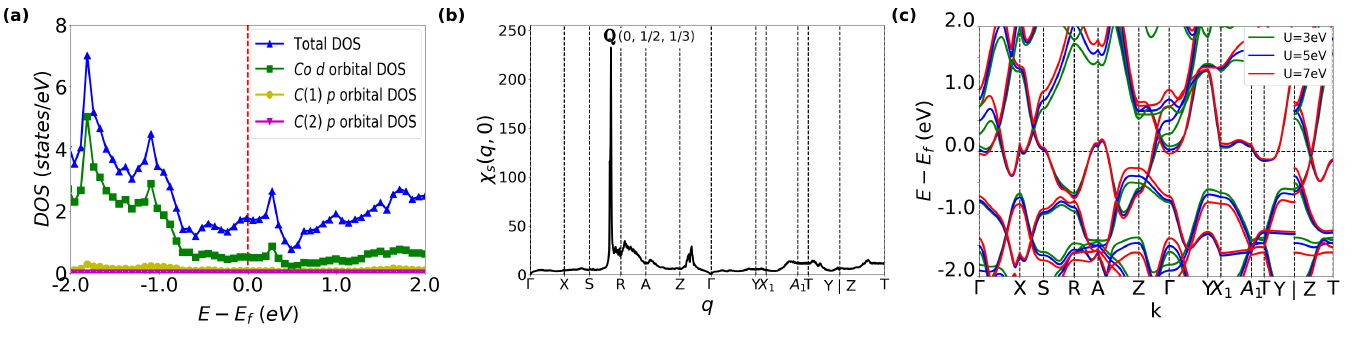}
\caption {(a) Compute partial density of states (pDOS) plotted for three relevant orbitals near the Fermi level. We clearly observe that Co-d orbitals dominate the low-energy states. (b) The real part of the RPA susceptibility (static) in the spin-channel is plotted to the same high-symmetric directions. We observe a distinctly sharp peak at {\bf Q} (0,1/2,1/3) in the susceptibility. This nesting peaks connect two Fermi surfaces across the k$_{z}$ = 0 plane. (c) Computed band structures for three different values of U. We notice a gradual shift of the bands, inducing a change in the FS area, but no change in the Fermi surface topology is observed in this wide range of $U$. U = 5 (middle plot) is used in the paper.}
\label{fig510}
\end{figure*}

From interacting Hamiltonian, we define the SC gap equation as
\begin{eqnarray}
\Delta_{\mathbf{k}}&=-\sum_{\mathbf{k^{'}}}V_{\sigma\sigma'}(\mathbf{k}-\mathbf{k^{'}})\langle c_{\mathbf{-k}'\sigma}c_{\mathbf{k}'\sigma'}\rangle,\\
&= -\sum_{\mathbf{k^{'}}}V_{\sigma\sigma'}(\mathbf{k}-\mathbf{k^{'}})\frac{\Delta_{\mathbf{k^{'}}}}{2\xi_{\mathbf{k^{'}}}} \tanh{\Big(\frac{\xi_{\mathbf{k^{'}}}}{k_BT_C}\Big)}
\end{eqnarray}
In the limit $T\to 0$ we have $\langle c_{-{\bf k}\sigma}c_{\mathbf{k}\sigma'}\rangle\to \lambda \Delta_{\mathbf{k^{}}}$, leading to an eigenvalue equation as
\begin{equation}
\Delta_{\mathbf{k}}=-\lambda \sum_{\mathbf{k^{'}}}V_{\sigma\sigma'}(\mathbf{k}-\mathbf{k^{'}})\Delta_{\mathbf{k^{'}}}.
\end{equation}
We solve this equation separately for spin singlet ($\sigma=-\sigma'$), and spin triplet ($\sigma=\sigma'$) cases. Since we restrict ourselves to the static limit, the above equation is solved for only Fermi momenta $\mathbf{k}_F$, $\mathbf{k}'_F$ in which $V_{\sigma\sigma'}(\mathbf{k}_F-\mathbf{k}'_F)$ becomes a $n\times n$ matrix with $n$ being the number of discrete Fermi momenta considered in a 3D BZ.

\subsection*{Pairing eigenvalue calculation}

We perform a pairing symmetry calculation using single-band Hubbard model. By carrying out a perturbative expansion of the Hubbard interaction, we obtain an effective pairing potential in the single and triplet pairing channels~\cite{kubo2007pairing, monthoux1991toward}
%
\begin{eqnarray}
\label{pairpot}
V_{\uparrow\downarrow}(\mathbf{q})&=&\frac{U}{2}\left[1+3U\chi^{s}(\mathbf{q})\right]+\frac{U}{2}\left[1-U\chi^{c}(\mathbf{q})\right],\nonumber\\
V_{\uparrow\uparrow}(\mathbf{q})&=&\frac{U}{2}\left[1-U\chi^{s}(\mathbf{q})\right]+\frac{U}{2}\left[1-U\chi^{c}(\mathbf{q})\right].\\
\end{eqnarray}
Here $\chi^{s/c}(\mathbf{q})$ are the spin and charge susceptibilities, decoupled within the random-phase approximation (RPA), as $\chi^{s/c}({\bf q}) = \chi^{0}({\bf q})/(1\mp U\chi^{0}({\bf q}))$ with $\chi^{0} ({\bf q})$ being bare static Lindhard susceptibility. For such a potential, the BCS gap equation (for details see Supplementary materials) becomes
\begin{equation}
\Delta_{\mathbf{k}}=-\lambda \sum_{\mathbf{k^{'}}}V(\mathbf{k}-\mathbf{k^{'}})\Delta_{\mathbf{k^{'}}}.
\label{Paireig}
\end{equation}
This is an eigenvalue equation with the pairing strength $\lambda$ denoting the eigenvalue. For $V>0$, a positive eigenvalue $\lambda$ can commence with the corresponding eigenfunction $\Delta_{\bf k}$ which changes sign as ${\rm sgn}[\Delta_{\bf k}]=- {\rm sgn}[\Delta_{\bf k'}]$ promoted by strong peak(s) in $V$ at ${\bf Q}=\mathbf{k}-\mathbf{k^{'}}$. Looking into the origin of $V$ in Eq.~\ref{pairpot}, we notice that $V$ inherits strong peaks from that in $\chi^{s/c}$, which is directly linked to the FS nesting feature embedded in $\chi^0$. In this way, the pairing symmetry of a system is intricately linked to the FS nesting instability. This theory of spin-fluctuation driven superconductivity consistently links between the observed pairing symmetry and FS topology in many different unconventional superconductors~\cite{monthoux1991toward, hirschfeld2011gap, ikeda2015emergent}.

For the present FS topology presented in Fig.~\ref{fig59}(c), we find that the nesting is dominant between the two FS pockets centered at the $k_z=0$, and $k_z=\pm \pi$ planes. This nesting is ingrained within the above pairing potential $V$, and naturally dictates a dominant pairing channel that changes sign between these two $k_z$-planes. In fact, our exact evaluation of the pairing eigenvalue gives the dominant pairing channel to be a spin-singlet pairing with $d_z$-wave pairing channel $\Delta_{\bf k}=\cos(k_z)$ with an eigenvalue of $\lambda\sim 0.2$. The computed pairing eigenfunction is shown in Fig.~\ref{fig510} on the FS in a color plot. The computed pairing channel does not acquire any in-plane anisotropy, and hence a nodal line on the $k_z=\pm \pi/2$. The experimental data fitted with the same nodal-like pairing channel indeed gives a better fit compared to other pairing functions.

This work paves the way for further studies of the large numbers of unexplored NCS compounds with the CeNiC$_2$ type structure for the hunt of the unconventional superconductor.
%

%% file: chapter6.tex
\chapter{Investigation of intermediate valence superconductor CeIr$_{3}$} 
\label{chapter:6}
\section{Introduction}
The strongly correlated electron systems of Ce, Yb, and U have attracted considerable attention in condensed matter physics, both theoretically and experimentally, due to the observation of heavy-fermion (HF) and valence fluctuation behavior, unconventional superconductivity, quantum criticality, and spin and charge gap formation~\cite{coleman2015introduction}. The great interest in heavy fermion systems originated with the identification of superconductivity in CeCu$_2$Si$_2$ with a $T_{C} =0.7$~K in 1979 by Steglich {\it et al.}~\cite{steglich1979superconductivity}. At that time, it was thought that magnetism and superconductivity would not occur simultaneously. Nevertheless, in CeCu$_2$Si$_2$, the 4$f$ electrons which give rise to the local magnetic moments also seem to be responsible for the unconventional superconductivity~\cite{smidman2018interplay}. Unconventional superconductivity was also reported in other Ce-based heavy fermion compounds including CeCoIn$_5$, which has a $T_{C}$ of 2.3 K~\cite{petrovic2001heavy, sidorov2002superconductivity}, and the noncentrosymmetric HF superconductor CePt$_{3}$Si~\cite{bauer2004heavy}, a system without a center of inversion in the crystal structure that exhibits a coexistence of antiferromagnetic order ($T_{N}=2.2$~K) and superconductivity ($T_{C}=0.75$~K). Usually, the conventional BCS theory of superconductivity does not apply to these exotic systems~\cite{BCS}. Heavy fermions have a diverse range of ground states, including superconductors such as UBe$_{13}$~\cite{mcelfresh1990structure} and UPt$_3$~\cite{de1984resistivity, trappmann1991pressure} both with unconventional superconducting ground states. There are many magnetic HF systems which exhibit unconventional superconductivity under applied pressure. For example, CeIn$_{3}$ ($T_{C} = 1.2$~K at 2.46~GPa)~\cite{knebel2001electronic}, CePd$_{2}$Si$_{2}$ ($T_{C} = 0.43$~K, at 30~GPa)~\cite{mathur1998magnetically}, CeRh$_{2}$Si$_{2}$ ($T_{C} = 0.35$~K at 0.9~GPa)~\cite{movshovich1996superconductivity, araki2002pressure} and Ce$TX_{3}$ ($T=$~Co, Rh, Ir, $X=$~Si and Ge; $T_{C} = 0.7$~-~1.3~K, 1~-~22~GPa)~\cite{muro1998contrasting,kimura2005pressure,kimura2007normal,sugitani2006pressure,okuda2007magnetic, settai2007pressure,knebel2009high,thamizhavel2005unique,kawai2008magnetic,honda2010pressure}. All these HF superconductors have very high upper critical fields, and some of them exhibit anisotropic behavior. Furthermore, it is reported that the superconductivity in CeIn$_{3}$ ($T_{C}=0.2$~K at a pressure of 2.5~GPa) and CeCoIn$_5$ has $d$-wave pairing symmetry, mainly induced by antiferromagnetic spin fluctuations, in a way that is very similar to the high-temperature cuprates. Strong interest in heavy fermions is also generated by the similarities seen in the phase diagrams of HF superconductors and high-temperature superconductors, including the cuprates and Fe-based materials~\cite{grauel1992tetravalency, si2010heavy, zhao2008structural, paglione2010high}, where spin fluctuations are also suggested to play an important role. 

\begin{figure*}
\centering
 \includegraphics[width=0.6\linewidth]{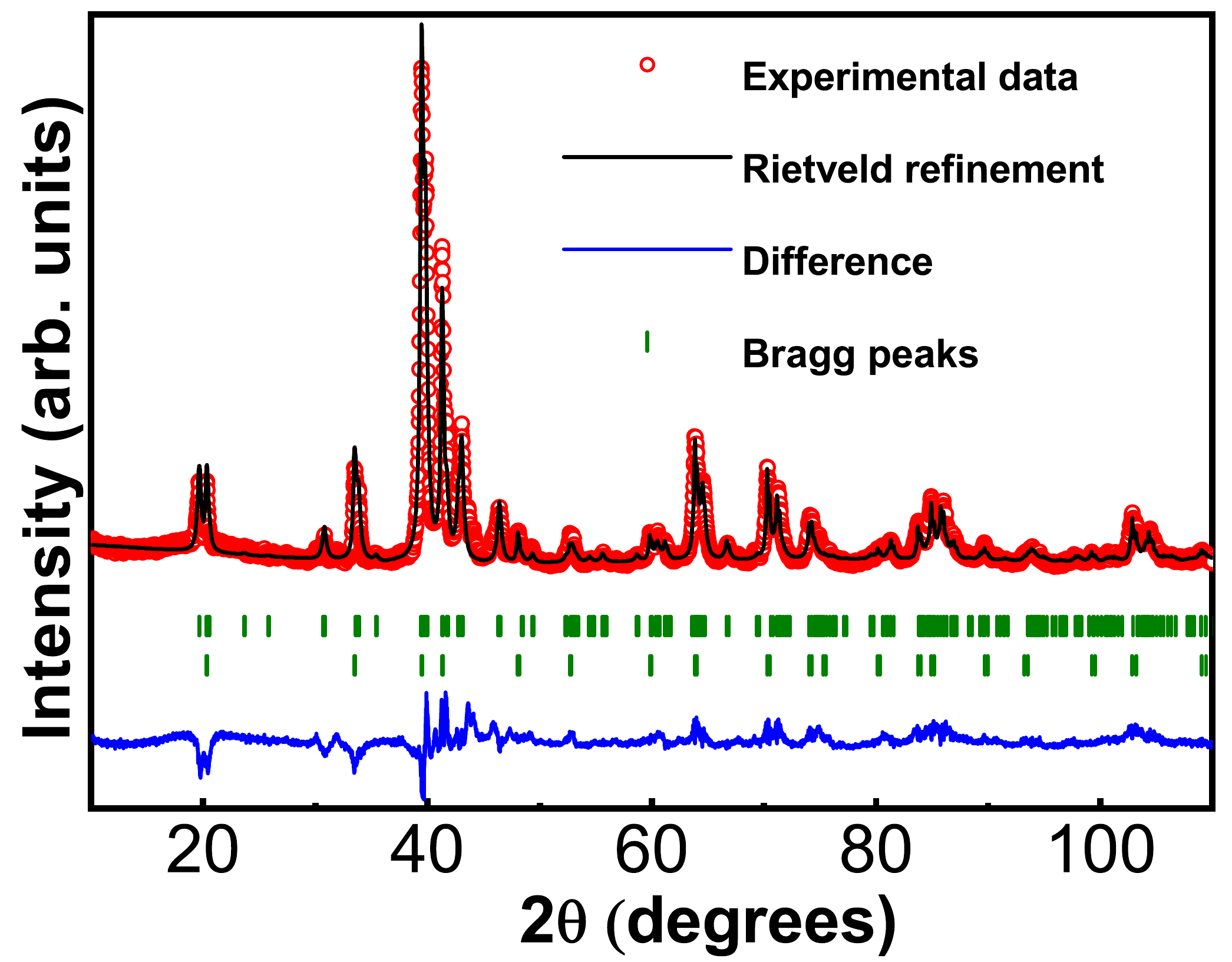}
\caption {Rietveld refinement of the powder X-ray diffraction pattern of CeIr$_{3}$. The data are shown as red circles and the result of the refinement as a solid line (black). We have used the rhombohedral phase (space group $R\bar{3}m$, No. 166) of CeIr$_{3}$ as the main phase and added CeIr$_{2}$ cubic phase (space group $Fd\bar{3}m$, No. 227) as an impurity phase. The vertical green bars show the Bragg peaks' positions, top for CeIr$_{3}$ phase and bottom for CeIr$_{2}$ phase.}
\label{fig61}
\end{figure*}

\begin{figure*}
\centering
 \includegraphics[width=0.6\linewidth]{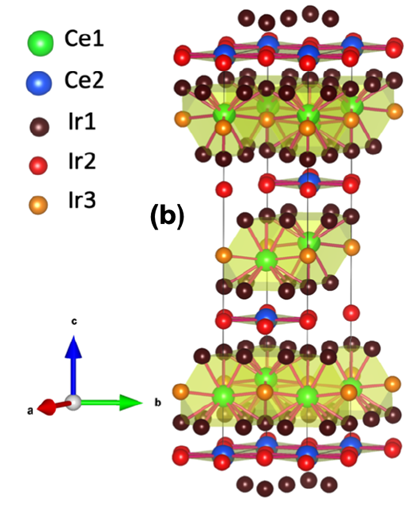}
\caption { Rhombohedral crystal structure of CeIr$_{3}$ where the Ce atoms are the bigger spheres, and the Ir atoms are the smaller spheres.}
\label{fig62}
\end{figure*}

Recently, $R$Ir$_{3}$ ($R =$~La and Ce) materials have attracted considerable attention both experimentally and theoretically due to the observation of superconductivity with strong spin-orbit coupling~\cite{haldolaarachchige2017ir, sato2018superconducting, bhattacharyya2019}. CeIr$_3$ forms in a PuNi$_3$-type rhombohedral crystal structure (Fig.~\ref{fig62}), space group $R\bar{3}m$ (166, $D^5_{{3d}}$ )~\cite{sato2018superconducting}. Sato {\it et al.}~\cite{sato2018superconducting} reported bulk type-II superconductivity in HF CeIr$_3$, with a $T_{C} = 3.4$~K which is the second-highest $T_{C}$ among the Ce-based HF compounds. The crystal structure consists of two non-equivalent Ce sites (Ce1 and Ce2) and three Ir sites (Ir1, Ir2, and Ir3) (Fig.~\ref{fig61}). Gornicka \textit{et al.}~\cite{gornicka2019ceir3} calculated the band structure of CeIr$_3$, which confirmed a non-magnetic ground state, with a small contribution from the Ce 4$f$ shell. It was reported that the density of states (DOS) at the Fermi surface principally arises from the 5$d$ states of the Ir atoms, suggesting that CeIr$_3$ is indeed an Ir 5$d$-band superconductor and that the 5$d$ electrons play a crucial role in the superconductivity. An X-ray photoelectron spectroscopy study reported that the Ce ions have a strong intermediate valence character in CeIr$_3$~\cite{gornicka2019ceir3}.

The Ce ion valence of 3.6 in CeIr$_3$ was estimated using the superconducting transition temperatures, $T_{C}$, of the pseudo-binaries of the isostructural compounds LaIr$_3$, CeIr$_3$, and ThIr$_3$~\cite{hakimi1985valence}. Furthermore, evidence of an intermediate valence, between $3^+$ and $4^+$, of the Ce ions in CeIr$_3$ comes from Vegard's law by plotting the volume versus covalent radius of the $R^{3+}$ metal in the $R$Ir$_3$ series. The volume increases monotonically with an increase in the radius, except for Ce, for which the unit cell volume is much smaller and comparable with the unit cell volume of GdIr$_3$ supporting the intermediate valence of Ce ion in CeIr$_3$~\cite{gornicka2019ceir3}.

The isostructural compound LaIr$_3$, with $T_{C} = 2.5$~K, is another of the few materials~\cite{haldolaarachchige2017ir, sato2018superconducting, bhattacharyya2019} with 5$d$-electrons that exhibits superconductivity. Here as well, the bands at the Fermi surface are dominated by the Ir 5$d$ states with spin-orbit coupling, without any contribution from the La-orbitals; a similar situation is observed for CeRu$_2$~\cite{hakimi1985valence}. Very recently, we have investigated the superconducting properties of LaIr$_3$ using transverse-field (TF) and zero-field (ZF) muon spin rotation and relaxation ($\mu$SR) measurements. Our TF-$\mu$SR measurements revealed a fully gapped isotropic $s-$wave superconductivity with a gap to $T_{C}$ ratio, 2$\Delta(0)/k_{{B}}T_{{C}} = 3.31$, which is smaller than the value expected from the BCS theory of 3.53, implying weak-coupling superconductivity~\cite{bhattacharyya2019}. Moreover, zero-field $\mu$SR measurements show there is no spontaneous magnetic field below $T_{C}$, which confirms that the time-reversal symmetry is preserved in LaIr$_3$~\cite{bhattacharyya2019}.
\begin{figure*}
\centering
 \includegraphics[width=0.6\linewidth]{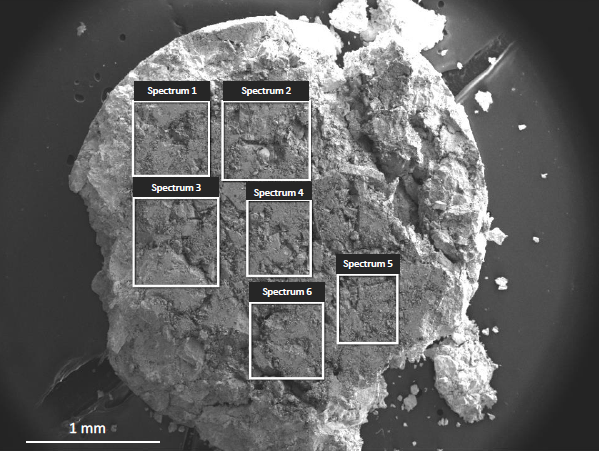}
\caption {Scanning electron microscope image of the polycrystalline CeIr$_{3}$ sample used for the EPMA. The white boxes indicate the regions probed. Table I shows the percentage of Ce and Ir, which is close to 1:3 in all six regions.}
\label{fig63}
\end{figure*}

In this chapter, I have presented the superconducting states of the mixed-valence metal CeIr$_3$ employing magnetization, heat capacity, and TF/ZF-$\mu$SR measurements. The temperature dependence of the magnetic penetration depth, determined by TF-$\mu$SR measurements, implies a fully gapped isotropic $s-$wave nature for the superconducting state. The ZF-$\mu$SR data show no evidence of any spontaneous internal fields developing at and below $T_{C}$ in the superconducting state, suggesting that time-reversal symmetry is preserved in the superconducting state of CeIr$_3$. Further, the very weak temperature dependence of the ZF-$\mu$SR relaxation rate below 3~K suggests the presence of weak spin fluctuations.

\section{Sample preparation}
A polycrystalline sample of CeIr$_3$ was prepared in a tetra arc furnace by arc melting stoichiometric quantities of the starting elements (Ce: 99.9~wt\%; Ir: 99.999~wt\%). The ingot was flipped and remelted five times, and the sample was quenched. The sample was subsequently annealed at 900~$^{\circ}$C for 6~days under a vacuum of $1 \times 10^{-4}$~Pa in a quartz ampoule. The sample was wrapped in tantalum (Ta) foil during the annealing. The sample was heated to 900~$^{\circ}$C and held at this temperature for 6~days and then quenched by switching off the furnace.

\begin{figure*}
\centering
 \includegraphics[width=0.9\linewidth]{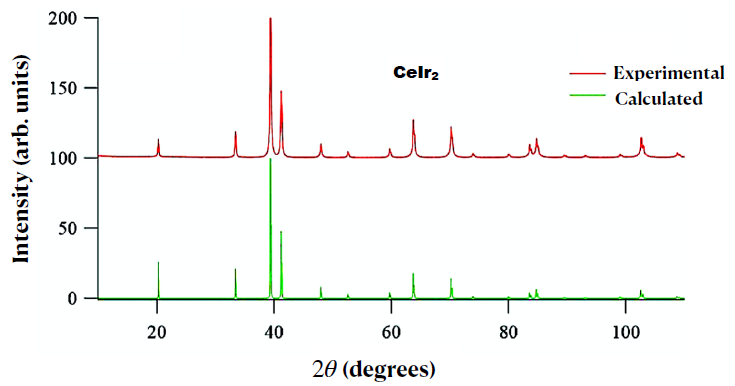}
\caption {(Upper panel). Experimental x-ray diffraction pattern of CeIr$_{2}$. (Lower panel). Simulated x-ray diffraction pattern of CeIr$_{2}$ using a cubic structure (space group Fd$\bar{3}$m, No. 227)}
\label{fig64}
\end{figure*} 

\section{X-ray diffraction}
The quality of the sample was verified through powder x-ray diffraction using a Panalytical X-Pert Pro diffractometer. Fig.~\ref{fig61} presents the powder x-ray diffraction (XRD) pattern and a Rietveld refinement of the data for our polycrystalline sample of CeIr$_{3}$ sample. CeIr$_{3}$ crystallizes in the PuNi$_{3}$-type rhombohedral structure with the space group $R\bar{3}m$. Analysis of the XRD data reveals the fit can be improved by adding a small quantity of cubic CeIr$_2$ (space group $Fd\bar{3}m$, No. 227) as an impurity phase, although overlap of the peaks for the two structures prevents a quantitative analysis. A schematic of the unit cell obtained from the Rietveld analysis of the XRD data of CeIr$_{3}$ is shown in the inset of Fig.~\ref{fig62}. The lattice parameters of the synthesized CeIr$_3$ sample are $a = 5.2943(2) $~\AA~and $c = 26.2134(1)$~\AA, $\alpha = \beta = 90^{\circ}$ and $\gamma = 120^{\circ}$ which are in agreement with a previous report~\cite{sato2018superconducting}. Electron probe microanalysis (EPMA) shows that the composition of the polycrystalline sample is 26(1) Ce: 74(1) Ir, which is close to the expected stoichiometry of the CeIr$_3$ phase, and no evidence for the presence of CeIr$_2$.
\begin{figure*}
\centering
 \includegraphics[width=0.9\linewidth]{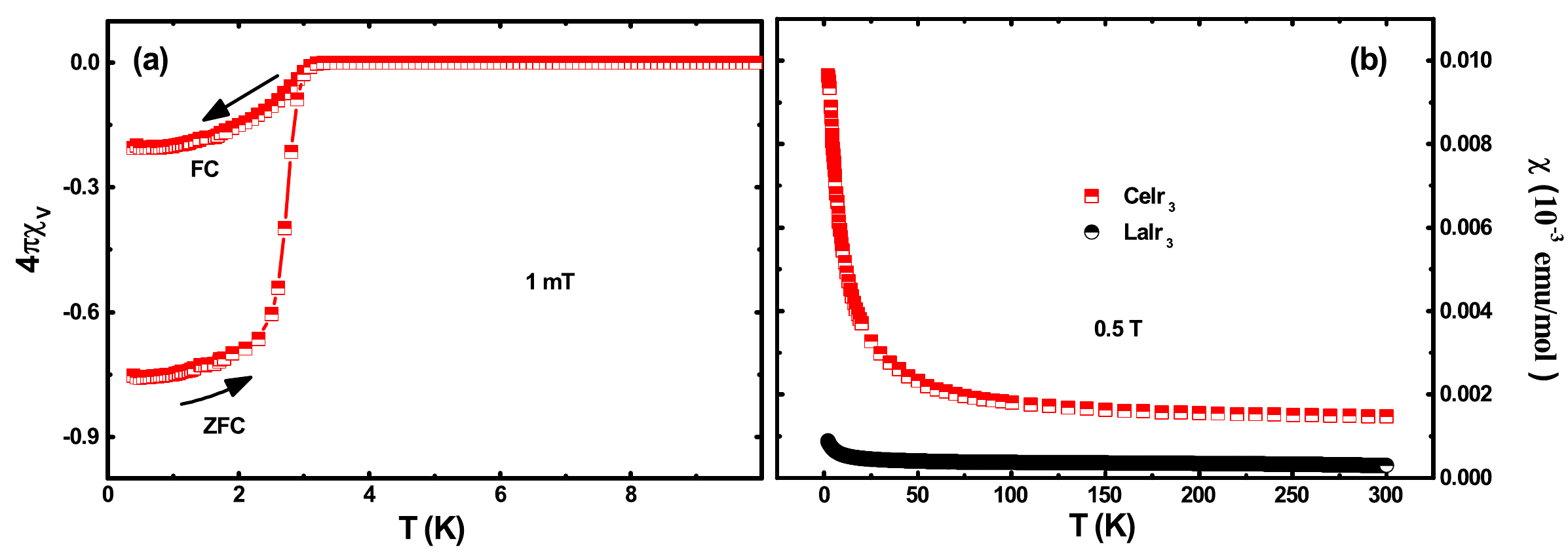}
\caption {Temperature dependence of the dc magnetic susceptibility of CeIr$_{3}$ in zero-field-cooled (ZFC) and field-cooled-cooling (FCC) mode. The inset shows the temperature dependence of the magnetic susceptibility of CeIr$_{3}$ (red squares) and LaIr3 (black circles) measured in a field of 0.5 T.}
\label{fig66}
\end{figure*}

\section*{Powder x-ray diffraction study of CeIr$_{2}$}
A polycrystalline sample of CeIr$_{2}$ was prepared in a tetra arc furnace by arc melting stoichiometric quantities of the starting elements (Ce: 99.9 wt\%; Ir: 99.999 wt\%). The ingot was flipped and remelted five times to improve the homogeneity. The quality of the sample was verified through powder x-ray diffraction using a Panalytical X-Pert Pro diffractometer. The powder x-ray diffraction (XRD) pattern of CeIr$_{2}$ is shown in Fig. ~\ref{fig64} along with the simulated XRD pattern for a cubic structure (space group Fd$\bar{3}$m, No. 227), with a lattice parameter a = 5:394 \AA. The excellent agreement between the experimental XRD pattern and the simulated pattern indicates the single phase nature of the CeIr$_{2}$ sample.  

\section{Electron probe microanalysis (EPMA)}
Electron probe microanalysis shows that the composition of our CeIr$_{3}$ polycrystalline sample is 26(1) Ce:74(1) Ir.), which is close to 1:3 (see Table \ref{EPMA}). Measurements of the average composition were made in the six regions indicated by the white boxes in Fig \ref{fig63}. Several points within each box were also examined. This analysis shows that there are no regions where the sample composition is close to CeIr$_{2}$.
\begin{table}
\caption{Composition of a polycrystalline sample of CeIr$_{3}$ determined using electron probe microanalysis. This analysis was carried out on part of the sample used in our $\mu$SR study. }
\label{EPMA}
\centering
\begin{tabular}{lcc}
\hline\hline
Spectrum & Ce (mol \%) & Ir (mol \%) \\ \hline
1  & 26.0(1) & 74.0(1)\\~\\
2 & 26.3 (1) & 73.7(1)\\~\\
3 & 26.3 (1) & 73.7(1)\\~\\
4 & 25.5 (1) & 74.5 (1) \\~\\
5 & 26.3 (1) & 73.7 (1) \\~\\
6 & 25.9 (1) & 74.1 (1) \\ ~\\

\hline\hline
\end{tabular}
\label{tab:61}
\end{table}

\section{Magnetization data}
\begin{figure*}
\centering
 \includegraphics[width=0.6\linewidth]{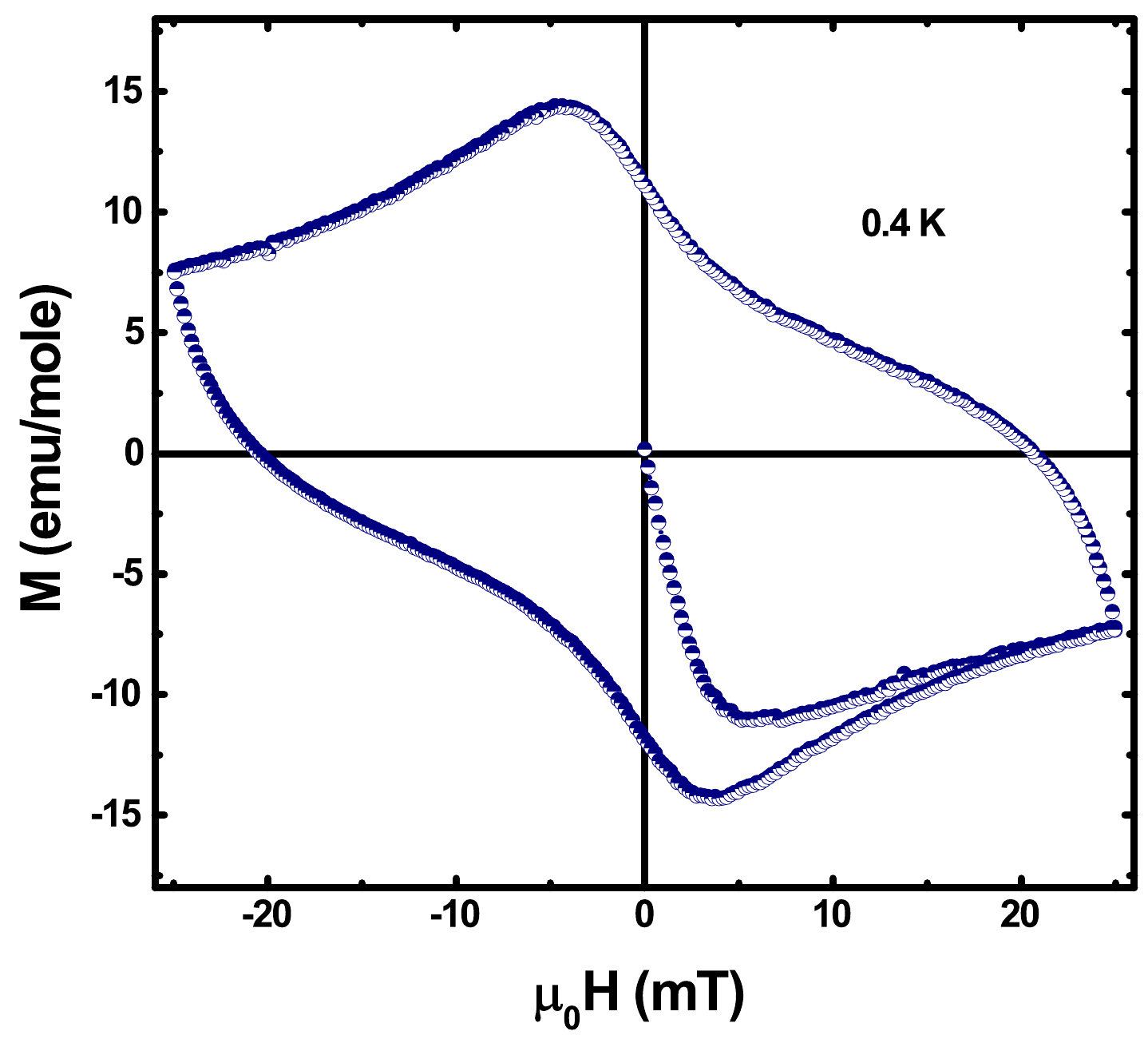}
\caption {Isothermal magnetic field dependence of the magnetization of CeIr$_{3}$ at 0.4 K.}
\label{fig67}
\end{figure*}

The temperature and field dependence of magnetization was measured using a Quantum Design Magnetic Property Measurement System SQUID magnetometer. Fig.~\ref{fig65}(a) presents the temperature dependence of the magnetic susceptibility $\chi\left(T\right)$ in zero-field-cooled (ZFC) state and field-cooled (FC) state which confirms the bulk type-II superconductivity at 3.1~K in CeIr$_3$.  Fig.~\ref{fig65}(b) shows the magnetic susceptibility measured at 0.5 T field up to 300~K for both CeIr$_3$ and LaIr$_3$. The susceptibility of CeIr$_3$ is higher than that of LaIr$_3$ and exhibits considerable temperature dependence below 25~K. This low-temperature rise could be attributed to a Curie tail from impurities. The high temperature (50 - 300~K), the weak temperature dependence of the susceptibility of CeIr$_3$ indicates the presence of strong hybridization between localized 4f-electron and conduction electrons and the mixed-valence of the Ce ions.  The isothermal magnetic field dependence of the magnetization at 0.4~K is shown in Fig.~\ref{fig67}. 

\section{Heat capacity analysis}
 Heat capacity down to 500~mK was measured using a Quantum Design Physical Property Measurement System with a $^3$He insert. Fig.~\ref{fig67}(a) shows the temperature dependence of the heat capacity $C_{{P}}$ at different applied magnetic fields. The inset in Fig.~\ref{fig67}(a) shows the temperature variation of heat capacity at zero applied magnetic field. A clear signature of a superconducting transition is observed at 3.1~K in $C_{{P}}\left(T\right)$ data. Another weaker transition in $C_{{P}}\left(T\right)$ is seen below 1.6~K. The heat capacity of single crystal  CeIr$_{3}$ shows only one transition at $T_{C} = 3.1$~K, as shown in Fig.~\ref{fig67}(b), and no sign of second transition~\cite{sato2018superconducting}, suggests that the second transition observed in the polycrystalline CeIr$_{3}$ near 1.6~K might be related to a superconducting impurity phase or a small variation in the Ir composition (i.e., inhomogeneous Ir composition, CeIr$_{3-\delta}$)~\cite{haldolaarachchige2017ir}. It is to be noted that LaIr$_3$ has a $T_{C} = 3.3$~K, while LaIr$_{2.8}$ has a $T_{C} = 2.75$~K~\cite{haldolaarachchige2017ir}. Sugawara {\it et al.}~\cite{sugawara1994single} reported superconductivity in CeIr$_2$ at 0.25~K. In order to confirm that the transition near 1.6~K does not arise from CeIr$_{2}$, we have synthesized a polycrystalline sample of CeIr$_{2}$ and carried out a powder XRD study and measured the temperature dependence of the heat capacity, $C_{{P}}\left(T\right)$, of this sample down to 400~mK. The results are given in the following section and confirm that the weak anomaly observed in the heat capacity data of CeIr$_3$ near 1.6~K is not due to CeIr$_{2}$. We also considered the possibility that the anomaly is due to superconducting CeIr$_{5}$ which has a reported $T_{C}$ of 1.8~-~1.9~K~\cite{tiede2012d}. However, there is no evidence for CeIr$_5$ in the XRD or magnetic susceptibility data of our polycrystalline CeIr$_{3}$ sample. The anomaly at 1.6~K, therefore, requires further investigation. The jump in the heat capacity of CeIr$_3$ is suppressed in a magnetic field of 6~T. The heat capacity data were fitted using $C_{{P}}\left(T\right)/T = \gamma +\beta T^2$ where $\gamma$ and $\beta$ are electronic specific heat coefficient and lattice specific heat coefficient, respectively. The least-squares fit yields $\gamma = 21.66(2)$~mJ/(mol-K$^2$), $\beta = 1.812(1)$~mJ/(mol-K$^4$), and then using $\beta = {n}\frac{12}{5}\pi^4R\Theta_D^{-3}$, where $R = 8.314$~J/mol-K is the universal gas constant, $n$ is the number of atoms per formula unit,we estimate that the Debye temperature $\Theta_D = 162(2)$~K, which is similar to the $\Theta_D$ values of isostructural ThIr$_3$ [= 169~K]~\cite{gornicka2019iridium} and DyIr$_3$ [155~K]~\cite{mondal2020identification}. Sato {\it et al.} have reported the heat capacity jump $\Delta C_{{e}}/\gamma T_{{C}} \sim 1.39(1)$ and 2$\Delta(0)$/$k_{{B}}T_{{C}}$ = 3.83(1)~\cite{sato2018superconducting} for a single crystal of CeIr$_3$, which is closer to the theoretical BCS limit of a weak-coupling superconductor (3.53). Both of these values suggest that CeIr$_3$ can be categorized as a weak-coupling superconductor.
 
 \begin{figure*}
\centering
 \includegraphics[width=0.6\linewidth]{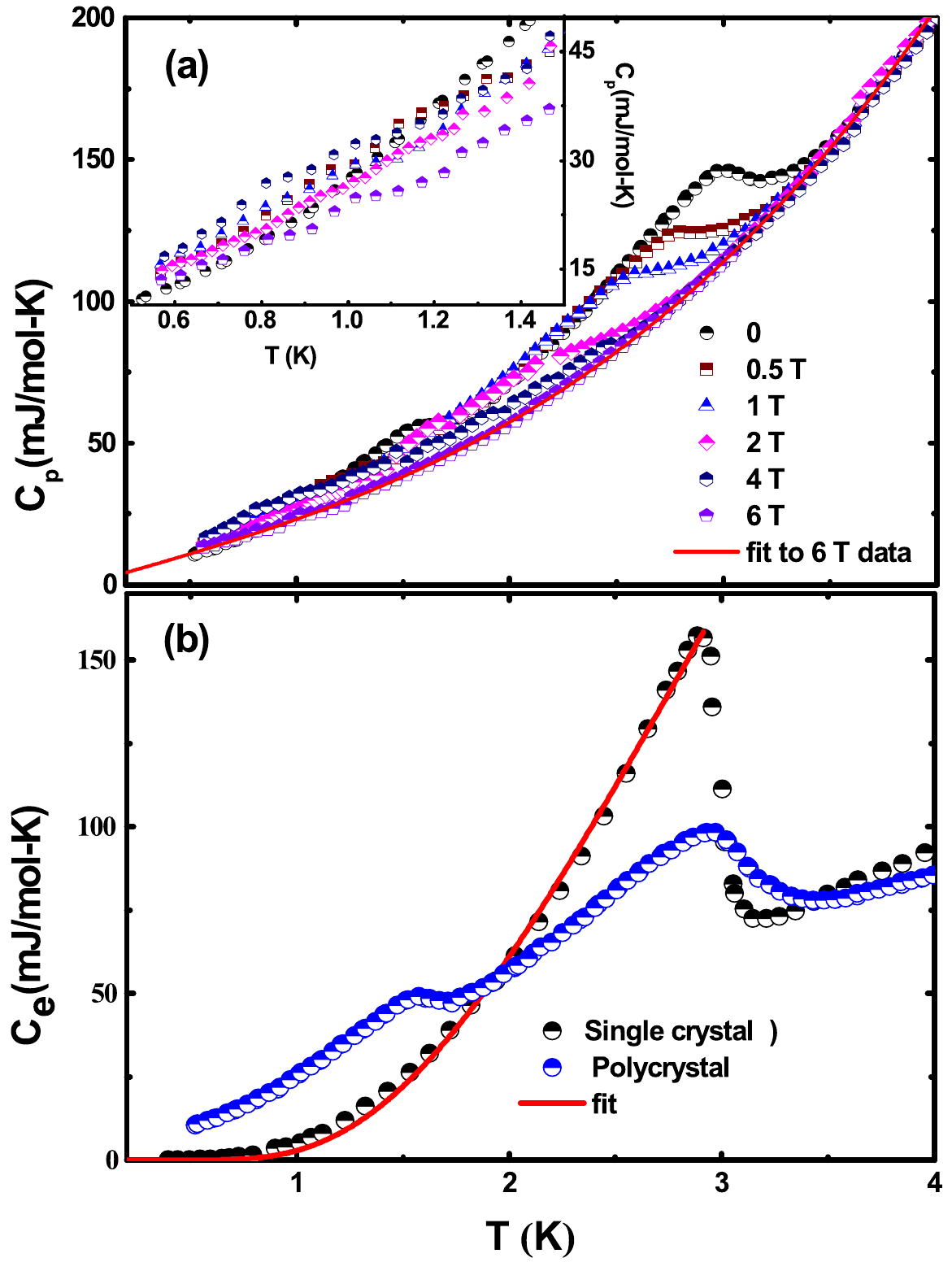}
\caption {(a)Temperature dependence of the heat capacity of CeIr$_{3}$ at different applied fields. The solid line shows a fit to the 6-T data. The inset shows the low-temperature heat capacity versus temperature in various applied magnetic fields on an expanded scale. (d) Electronic heat capacity for CeIr$_{3}$ single crystal from Ref.~\cite{sato2018superconducting} presented here for comparison with the data for polycrystalline CeIr$_{3}$. The solid red line represents a fit to fully gapped superconductivity.}
\label{fig68}
\end{figure*}

For comparison, in Fig.~\ref{fig67}(b) we have plotted the heat capacity of CeIr$_{3}$ single crystal along with our data for polycrystalline CeIr$_{3}$. 
From the exponential temperature dependence of $C_{{e}}$ for single crystal CeIr$_{3}$, (see Fig.~\ref{fig67}(b)), we obtain 2$\Delta(0)$/$k_{{B}}T_{{C}}=3.81$.  
 \begin{figure*}
\centering
 \includegraphics[width=0.6\linewidth]{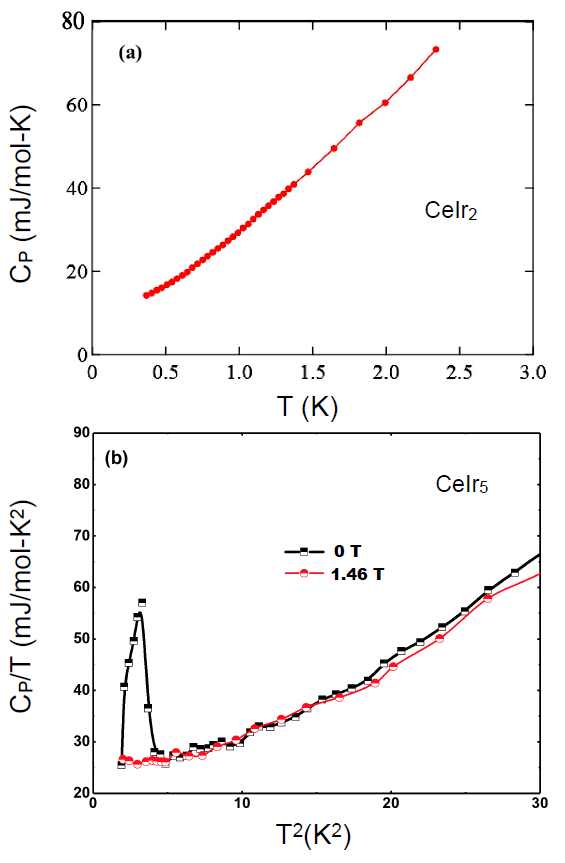}
\caption {(a) Temperature dependence of the heat capacity of CeIr$_{2}$ in zero applied field. (b) Temperature dependence of the heat capacity of CeIr$_{5}$ in zero and 1.46 T, taken from Ref.~\cite{Tang}.}
\label{fig65}
\end{figure*}

\subsection*{Heat capacity of CeIr$_{2}$}
The temperature dependence of the heat capacity, C (T), of CeIr$_{2}$ was measured down to 400 mK using a Quantum Design Physical Property Measurement System with a 3He insert. Fig.~\ref{fig65}(a) shows the temperature dependence of the heat capacity of CeIr$_{2}$ between 400 mK and 2.5 K in zero applied field. The heat capacity decreases with decreasing temperature with no sign of superconductivity observed down to the lowest temperature measured, although the non-zero value for C extrapolated to T = 0 K suggests a superconducting transition may occur in CeIr$_{2}$ at some temperature below 400 mK. This in agreement with the $T_{C}$ = 0.25 K for CeIr$_{2}$ reported by Sugawara et al~\cite{sugawara1994single}.

\subsection*{Heat capacity of CeIr$_{5}$}
The onset of superconductivity in CeIr5 is reported to occur at 1.8 - 1.9 K~\cite{tiede2012d}. The heat capacity of CeIr$_{5}$ taken from Ref. 6 is shown in Fig. \ref{fig65}(b).

\section{TF-$\mu$SR analysis}

\begin{figure*}
\centering
 \includegraphics[width=0.6\linewidth]{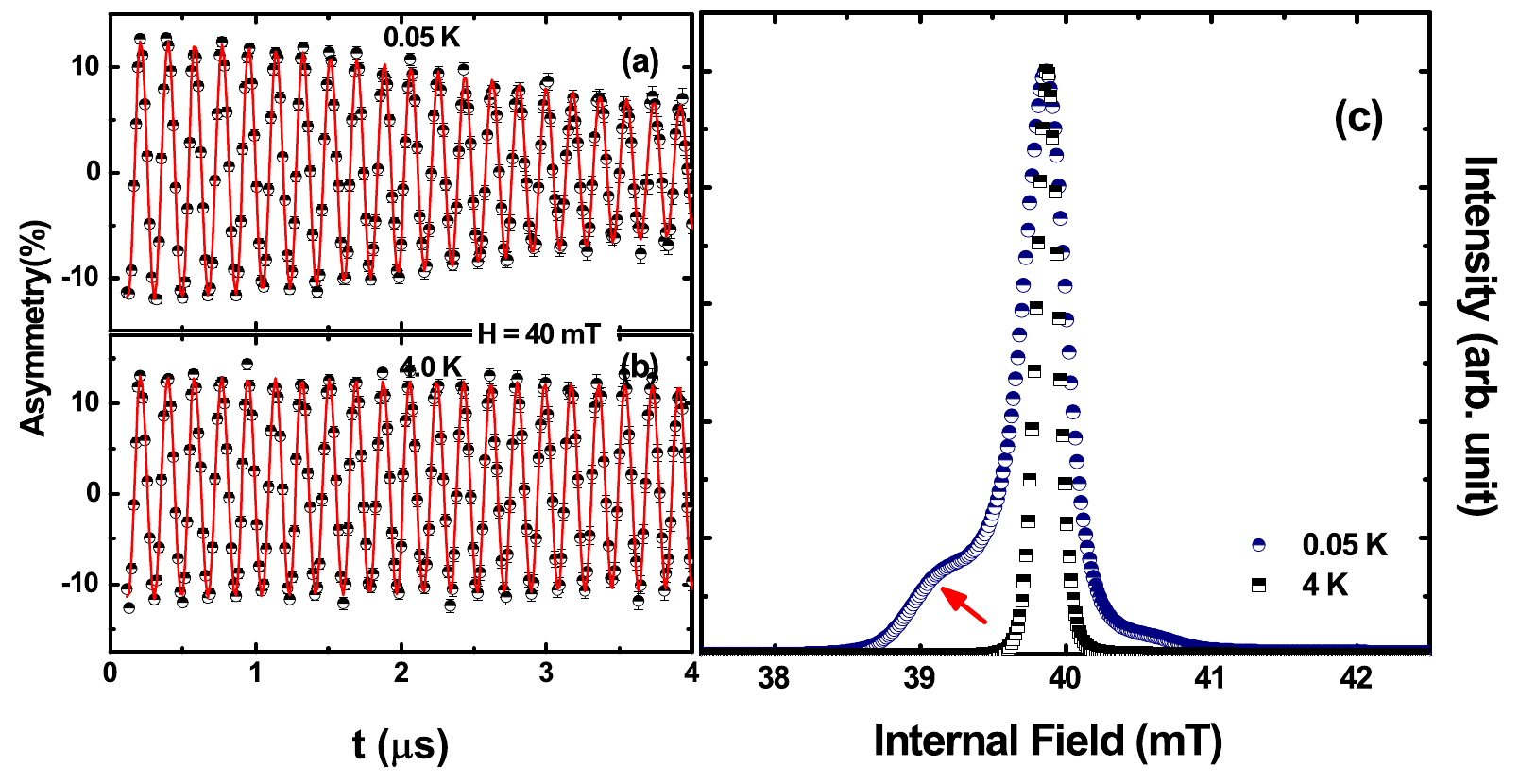}
\caption {TF-$\mu$SR spin precession signals for CeIr$_{3}$ collected in a transverse magnetic field of $\mu_{0}$H = 40mT. Asymmetry vs time in (a) the superconducting state at 0.05 K and (b) the normal state at 4.0 K. Solid lines represent fits to the data using Eq.\ref{MuonFit1}. (c) ) display the corresponding maximum entropy spectra (below and above T$_{C}$).}
\label{fig69}
\end{figure*}

The TF-$\mu$SR asymmetry spectra measured in an applied magnetic field of 40~mT are displayed in Figs.~\ref{fig69} (a) and (b). The data in Fig.~\ref{fig69} were taken at the base temperature in the superconducting state and in Fig.~\ref{fig69}(b) at a higher temperature, well into the normal state. At $T\ge T_{{C}}$, the muon asymmetry oscillates with minimal damping, suggesting that the internal field distribution is extremely uniform. On the other hand, the asymmetry spectrum measured at $T\le T_{{C}}$ shows an increase in damping, suggesting an inhomogeneous field distribution due to the vortex state. The corresponding maximum entropies are shown in Fig. \ref{fig69} (c). To obtain quantitative information about the superconducting state in CeIr$_3$, we first tried to analyze the TF-$\mu$SR data recorded at various temperatures using two Gaussian components, one to account for the CeIr$_3$ phase and another to account for the impurity phase. However, the two component model gave unphysical values for the parameters, and the fit did not converge. We, therefore, fitted our TF-$\mu$SR data using a single Gaussian model~\cite{bhattacharyya2018brief, adroja2017multigap, bhattacharyya2019evidence, anand2014physical} given by,
\begin{equation}
G_{x}(t) = C_{1}\cos(\omega_{1}t+\Phi)\exp\bigg(\frac{-\sigma^{2}t^{2}}{2}\bigg)+C_{2}\cos(\omega_{2}t+\Phi),
\label{MuonFit1}
\end{equation}
\noindent where $C_{i}$ and $\omega_{i}$ ($i = 1$, 2), are the transverse-field asymmetries and the muon spin precision frequencies that arise from the sample and the silver sample holder (this could also include the impurity phase), and $\Phi$ and $\sigma$ are a phase factor and total Gaussian depolarization rate, respectively. During the fitting $C_{2}$ was fixed at 35\%, its low-temperature value and the asymmetry spectra were then fit by varying the value of $C_{1}$, which is nearly independent of temperature. The phase, $\Phi$, was also fixed to the value obtained at low temperatures. Figures.~\ref{fig69}(a) and ~\ref{fig69}(b) also include fits to the data (the solid red lines) using Eq.~\ref{MuonFit1}, and show a good correspondence between the experimental and the calculated asymmetry spectra.

The values of $\sigma$ determined from the fits consist of two parts; one part comes from the superconducting signal, $\sigma_{sc}$, and the other part is the nuclear magnetic dipolar contribution, $\sigma_{{nm}}$, which is taken to be constant over the entire temperature range studied.
The superconducting depolarization rate $\sigma_{{sc}}$ is then calculated using $\sigma_{{sc}} = \sqrt{\sigma^{2}-\sigma_{{nm}}^2}$. The temperature variation of $\sigma_{{sc}}$ shown in Fig.3(c). It is to be noted that there is no clear feature in the $\sigma_{{sc}}$ data at 1.6 K (Fig.3c), where the heat capacity exhibits second anomaly. The $\sigma_{{sc}}$ is modeled using a standard expression within the local London approximation~\cite{bhattacharyya2017superconducting,adroja2017multigap} with 

\begin{eqnarray}
\label{MuonFit2}
\frac{\sigma_{{sc}}\left(T\right)}{\sigma_{{sc}}\left(0\right)} &=& \frac{\lambda^{-2}\left(T,\Delta_{0}\right)}{\lambda^{-2}\left(0,\Delta_{0}\right)} \nonumber \\
 &=& 1+\frac{1}{\pi}\int_{0}^{2\pi}\int_{\Delta\left(T\right)}^{\infty}\left(\frac{\delta f}{\delta E}\right) \frac{E{d}E{d}\phi}{\sqrt{E^{2}-\Delta^2\left(T,\phi\right)}},
\end{eqnarray}

\noindent where $f = [1+\exp(E/k_{B}T)]^{-1}$ is the Fermi function, $\phi$ is the azimuthal angle in the direction of Fermi surface, and $\Delta_{}(T,\phi) = \Delta_{0}\delta(T/T_{C})$g$(\phi)$. $\Delta_{0}$, the gap value at zero temperature, is the only adjustable parameter. The temperature dependence of the gap can be approximated by $\delta\left(T/T_{{C}}\right) = \tanh\left[1.82\left[1.018\left(T_{{C}}/T-1\right)\right]^{0.51}\right]$, and g$\left(\phi\right)$ gives the angular dependence of the gap function where $\phi$ is the polar angle for the anisotropy. The spatial dependence g$(\phi$) is substituted by (a) 1 for an $s$-wave gap, and (b) $\vert\cos(2\phi)\vert$ for a $d$-wave gap with line nodes~\cite{annett1990symmetry,pang2015evidence}. 

\begin{figure*}
\centering
 \includegraphics[width=0.6\linewidth]{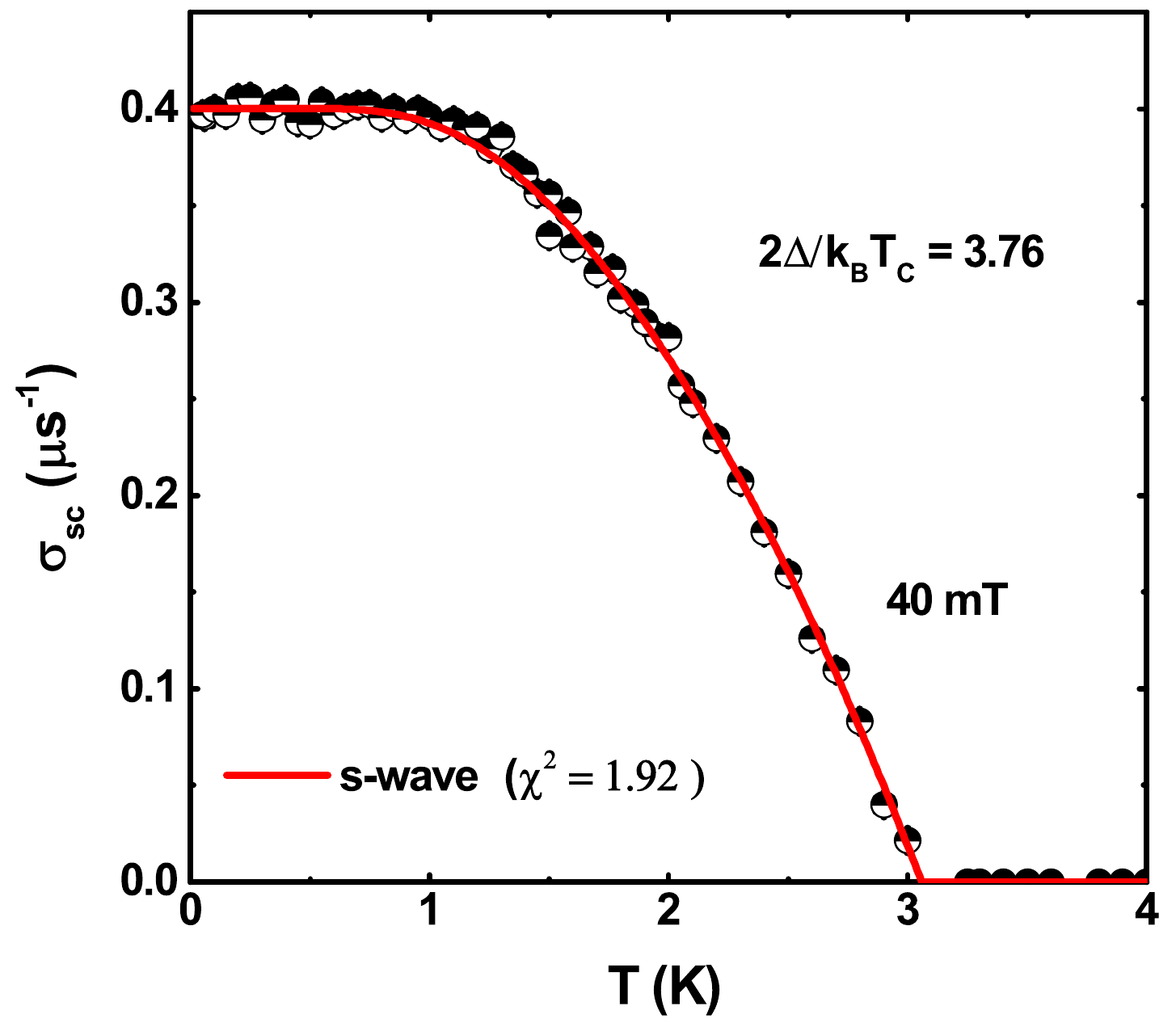}
\caption {Temperature variation of the Gaussian superconducting relaxation rate $\sigma_{sc}$(T ). The solid line is a fit to the data using an isotropic, fully gapped s-wave model using Eq. ~\ref{MuonFit2}.}
\label{fig610}
\end{figure*}

A conventional isotropically gapped model describes the data very well, as shown by the solid red line in Fig.~\ref{fig610}. Using this isotropic model, the refined critical temperature is $T_{C} = 3.1$~K and the gap to $T_{C}$ ratio of 2$\Delta(0)/k_{B}T_{{C}} = 3.76(3)$, is close to the value of 3.53 expected from a weak-coupling BCS theory. This value is in agreement with the heat capacity data for single crystal CeIr$_3$. 

Using the TF-$\mu$SR results, the other superconducting parameters characterizing the superconducting ground state of CeIr$_3$ can be evaluated. For a triangular lattice~$\sigma_{sc}^{2} = \frac{0.00371 \times \phi_{0}^2}{\lambda^{4}}$ where $\phi_{0}$ is the flux quantum number 2.07$\times$ 10$^{-15}$ T m$^{2}$ and $\gamma_{\mu}$ is the muon gyromagnetic ratio, $\gamma_{\mu}/2\pi$ = 135.5 MHz T$^{-1}$. Using this relation we have estimated the magnetic penetration depth, $\lambda(0) = 435(2)$~nm. The London theory~\cite{sonier2000musr} gives the relation between microscopic quantities, $\lambda$ (or $\lambda_{{L}}$), effective mass, $m^{*}$, and the superconducting carrier density, $n_{s}$; $\lambda_{{L}}^2=\lambda^2 = \frac{m^{*}c^{2}}{4\pi n_{s}e^{2}}$, here $m^{*} = \left(1+\lambda_{e-ph}\right)m_{e}$, where  $\lambda_{{e-ph}}$ is the electron-phonon coupling constant and ${m_{e}}$ is an electron mass. Using McMillan's relation~\cite{mcmillan1968transition}, $\lambda_{{e-ph}}$ can be determined using

\begin{equation}
\label{McMillanEq}
\lambda_{e-ph} = \frac{1.04+\mu^{*}\ln(\Theta_{D}/1.45T_{C})}{(1-0.62\mu^{*})\ln(\Theta_{D}/1.45T_{C})-1.04},
\end{equation}

\noindent where $\Theta_{{D}}$ is the Debye temperature. Assuming a repulsive screened Coulomb parameter $\mu^{*} = 0.13$~\cite{allen1999handbook}, we have estimated $\lambda_{{e-ph}} = 0.57(2)$. This value of $\lambda_{{e-ph}}$ is larger than 0.02 to 0.2 observed for many Fe-based superconductors (11- and 122-families) and cuprates (YBCO-123)~\cite{an2013electron}, but smaller than 1.38 for LiFeAs~\cite{kordyuk2011angle}, 1.53 for PrFeAsO$_{0.60}$F$_{0.12}$~\cite{bhoi2008resistivity} and 1.2 for LaO$_{0.9}$F$_{0.1}$FeAs~\cite{gang2008nodal}. Given CeIr$_3$ is a type II superconductor, using the value of $\lambda_{{e-ph}}$ estimated above and $\lambda_{{L}}$, we find the effective-mass enhancement $m^{*} = 1.69(1) m_{{e}}$ and superconducting carrier density $n_{{s}} = 2.5(1) \times 10^{27}$~{carriers}$~m^{-3}$. The superconducting parameters of CeIr$_3$ and LaIr$_3$ are listed together in Table~\ref{Comparison}.

 \begin{figure*}
\centering
 \includegraphics[width=0.6\linewidth]{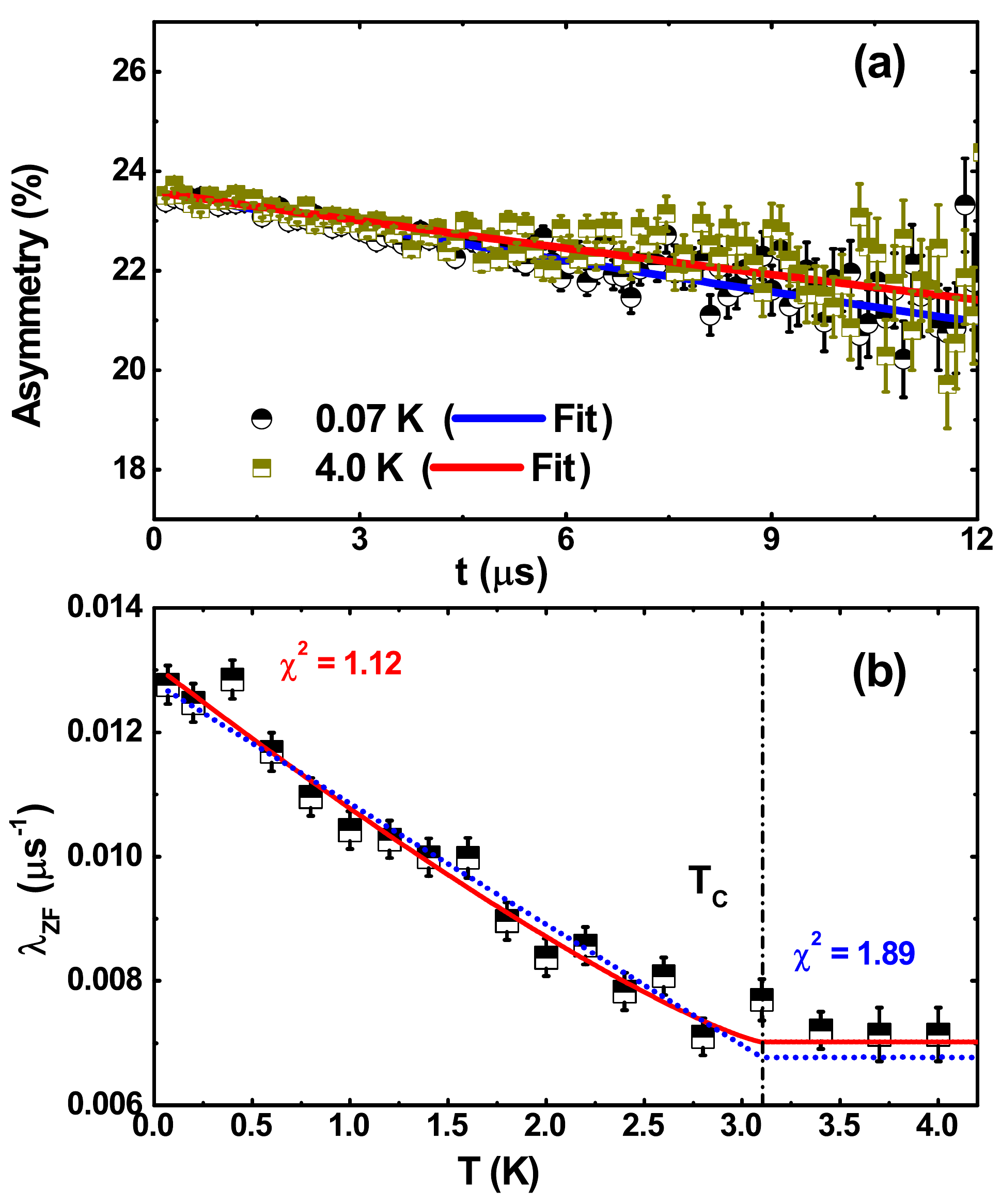}
\caption {(a) Zero-field $\mu$SR asymmetry spectra for CeIr$_{3}$ collected at 0.07 K (black circles) and 4.0 K (dark yellow squares) together with solid lines that are least-squares fits to the data using Eq.\ref{MuonFit3}. (b) Temperature dependence of the zero-field muon relaxation rate. The solid red line indicates a fit made to a power-law behavior using $\lambda_{ZF}(T ) = \lambda_{ZF}(0)[1 - (T/T_{C})^\alpha]^\beta + a_{1}$ with $T_{C}$ fixed at 3.1 K (see text).}
\label{fig611}
\end{figure*}   

\begin{table}
\caption{Superconducting parameters of CeIr$_{3}$ and LaIr$_{3}$. The parameters values of LaIr$_{3}$ comes from Ref.~\cite{bhattacharyya2019} }
\label{Comparison}
\centering
\begin{tabular}{lcc}
\hline\hline
Parameter (units) & CeIr$_3$ & LaIr$_3$ \\ \hline
$T_{C}$~(K)   & 3.1 & 2.5 \\~\\
$\mu_{0}H_{C1}$~(mT) & 5.1(2)& 11.0(2)\\~\\
$\mu_{0}H_{C2}$~(T) & 4.65(3) & 3.84(2)\\~\\
$\gamma$(0)~(mJ/mol K$^{2}$) & 21.66(2) &15.32(3) \\~\\
$\Theta_{D}$~(K) & 162(2) & 430(4) \\~\\
$\Delta C/\gamma T_{C}$ & 1.39(1) & 1.0(2) \\ ~\\
2$\Delta$/$k_{B} T_{C}$ & 3.76(3) & 3.31(1) \\~\\
$\lambda$~(nm) & 435(2) & 386(3) \\~\\
$\lambda_{e-ph}$ & 0.57(2) & 0.53(3) \\~\\
$n_{s}$~($\times$ 10$^{27}$ carriers/m$^{3})$ & 2.5(1) & 2.9(1)\\ ~\\
\hline\hline
\end{tabular}
\end{table}

\section{ZF-$\mu$SR analysis}

ZF-$\mu$SR muon asymmetry spectra above (dark yellow) and below (black) $T_{C}$, that are representative of the data collected are shown in Fig.~\ref{fig611}(a). Both spectra exhibit a slow and almost indistinguishable exponential relaxation. Fits to the ZF-$\mu$SR spectra at several temperatures between 0.07 and 4.0~K were made using the Lorentzian function~\cite{adroja2015superconducting, adroja2017multigap, adroja2017nodal, bhattacharyya2017superconducting,bhattacharyya2019investigation},

\begin{equation}
\label{MuonFit3}
G_{{z}}(t) =C_{{0}} \exp ({-\lambda_{ZF} t})+C_{{bg}},
\end{equation}

\noindent where $C_{{0}}$, $C_{{bg}}$ and $\lambda_{ZF}$ are the total initial asymmetry from muons probing the sample, the asymmetry arising from muons landing in the silver sample holder, and the electronic relaxation rate, respectively. The parameters $C_{{0}}$, and $C_{{bg}}$ are found to be temperature independent. The zero-field-$\mu$SR measurements reveal the relaxation rate between 0.07 and 4~K is slightly temperature dependent (see Fig.~\ref{fig611}(b)), suggesting the presence of weak spin-fluctuations. This effect is not seen in LaIr$_3$, which suggests that the spin fluctuations originate from the Ce moments that are in an intermediate valence state. There is no clear change in $\lambda_{ZF}$ as the samples cool though $T_{C}$ indicating that the time-reversal symmetry is likely preserved in CeIr$_{3}$. If the temperature dependence of the ZF relaxation was due to extrinsic impurities, the relaxation ought to saturate below some temperature, independently of  T$_{C}$, or go through a maximum and then decrease. The absence of this behaviour supports the suggestion that the temperature dependence of the ZF relaxation is an intrinsic property of the CeIr$_3$ phase. Further support on the intrinsic nature of the ZF relaxation rate in CeIr$_3$ comes when we compare the value of ZF relaxation observed in YBCO, which exhibits a very similar change in the relaxation rate below the pseudogap temperature ~\cite{zhang2018discovery}. We have analysed the ZF relaxation rate based on a power law behaviour, $\lambda_{ZF}(T) = \lambda_{ZF}(0)[1-(T/T_{C})^{\alpha}]^{\beta}+a_{1}$, here $\alpha$  and $\beta$ are the critical exponents and $a_{1}$ is the temperature independent relaxation rate above Tc. We have  $\alpha$=1 and  Tc = 3.1 K (from magnetization measurement). The value of the fit parameters obtained are $\beta$= 1.23(9), $\lambda_{ZF}(0)$= 0.0061(2) and $a_{1}$=0.0071(1). The fit is shown by a red solid line in Fig.4b. It is to be noted that when we are allowed to vary $\alpha$  it value remains close to 1, but gave a larger error. A very similar analysis was also performed for ZF-$\mu$SR relaxation rate for UPt$_3$ single crystals by Luke et al., ~\cite{luke1993muon} and the value of the parameters reported are (sample dependent) $\alpha$=0.89-1 and $\beta$=1.53-2.1. In the case of  UPt$_3$, the ZF  relaxation rate exhibits saturation at the lowest temperature, while in CeIr$_3$ the relaxation rate increase almost linearly down to the lowest temperature.\\

\section{Summary}

In summary, we have examined the superconducting properties, including the superconducting ground state, of CeIr$_3$. Magnetic susceptibility measurements show that CeIr$_{3}$ is a bulk type-II superconductor with $T_{C} = 3.1$~K. The heat capacity of polycrystalline CeIr$_{3}$ shows the superconducting transition near 3.1~K and a second weaker anomaly near 1.6~K. Given that the heat capacity of CeIr$_{3}$ single crystal exhibits only one transition near $T_{C} = 3.1$~K ~\cite{haldolaarachchige2017ir} and no peak observed in the heat capacity of CeIr$_{2}$ between 400~mK and 2.5~K, the second transition near 1.6~K could be associated with some variation in Ir content throughout the polycrystalline sample and needs further investigation. The temperature dependence of the ZF-$\mu$SR relaxation rate confirmed the preservation of time-reversal symmetry below $T_{C}$ and suggests the presence of weak spin fluctuations in CeIr$_{3}$. Instead, we suggest this relaxation, which may be slightly enhanced below $T_{C}$, is due to the presence of weak spin fluctuations. This effect is not seen in LaIr$_{3}$, which suggests that the spin fluctuations originate from the Ce moments that are in an intermediate valence state. Transverse-field $\mu$SR measurements reveal that CeIr$_3$ exhibits an isotropic fully gapped $s$-wave type superconductivity with a gap to $T_{C}$ ratio, 2$\Delta(0)/k_{{B}}T_{{C}} = 3.76$, compared to the expected BCS value of 3.53 suggesting weak-coupling superconductivity. The $s$-wave pairing symmetry observed in both LaIr$_3$~\cite{bhattacharyya2019}, a material with no 4$f$-electrons and CeIr$_{3}$, with less than one 4$f$-electron, indicates that the superconductivity is controlled by the Ir-$d$ bands near the Fermi level in both the compounds.  

%% file: chapter7.tex
\chapter{Quantum fluctuations in the non-fermi liquid system CeCo$_{2}$Ga$_{8}$} 
\label{chapter:7}
\section{Introduction}

\subsection*{Feature of Fermi liquid theory}
The so-called free electron model, in which non-interacting electrons travel freely in a positively charged environment, underpins most of the present understanding of basic metals. Despite its simplicity, such a model may accurately predict many metallic transport characteristics long before quantum theory was developed. Using this model one can successfully derive the Ohm- and Wiedemann-Franz laws. Later Sommerfeld treated the electrons using a quantum mechanical approach governed by Scr\"{o}dinger's equation, known as the Fermi gas model. The gas of non-interacting electrons obeys the Fermi Dirac statistics. Following the Pauli-exclusion principle, the electrons fill up the available states up to the Fermi energy, $E_{F} = \frac{\hbar^2}{2m}k_{F}^2$, which is represented in k-space by the so-called Fermi surface. The density of state (DOS) i.e. the number of energy states per unit energy per unit volume in 3-dimension is given by:
\begin{equation}
D(E) = \frac{dN}{dE}= \frac{V}{2\pi^2}\left(\frac{2m}{\hbar^2}\right)^{3/2}\sqrt{E}
\end{equation}

However, in real cases, there is a large number of particles ($\sim$ 10$^{23}$), and they interact with each other. This typical problem is addressed by Landau and this theory is known as Fermi liquid theory. This theory rests upon a very important concept called `Adiabatic continuation'. So here is a correspondence between the non interacting particles, which are known as quasi particles. These quasi particles are known as 'Landau quasi particle', are characterized by renormalized mass m$^{*}$ and the energy of the quasi particle depends on the other quasi particles in the system. Landau had taken into by a function called as `F function', which give rise to the Hamiltonian known as Landau Hamiltonian:
\begin{equation}
H = \sum_{k,\sigma}\epsilon_{k}\delta n_{k,\sigma} + \sum_{k,k',\sigma,\sigma{'},n,n'} f_{nk,n'k'}\delta n_{k \sigma}\delta n_{k'\sigma'}
\end{equation} 
Here one thing is crucial, the quasiparticles are long lived, this is only possible at low temperature and near the Fermi surface i.e.
\begin{equation}
\tau \propto \frac{1}{E-E_{F}}
\end{equation}

The Fermi liquid theory predicts 
\begin{eqnarray}
C &=& \frac{m^{*}k_{F}k_{B}^{2}}{3\hbar^{2}}T\\
\chi &=& \frac{\mu_{0}\mu_{B}^{2}m^{*}k_{F}}{\pi^{2}\hbar^{2}} \\
\rho & =& \rho_{0}+ AT^{2}
\end{eqnarray} 
for the specific heat, the magnetic susceptibility, and the resistivity, respectively.

This is the fundamental theory for the study of simple metals. It also holds in extreme cases such as heavy-fermion systems and high-$T_{C}$ superconductors, where electronic correlations are significantly greater.

\subsection*{Non Fermi Liquid}
\noindent Searching for quantum critical point (QCP) is a great challenge in strongly correlated materials since it only emerges at zero temperature by varying a control parameter such as magnetic field, pressure, and chemical doping or alloying etc~\cite{lohneysen2007fermi,shibauchi2014quantum,fischer2005field}. Heavy fermion (HF) materials exhibit many exotic states in the vicinity of a magnetic QCP, including non-Fermi- liquid (NFL) and unconventional superconductivity~\cite{lohneysen2007fermi,georges1996dynamical,varma1976mixed,coleman2007handbook,riseborough1992theory,stewart2001non}. At QCP, the quantum fluctuations dominate over the thermal fluctuations that break the predictions of well-known Landau-Fermi-liquid behavior~\cite{pomeranchuk1958stability,baym2008landau}, and hence the system exhibits NFL behavior. The nature of quantum fluctuations and the development of magnetic correlations will depend on the dimensionality of the systems, and hence it is very important to investigate the effect of dimensionality on the QCP/NFL. So far, most Ce-based QCP/NFL systems investigated are two-dimensional (2D) or 3D, and there are no reports on 1D Ce-based NFL systems~\cite{stewart2001non,krellner2011ferromagnetic,steppke2013ferromagnetic}. Furthermore, the physical properties of Ce-based compounds at low temperature exhibit Fermi liquid behavior predicted by Landau theory~\cite{landau1992hydrodynamic}. For example, at low-temperature electrical resistivity $\rho \sim T^{2}$, heat capacity $C \sim T$ and dc magnetic susceptibility independent of temperature~\cite{schofield1999non,stewart2001non,gegenwart2008quantum,si2010heavy,stockert2011unconventional}. Interestingly some of the Ce- and Yb- based materials deviate from conventional Fermi liquid behavior to so call NFL behavior, which can be tuned from an antiferromagnetic ground state to zero-temperature QCP, where quantum fluctuations are responsible for NFL behavior. In case of NFL compounds $\rho$ $\sim T^{n}$ (1 $\leq$ n $<$ 2),~$C/T$ $\sim$ $-\ln T$ $C/T$ $\sim$ $a - bT^{1/2}$ and $\chi$ $\sim$ $-\ln T$ or $\chi$ $\sim$ $T^{-p}$ $(p < 1)$~\cite{schofield1999non,stewart2001non,gegenwart2008quantum,si2010heavy,stockert2011unconventional}. 

\begin{figure*}
\centering
 \includegraphics[width=0.6\linewidth]{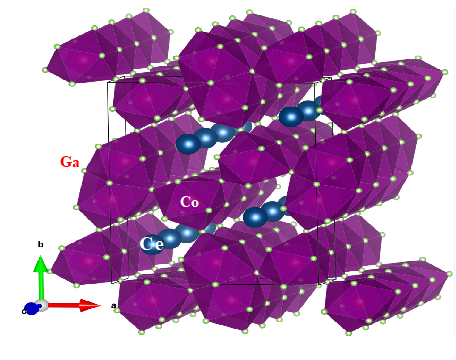}
\caption {Represent the crystal structure of CeCo$_{2}$Ga$_{8}$, showing the quasi-1D chains of Ce atoms along the $c$-axis. The individual unit cell holds four Ce atoms.}
\label{fig71}
\end{figure*}

\noindent To accommodate deeper insight toward the specific nature of the QCP, both theoretical and experimental efforts have been made a lot in recent years. Still, most of them focus on the quasi-2D or 3D HF~\cite{stewart2001non,krellner2011ferromagnetic,steppke2013ferromagnetic}. Considering dimensionality is a fundamental component in defining the unique NFL attributes in these materials, lower dimension anticipates a substantial magnetic frustration parameter, and it is imperative to search QCP in quasi-1D HF compounds, whose science can be further easier approximated by the density matrix renormalization group method as well as mean-field approximation~\cite{schollwock2005density}. QCP has been observed in CeCu$_{6-x}$T$_{x}$ (T = Au, Ag)~\cite{von1996non,heuser1998inducement}, in which the heavy NFL at x$_{c}$ = 0.1 at an ambient pressure is driven to a magnetically ordered state via further doping, in antiferromagnetic ordered HF compounds such as CeIn$_{3}$~\cite{shishido2010tuning,gor2006antiferromagnetism} or CePd$_{2}$Si$_{2}$ ~\cite{mathur1998magnetically}. To date, there exist hardly a few undoped or stoichiometric materials which exhibit NFL states at ambient pressure such as UBe$_{13}$~\cite{ramirez1994nonlinear}, CeNi$_{2}$Ge$_{2}$~\cite{steglich1997non} and CeCu$_{2}$Si$_{2}$~\cite{steglich1997non}, CeRhBi~\cite{sasakawa2005non}.

\begin{figure*}
\centering
 \includegraphics[width=0.6\linewidth]{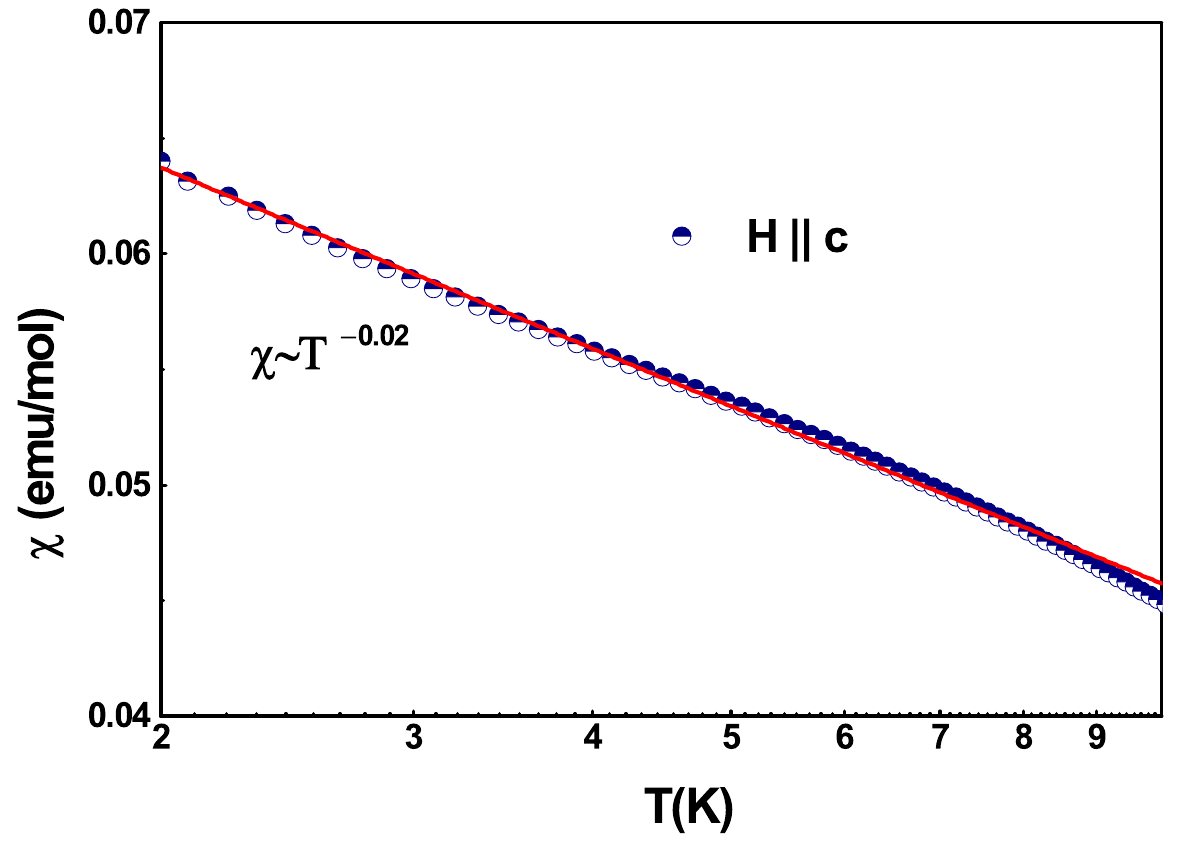}
\caption {Magnetic susceptibility as a function of temperature along the parallel to $c$-axis ($\chi \sim -\ln{T}$). The solid line shows the power law fit.}
\label{fig72}
\end{figure*}

\noindent As chemically disorder NFL states remain challenging to understand by theoretical models, it is highly desirable to examine stoichiometric and homogeneous systems, which are prototypical materials for theoretical modeling. The recently discovered quasi-1D Kondo lattice system CeCo$_{2}$Ga$_{8}$, which crystallizes in the YbCo$_{2}$Ga$_{8}$-type orthorhombic structure, provides us with a rare opportunity to scrutinize a QCP at ambient pressure~\cite{wang2017heavy}. The onset of coherence is about $T^* \sim$ 20 K, and no sign of superconductivity is found down to 0.1 K. Furthermore, 1D spin-chain behavior is also clear from susceptibility data~\cite{wang2017heavy}, and density functional computations predict flat Fermi surfaces originating from the 1D $f$-electron bands along the $c$-axis. NFL state develops in a wide temperature region, as apparent from the pressure dependence resistivity data~\cite{wang2017heavy}. All these facts firmly intimate CeCo$_{2}$Ga$_{8}$ is naturally positioned in the proximity of a magnetic QCP~\cite{wang2017heavy}. Nevertheless, the $T$-linear resistivity and a logarithmically divergent-specific heat is expected in the 2D antiferromagnetic QCP from the conventional Hertz-Millis theory~\cite{hertz1976quantum,millis1993effect}, but not in a quasi-1D Kondo lattice system. Interestingly in the case of CeCo$_{2}$Ga$_{8}$, anisotropic magnetic attributes well explained utilizing crystal field theory, and the ratio of the exchange interaction is $|J_{ex}^c/J_{ex}^{a,b}|\sim$ 4-5~\cite{cheng2019realization}. Our investigation on CeCo$_{2}$Ga$_{8}$ includes electrical resistivity $\rho$(T), dc susceptibility $\chi(T)$, heat capacity $C_\mathrm{P}$(T) data used for characterization of CeCo$_{2}$Ga$_{8}$, and muon spin relaxation ($\mu$SR) measurement to study the low energy spin dynamics. Our microscopic examination confirms the stoichiometric CeCo$_{2}$Ga$_{8}$ compound exhibits NFL ground state without any doping. It is interesting to note that NFL ground state without doping is rare in Ce-based compounds, only a few CeRhBi~\cite{sasakawa2005non}, CeRhSn~\cite{ho2004non}, CeInPt$_{4}$~\cite{malik1989heavy,hillier2007understanding} located in this group.

\begin{figure*}
\centering
 \includegraphics[width=0.6\linewidth]{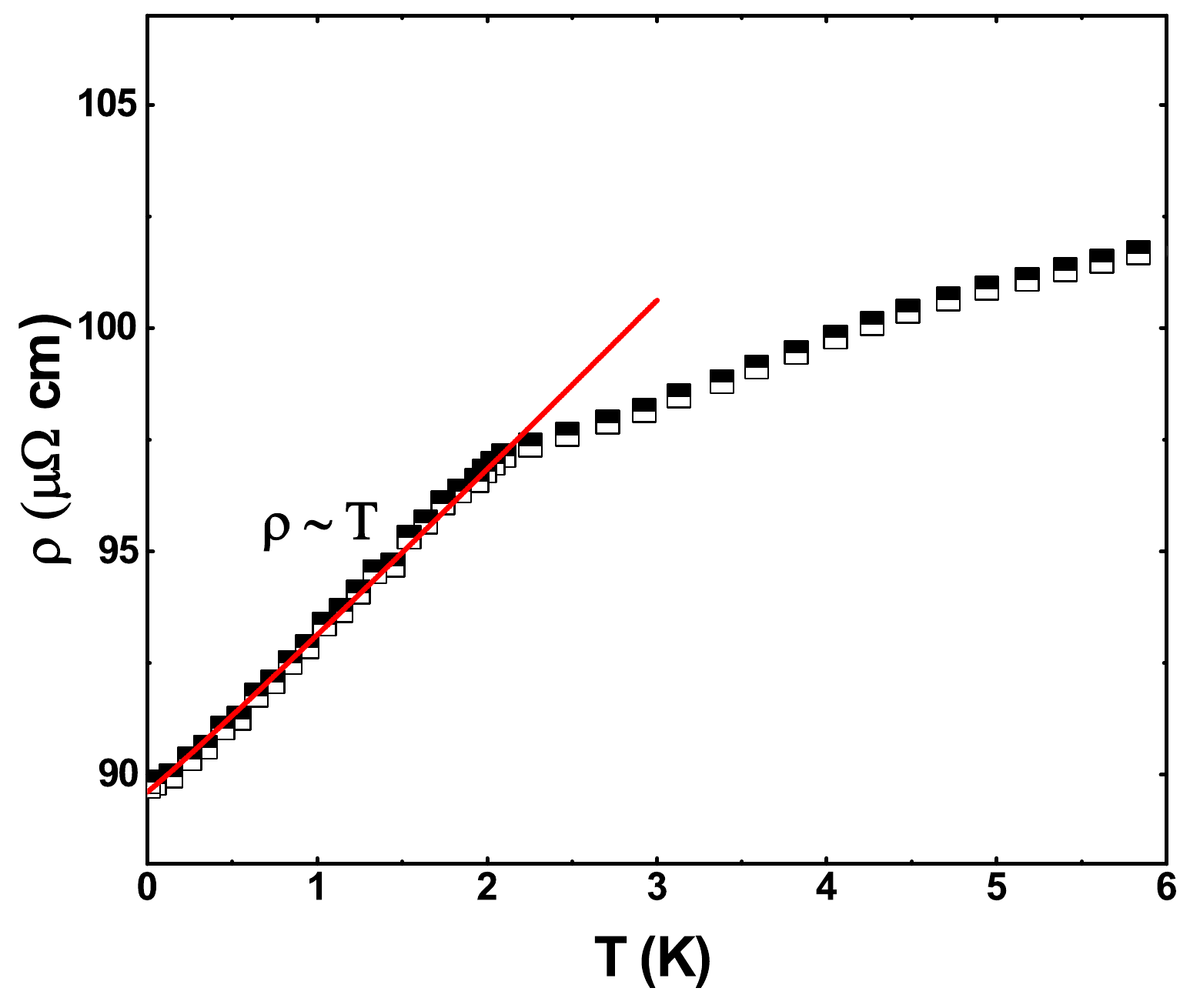}
\caption {Low temperature region of resistivity ($\rho \sim T$).}
\label{fig73}
\end{figure*}

\section{Single crystal preparation and crystal structure}
\noindent For the present investigation, a high-quality single crystals of CeCo$_{2}$Ga$_{8}$ were grown to employ the Ga-flux method. The complete growth method can be found in Ref.~\cite{wang2017heavy}. Figs.~\ref{fig71} represent the orthorhombic structure (space-group $Pbam$, No. 55) of CeCo$_{2}$Ga$_{8}$, showing the quasi-1D chains of Ce atoms along the $c$-axis. The individual unit cell holds four Ce atoms.

\section{Magnetic properties}
Temperature-dependent magnetic susceptibility [$\chi(T)$] was measured using a Quantum Design Magnetic Property Measurement System (MPMS-SQUID) with the applied field parallel to the $c$-axis. The temperature variation of susceptibility measured in a field cool condition with 50 mT applied field parallel to the $c-$ axis is manifested in Fig.~\ref{fig72}. $\chi(T)$ manifests a power-law response suggesting CeCo$_{2}$Ga$_{8}$ placed near the quantum phase transition~\cite{wang2017heavy}. High temperature Curie-Weiss fit yields an effective magnetic moment $p_{eff}$ = 2.74 $\mu_{B}$, which is larger compare to free Ce$^{3+}$-ion value (2.54 $\mu_{B}$), may indicate very  weak magnetic contribution from the Co ion in CeCo$_{2}$Ga$_{8}$. This peculiarity is well-known in Ce-based heavy Fermion systems, for instance, CeCoAsO~\cite{sarkar2010interplay} and CeCo$_2$As$_2$~\cite{stevens1952matrix}.

\section{Transport properties}
\begin{figure*}
\centering
 \includegraphics[width=\linewidth]{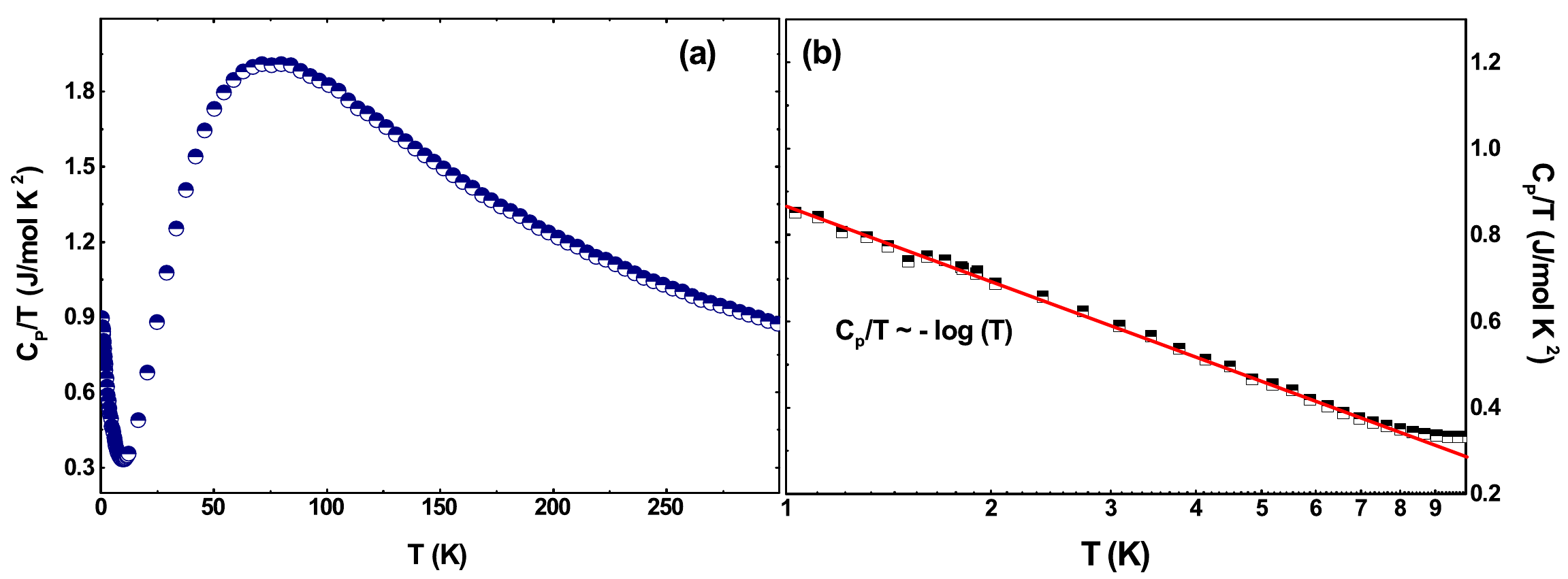}
\caption { (a) Low temperature dependence heat capacity divided by temperature $C_{p}/T$ plotted in semi-logarithmic scale in zero applied field ($C_{p}/T \sim -\ln{T}$). (d) Display $C_{p}/T$ as a function over a large temperature range.}
\label{fig74}
\end{figure*}

  Electrical resistivity [$\rho(T)$] and heat capacity [$C_p(T)$] measurements were done using a Physical Property Measurement System (PPMS) with He$^{3}$ cryostat. In low temperature limit as shown in Fig.~\ref{fig73}, 0.1 K $\leq T \leq$ 2 K, $\rho(T)$ varies linearly with temperature ($\rho \sim$ $T$), characteristics of a NFL state. Heat capacity varies logarithmically $C_{p}/T$ $\sim-\ln(T)$ in the low $T$ limit as shown in Fig.~\ref{fig74}. It is clear from the inset of \ref{fig74} that the heat capacity exhibits a broad maximum near 90 K, which can the attributed to the Schottky anomaly arising from the crystal field effect in the presence of the Kondo effect. All of these attributes of CeCo$_{2}$Ga$_{8}$ are pretty similar to quasi 1D NFL ground state and lead to further examination using a microscopic technique such as muon spin relaxation ($\mu$SR) measurement.

\par

\section{Zero field muon spin relaxation}
 
\begin{figure*}
\centering
 \includegraphics[width=0.9\linewidth]{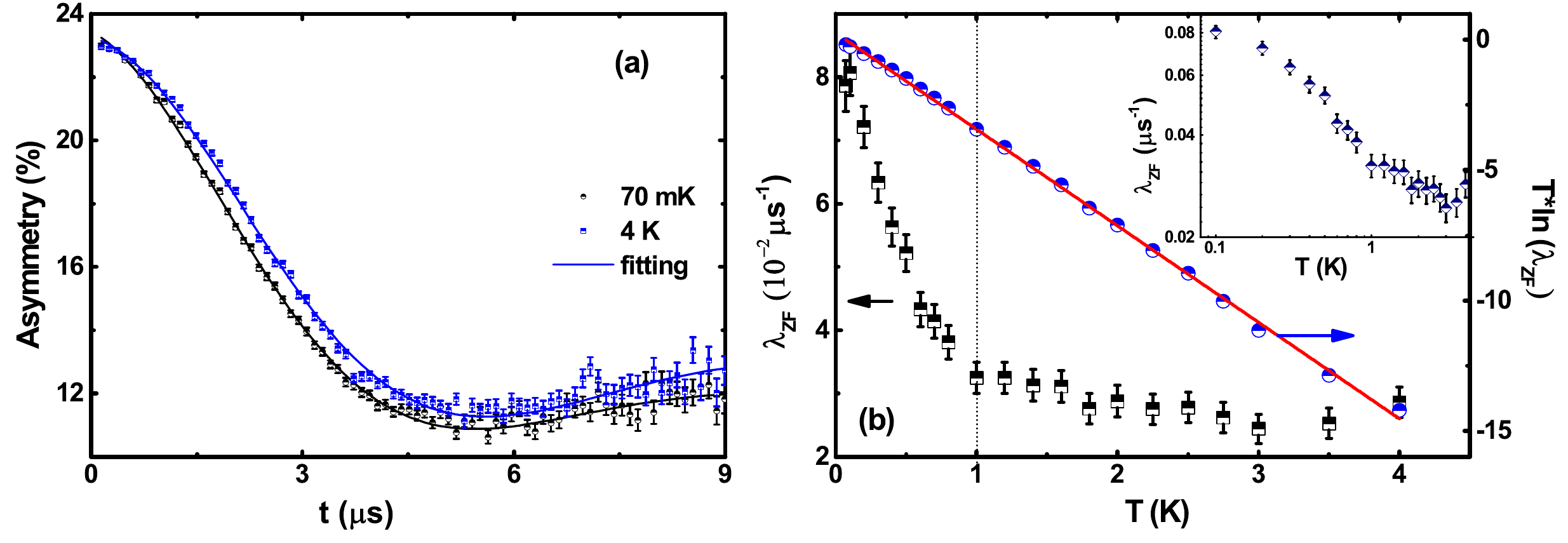}
\caption {a) Time-dependent zero-field $\mu$SR spectra of CeCo2Ga8 collected at 70 mK (black symbols) and 4 K (blue symbols). Solid lines are the least-squares fit applying Eq. ~\ref{ZF} (b) The left y-axis plots the electronic contribution of the muon relaxation rate $\lambda_{ZF}$. A clear signature of quantum fluctuations is seen below 1 K, confirming the NFL state as shown by the bulk properties. The right axis demonstrates the Arrhenius behavior of $\lambda_{ZF}$. The line is the least-squares fit of the data, as presented in the text. Inset: A log-log plot of $\lambda_{ZF}$ vs temperature.}
\label{fig75}
\end{figure*}

\noindent To probe the NFL state as seen from electrical, thermal and magnetic measurements at low $T$, we did ZF/LF-$\mu$SR measurements. The ZF depolarization reveals the sum of the local responses of muons embedded at different stopping sites in CeCo$_{2}$Ga$_{8}$. ZF-$\mu$SR muon asymmetry spectra of CeCo$_{2}$Ga$_{8}$ at $T$ = 70 mK (black) and $T$ = 4 K (blue), that are representative of the data collected, are shown in Fig.~\ref{fig75}(a). Both the 4 K and 70 mK data in Fig.~\ref{fig75}(a) reveal the same value of the initial asymmetry at t=0 along with the lack of oscillations confirm the absence of long range magnetic ordering in CeCo$_{2}$Ga$_{8}$ down to 70 mK. This precludes any argument for static magnetism in the sample~\cite{suzuki2009evidence}. Hence the moderate increase of the relaxation upon cooling from high temperature reflects only a slowing down of the electronic spin dynamics. Fits to the ZF-$\mu$SR data at different temperatures were done employing a Gaussian Kubo-Toyabe function multiplied by an exponential decaying function~\cite{bhattacharyya2019evidence}, 

\begin{equation}
P_{z}(t)=A_{1}[\frac{1}{3}+\frac{2}{3}(1-\sigma^{2}_{KT}t^{2})\exp^{(\frac{-\sigma^{2}_{KT}t^{2}}{2}}]\exp{(-\lambda_{ZF} t)} + A_{bg}
\label{ZF}
\end{equation} 

\noindent where $\lambda_{ZF}$ is the ZF relaxation rate arising due to the local moment, $A_{1}$ and $A_{bg}$ are the asymmetries were originating from sample and background, respectively. $A_{bg}$ was determined from high-temperature ZF, which was kept fixed for the analysis. $\sigma_{KT}$ is the nuclear contribution that emerges from the Gaussian distribution of the magnetic field at the muon site. The relaxation term $\exp$(-$\lambda_{ZF} t$) is the magnetic contribution that comes from the fluctuating electronic spins, which provides information about the low energy spin dynamics of CeCo$_{2}$Ga$_{8}$. As shown in the left panel of Fig.~\ref{fig75}(b), the $T$ variation of $\lambda_{ZF}$ sharply increases below 1 K, indicating the development of the NFL state as evidence from bulk properties. Above 1 K, $\lambda_{ZF}$ decreases with increasing temperature. The right panel of Fig.~\ref{fig75}(b) represents the Arrhenius like behavior of $\lambda_{ZF}(T)$, i.e., follows the form, $\lambda_{ZF} = \lambda_{0}\exp (-\frac{E_{a}}{k_BT})$, where $E_{a}$ and $k_B$ are the activation energy and Boltzmann constant respectively. This confirms that the low $T$ spin dynamics of CeCo$_{2}$Ga$_{8}$ is a thermally activated with $E_{a}$ = 2.3 mK which is similar as observed for CeInPt$_{4}$~\cite{hillier2007understanding} and CeRhBi~\cite{anand2018zero} with $E_{a}$ values are 2.9 mK and 140 mK, respectively. It is an open question why the temperature dependence relaxation of these three stoichiometry compounds exhibits Arrhenius behavior in NFL state (as T$\to$ 0), while that of chemically disordered NFL systems exhibits power law behavior~\cite{adroja2008muon}. We also plotted $\lambda_\mathrm{ZF}$(T) data of CeCo$_{2}$Ga$_{8}$ in log-log plot [Inset of Fig.~\ref{fig75} (b)] to see power law behavior, but the data did not follow the power law-behavior. 

\par

\section{Longitudinal field muon spin relaxation}
\begin{figure*}
\centering
 \includegraphics[width=0.9\linewidth]{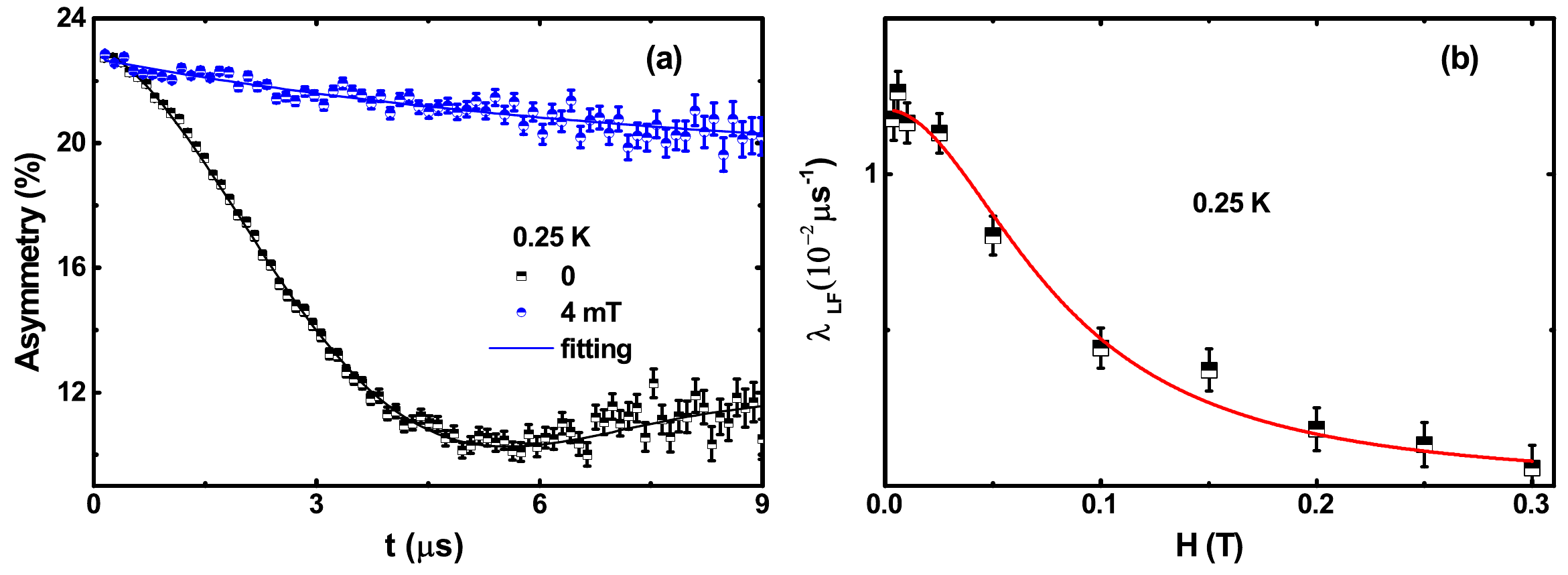}
\caption {Time-dependent longitudinal-field muon asymmetry spectra of CeCo$_{2}$Ga$_{8}$ measured at T = 0.25 K in zero field and at 40 mT. Solid lines are the least-squares fit to the raw data using Eq.~\ref{ZF}. (d) Longitudinal-field dependence of the muon relaxation rate $\lambda$ for CeCo$_{2}$Ga$_{8}$ at 0.25 K. The solid red line is the least-squares fit using Eq.~\ref{eq2}.}
\label{fig76}
\end{figure*}

\noindent A longitudinal field of just 40 mT removes any relaxation due to the spontaneous field and is adequate to decouple muons from the relaxation channel, as presented in Fig.~\ref{fig76}(a) at 250 mK. Once muon is decoupled from the nuclear moments, the spectra can be best-fitted using~\cite{hayano1979zero}, $G_{z(t)}=A_{1}\exp{(-\lambda_{LF}t)}+A_{bg}$. $\lambda_{LF}$ decreases rapidly at low $H$ and saturates at a high $H$. $\lambda_{LF}(H)$ can be adequately expressed by the standard description description given by the Redfield formula~\cite{bhattacharyya2014mu},

\begin{equation}
\lambda = \lambda_0+\frac{2\gamma^2_{\mu}<H_{l}^2>\tau_C}{1+\gamma^2_{\mu}H^2\tau_C}
\label{eq2}
\end{equation}

\noindent where $\lambda_0$ is the field independent depolarization rate, $<H_{l}^2>$ is the time-varying local field at muon sites due to the fluctuations of Ce 4$f$ moments. $H_{l}$ is applied longitudinal field and  the correlation time $\tau_\mathrm{C}$ is related to the imaginary component of the $q$-independent dynamical susceptibility, $\chi''(w)$ through the fluctuation-dissipation theorem~\cite{toll1956causality}. The red line in Fig.~\ref{fig76}(b) represents the fit to the $\lambda_{LF}(H)$ data. The calculated parameters are $\lambda_0$ = 0.19(1) $\mu s^{-1}$, $<H_{l}^2>$ = 1.3(1) mT, and $\tau_c$ = 3.1(6)$\times$10$^{-8}$ s. The value of the time constant of CeCo$_{2}$Ga$_{8}$ unveils a slow spin dynamics, which is originated by the quantum critical fluctuations at low $T$. A similar values are observed for CeRhBi~\cite{anand2018zero}, $\lambda_0$ = 0.17(1) $\mu s^{-1}$, $<H_{l}^2>$ = 1.5(1) mT, and $\tau_c$ = 4.2(6)$\times$10$^{-8}$ s. 

\section{Conclusion}

\noindent In conclusion, we have presented the magnetization, resistivity, heat capacity, and ZF/LF muon spin relaxation measurements on the quasi one-dimensional CeCo$_{2}$Ga$_{8}$ compound. The linear behavior of $\rho(T)$ and logarithmic divergence of $C_p(T)/T$ imply an NFL ground state of CeCo$_{2}$Ga$_{8}$. Moreover, the increase in the ZF relaxation rate $\lambda_{ZF}$ below 1 K and the Arrhenius like behavior of CeCo$_{2}$Ga$_{8}$ suggest NFL ground state, which is quite similar to that seen in other NFL systems, CeRhBi~\cite{anand2018zero} and CeInPt$_{4}$~\cite{hillier2007understanding}. Our ZF $\mu$SR measurements confirm the absence of long range magnetic ordering down to 70 mK. Furthermore, the longitudinal field dependence $\mu$SR study provides information on the spin fluctuations rate and the width of field distribution at the muon sites. The observed quantum fluctuations below 1 K in the undoped CeCo$_{2}$Ga$_{8}$ compound makes it a prototype material to investigate the low $T$ quantum fluctuations in low dimensional NFL systems and other Ce128 counterparts with YbCo$_{2}$Ga$_{8}$-type structure. This work will pave the way in our understanding of NFL in 1D systems both theoretically and experimentally.


%% file: chapter8.tex
\chapter{Summary and Conclusion} \label{chapter:8}
For both experimentalists and theoreticians, the area of superconductivity offers one of the broadest and most diversified playgrounds. The discovery of high-temperature superconductors ushered a new age of research marked by the pursuit of the unconventional superconductor, reigniting the search for room-temperature superconductivity. Surprising discoveries and advancements have happened throughout the history of this area, dating back to Onnes' initial unexpected observation in 1911. It is a noble objective to research these materials for pure knowledge, as it is only through following these interests that the next ground-breaking discovery will be discovered.

Condensed matter physics is a highly diverse area of research that is often driven by the search for new materials with unexpected properties. In this thesis, I concentrate on two of its most active areas of interest in recent years, namely, the superconducting gap in novel superconductors and quantum fluctuation in heavy fermion compounds. In particular, I have investigated a range of novel materials using a variety of complementary experimental techniques such as low temperature resistivity, magnetization, heat capacity, and muon spin rotation and relaxation measurements. These experiments have been motivated by the desire for improving our fundamental understanding of these materials, as well as the prospect of advancing a range of technological applications. The key results are summarised in the following.

The first study in chapter \ref{chapter:4} focuses on the TIrX family of ternary equiatomic superconductors, where T = Hf, Zr. These materials exhibit superconductivity in 3.6 K (Hf) and 1.7 K (Zr), respectively, and have upper critical field 2.2 T (Hf) and 0.6 T (Zr). These materials are important as they are a structure to investigate the role of spin-orbit (SO) coupling in superconductivity, which has not been so well studied in these systems I find that these superconductors have conventional BCS gap structure. The value of 2$\Delta(0)/k_\mathrm{B}T_\mathrm{C}$ = 3.38 obtained from the $s$-wave gap fit, suggests weak-coupling BCS-type superconductivity in HfIrSi, whereas the obtained gap value is 2$\Delta(0)/k_\mathrm{B}T_\mathrm{C}$ = 5.10, which suggest ZrIrSi to be a strongly coupled BCS superconductor. The ZF-$\mu$SR measurements reveal no sign of any spontaneous field appearing below T$_{C}$ which suggests that TRS is preserved in both the compounds. {\it Ab initio} electronic structure calculation indicates BCS superconductivity, which supports our experimental results. The present results pave the way to develop a realistic theoretical model to interpret the origin of superconductivity in ternary systems.

In chapter \ref{chapter:5}  I have investigated the gap symmetry of ThCoC$_{2}$ using transverse field $\mu$SR. The $\mu$SR analysis confirmed that a nodal line $d-$wave model fits better than a single or two gaps isotropic $s-$wave model to the observed temperature dependence of the penetration depth. This finding is in agreement with the field dependent heat capacity, $\gamma(H) \sim \sqrt{H}$. The upper critical field ($H_{C2}$) vs temperature phase diagram confirms a positive curvature, signature of unconventional superconductivity.  The finding of the nodal line gap is further supported through our theoretical calculations. ZF$\mu$SR measurement reveals no appearance of the spontaneous magnetic field in the superconducting ground state, confirming the time reversal symmetry is preserved. 

Next in chapter \ref{chapter:6} the superconducting properties, including the superconducting ground state, of CeIr$_3$ is thoroughly investigated. Magnetic susceptibility measurements show that CeIr$_{3}$ is a bulk type-II superconductor with $T_{C} = 3.1$~K. The heat capacity of polycrystalline CeIr$_{3}$ shows the superconducting transition near 3.1~K and a second weaker anomaly near 1.6~K. Given that the heat capacity of CeIr$_{3}$ single crystal exhibits only one transition near $T_\mathrm{C} = 3.1$~K and no peak observed in the heat capacity of CeIr$_{2}$ between 400~mK and 2.5~K, the second transition near 1.6~K could be associated with some variation in Ir content throughout the polycrystalline sample and needs further investigation. The temperature dependence of the ZF-$\mu$SR relaxation rate confirmed the preservation of time-reversal symmetry below $T_\mathrm{C}$ and suggests the presence of weak spin fluctuations in CeIr$_{3}$. Transverse-field $\mu$SR measurements reveal that CeIr$_3$ exhibits an isotropic fully gapped $s$-wave type superconductivity with a gap to $T_{C}$ ratio, 2$\Delta(0)/k_{\mathrm{B}}T_{\mathrm{C}} = 3.76$, compared to the expected BCS value of 3.53 suggesting weak-coupling superconductivity. 

Finally, in chapter \ref{chapter:7} I have presented the magnetization, resistivity, heat capacity, and ZF/LF muon spin relaxation measurements on the quasi one-dimensional CeCo$_{2}$Ga$_{8}$ compound. The linear behavior of $\rho(T)$ and logarithmic divergence of $C_p(T)/T$ imply an NFL ground state of CeCo$_{2}$Ga$_{8}$. Moreover, the increase in the ZF relaxation rate $\lambda_{ZF}$ below 1 K and the Arrhenius like behavior of CeCo$_{2}$Ga$_{8}$ suggest NFL ground state, which is quite similar to that seen in other NFL systems, CeRhBi and CeInPt$_{4}$. Our ZF $\mu$SR measurements confirm the absence of long range magnetic ordering down to 70 mK. Furthermore, the longitudinal field dependence $\mu$SR study provides information on the spin fluctuations rate and the width of field distribution at the muon sites.

The superconducting pairing mechanism of different superconductors including the superconductors of this thesis are presented in the following tables. The part of the table is adapted from Ref. ~\cite{bhattacharyya2018brief}:

\begin{table*}[b]
\caption{The recent research work on superconducting order parameters and pairing symmetries of iron based materials was examined via different techniques from various international research groups. Theoretically in the weak coupling limit $2\Delta/k_{\rm B}T_{\rm c}\approx 3.53$ for $s$-wave gap, and $2\Delta/k_{\rm B}T_{\rm c}\approx 4.28$ for $d$-wave. Remark on acronyms used for methods in Table: TDO stands for tunnel diode oscillator, TC for thermal conductivity, HC for heat capacity, STM for scanning tunnelling microscopy, NQR for nuclear quadrupole resonance and QPI for quasiparticle interference.} 
\label{FeAsTable}
\noindent\hrulefill

\smallskip\noindent
\resizebox{\linewidth}{!}{%
\begin{tabular}{l c c c c c r }
\hline
\hline
Compounds & $T_c$ (K) & $2\Delta/k_{\rm B}T_{\rm c}$ & Pairing state & Nodeless/nodal & Techniques \\
\hline
\bf{ Fe-based}\\
\hline
LaFePO &6, 7.5 & & $s_{+-}$ & Nodal &TC/TDO \\

LiFeAs &18 & & Anis. $s$-wave & Nodeless &ARPES/QPI\\

LiFeP & 4.5 & & & Nodal &TDO \\

LiFeAsO$_{1-x}$F$_x$ &18 &4.2, 1.1 & $s_{+-}$ & Two-gap & $\mu$SR \\

FeS & 4.04 & 4.54, 2.47 & $s+d$-wave & Nodal, two-gap & $\mu$SR \\

FeSe & 8.5&3.69, 1.64 & Iso. \& Anis. $s$-wave & Nodeless, two-gap & HC/STM \\

FeSe$_{0.85}$&8.3 & 4.49, 1.07 &$s+s$ &Nodeless &$\mu$SR \\

& & 6.27 &Anis. $s$-wave &Nodeless &$\mu$SR \\

FeTe$_{0.55}$Se$_{0.45}$&14.5 & &$s$-wave &Nodeless &ARPES \\

KFe$_2$As$_2$ & 3.8 & & $d$-wave & Nodal &TC/TDO/$\mu$SR \\

 & & & $s$-wave & Nodal &ARPES \\

RbFe$_2$As$_2$ & 2.5 &4.55, 1.39 &$s+s$ & Nodeless & $\mu$SR\\

& & &$d$ & Nodal & TC \\

CsFe$_2$As$_2$ &1.8 & &$d$& Nodal & TC/HC \\

Ba$_{0.6}$K$_{0.4}$Fe$_2$As$_2$ &37 &7.5, 3.7 & $s$ & Nodeless, two gap &ARPES \\

& 38 &7.3, 4.1 & $s$ & Nodeless, two gap &$\mu$SR \\

BaFe$_2$(As$_{0.7}$P$_{0.3}$)$_2$ &30 & & $s$ &Nodal &ARPES/TC \\

Ba(Fe$_{0.77}$Ru$_{0.23}$)$_2$As$_2$ &17 & & & Nodal&TC \\

Ba(Fe$_{2-x}$Co$_x$)As$_2$& 22.1& 3.77, 1.57 & $s$ & Nodeless & $\mu$SR \\

CaKFe$_4$As$_4$ &34.3 &5.82, 1.69 & $s+s$ & Nodeless, two gap & $\mu$SR/TDO/STM \\

KCa$_2$Fe$_4$As$_4$F$_2$ & 33.4 &7.03, 1.28& $s+d$ &Line nodes &$\mu$SR \\

& & 10.14, 1.19&$d+d$ & Line nodes &$\mu$SR \\

RbCa$_2$Fe$_4$As$_4$F$_2$ &29.2& 6.48, 0.70 & $s+s$& &$\mu$SR \\

& &6.42, 0.73 & $s+d$ & Nodal &$\mu$SR \\

CsCa$_2$Fe$_4$As$_4$F$_2$ &28.3 & 6.15, 1.24& $s+d$&Line nodes &$\mu$SR \\

ThFeAsN &28.1 & 4.29, 0.25 & $s+d$, $s+s$&Nodal and nodeless &$\mu$SR/HC \\
\end{tabular}}
\end{table*}

\begin{table*}[t]

\caption {The recent research work on superconducting order parameters and pairing symmetries of chromium based, Caged type, Noncentrosymmetric, Ru-based and HFSCs materials was examined via different techniques from various international research groups. Remark on acronyms used for methods in Table: TDO stands for tunnel diode oscillator, TC for thermal conductivity, HC for heat capacity, STM for scanning tunnelling microscopy, NQR for nuclear quadrupole resonance and QPI for quasiparticle interference.} 
\label{FeAs1Table}
\noindent\hrulefill

\smallskip\noindent
\resizebox{\linewidth}{!}{%
\begin{tabular}{l c c c c c  }

\hline
\hline
Compounds & $T_c$ (K) & $2\Delta/k_{\rm B}T_{\rm c}$ & Pairing state & Nodeless/nodal & Techniques  \\
\hline

\bf{Cr based}\\
\hline
CrAs & 1.5 K (1.09 GPa) & & & & \\
K$_2$Cr$_3$As$_3$ &5.8 &6.4 & $d$& Line nodes &$\mu$SR/TDO \\
Cs$_2$Cr$_3$As$_2$ & 2.1&6.0 & $d$ & Line nodes &$\mu$SR \\
Rb$_2$Cr$_3$As$_3$ & 4.8 &4.2 & & Point nodes& NQR \\
\hline
\bf{Caged type}\\
\hline
PrOs$_{4}$Sb$_{12}$ & 1.8 & 4.2 & $s$ & Nodeless & $\mu$SR/STM  \\
PrPt$_{4}$Ge$_{12}$ & 7.9 &4.58&$D$ & point node & $\mu$SR/HC \\
Lu$_{5}$Rh$_{6}$Sn$_{18}$ & 4.0 &4.4 &$s$ & Nodeless & $\mu$SR \\
Y$_{5}$Rh$_{6}$Sn$_{18}$ & 3.0 &3.91 &$s$ & Nodeless & $\mu$SR \\
Sc$_{5}$Rh$_{6}$Sn$_{18}$ & 4.4 &5.3 &$s$ & Nodeless & $\mu$SR \\
\hline
\bf{Noncentrosymmetric}\\
\hline
ThCoC$_{2}$ & 2.3 &7.8 &$d$ & Nodal & $\mu$SR \\
LaNiC$_2$ & 2.7 & 3.34 & $s$ & Nodeless & HC/NQR\\
LaNiGa$_2$ & 2.1 & 4.08, 2.58 &$s+s$& Two gap, nodeless & HC \\
CePt$_3$Si & 0.75 & & & Line nodes& NMR \\
Li2Pd3B & 8.0 & 2.2 &$s$& Nodeless & NMR \\
Re$_{6}$Zr & 6.75 & 4.2 & $s$ & Nodeless, single gap & HC/$\mu$SR \\
YPtBi & 0.77 & & & Line Node& penetration depth \\
La$_{7}$Ir$_{3}$ & 2.25 & 3.81 & $s$ & Nodeless & $\mu$SR  \\
\hline
Sr$_2$RuO$_4$ & 1.48 & 8.0 & $p$ & & STM \\
\hline
{\bf HF SC}\\
\hline
CeCu$_2$Si$_2$ & 0.6 & 4.0,1.5 & $d+d$ & Fully gapped & TDO  \\
UBe$_{13}$ &0.94 & 4.2 & $p$ & &  \\
CeCoIn$_{5}$ & 2.3&3.1 & $d$ & &spin resonance \\
CeRhIn$_{5}$ & 2.1 (1.6 GPa)& &$d$& Nodal &ARPES \\
UPt$_{3}$ & 0.54 & & &Line Node & NMR  \\
URu$_{2}$Si$_{2}$ & 1.5 & &$d$ & Two gap &polar kerr effect \\
UPd$_{2}$Al$_{3}$ & 2.0 & 5.6 & &Nodeless & Neutron resonanance/NMR  \\
UNi$_{2}$Al$_{3}$ & 1.1 & &&& \\
\hline
{\bf This thesis}\\
\hline
ZrIrSi & 1.7 & 5.1 & $s$ & Fully gapped & $\mu$SR \\
HfIrSi & 3.6 & 3.38  & $s$ & Fully gapped & $\mu$SR \\
ThCoC$_{2}$ & 1.7 & 7.8 & $d$ & Line nodes & $\mu$SR \\
CeIr$_{3}$& 3.1 & 3.76 & s & Fully gapped & $\mu$SR \\
Zr$_{5}$Pt$_{3}$C$_{0.5}$ & 3.8 & 3.84 & s & Fully gapped & $\mu$SR \\
\hline
\hline
\end{tabular}}
\end{table*}


%% file: biblio.tex
\singlespacing